\documentclass{elsart}
\usepackage{epsfig}
\usepackage{graphicx}
\usepackage{amssymb}
\usepackage{citesort}
\usepackage{psfrag}
\begin{document}
\def\LL  {\Lambda ^{\!\!\scriptscriptstyle-1}}
\newcommand{\B}[1]{\mbox{\boldmath $#1$}}
\newcommand{\C}[1]{\mathcal #1}
\newcommand{\rb}{Rayleigh-B\'enard }
\newcommand{\Br}{\B{r}}
\def\Obr{\overline{\Br}}
\newcommand{\KO}{\hat{\mathcal{K}}}
\newcommand{\KM}{{\Bbb K}}
\newcommand{\PM}{{\Bbb P}}
\newcommand{\MM}{{\Bbb M}}
\newcommand{\UM}{{\Bbb U}}
\newcommand{\NM}{{\Bbb N}}
\newcommand{\VecII}[2]{\left( \begin{array}{c} #1 \\ #2 \end{array} \right)}
\newcommand{\MatII}[4]{\left( \begin{array}{cc} #1 & #2 \\
                                                #3 & #4 \end{array}\right)}
\newcommand{\BR}{\B{R}}
\newcommand{\FT}{{\hat F}}
\newcommand{\BB}{\B{B}}
\newcommand{\Bx}{\B{x}}
\newcommand{\By}{\B{y}}
\newcommand{\Bk}{\B{k}}
\newcommand{\Bq}{\B{q}}
\newcommand{\Bp}{\B{p}}
\newcommand{\f}{\B{f}}
\renewcommand{\v}{\B{v}}
\newcommand{\Bv}{\B{v}}
\newcommand{\Bu}{\B{u}}
\newcommand{\Bt}{\B{t}}
\newcommand{\Bn}{\B{n}}
\newcommand{\Be}{\B{e}}
\newcommand{\be}{\begin{equation}}
\newcommand{\ee}{\end{equation}}
\newcommand{\bea}{\begin{eqnarray}}
\newcommand{\eea}{\end{eqnarray}}
\renewcommand{\k}{{\bf k}}
\newcommand{\x}{{\bf x}}
\renewcommand{\r}{{\bf r}}
\def\Or {\overline{r}}
\def\dv{{\delta_r v}}
\def\cP{{\mathcal P}}
\def\cT{{\mathcal T}}
\def\cS{{\mathcal S}}
\def\cC{{\mathcal C}}
\def\Btensor{ B_{n,jm}^{\alpha_1 \ldots \alpha_{n}} }
\def\Ctensor #1#2{ \delta^{\alpha_{#1}\alpha_{#2}}
B_{n-2,jm}^{ \stackrel{ \mbox{no } #1 ,#2} {
\overbrace{_{\alpha_1 \ldots \alpha_{n}} } }}}
\renewcommand{\a}{\alpha}
\renewcommand{\e}{\epsilon}
\newcommand{\eb}{\bar \epsilon}
\renewcommand{\b}{\beta}
\renewcommand{\l}{\ell}
\newcommand{\lp}{\left(}
\newcommand{\rp}{\right)}
\newcommand{\la}{\langle}
\newcommand{\ra}{\rangle}
\newcommand{\ut}{\underline{\theta}} \newcommand{\uu}{\underline{u}}
\newcommand{\ud}{\underline{\Delta}} \newcommand{\PO}{\hat{\mathcal{P}}}
\newcommand{\OO}{\hat{\mathcal{O}}}
\newcommand{\LM}{{\Bbb M}}
\newcommand{\OM}{{\Bbb O}}
\newcommand{\PRO}{\hat{\mathcal{P}}_{\mathrm{R}}}
\newcommand{\PLO}{\hat{\mathcal{P}}_{\mathrm{L}}}
\newcommand{\Eq}[1]{Eq.~(\ref{#1})}
\newcommand{\Fig}[1]{Fig.~\ref{#1}}
\newcommand{\EqDef}{\stackrel{\mathrm{def}}{=}}
\newcommand{\AO}{\mathcal{A}}
\newcommand{\BO}{\mathcal{B}}
\newcommand{\CO}{\mathcal{O}}
\newcommand{\DO}{\mathcal{D}}
\newcommand{\sepBR}[3]{\begin{picture}(9,1)
   \put(#2,#3){\line(1,0){#1}\line(0,1){.25}}\end{picture}}
\newcommand{\sepTL}[3]{\begin{picture}(9,1)
   \put(#2,#3){\line(0,-1){0.25}}
   \put{#2,#3}{\line(1,0){#1}}\end{picture}}
\renewcommand{\S}{{\mathcal S}}
\newcommand{\Even}{{\mbox{\tiny Even}}}
\newcommand{\unitr}{{\mbox{\boldmath$\hat r$}}}
\newcommand{\unitR}{{\mbox{\boldmath$\hat R$}}}
\def\unitz{{\mbox{\boldmath$\hat z$}}}
\newcommand{\SO}{$SO(3)$}
\begin{frontmatter}
\title{Anisotropy in Turbulent Flows  and in Turbulent Transport}
 \author{Luca Biferale}
\address{Dept of Physics and INFM, University of Rome ``Tor
Vergata'', Via della Ricerca Scientifica 1, 00133 Roma, Italy}
 \author{Itamar Procaccia}
\address{Dept. of Chemical Physics, The Weizmann Institute of Science,
Rehovot 76100, Israel}

\begin{abstract}
The problem of anisotropy and its effects on the statistical theory of
high Reynolds-number (Re) turbulence (and turbulent transport) is
intimately related and intermingled with the problem of the
universality of the (anomalous) scaling exponents of structure
functions. Both problems had seen tremendous progress in the last five
years. In this review we present a detailed description of the new
tools that allow effective data analysis and systematic theoretical
studies such as to separate isotropic from anisotropic aspects of
turbulent statistical fluctuations.  Employing the invariance of the
equations of fluid mechanics to all rotations, we show how to
decompose the (tensorial) statistical objects in terms of the
irreducible representation of the SO($d$) symmetry group (with $d$
being the dimension, $d=2$ or 3). This device allows a discussion of
the scaling properties of the statistical objects in well defined
sectors of the symmetry group, each of which is determined by the
``angular momenta" sector numbers $(j,m)$.  For the case of turbulent
advection of passive scalar or vector fields, this decomposition
allows rigorous statements to be made: (i) the scaling exponents are
universal, (ii) the isotropic scaling exponents are always leading,
(iii) the anisotropic scaling exponents form a discrete spectrum which
is strictly increasing as a function of $j$. This emerging picture
offers a complete understanding of the decay of anisotropy upon going
to smaller and smaller scales. Next we explain how to apply the SO(3)
decomposition to the statistical Navier-Stokes theory. We show how to
extract information about the scaling behavior in the isotropic
sector.  Doing so furnishes a systematic way to assess the
universality of the scaling exponents in this sector, clarifying the
anisotropic origin of the many measurements that claimed the
opposite. A systematic analysis of Direct Numerical Simulations (DNS) of the
Navier-Stokes equations and of experiments provides a strong support
to the proposition that also for the non-linear problem there
exists foliation of the statistical theory into sectors of the
symmetry group. The exponents appear universal in each sector, and
again strictly increasing as a function of $j$. An approximate
calculation of the anisotropic exponents based on a closure theory is
reviewed. The conflicting experimental measurements on the decay of
anisotropy are explained and systematized, showing agreement with the
theory presented here.
\end{abstract}
\end{frontmatter}
\tableofcontents
%
%
\section{Introduction}
\label{chap:introduction}
The statistical theory of fluid turbulence is concerned with
correlation functions of the turbulent velocity vector field $\B
u(\B x,t)$ where $\B x$ is the spatial position and $t$ the time
\cite{my75}. Since the velocity field is a vector, multi-point and
multi-time correlation functions are in general tensor functions
of the vector positions and the scalar times. Naturally such
functions have rather complicated forms which are difficult to
measure and to compute. Consequently, almost from its very
beginning, the statistical theory of turbulence was
discussed in the context of an isotropic and homogeneous model.
The notion of isotropic turbulence was first introduced by
G.~I.~Taylor in 1935 \cite{tay35}. It refers to a turbulent
flow, in which the statistical averages of every function of the
velocity field and its derivatives with respect to a particular
frame of axes is invariant to any rotation in the axes. This is a
very effective mathematical simplification which, if properly
used, can drastically reduce the mathematical complexity of the
theory. For this reason, it was very soon adopted by others, such
as T.~D.~K\'arm\'an and L.~Howarth \cite{kar38} who derived the
K\'arm\'an-Howarth equation (see below), and A.~N.~Kolmogorov
\cite{kol41,fri95} who derived the 4/5 law (re-derived
below). In fact, most of the theoretical work in turbulence in the
past sixty years was limited to the isotropic model.

Experimentally, however, we know that isotropy holds only as an
approximation with a varying degree of justification. In all realistic
flows there always exists some anisotropy at all scales; the
statistical properties of the velocity field are effected by the
geometry of the boundaries or the driving mechanism, which are never
rotationally invariant
\cite{hin75,she00,lum67,sad94,kur00,lum65,tav81,kur00a,fer00}. Therefore, a
realistic description of
turbulence cannot be purely isotropic and must contain some
anisotropic elements. Yet the problem is that once we take anisotropy
into account, we face a drastic increase in the complexity of the
theory. The number of variables that is needed to describe the common
statistical quantities, such as correlation functions and structure
functions of the velocity field, increases a lot. For example, under
shear there is a characteristic length scale, which can be constructed from
the
typical velocity and the typical shear \cite{lum67,bif02}.
This length has to be
considered in order to distinguish those scales where the turbulent
evolution is mainly dominated by the inertial effects of fluid
mechanics or by the direct input of energy due to the anisotropic
shear \cite{ben02,cas00,tos99,tos00}. Similarly,
all dimensional estimates acquire a significant
degree of ambiguity because of the proliferation of different
dimensional quantities related to the parameters of anisotropy. As a
consequence of these inherent difficulties the existing anisotropic
effects were simply ignored in many of the experimental and
simulational studies of statistical turbulence.  This attitude gave
rise to ambiguous assessments of important fundamental issues like the
universality of the scaling exponents in turbulence.

The standard justification for ignoring anisotropic effects is
that the basic phenomenology, since the pioneering works of
Kolmogorov \cite{kol41,fri95}, predicts  a {\it
recovery of isotropy} at sufficiently small scales of the
turbulent flows. Nevertheless, both recent experimental works and
theoretical analysis suggested that the actual rate of recovery
is much slower than predicted by simple dimensional analysis,
pointing out even the possibility that some anisotropic
correlation function, based on velocity gradients, stays $O(1)$
for any Re \cite{pum95,pum96,she00,sch02}. In order
to settle this kind of problems, theoretically or experimentally,
it is crucial to possess systematic tools to disentangle
isotropic from anisotropic fluctuations and to distinguish among
different kinds of anisotropic fluctuations. Thus a central
challenge in the theory of anisotropy in turbulence is
the construction of an efficient mathematical language to
describe it.  Without a proper description, the complexities of
the formalism can soon obscure the physical content of the
processes that we wish to study.

The problem of anisotropy is not disconnected from the other
fundamental problem which has to do with the nature of
universality in turbulence. By universality, we mean the tendency
of different turbulent systems to show the same small-scales
statistical behavior when the measurements are done far away from
the boundaries. Consider, for example, the longitudinal two-point
structure function
\begin{equation}
  S^{(2)}(\Br) \equiv \left< \delta u_{\l}^2(\Bx, \Br,t) \right> \ , \quad
  \delta u_{\l}(\Bx,\Br,t) \equiv
  \hat{\Br}\cdot\big[\B{u}(\Bx+\Br,t)-\B{u}(\Bx,t)\big] \ ,
\label{defS2long}
\end{equation}
with $\hat{\Br}$ being the unit vector in the direction of $\Br$, and
$\left< \cdot \right>$ stands for an appropriate ensemble
average.  This function shows essentially the same dependence on the
separation vector $\Br$, whether it is measured in the atmospheric
boundary layer, in a wind tunnel or in a DNS, provided
it is measured for sufficiently small separations and far from the
boundaries. This high-degree of universality cannot be expected if
anisotropic fluctuations were the dominant contributions to the
two-point structure functions. Different boundary conditions and
different forcing mechanism necessarily introduce different
large-scale anisotropies in the flow, which would translate to
different small-scale anisotropic fluctuations.  Small-scale
universality can be achieved only if anisotropic fluctuations are
sub-leading with
respect to the isotropic fluctuations. In the following, we also
discuss which aspects of the anisotropic fluctuations are {\it universal}
and
which are not.  We will see that some aspects of the anisotropic
fluctuations
depend on the boundary conditions while other aspects do
not. In fact we will show that scaling exponents are expected to be
universal whereas amplitudes depend on the boundary conditions.

In the last 5 years, a tremendous progress in the understanding of
the two aforementioned problems, i.e., finding a mathematical
language that properly describes anisotropic turbulence and its universal
properties  has been achieved. Not surprisingly, the two
problems are closely related, as it often happens in physics - a
problem becomes considerably simpler if described in the proper
mathematical language.
The technical core of these recent achievements is the SO($3$)
decomposition \cite{ara99b}. This tool enjoys the advantages of being
mathematically simple, yet very powerful and systematic. By using
it, many of the mathematical complexities of dealing with
anisotropy in turbulence and in other hydrodynamic problems are
greatly simplified.  The principal idea is to represent the main
statistical observables, such as structure functions and
correlation functions, in terms of their projections on the
different $(j, m)$ sectors of the group of rotations.
It can be applied to all the
statistical quantities in turbulence, creating a detailed profile
of the effects of anisotropy.  Additionally, and perhaps more importantly,
the
SO($3$) decomposition reveals some new universal properties of
fully developed turbulence. It is expected that each sector of the
SO($3$) group has its own universal exponents. In particular, it is shown
that the exponents associated with the anisotropic sectors are
larger than the isotropic exponents, in accordance with
the isotropization of the statistics as smaller and smaller
scales are observed.

As already mentioned, the SO($3$) decomposition is useful also to
investigate isotropic and anisotropic fluctuations in other
hydrodynamic problems. In particular we will focus on the case
of scalar and vector quantities passively advected by a turbulent velocity
field. In
these cases, one may often elevate the phenomenological assumptions
made for turbulent anisotropic fluctuations to the status of rigorous
statements \cite{fal01}. By using a systematic decomposition in
different sectors of the SO($3$) group one may show that passive
scalars, advected by stochastic self-similar Gaussian velocity
fields, always possess isotropic {\it leading} small-scale
fluctuations. Moreover, one may quantitatively distinguish among
different kinds of anisotropies, assessing their rate of decay by
going to smaller and smaller scales. It turns out that the rate
of {\it recovery of isotropy} is typically much slower than
expected on the basis of dimensional analysis. Moreover, all
different anisotropic fluctuations decay in a self-similar way
but with different rates; the scaling exponents being universal,
while prefactors are non-universal \cite{ara00}. The very same
can be rigorously proved for the passive advection of vector-like
quantities, as for the case of magnetic fields when the feedback
on the velocity evolution due to the Lorentz force is neglected.
There, the vector nature of the transported quantity leads to an
even richer, and more complex, list of possible anisotropic
fluctuations \cite{ara00a,lan99}. Another important problem which we address
in
detail is the case of the passive advection of a vector-like
incompressible quantity, i.e.  a {\it passive vector with
pressure} \cite{ara01}. Although without any counterpart in nature, such a
system is particularly interesting because it can be seen, for
some aspects, as the closest {\it linear} approximation to the
non-linear Navier-Stokes evolution.  For example, it allows to
study in a systematic way some  problems
connected to the convergence of integrals involving the
pressure term.  Similar technical problems arise
also in the analysis of both isotropic and anisotropic multi-point
velocity correlations in Navier-Stokes equations.

Of course, a significant part of this review will be devoted to
applications of the theoretical and technical tools to physical
experimental data \cite{ara98,kur00,kur00a,she02,she02a,wander,iacob04}
and numerical data sets \cite{ara99,bif01,bif01a,bif02,ish02,sch00}.
In order to exploit the
entire potentiality of the SO($3$) decomposition one needs to measure
the whole velocity field, $\B u(\B x)$, in a 3 dimensional volume.  This is
because in order to disentangle different projections on different
sectors one needs to integrate the given correlation function against
the proper eigenfunction of the rotation group on the 3$d$ sphere of
radius $r$.  By doing that, the exact projection on each different
sector of the SO($3$) decomposition is under control, with the only
practical limitations for reaching highly anisotropic sectors being
the lack of resolution of highly fluctuating angular properties.  At
the present stage of experimental capabilities the exact decomposition
can be carried out explicitly only in data sets coming from DNS.
Here the velocity field in the
whole testing volume is available. For experimental data, the best way
to exploit the SO($3$) decomposition is to either select observables
with vanishing isotropic components, in order to focus directly on
anisotropic sectors, or to perform a multi-sector analysis, i.e.  to
fit simultaneously the isotropic and anisotropic components.

The review is organized as follows. Sect. \ref{historyiso} offers a
historical review of isotropic turbulence. We present a modern derivation
of the exact results pertaining to the 3'rd order structure function
and the celebrated 4/5 law.  We review the standard theory for all
correlation functions, and discuss the experimental difficulties with
the isotropic theory. These difficulties included apparent persistence
of anisotropies into the small scales for high Re,
apparent location dependent scaling exponents etc. In Sect.
\ref{history-aniso}
we review the history of attempts to deal with anisotropy.  In Section
\ref{chap:modern}, the technical basis of the SO($3$) decomposition is
introduced focusing on the particular statistical problems of
anisotropic fluctuations discussed in the previous section.  Then, in
Sect. \ref{chap:analytical} we switch to study {\it exactly solvable}
hydrodynamic problems with emphasis on either those aspects
peculiar to each different model and to those features in common
with the  non-linear Navier-Stokes case. Among the common aspects
we cite the possibility to study in these models in full details the
{\it foliation} of the equations of correlation functions in different
anisotropic sectors; the universality of isotropic and anisotropic
exponents; the hierarchical organization of exponents --leading to
{\it recovery of small-scales isotropy}.  At the end of this section
we present closure results for two-point turbulent structure function
in the anisotropic sectors $j=2,4,6$.  In Sect.
\ref{chap:experiment}, the utility of this language is demonstrated by
discussing experimental data in atmospheric boundary layer and on
homogeneous-shear flows.  In Sect.
\ref{chap:numerics},
we present the analysis of anisotropy in DNS of
typical strongly anisotropic flows. Two cases are discussed in depth:
channel flows and random Kolmogorov flows, the latter being homogeneous
flows
stirred at the large scales.  Sect.
\ref{chap:conclusions} presents a  summary and conclusions.
Technical details are collected in the appendices.
%
%
\section{Historical Review: Isotropic Turbulence}
\label{historyiso}
In the first two sections we present a historical review. We start
with the model of homogeneous isotropic turbulence, and then turn to
previous attempts to treat theoretically  anisotropy in
turbulence.
\subsection{Homogeneous and Isotropic Turbulence}
The Navier-Stokes equations for the velocity field are invariant to
all rotations:
\begin{eqnarray}
\frac{\partial \Bu(\Bx,t)}{\partial t} +\left[\Bu(\Bx,t)\cdot
\B\nabla\right]\Bu(\Bx,t)&=&
-\B\nabla p(\Bx,t) +\nu \Delta^2 \Bu(\Bx,t) \ , \label{NS}\\
\B \nabla \cdot \Bu(\Bx,t) &=&0,\nonumber
\end{eqnarray}
with $p(\Bx,t)$ and $\nu$ being the pressure and kinematic viscosity
respectively.  Since the gradient and Laplacian operators are both
rotationally invariant, the rotation symmetry of the equation can be
broken only by anisotropic forcing terms or anisotropic boundary
conditions. Rather naturally then the statistical theory of turbulence
was mostly developed in the framework of \emph{isotropic turbulence}
\cite{tay35}.
The central idea of this approximation is that
the statistical average of any function of the velocity
components in any coordinate system  is unaltered
if this coordinate system  is rotated or reflected in any manner.
The assumption of isotropy was  widely adopted. In 1938,
K\'arm\'an and Howarth \cite{kar38} used it  to explore the
second- and third-order correlation-functions of the velocity field. Their
use of tensor notation was more elegant and compact than that used by
Taylor. It enabled them to derive some constraints on these correlation
functions and express them in terms of a few scalar functions. For example,
for the second-order correlation-function in homogeneous turbulence
\begin{equation}
\label{def:KA-C2}
  C^{\alpha\beta}(\Br,t)
    \equiv \left< u^\alpha(\Bx+\Br,t) u^\beta(\Bx,t) \right> \ ,
\end{equation}
they used the representation
\begin{equation}
\label{eq:KA-C2}
  C^{\alpha\beta}(\Br,t) = [f(r,t)-g(r,t)]\hat{r}^\alpha\hat{r}^\beta
    + g(r,t)\delta^{\alpha\beta} \ ,
\end{equation}
and then derived a linear differential relation between $f(r,t)$ and
$g(r,t)$ using the solenoidal condition of $\partial_\alpha
C^{\alpha\beta}(\Br,t)=0$,
$$
  2f(r,t) - 2g(r,t) = -r\frac{\partial f(r,t)}{\partial r} \ .
$$
This means that under the assumption of isotropy, and using the
solenoidal condition, the second-order correlation-function can be
written in terms of one scalar function instead of nine.
Similarly, K\'arm\'an and Howarth analyzed the third-order correlation
function by representing it as an isotropic tensor and then reducing the
number of scalar functions using the solenoidal condition. They were
also able to connect it to the second-order correlation function in
decaying turbulence using the Navier-Stokes equations. These computations
have
since found their way into  every standard text-book on the
statistical theory of turbulence.

The mathematical representation of isotropic turbulence has
reached its most elegant and powerful form in a paper by
H.~P.~Robertson from 1940 \cite{rob40}.
Robertson provided a systematic way to represent isotropic
tensors using the theory of invariants.  For example, to derive the general
representation (\ref{eq:KA-C2}) in the stationary case using Robertson's
method, we
consider the \emph{scalar} function
$$
  C(\B{a}, \B{b}, \Br) \equiv C^{\alpha\beta}(\Br)a_\alpha b_\beta \ ,
$$
with $\B{a}$ and $\B{b}$ being two arbitrary vectors. If
$C^{\alpha\beta}(\Br)$ were an isotropic tensor, $C(\B{a}, \B{b},
\Br)$ would preserve its functional form upon an arbitrary
(simultaneous) rotation of the three vectors $\Br, \B{a}, \B{b}$.
Using invariant theory, Robertson deduced that $C(\B{a}, \B{b},
\Br)$ must be a function of the six possible scalar products
$(\Br\cdot\Br)$, $(\Br\cdot\B{a})$,... and of the determinant
$[\Br\B{a}\B{b}]\equiv\epsilon_{\mu\alpha\beta}r^\mu a^\alpha
b^\beta$. Additionally, by definition, it must be a bilinear
function of $\B{a}$ and $\B{b}$ and therefore must have the
following form:
$$
  C(\B{a}, \B{b}, \Br) = A(r)(\Br\cdot\B{a})(\Br\cdot\B{b})
    + B(r)(\B{a}\cdot\B{b}) + C(r)[\Br\B{a}\B{b}] \ ,
$$
where $A(r)$, $B(r)$ and $C(r)$ are arbitrary functions. Finally,
recalling that $C(\B{a}, \B{b}, \Br)$ is the contraction of
$C^{\alpha\beta}(\Br)$ with $\B{a}$ and $\B{b}$, we find that
$$
 C^{\alpha\beta}(\Br) = A(r)r^\alpha r^\beta + B(r)\delta^{\alpha\beta}
    + C(r)\epsilon^{\mu\alpha\beta}r_\mu \ .
$$
If we further demand $C^{\alpha\beta}(\Br)$ to be invariant to
improper rotations as well (i.e., rotations plus reflections), we
can drop the skew-symmetric part $\epsilon^{\mu\alpha\beta}$,
thus retaining a representation which is equivalent to
\Eq{eq:KA-C2}.
\subsection{The 4/5 law in Isotropic Turbulence and its Generalization}
By using the isotropic representation of the third-order
correlation function, in 1941 Kolmogorov proved the ``four-fifth law''
well inside the inertial range of a fully developed turbulence. This law
pertains
to the third order moment of longitudinal velocity differences,
stating that in homogeneous, isotropic and stationary turbulence,
in the limit of vanishing kinematic viscosity $\nu\to 0$
$$
\left<[\delta u_l(\B x,\Br,t)]^3\right> = -\frac{4}{5}
\eb r \ , $$
where $\eb$ is the mean
energy flux per unit time and mass, $\eb\equiv
\nu\left<|\nabla_\alpha u_\beta|^2\right>$. The fundamental assumption
needed to derive this law is  the so-called ``dissipation anomaly''
which means that the dissipation  is finite in
the limit $\nu\to 0$. As noted in \cite{fri95}, ``this is one of the
most important results in fully developed turbulence because it
is both exact and nontrivial. It thus constitutes a kind of
`boundary condition' on theories of turbulence: such theories, to
be acceptable, must either satisfy the four-fifth law, or
explicitly violate the assumptions made in deriving it".

To demonstrate how isotropy helps in deriving this result, we
present a re-derivation in which we will obtain an additional
exact relation that appears to have the same status as the
four-fifth law, pertaining to homogeneous, stationary and
isotropic turbulence with helicity \cite{chk96,lvo97a}. Defining the
velocity $\Bv(\Bx,t)$ as
 $\B v(\Bx,t)\equiv \B u(\Bx,t)-\left<\B u\right>$
we consider the simultaneous 3rd order tensor
correlation function which depends on two space points:
\begin{equation}
J^{\alpha,\beta\gamma}(\Br)\equiv
 \left<v^\alpha(\Bx+\Br,t)v^\beta(\B x,t) v^\gamma(\B x,t)\right> \ .
\label{defJ}
\end{equation}
We show that in the limit $\nu\to 0$, under the same assumption
leading to the fourth-fifth law, this correlation function reads
\cite{lvo97a}
\begin{equation}
J^{\alpha,\beta\gamma}(\Br)=-{\eb \over
10}(r^\gamma\delta_{\alpha\beta}
+r^\beta\delta_{\alpha\gamma}-{2\over
3}r^\alpha\delta_{\beta\gamma}) - {h\over 30}
(\epsilon_{\alpha\beta\delta}r^\gamma+\epsilon_{\alpha\gamma\delta}r^\beta)
r^\delta \ , \label{result}
\end{equation}
where $\delta_{\alpha\beta}$ is the Kronecker delta and
$\epsilon_{\alpha\beta\gamma}$ is the fully antisymmetric tensor.
The quantity $h$ is the mean dissipation of helicity per
unit mass and time,
$$
h \equiv \nu \left<(\nabla^\alpha u^\beta)(\nabla^\alpha[\B
\nabla\times \B u]^\beta)\right> \ , $$
where repeated indices are summed upon. In the derivation below it assumed
that $h$ remains constant when $\nu \rightarrow 0$ in the same spirit of the
dissipation anomaly \cite{eyink1,eyink2,bif98c}. The first term in Eq.
(\ref{result}) is just the 4/5 law. The new part of result
(\ref{result}) can be also displayed in a form that depends on
$h$ alone by introducing the longitudinal and transverse
parts of $\B u$: the longitudinal part is $\B u_l\equiv \B r(\B
u\cdot \B r)/r^2$ and the transverse part is $\B u_t \equiv \B
u-\B u_l$. In terms of these quantities we
can present a ``two fifteenth law"
\begin{equation}
\left<[\delta\B u_l(\B x,\B r,t)]\cdot [\B u_t(\B r+\Bx,t)
\times \B u_t(\B x,t)]\right> =\frac{2}{15}h r^2 \ .
\label{short}
\end{equation}
We note that this result holds also when we replace $\B u$ by $\B
v$ everywhere.

To derive the result (\ref{result}) we start from the correlation
function $J^{\alpha,\beta\gamma}(\B r)$ which is symmetric with
respect to exchange of the indices $\beta$ and $\gamma$ as is
clear from the definition. In an isotropic homogeneous medium with
helicity (no inversion symmetry), the most general form of this
object is \cite{lvo97a}:
\begin{eqnarray}
&&J^{\alpha,\beta\gamma}(\B
r)=a_1(r)[\delta_{\alpha\beta}r^\gamma+\delta_{\alpha\gamma}
r^\beta+\delta_{\beta\gamma}r^\alpha] \label{general}
+\tilde
a_1(r)[\delta_{\alpha\beta}r^\gamma+\delta_l{\alpha\gamma}
r^\beta-2\delta_{\beta\gamma}r^\alpha]\nonumber \\
&&+b_2(r)[\epsilon_{\alpha\beta\delta}
r^\gamma+\epsilon_{\alpha\gamma\delta} r^\beta]r^\delta
+a_3(r)[\delta_{\alpha\beta}r^\gamma+\delta_{\alpha\gamma}
r^\beta+\delta_{\beta\gamma}r^\alpha-5r^\alpha r^\beta
r^\gamma/r^2] \ . \nonumber
\end{eqnarray}
This general representation is invariant to the choice of
orientation of the coordinates.
Not all the coefficients are independent for incompressible
flows. Requiring $\partial J^{\alpha,\beta\gamma}(\B r)/\partial
r^\alpha=0$ leads to two relations among the coefficients:
$$
\Big({d\over dr}+{5\over r}\Big)a_3(r)={2\over 3}{d\over
dr}\big[a_1(r)
+\tilde a_1(r)\big]  \ , \qquad
\Big({d\over dr}+{3\over r}\Big)\big[5a_1(r)-4\tilde
a_1(r)\big]=0 \ . $$
As we have two conditions relating the three coefficients
$a_1,\tilde a_1$ and $a_3$ only one of them is independent.
Kolmogorov's derivation related the rate of energy dissipation to
the value of the remaining unknown. Here the coefficient $b_2$
remains undetermined by the incompressibility constraint; it will
be determined by the rate of helicity dissipation.

Kolmogorov's derivation can be paraphrased in a simple manner.
Begin with the second order structure function $\tilde S^{(2)}(r)\equiv
\la |\Bu (\B x +\B r)- \Bu(\B x)|^2 \ra$.
Computing the rate of change of this (time-independent) function
from the Navier-Stokes equations (\ref{NS}) we find
\begin{equation}
0={\partial \tilde S^{(2)}(r)\over 2\partial t}=-{\mathcal
D}^{(2)}(r)-2\eb+\nu\nabla^2 \tilde S^{(2)}(r) \ , \label{bal}
\end{equation}
where ${\mathcal D}^{(2)}(r)$ stems from the nonlinear term $(\B u\cdot\B
\nabla)\B u$ and as a result it consists of a correlation
function including a velocity derivative. The conservation of
energy allows the derivative to be taken outside the correlation
function:
\begin{equation}
{\mathcal D}^{(2)}(r) \equiv {\partial\over \partial r^\beta} \langle
u^\alpha(\B x,t)u^\alpha(\B x+\B r,t)
\big[u^\beta(\Bx,t)-\!u^\beta(\B x+\B r,t)\big]\rangle \ . \label{D2}
\end{equation}
In terms of the function of Eq. (\ref{defJ}) we can write
\begin{equation}
{\mathcal D}^{(2)}(r) =  {\partial\over \partial
r^\beta}\Big[J^{\alpha,\beta\alpha}(\B r,t)-
J^{\alpha,\beta\alpha}(-\B r,t)\Big] \ . \label{relate}
\end{equation}
Note that Eq. (\ref{defJ}) is written in terms of $\B v$ rather
than $\B u$, but using the incompressibility constraint we can
easily prove that Eq. (\ref{D2}) can also  be identically written
in terms of $\B v$ rather than $\B u$. We proceed using Eq.
(\ref{general}) in Eq. (\ref{relate}), and find
\begin{equation}
{\mathcal D}^{(2)}(r) =  2{\partial\over \partial
r^\beta}r^\beta\big[5a_1(r)+2\tilde a_1(r)\big] \ . \label{D2a1}
\end{equation}
For $r$ in the inertial interval, and for $\nu\to 0$, we can read
from Eq. (\ref{bal}) ${\mathcal D}^{(2)}(r)=-2\eb$ and therefore
have the third relation that is needed to solve all the three
unknown coefficients. A calculation leads to
$$
a_1(r)=-2\eb/45\ , \quad \tilde a_1=-\eb/18\ ,
\quad a_3=0\ . $$
The choice of the structure function $\tilde S_2(r)$ leaves the
coefficient $b_2(r)$ undetermined, and another correlation
function is needed in order to remedy the situation. Since the
helicity is $\B u\cdot [\B \nabla\times \B u]$, we seek a
correlation function which is related to the helicity of eddys of
scale of $r$:
$$
T^{(2)}(r)\equiv \langle \big[\B u(\B r+\B x,t)-\B u(\B x,t)\big]
\nonumber \cdot \big[\B \nabla\times \B u(\B x+\B r,t)-\B
\nabla\times \B u(\B x,t)\big]\rangle \ . $$
Using the Navier-Stokes equations to compute the rate of change
of this quantity we find
\begin{equation}
0={\partial T^{(2)}(r)\over 2\partial t}=-G^{(2)}(r)-2h
-\nu\nabla^2T^{(2)}(r) \ , \label{bal2}
\end{equation}
which is the analog of (\ref{bal}), and where
\begin{eqnarray}
G^{(2)}(r)&=&\left\{\langle\B u(\B x,t)\cdot\left[\B \nabla_r\times\left[\B
u(\B x+\B r,t)\times \left[\B \nabla_r \times \B
u(\B x+\B r,t)\right]\right]\right]\rangle\right\} \nonumber\\&+&\{{\rm
term}~\B r\to -\B
r\} \ . \label{G2}
\end{eqnarray}
The conservation of helicity allows the extraction of two
derivatives outside the correlation functions. The result can be
expressed in terms of the  definition (\ref{defJ}):
$$
G^{(2)}(r)={\partial\over \partial R^\lambda}{\partial\over \partial
r^\kappa}
\epsilon_{\alpha\lambda\mu}\epsilon_{\mu\beta\nu}\epsilon_{\nu\kappa\gamma}
\big[J^{\alpha,\beta\gamma}(\B r)+J^{\alpha,\beta\gamma}(-\B
r)\big] \ . $$
Substituting Eq. (\ref{general}) we find
$$
G^{(2)}(r)=2{\partial^2\over \partial r^\lambda \partial
r^\kappa}b_2(r)\big[r^\lambda r^\kappa
-\delta_{\lambda\kappa}r^2\big] \ , $$
which is the analog of Eq. (\ref{D2a1}). Using Eq. (\ref{bal2})
in the inertial interval in the limit $\nu\to 0$ we find the
differential equation
$$
r^2{d^2 b_2(r)\over dr^2}+9r{db_2(r)\over dr} +15 b_2(r)=-{h
\over 2} \ . $$
The general solution of this equation is
$b_2(r) = -h / 30+\alpha_1 r^{-5}+\alpha_2 r^{-3}$.
Requiring finite solutions in the limit $r\to 0$ means that
$\alpha_1=\alpha_2=0$. Accordingly we end up with Eq.
(\ref{result}).
The moral of this example is that {\em even in isotropic and
homogeneous systems there exist sub-leading terms} which can
become dominant for specially selected objects like (\ref{short}).
Once anisotropy exists, there are many more (in fact infinitely
many) sub-leading contributions that need to be assessed
carefully. Similar results for slightly different correlation functions
have also been found in \cite{kur03,pol03}
\subsection{Kolmogorov's Theory for 2'nd, 4'th, and Higher Order Structure
Functions}
\label{K41forSp}
Unfortunately, the exact result pertaining to the 3'rd order
structure function is rather unique. The moment we consider 2'nd,
4'th or higher order correlation functions there is no exact
result for the scaling exponents. Kolmogorov's 1941 theory states
that, for Re large enough, small-scales turbulent
fluctuations should recover isotropy and homogeneity (if measured
far enough from boundaries) and should possess universal scaling
properties depending only on the mean energy flux, $\eb$.
For homogeneous and isotropic ensembles one defines
\begin{equation}
  S^{(n)}(r) \equiv \left< \delta u_{\l}^n(\Bx, \Br,t)
  \right>\ =  (\eb r)^{n \over 3}f^{(n)}({r \over L_0},{\eta
\over r}), \label{eq:k41a} \label{Sp}
\end{equation}
where $L_0$ and $\eta$ are the integral length scale and the
viscous scale, respectively. The function $f^{(n)}$ is supposed to be
well behaved in the limit of infinite Reynolds numbers for fixed
separation, $r$:
  $\lim_{x,y \rightarrow 0}  f^{(n)}(x,y) =$ const. In this limit,
the celebrated K41  scaling prediction for structure
functions in the inertial range,  $ \eta \ll r \ll L_0$, follows:
\begin{equation}
S^{(n)}(r) \sim  C^{(n)} (\eb r)^{\zeta^{(n)}}, \qquad \mbox{with}\;\;
\zeta^{(n)} = \frac{n}{3}.  \label{eq:k41}
\end{equation}
 In (\ref{eq:k41}) the constants
 $C^{(n)}$ depend only on the large scale
properties.  Because of stationarity, the mean energy flux in
Eq. (\ref{eq:k41}) can be equally taken to be the mean energy input
or the mean energy dissipation.

Kolmogorov's theory goes beyond the scaling prediction
(\ref{eq:k41}). For example, any non-vanishing $p$th order
structure functions, including purely transversal
   and mixed
longitudinal-transversal velocity increments, must possess the
same scaling exponents:
\begin{equation}
  S^{(n,m)}(r) \equiv \left< \delta u_{\l}^{n}(\Br) \delta
  u_t^{m}(\Br) \right>
\sim C^{(n,m)}(\eb r)^{\frac{(n+m)}{3}}, \label{eq:mixed}
\end{equation}
where $p=n+m$, $ \delta  \Bu_t(\Br) \equiv \delta  \Bu(\Br)- \delta
u_{\ell}(\Br)\hat{\Br}$,
and $\delta u_t(\Br)$ is one of the components of the two-dimensional
transverse velocity difference.
Notice that due to the assumption of isotropy, only even combinations
of transversal increment in (\ref{eq:mixed}) have a non vanishing
average.  It is also not difficult to extend the K41 reasonings to
describe also correlation functions at the viscous scales,
i.e. observables based on gradients statistics
\cite{fri90,ben91}.
\subsection{Experimental Difficulties with the Isotropic Theory}
\label{expcheck}

On the whole, experimental tests of Kolmogorov's theory ran into increasing
difficulties
when the data were analyzed with greater detail. The first systematic
attempt to check the isotropic scaling (\ref{eq:k41}) for high
Re number turbulence was
\cite{ans84}. These authors performed
 a high statistical test of K41 theory by going beyond
the usual two-point correlations. They measured  structure
functions of higher order, reaching good evidence that there exist
{\it anomalous} deviations from the scaling exponents (\ref{eq:k41}).  Their
data substantiate a power-law behavior with $\zeta(n) \ne n/3$.  At
that time, and for many year later, the situation was very
controversial both theoretically and experimentally. There appeared
many claims that the observed deviations were due to sub-leading
finite-Reynolds effects. One should not underestimate the difficulties
of getting reliable estimates of the scaling exponents.  First, one
must expect finite Reynolds numbers corrections which may {\it
strongly } reduce the inertial range  where scaling laws are
expected or introduce anisotropic corrections to the isotropic K41
predictions. Both effects are usually present in all experiments and
numerical simulations. Nowadays, state-of-the-art experiments of
turbulence in controlled geometries reach a maximum Re numbers
measured on the gradient scales, $\lambda^{-1} = \langle | \B\nabla \Bu|
\rangle
/\langle |\Bu| \rangle$, of $R_{\lambda}
\sim 5000$
where $R_{\lambda} = \frac{\lambda U}{\nu}$ and $U$ is the typical
large scale velocity.  In atmospheric flows $R_{\lambda}$ can be
as high as $R_{\lambda }\sim 20000$ but at the expense of high
anisotropy. More complex is the situation of  DNS where the best resolution
ever reached up to now is
$4096^3$ \cite{kan03}, corresponding to a $R_{\lambda}
\sim 1100$. DNS allow a minimization of the anisotropic corrections,
by implementing periodic boundary conditions
and fully isotropic forcing, something which is not
 experimentally feasible.
However, also in DNS the discrete symmetries
induced
by the finite lattice-spacing do not allow for  perfect isotropic
statistics. We thus either have high-Reynolds-numbers experiments
which are strongly perturbed by anisotropic effects, or DNS isotropic
flow at moderate Reynolds numbers. Therefore, one has to face the
problem of how to disentangle isotropic from anisotropic fluctuations
and how to extract information on the asymptotic scaling with a finite
--often short-- inertial-range extension. Only recently, after many
experimental and numerical confirmations of the results of
\cite{ans84},
the situation became clearer \cite{ben93b}. We may affirm now with
some degrees of certitude that the isotropic scaling exponents are
{\it anomalous}, the K41 prediction $\zeta(n) = n/3$ is wrong, except
for $n=3$ which is fixed to be $\zeta(3)=1$ by the exact $4/5$ law.
Moreover, the possibility to show analytically the existence of
anomalous scaling in turbulent advection
\cite{fal01}, definitely eliminated those arguments
supporting the {\it impossibility} to have a Re-independent
anomalous scaling in any hydrodynamic system. From a phenomenological point
of view, it is easy to extend the K41
theory such as to include  anomalous scaling.
Already Kolmogorov noticed, after Landau's criticism in 1962, that it is
unrealistic to expect the isotropic inertial range fluctuations to
depend only on the {\it mean} energy dissipation, $\eb$.
 Kolmogorov proposed in 1962 \cite{kol62} to employ the
coarse-grained energy dissipation over a box of size $r$,
\be
\label{eq:coarse}
\tilde
\epsilon(r,\Bx) = {1\over r^3}
\int_{|y| < r} d\By \eb(\Bx +\By),
\ee
 to
match the correct dimensions of structure functions in
(\ref{eq:k41a}), the so-called Refined Kolmogorov Hypothesis:
\begin{equation}
 S^{(n)}(r) = C^{(n)} \la \tilde \epsilon^{n\over 3}(r)\ra r^{n \over
 3}.
\label{eq:multifractal}
\end{equation}
This hypothesis connects the deviation from the K41
prediction, $\zeta(n) - n/3 = \tau(n/3)$, to the anomalous scaling of
the coarse-grained energy dissipation : $\la \tilde \epsilon^{n\over
3}(r)\ra
\sim r^{\tau(n/3)}$.
 Anomalous scaling of isotropic structure functions is therefore
 connected to the multifractal properties of the three dimensional
 energy dissipation field \cite{fri95}.  It should be noted however
 that the Refined Kolmogorov Hypothesis related inertial range scaling
 to scaling of dissipative quantities, and delicate issues connected
 to small distance expansions and fusion rules are being disregarded
 here \cite{lvo96,ben98}. At any rate, the relation presented by
 Eq.(\ref{eq:multifractal}) did not advance the calculation of the
 scaling exponents beyond crude phenomenology.
\subsection{Persistence of Anisotropies}
\label{sec:persis}
A central issue of K41 phenomenology is the assumption of {\it
return-to-isotropy} for smaller and smaller scales.  Recently this
assumption had been put to test in experiments and simulations
\cite{pum96,gar98,kur00,bif01}.  A useful experimental
set-up to test the return to isotropy is a {\it homogeneous} shear
flow
\cite{hin75} where the large-scale
mean-velocity has a linear profile: $\B{V} = (V_0 y,0,0)$.  The shear
is given by ${\mathcal S}_{ij} = \partial_i V_j = \delta_{iy}
\delta_{jx} V_0$. We thus have
a homogeneous but anisotropic flow, close to the ideal case for
studying the influence of large scale anisotropies on the small scale
statistics.  ``Small scales" are defined here in comparison to the
characteristic shear length, $L_{\mathcal S} = \eb^{1/3}/{\mathcal S}$;
for $ r \ll L_{\mathcal S}$ we may expect that anisotropic fluctuations
are sub-leading with respect to the isotropic ones. The case $ r \gg
L_{\mathcal S}$ is of interest in situations where the shear is very
intense, as very close to the walls in bounded flows. In such cases we
expect a dramatic change  from the K41 phenomenology
\cite{ben02,tos99,tos00}.
Fortunately, it is not that difficult to design experiments or DNS
possessing an almost perfect linear profile with homogeneous shear
\cite{pum96,she00,gar98,ben02}.
A popular way to measure small-scales anisotropies is to focus on the
Re dependence of isotropic and anisotropic statistical
observables built in terms of velocity gradients. For example, due to
the symmetries of the mean flow, gradients of the stream-wise
component in the shear direction, $\partial_y u_x$, may have a skewed
distribution only due to the anisotropic fluctuations; they have a
symmetric PDF in a perfectly isotropic flow.  A natural measure of the
residual anisotropy at small scales as a function of  Re
 is the mixed generalized skewness based on gradients:
\begin{equation}
M^{(2n+1)}(R_{\lambda})
\frac{ \la (\partial_y u_x)^{2n+1} \ra}{\la (\partial_y u_x)^2
\ra^{\frac{2n+1}{2}}}.
\label{eq:der_ske}
\end{equation}
These objects vanish in isotropic ensembles. Of course, at finite
Reynolds numbers one expects that the large-scale anisotropy
introduced by the shear still remains, even on the gradient scale.
Therefore, the rate of decay of (\ref{eq:der_ske}) as a function of
Re is a quantitative indication of the rate of decay of
anisotropy at small scales.  In the next section we review Lumley's
dimensional arguments \cite{lum67} for anisotropic fluctuations, which
predicts:
\begin{equation}
M^{(2n+1)}(R_{\lambda}) \sim R^{-{1 \over 2}}_{\lambda},\qquad \forall \;\;
n.
\label{eq:der_ske_mf}
\end{equation}
In fact, both numerical
\cite{pum96,pum95} (at low Reynolds numbers) and experimental
tests (up to $R_{\lambda} \sim 1000$) showed a clear disagreement with
the dimensional prediction (\ref{eq:der_ske_mf}). For example in
\cite{she00} the
authors quote a  decay in agreement with the prediction for
$M^{(3)}(R_{\lambda})$, an almost constant
behavior as a function of Re for the fifth order,
$M^{(5)}(R_{\lambda}) \sim O(1)$ and an {\it increasing} behavior for
the seventh order $M^{(7)}(R_{\lambda}) \sim R^{+0.63}_{\lambda}$ !
These results have cast a severe doubt
on the fundamental assumption of the K41 theory.  Similar results, with even
more
striking contradictions with the hypothesis of the return-to-isotropy, have
been measured in the problem of passive scalar fluctuations, $\theta = T -
\la T \ra $, advected by an isotropic velocity field in the presence of a
mean homogeneous scalar gradient, $\B \nabla \la T \ra = (g,0,0)$.
The equation of motions for the passive advected field in this case
are: $$
\partial_t \theta + \Bu \cdot \B \nabla \theta =
g u_x + \chi \partial^2 \theta.
$$
Both experimental and numerical data show a strong
disagreement with the prediction that generalized skewness of
temperature gradients becomes smaller upon increasing
Reynolds and Peclet numbers \cite{war00,gib77,ant78,sre91,p94,cel01}.

We will show below how the analysis based on SO(3) decomposition and
its theoretical consequences settles this puzzle of strong {\it
persistence of anisotropies} \cite{bif01}. In fact, contrary to
what appears, the K41 phenomenology with its assumption of {\it
return-to-isotropy} and the above experimental results are not at all
in contradiction (see section \ref{pers_ani}).
\subsection{Longitudinal and  Transversal Isotropic Structure Functions}
\label{sec:long-tran}
Another debated issue concerning the K41 phenomenology and its
multifractal generalization (\ref{eq:multifractal}) has to do with the
observed discrepancies between the scaling properties of longitudinal,
transversal, and mixed longitudinal-transversal structure functions in
(supposedly) isotropic fully developed turbulence
\cite{she02,che97,dhr97,got02,wat99}.  As previously
stated, K41 theory, for isotropic flows, predicts the same scaling
behavior, in the limit of high Re, independent of
the Cartesian components of the velocity increments in the structure
functions. For a given order $p=m+n$ only prefactors
in (\ref{eq:mixed}) may depend on the particular choice of $n$ and $m$.
Let us denote with
$\zeta^{(n,m)}$ the scaling exponent of the mixed structure function
(\ref{eq:mixed}) made of $n$ longitudinal increments and of $m$
transversal increments in a isotropic ensemble:
$$
  S^{(n,m)}(r)\sim C^{(n,m)} r^{\zeta^{(n,m)}}.
$$
For $p<4$ the issue does not exist; due to the incompressibility
constraint all second and third order longitudinal or transversal
structure functions have the same scaling in a isotropic ensemble. For
$p>3$ many experiments and numerical simulations found that
$\zeta^{(n,m)} < \zeta^{(n',m')}$ if $n<n'$ and $m>m'$ when $n+m=n'+m'$.  It
appears that with increasing $m$ the
scaling exponents reduce (the signal is more intermittent). The
largest difference for a structure function of order $p$ is therefore
achieved when we compare the purely longitudinal scaling $\zeta(p,0)$
with the purely transversal scaling $\zeta(0,p)$.  Other experimental
data suggests the possibility of a {\it slow} tendency of the
longitudinal and transversal scaling exponents to coalesce for
increasing Re \cite{zho99,nou97,cam96}.

We will argue below that the experimental measurements of different
exponents stems from anisotropic corrections that affect differently
the longitudinal and transverse components. In other words, by not removing
the
anisotropic contributions, one cannot expect pure power-law behavior.
  The situation is more
complex for the analysis of data from DNS. There, one may implement
highly isotropic forcing and boundary conditions, such that in most
cases any residual anisotropic effects may safely be neglected even at
moderate Re numbers. On the other hand, state-of-the-art
numerical simulations are still strongly limited in the maximum
Re  achievable. Only very recently reliable data with
high-statistics became available at resolution $1024^3$
\cite{got02}, while most of the previous DNS where limited to  lower
resolutions. At resolution of $1024^3$ one reaches a moderate
$R_{\lambda} \sim 400$, far below many experiments. Because of the
consequent limited extension of the inertial range, such DNS did not
resolve the puzzle of longitudinal {\it vs} transversal scaling. The
numerical results oscillate between evidence for different scaling
properties and for its opposite
\cite{che97,got02,bor97,ker01,gro98b}.  The issue is complicated
by the fact that longitudinal and transversal structure functions
possess different finite Re effects. For example, in
\cite{got02} it was shown that structure functions
of different order have different dependence on the viscous cut-off;
this introduces ambiguity in defining a common inertial range where
power law is expected. We thus propose that until high resolution
isotropic measurements became available, all evidence for different
scaling exponents for longitudinal and transverse structure functions
should be considered with suspicion.
\subsection{Position Dependent Scaling Exponents}
\label{sec:position-dep}
In some inhomogeneous
simulations and experiments it was claimed that the measured scaling
exponents depended on the point of measurement within the flow domain
\cite{gau98,sad94,tos99,tos00,ono00,str96}. If true, such finding would
deal a death blow to the idea of universality of the scaling exponents
in turbulence. It should be stressed that in all the examples where
such findings were reported the flow contained strong anisotropic and
inhomogeneous components and/or the scaling range was not sufficient
to actually present direct log-log plots for the structure functions
vs. $r$. In some of these cases, scaling was extracted by using the
method called ``Extended Self Similarity" (ESS) \cite{ben93b}; the use
of this method can be dangerous in presence of anisotropic and
inhomogeneous effects.  Whenever strong anisotropies are present one
has to distinguish among two scaling ranges. At scales larger than the
shear length, the energy cascade mechanism of the Kolmogorov theory is
overwhelmed by shearing effects \cite{ben02,tos00}.  Only
for scales smaller than the shear length the meaning of anisotropic
corrections to the isotropic K41 scaling theory is well posed.  We
argue below that the reported position-dependent isotropic exponents
in the latter case stem from anisotropic components which appear with
different amplitudes at different points in the flow. The different
``exponents" that were measured were not real exponents but the result
of a crossover between the isotropic and anisotropic corrections.
Once the data is projected onto the isotropic sector the leading
exponents become position independent as expected.
\section{Historical Review: Attempts at  Anisotropy}
\label{history-aniso}
\subsection{Bachelor's Approach}
The first systematic approach to anisotropy in turbulence was
suggested by G.~K.~Batchelor in 1946 \cite{bat46}.  Batchelor did
not attempt to describe the most general form of anisotropy in
turbulence, but instead confined himself to the easier case of
\emph{axisymmetric turbulence}. In axisymmetric turbulence
 the mean value of any function of the velocity field and its
 derivatives is invariant to all rotations of the axes in a given
 direction. Therefore, the anisotropy in axisymmetric turbulence is
 induced by a single direction in space.  We denote this symmetry
 axis by the unit vector $\Bn$. Being the easiest case of anisotropic
 turbulence, axisymmetric turbulence was the main model for
 studying anisotropy in subsequent years.

Batchelor used the invariant theory in order to take the anisotropy
vector $\Bn$ into account in the tensor representations. His method
is simple: add the vector $\Bn$ to the list of vectors in Robertson's
method. For example, suppose we wish to construct an axisymmetric
representation of the second-order correlation function
$C^{\alpha\beta}(\Br)$ defined in (\ref{def:KA-C2}). Then, just as in
the isotropic case, we create a scalar function by contracting the two
indices of $C^{\alpha\beta}(\Br)$ with two arbitrary vector $\B{a}$
and $\B{b}$, with the difference that now we assume that the resultant
scalar function depends on the unit vector $\Bn$ as well as on the
other vectors $\Br$, $\B{a}$ and $\B{b}$. We therefore look for an
invariant representation of the scalar function $C(\Br, \Bn, \B{a},
\B{b})$, which  depends only  on the different scalar products
$\Br\cdot\Br$, $\Br\cdot\Bn$, $\Br\cdot\B{a}\ $,... and the various
determinants $[\Br\Bn\B{a}]$, $[\Br\Bn\B{b}]\ $,...  For some reason,
Batchelor decided to ignore the skew-symmetric parts and considered
only the scalar products. Using the fact that $C(\Br, \Bn, \B{a},
\B{b})$ is a bilinear function of $\B{a}$ and $\B{b}$, Batchelor found
that
$$
  C^{\alpha\beta}(\Br) = A r^\alpha r^\beta + B \delta^{\alpha\beta} +
  C n^\alpha n^\beta + D n^\alpha r^\beta + E r^\alpha n^\beta \ ,
$$
where $A,B,C,D,E$ are functions of the amplitude $r$ and of the scalar
product $\hat{\Br}\cdot\Bn$.  Notice that in this expansion the number
of unknowns has grown from two to five with respect to the isotropic
expansion. It would have been nine, had we taken the skew-symmetric
parts into account. Indeed, a prominent characteristic of anisotropic
representations is that they are far more complex than their isotropic
counterparts.

Using this sort of representations, Batchelor was able to generalize
K\'arm\'an-Howarth results to the case of axisymmetric turbulence. That
is, after representing the second- and third-order correlation-functions in
terms of few scalar functions, Batchelor used the solenoidal condition
and the Navier-Stokes equations to derive some linear differential
relations among them.
\subsection{Chandrasekhar and Lindborg's Approaches}
A somewhat more elegant approach to axisymmetric turbulence was
offered a few years later by S.~Chandrasekhar
\cite{cha50}. Chandrasekhar's treatment is similar to Batchelor's
in following Robertson's work \cite{rob40}. Chandrasekhar
took advantage of the skew-symmetric tensor
$\epsilon^{\mu\alpha\beta}$ for creating a representation of
solenoidal axisymmetric tensors. He noticed that the curl of an
axisymmetric tensor automatically satisfies the solenoidal
condition. Therefore, by representing the second- and third-
correlation-functions as a curl of auxiliary tensors, Chandrasekhar
automatically solved the solenoidal equations, and was left with the
dynamical equations (which are derived from the Navier-Stokes
equation) only. Chandrasekhar's dynamical equations are considerably
simpler than those of Batchelor. Nevertheless, they are still very
complicated and this, perhaps, explains why there was no serious
attempt to continue Chandrasekhar's work in subsequent years.

In 1995 there was another attempt to formulate the kinematics of
homogeneous axisymmetric turbulence in \cite{lin95}.  The
representation in this paper was ``experimentally oriented'', in the
sense that the scalar functions that are used can be measured directly
in experiment. To accomplish that, one defines two auxiliary unit
vectors (that were also used in \cite{her74}): $
  \Be_1(\Br) \equiv \Bn \times \Br/|\Bn\times\Br| \ , \quad \Be_2(\Br)
  \equiv \Be_1(\Br) \times \Bn \ $.
The triplet $(\Bn, \Be_1, \Be_2)$ is an orthonormal basis of
$\mathbb{R}^3$. But since it is made out of $\Br$ and $\Bn$ using
invariant operations (i.e., vectorial products), it is invariant to
simultaneous rotations of $\Br$ and $\Bn$, and thus it is invariant to
rotations of $\Br$ alone around $\Bn$ (because in these rotations
$\Bn$ remains fixed anyhow). Therefore any tensor that is built from
these unit vectors and the products $\Br\cdot\Br$, $\Br\cdot\Bn$ is
necessarily an axisymmetric tensor. For example, in order to represent
the second-order correlation-function $C^{\alpha\beta}(\Br)$, one writes
\begin{eqnarray}
  C^{\alpha\beta}(\Br) &=& R_1 n^\alpha n^\beta +R_2 e_2^\alpha
  e_2^\beta + R_3 e_1^\alpha e_1^\beta +R_4 \big[n^\alpha e_2^\beta +
  n^\beta e_2^\alpha\big] \nonumber \\ &&+\ Q_1 \big[n^\alpha e_1^\beta +
  n^\beta e_1^\alpha\big] + Q_2 \big[e_2^\alpha e_1^\beta + e_2^\beta
  e_1^\alpha\big] \ ,\nonumber
\end{eqnarray}
where the six scalar functions $R_1,{\ldots}, R_4$ and $Q_1, Q_2$ are
functions of $\mu \equiv \Br\cdot\Bn$ and $\rho \equiv |\Br \times
\Bn|$.
  Notice that this representation takes into account the
  skew-symmetric part of $C^{\alpha\beta}(\Br)$ using the scalar
  functions $Q_1$ and $Q_2$.

The major advantage in this representation is that the scalar
functions have an immediate interpretation in terms of measurable
quantities. With respect to the example above, if $(u,v,w)$ are the
velocity components in the direction of $(\Bn, \Be_2, \Be_1)$
respectively, then due to the orthonormality of triplet $(\Bn, \Be_2,
\Be_1)$, we get
\begin{eqnarray}
&& R_1 = \left<u(\Bx)u(\Bx+\Br)\right> \ , \quad R_2 =
\left<v(\Bx)v(\Bx+\Br)\right> \ ,\nonumber  \\ && R_3 =
\left<w(\Bx)w(\Bx+\Br)\right> \ , \quad R_4 =
\left<u(\Bx)v(\Bx+\Br)\right> \ , \nonumber \\ && Q_1 =
\left<u(\Bx)w(\Bx+\Br)\right> \ , \quad Q_2 =
\left<v(\Bx)w(\Bx+\Br)\right> \ .\nonumber
\end{eqnarray}
Next one uses the solenoidal condition
 to derive linear differential relations between
the scalar functions. One can also consider the triple correlation
function and the velocity-pressure correlation function in the very
same method. This way one derives a representation for the dynamical
equation of $C^{\alpha\beta}(\Br)$.  Mathematically speaking, this
representation is no better than Batchelor's representation, and may
even be considered worse than Chandrasekhar's. This is because there
is no reduction in the number of scalar functions and there is no
simplification in the resulting equations. The motivation for this
representation is experimental without compelling physical or
mathematical contents.
\subsection{Case Specific Approaches}
The above mentioned works  can be viewed as
\emph{systematic} attempts to deal with the problem of anisotropic
turbulence, where a general method to describe the anisotropic (or,
more exactly, axisymmetric) quantities in turbulence is suggested. In
that respect they differ from most research that followed on the
subject, which was usually confined to a particular model or a
specific problem related to anisotropy.
\subsubsection{Temporal Return to Isotropy}
 One such problem is the
temporal  return  to isotropy in which one tries to understand the
mechanisms that drive a decaying turbulence which is initially
anisotropic into being statistically isotropic. As in the works of
Batchelor and Chandrasekhar, the statistics was usually assumed to be
spatially homogeneous to simplify the problem. The theoretical
attempts to explain this phenomenon can be roughly divided into two
groups. The first group, which was initiated by  Rotta
\cite{rot51} in 1951, consists of attempts to model the decay of
anisotropy using \emph{one-point} closures. In that framework, one
usually considers the dynamical equation of the Reynolds stress, which
is the \emph{same-point} correlation-function of the velocity field
$
  C^{\alpha\beta}(t) \equiv \left<u^\alpha(\Bx,t) u^\beta(\Bx,t)\right> \ .
$
In a homogeneous, decaying turbulence, this correlation obeys the following
equation
$$
  \partial_t C^{\alpha\beta} = -\eb \phi^{\alpha\beta}
    - \frac{2}{3}\eb\delta^{\alpha\beta} \ ,
$$
with
\begin{equation}
\label{eq:return-to}
  \eb \phi^{\alpha\beta} \equiv \left<u^\alpha\partial^\beta p\right>
    + \left<u^\beta\partial^\alpha p\right>
    + 2\left[\nu\left<\partial_\mu u^\alpha \partial^\mu u^\beta\right> -
    \frac{1}{3}\eb\delta^{\alpha\beta}\right] \ .
\end{equation}
 Notice that $C^{\alpha\beta}$ is a second-rank $\Br$-independent
 tensor that contains an isotropic part (which is its trace) and an
 anisotropic (traceless) part.  This explains the motivation behind
 the definition of $\phi^{\alpha\beta}$, which is to capture the
 anisotropic part of the decay rate of $C^{\alpha\beta}$. To solve
 \Eq{eq:return-to}, one must model the $\phi^{\alpha\beta}$
 tensor. Usually this has been done on a phenomenological basis. A
 systematic treatment to this problem was offered by in
 \cite{lum77}. In that paper, the authors suggested that
 $\phi^{\alpha\beta}$ should depend on time implicitly through
 $C^{\alpha\beta}$, $\eb$ and $\nu$. Additionally, since
 $\phi^{\alpha\beta}$ is a dimensionless tensor, it must depend only
 on dimensionless parameters. There are six such independent
 dimensionless quantities. The authors chose to represent them in a
 way which isolates the property of anisotropy from other properties,
 and form the tensor
$$
  b^{\alpha\beta} \equiv C^{\alpha\beta}/q^2 -
\frac{1}{3}\delta^{\alpha\beta}
  \ , \quad q^2 = C^{\alpha \alpha} = \left<\Bu^2(\Bx)\right> \ .
$$
The tensor $b^{\alpha\beta}$ is proportional to the anisotropic, traceless
part of $C^{\alpha\beta}$ and hence contains five independent components.
It is often denoted as the ``Reynolds stress anisotropy'' or simply as the
``anisotropy tensor''. This tensor has become a central measure for
anisotropy in turbulence and has been used extensively in experimental and
numerical analysis of anisotropy in turbulence. The sixth component was
defined to be proportional to the isotropic part of $C^{\alpha\beta}$ (the
energy) by
$ R_l \equiv \frac{q^4}{9\eb\nu}
$. With these dimensionless quantities, $\phi^{\alpha\beta}$ can be written
as
$
  \phi^{\alpha\beta} = \phi^{\alpha\beta}(\B{b}, R_l) \ $.
They further simplified that expression by noticing that
if $\phi^{\alpha\beta}$ depends solely on $\B{b}$ and $R_l$ then it must
depend on them in an \emph{isotropic} manner, since any anisotropic
dependence necessarily means that $\phi^{\alpha\beta}$ also depends on
the boundary conditions. To represent this isotropic dependence
explicitly, they used the invariant theory \cite{lum70} and
introduced the second- and third- principal-invariants of the
traceless tensor $\B{b}$:
$$
  II \equiv Tr[\B{b}^2] \ , \quad III \equiv Tr[\B{b}^3] \ .
$$
According to the invariant theory, $\phi^{\alpha\beta}$ can be most
generally written as
\begin{equation}
  \phi^{\alpha\beta} = \beta(II,III, R_l) b^{\alpha\beta} +
  \gamma(II,III, R_l) \Big[b^{\alpha\mu}b^{\mu\beta} -
  \frac{1}{3}II\delta^{\alpha\beta}\Big] \ ,
\end{equation}
and the problem is reduced to determining the functional form of
$\beta(II,III,R_l)$ and $\gamma(II,II,R_l)$. Based on this formalism,
there have been many attempts to model the functions $\beta(II,III,
R_l)$ and $\gamma(II,III, R_l)$ to match experimental results
\cite{chu95}. For example, Rotta's model is considered a linear
model for the anisotropy decay because he used
$$
  \beta(II,III,R_j) = C_1 \approx 3.0 \ , \qquad
  \gamma(II,III,R_j) = 0 \ .$$
Consequently, the decay of the anisotropy tensor is given by
$$
  \partial_t b^{\alpha\beta} = -(\eb/q^2)(C_1-2)b^{\alpha\beta} \,
$$
which is a linear equation in $b^{\alpha\beta}$, provided that the
isotropic quantities $\eb, q^2$ are independent of
$b^{\alpha\beta}$.  This sort of equation predicts that
$b^{\alpha\beta}(t)$ is proportional to $b^{\alpha\beta}(t=0)$ and
that every component of the tensor decays at the same
rate. Experimentally however linearity is not supported. For example,
it has been observed experimentally that the return to isotropy is
relatively rapid, at least at the beginning of the process, when the
invariant $III$ is negative, whereas the return to isotropy is fairly
slow in the case where the invariant is positive \cite{gen80}.

The other line of research that was used to study the problem of the
return-to-isotropy consists of attempts to model the decay with
\emph{two-point closures}. In these models one considers the different
correlation functions of the velocity field across a separation vector
$\Br$, instead of using the same-point correlations as in the
one-point closures. The mathematical structure here is usually much
more complicated than that of the one-point closures, but in return,
the two-points models often provide a deeper understanding of the
physics involved.

An important example for such a model is given by \cite{her74}. In
this work the author used the direct-interaction approximation (DIA)
\cite{kra59} to study the decay of an axisymmetric turbulence into
isotropy. The DIA is a well-known truncation of the renormalized
perturbation theory for turbulence. The perturbation is done in the
interaction strength parameter (which is set to unity in the end), and
is truncated at the second-order - i.e., at the direct-interaction
terms.  The calculations are done in Fourier space. They result in two
coupled equations for the time evolution of the two-time, second-order
correlation-function $C^{\alpha\beta}(\Bk,t,t') \equiv \left<
  u^\alpha(\Bk,t)u^\beta(-\Bk,t')\right>$ and the response
function $G^{\alpha\beta}(\Bk,t,t')$. The latter is defined as the
average of the change in the velocity field at time $t$ as a result of
an infinitesimal change in the forcing at time $t'$.
The equations for
$C^{\alpha\beta}(\Bk,t,t')$ and $G^{\alpha\beta}(\Bk,t,t')$ determine
their time evolution. They are nonlinear, nonlocal
integro-differential equations and are therefore very hard to deal
with. In the isotropic case the equations can be considerably
simplified by noting that both $C^{\alpha\beta}(\Bk,t,t')$ and
$G^{\alpha\beta}(\Bk,t,t')$ must satisfy the solenoidal condition,
which in Fourier space means that both tensors must vanish once we
contract any of their indices with the vector $\Bk$. It is easy to see
that under such a condition, $C^{\alpha\beta}(\Bk,t,t')$ and
$G^{\alpha\beta}(\Bk,t,t')$ can be represented in terms of one scalar
function:
\begin{equation}
  C^{\alpha\beta}(\Bk,t,t') = c(k,t,t')D^{\alpha\beta}(\hat{\Bk}) \ ,
  \quad
  G^{\alpha\beta}(\Bk,t,t') = g(k,t,t')D^{\alpha\beta}(\hat{\Bk}) \ ,
\end{equation}
where
\begin{equation}
\label{eq:projection-D}
  D^{\alpha\beta}(\hat{\Bk}) \equiv \delta^{\alpha\beta} -
  \hat{k}^\alpha\hat{k}^\beta \ .
\end{equation}
When turning to the axisymmetric case, the representation of the tensors
become much more complex. Instead of one scalar function for each tensor,
two functions must be used corresponding to the two scalar functions that
where used by Batchelor \cite{bat46} and Chandrasekhar \cite{cha50} to
describe the second-order correlation-function in real space. One  uses
the separating vector $\Bk$ and the anisotropy unit-vector $\Bn$ to create
two unit vectors which are orthogonal to $\Bk$:
\begin{equation}
  \Be_1(\Bk) \EqDef \Bk \times \Bn/|\Bk\times\Bn| \ , \quad
  \Be_2(\Bk) \EqDef \Bk\times(\Bk \times \Bn)/|\Bk\times(\Bk\times\Bn)| \ .
\end{equation}
With these vectors, $C^{\alpha\beta}(\Bk, t,t')$ was written as
\begin{equation}
\label{eq:Herring-rep}
  C^{\alpha\beta}(\Bk, t, t') = c_1(\Bk,t,t')\Be_1^\alpha\Be_1^\beta +
  c_2(\Bk,t,t')\Be_2^\alpha\Be_2^\beta \ ,
\end{equation}
and $G^{\alpha\beta}(\Bk,t,t')$ was written in a similar way using the
functions $g_1(\Bk,t,t')$ and $g_2(\Bk,t,t')$. To parameterize the angular
dependence of the scalar functions, one expands the four scalar
functions in terms of Legendre polynomials,
$$
  c_i(\Bk,t,t') = \sum_j c_{i,j}(k,t,t')P_j(\hat{\Bk}\cdot\Bn)
  \ ,\quad
  g_i(\Bk,t,t') = \sum_j g_{i,j}(k,t,t')P_j(\hat{\Bk}\cdot\Bn) \ ,
$$
obtaining an infinite set of coupled equations for the infinite set of
functions $c_{i,j}(k,t,t')$ and $g_{i,j}(k,t,t')$. These
equations were solved numerically after truncating all the $j>0$
part of the expansion. Doing so,one finds the time evolution of
$c_{i,0}(k,t,t')$ and $g_{i,0}(k,t,t')$ and connects them to the
physical observables of the one-point closures, such as Rotta's
constant. The  conclusions (partially numerical and partially obtained
after a long series of uncontrolled approximations) were that the
return-to-isotropy is much stronger at small scales (large $k$) and
that in some classes of initial conditions, the return-to-isotropy is
indeed a linear phenomenon.

This calculation was soon revised in \cite{sch76}. In this
paper, the authors compared the DIA calculation to the results of a
numerical simulation of a homogeneous and axisymmetric
turbulence. This time, however, the Legendre polynomials expansion in
the DIA calculation was extended to include also the $j=2$
components. Their conclusions were that the DIA calculation that
included the $j=2$ parts were in a good agreement with the numerical
simulation, especially the $j=0$ parts, provided that the initial
anisotropy was small. Additionally, the authors found that the
previous calculation \cite{her74}, which considered the $j=0$ part
only, was quite inadequate to describe the process of
return-to-isotropy - even in the case of weak anisotropy.

Other attempts to study the problem of return-to-isotropy in an
axisymmetric turbulence used two-point closures. For example,
eddy-damped quasinormal Markovian approximation (EDQNM) \cite{ors70,ors77}
 have been used in \cite{nak87}. In this closure scheme,
one approximates the fourth order  cumulants of the velocity field by a
linear damping term of the third-order correlation-function of the
velocity field (the eddy dumping).  Additionally, a
``Markov'' assumption is used that allows one to integrate
the history integral in the equations and retain an equation for the
\emph{same time} second-order correlation-function
$C^{\alpha\beta}(\Bk,t)$. As in the DIA model, this equation is
formulated in Fourier space and is both nonlinear and nonlocal. To
parameterize the axisymmetric correlation function in Fourier space,
the authors  used the following representation
$$
  C^{\alpha\beta}(\Bk, t)
  =
  \frac{1}{4\pi k^2}\big[D^{\alpha\beta}(\hat{\Bk})E(k,\mu,t) + n^\tau
  n^\sigma D^{\alpha\tau}(\hat{\Bk})D^{\beta\sigma}(\hat{\Bk})
  F(k,\mu,t)\big] \ ,
$$
where here $ \mu \equiv \Bk\cdot\Bn $ and $D^{\alpha\beta}(\hat{\Bk})$
is defined in \Eq{eq:projection-D}. This representation incorporates
the solenoidality condition and is more elegant than
(\ref{eq:Herring-rep}) in the sense that the isotropic case is very
easily recovered once we set $F(k,\mu,t)=0$ and let $E(k,\mu,t)$
become $\mu$-independent. Plugging this expansion to the dynamical
equation of $C^{\alpha\beta}(\Bk,t)$, the authors obtained two coupled
equations for $E(k,\mu,t)$ and $F(k,\mu,t)$ which they solved
numerically for ``medium'' and ``strong'' anisotropic initial
conditions. Their results indicate that in the medium anisotropy
cases, Rotta's constant approaches a constant of the order of unity,
qualitatively agreeing with Rotta's model and with the results
\cite{her74}. On the other hand, in the strong anisotropy case this
constant does not show any saturation, indicating the failure of
Rotta's model. Additionally, their results support the idea that the
decay isotropy strengthens when the $III$ invariant is negative.
\subsection{Dimensional Analysis in the Presence of Strong Shear}
\label{sec:lumley}
 An important discussion of the effects of strong shear on the energy
 spectrum was presented in \cite{lum67}.  In this paper the author
 included anisotropic corrections in to the K41 framework extending
 the phenomenological dimensional reasonings leading to
 (\ref{eq:k41a}). He considered the dependence on {\it anisotropic}
 mean observables, like the large-scale Shear proportional to the
 large-scale mean gradient: ${\mathcal S} \propto
\partial \la V \ra$:
$$
  S^{(n)}(\Br) = (\eb r)^{n \over 3}f^{(n)}({r \over L_0},{\eta \over r},
{\mathcal S}).
$$
By further assuming that anisotropic corrections are ``small'' and
{\it analytic} in the intensity of the shear ${\mathcal S}$, he proposed
the following form for the anisotropic correction to the isotropic
two-point longitudinal structure functions, in the inertial range
\cite{lum67}:
\begin{equation}
 S^{(2)}(\Br) \sim C^{(2)}(\eb r)^{2 \over 3} + D^{(2)}(\hat{\Br}){\mathcal
S} r^{4
\over 3}
\label{eq:Lumley}
\end{equation}
where the coefficient $D^{(2(}(\hat{\Br})$ takes into account the
 dependence on the direction $\hat{\Br}$ in
the anisotropic term.  The counterpart of (\ref{eq:Lumley}) for the
spectrum and co-spectrum in Fourier space is:
\begin{equation}
 \la k^2 u_i(\Bk)u_l(-\Bk) \ra \sim k^{-{5\over 3}}(\delta_{il}
 -{k_ik_l
\over k^2}) + A_{il} k^{-{7\over 3}}
\label{eq:k41F}
\end{equation}
where the first term on the RHS is the isotropic K41 scaling and the
second term is the anisotropic contribution with $A_{il}$ being a
traceless matrix depending on the details of the large scale shear.\\
In the past, most of  the measurements of the anisotropic contributions to
$S^{(2)}(r)$
concentrated on the Fourier representation (\ref{eq:k41F}),
\cite{sad94,tav81,sou95,lum65}.
In \cite{sad94} the authors showed
that the prediction (\ref{eq:Lumley} \ref{eq:k41F})
is well verified in a wind tunnel
flow. Later, many other experiments have confirmed this result in
different experimental situations (see for example the recent results
for an homogeneous shear in \cite{she00}). Only recently, a more extensive
study of anisotropies has been carried out, considering also
higher order statistical objects \cite{she00,she02,kur00,ara98}.  The
situation became immediately less clear:
the prediction (\ref{eq:k41F}) is not the end of the story (see below the
section on anomalous scaling for anisotropic fluctuations).
 We show later that in the jargon of the SO(3) decomposition the
anisotropic part of a spherically averaged and solenoidal second-rank
tensor is made  from $j=2$ contribution only,
 for this reason the
dimensional analysis is often viewed as predicting a $4/3$ exponent
for the $j=2$ sector of the second-order structure function. This result
was later derived by several authors in terms of Clebsch variables,
but again by  dimensional reasoning
\cite{yak94,gro94,fal95}. Another, more systematic attempt
to derive  the scaling behavior of the second order structure function
in a weakly anisotropic turbulent flow was presented in \cite{gro00}
within a variable scale mean field theory. In that paper the authors
reached the conclusion that all anisotropic contribution to the second order
structure function must scale $\sim r^{4/3}$. To reach this result the
authors
had to simplify  the tensorial structure of the equations for the second
order correlation functions; we argue below that this uncontrolled
simplification 
biased the estimate of the anisotropic exponents.
\section{The Modern Approach to Anisotropy}
\label{chap:modern}
In the past 10 years, the subject of anomalous scaling in turbulence
has gained a great deal of attention, as it became more and more
accepted that in the infinite Re limit, the scaling
exponents of the structure functions in the inertial range do not
conform with the classical prediction of the Kolmogorov theory.
 The numerical values of these exponents, as
well as the physical mechanism which is responsible for the anomalous
scaling, have been the target of an extensive experimental, numerical,
and theoretical research.

On the theoretical side, important progress was made by studying
Kraichnan's model of passive scalar advection \cite{kra68}. This model
describes the advection of a passive scalar field by a synthetic,
solenoidal velocity field with a Gaussian, white-in-time statistics. The
linearity of the equations for the passive scalar field and
the white-in-time statistics of the velocity field make it possible to
write down a closed set of equations for the same-time correlation
functions of the passive scalar \cite{kra68}. In
 \cite{gaw95,che95}, it was shown that
 the solution of these equations can lead to anomalous scaling. The
 key point is that the homogeneous solutions of these equations are
 scale invariant with nontrivial anomalous scaling exponents, which
 are different from the dimensional scaling exponents that
 characterize the inhomogeneous, ``forced'' solutions. Being usually
 smaller than the dimensional scaling exponent, the anomalous
 exponents dominate the small scales statistics of the passive scalar
 field. The homogeneous solutions are commonly referred to as ``zero
 modes'', and have been calculated to first order perturbatively in
 refs.~\cite{gaw95,che95} for the fourth order structure
 function and for all even structure functions in
 ref.~\cite{ber96}. Exact computer assisted calculations of the
 exponents of the third order structure functions were presented in
 \cite{gat}.   Besides suggesting an elegant mechanism for anomalous
 scaling, Kraichnan's model also provided an example in which the
 scaling of the anisotropic parts of structure functions is different
 from the isotropic scaling. In the paper \cite{fai96} it was
 shown how such a thing can happen, by expanding the second-order
 structure function of the passive field in terms of spherical
 harmonics $Y_{j,m}(\hat r)$. It was found that this expansion
 leads to a set of decoupled $j$-dependent equations for the expansion
 prefactors. These equations can be easily solved by a power law whose
 exponent is an increasing function of $j$. These exponents are
 universal in the sense that they are independent of the forcing and
boundary
 conditions.

The authors in \cite{fai96} also noticed that the fact that the anisotropic
exponents are
higher than the isotropic exponent neatly explains the isotropization of
the statistics as smaller and smaller scales are probed.
Based on this example, it was suggested in
\cite{lvo96c} that a similar mechanism may exist in a Navier-Stokes
turbulence. The authors  expanded the second-order structure function
in terms of spherical harmonics
\begin{equation}
  S^{(2)}(\B r) = \sum_{j,m} S^{(2)}_{j m}(r) Y_{j m}(\hat{\Br}) \ ,
\label{SO3dec}
\end{equation}
and argued that in the case of weak anisotropy, one can linearize the
equations for the anisotropic corrections of the second-order structure
function around the isotropic solution. In such a case, the kernel of the
linearized equation is invariant under rotations (isotropic), and as a
result the equations for the different $(j,m)$ components decouple, and
are $m$-independent - much as in the case of the second-order structure
function in Kraichnan's model. In a scale-invariant situation, this leads
to anisotropic, $j$-dependent exponents
$$
  S^{(2)}_{j m}(r) \sim (\eb r)^{2/3}\left(\frac{r}{L}\right)^{\delta_j}
    \sim r^{\zeta^{(2)}_j} \ . $$
If one accepts that homogeneous turbulence enjoys universal
statistics in the inertial range, then the kernel of the above
linearized equation is universal, and consequently so are the
anisotropic scaling exponents $\zeta^{(2)}_j$. All of these
statements could not have been proved rigorously (and still
haven't been proved rigorously), yet they offered a new approach
to understanding anisotropy in turbulence, an approach that is
explored in the rest of this review.
%
\subsection{Mathematical Framework}
Experiments in fluid turbulence are usually limited to the measurement
of the velocity field at one single spatial point as a function of
time. This situation has begun to improve recently, but still much of
the analysis of the statistical properties of Navier-Stokes turbulence
 is influenced by this tradition: the Taylor hypothesis \cite{tay38} is
used to justify the identification of velocity differences at different
times with differences of longitudinal velocity components across a
spatial length scale $r$. Most of the available statistical
information is therefore about properties of longitudinal two-point
differences of the Eulerian velocity field and their moments. Recent
research \cite{lvo96} has pointed out the advantages of
considering not only the longitudinal structure functions, but
tensorial multi-point correlations of velocity field differences
$$
{\B w}({\B x},{\B x}^{\prime },t)\equiv {\B u}({\B x}^{\prime },t)-{\B
u}({\B x},t),  $$
given by
\begin{equation}
F^{\alpha_1\dots\alpha_n}({\B x}_{1},{\B x}_{1}^{\prime},t_{1};;\dots ;{\B
x}_{n},{\B x}_{n}^{\prime },t_{n})
=\langle w^\alpha_1({\B x}_{1},{\B x}_{1}^{\prime },t_{1})
\dots w^\alpha_n({\B x}_{n},{\B x}_{n}^{\prime},t_{n})\rangle
\label{defFtilde}
\end{equation}
where all the coordinates are distinct. When the coordinates fuse to
yield time-independent structure functions depending on one separation only,
these are the so-called tensorial structure functions,  denoted as
\begin{equation}
S^{\alpha_1 \dots \alpha_n}({\B r})\equiv \langle \lbrack u^{\alpha_1
}({\Bx}+{\Br})-u^{\alpha_1 }({\B x})]
\cdots[u^{\alpha_n }({\B x}+{\B r})-u^{\alpha_n }({\B x})]
\rangle \ .  \label{Sn}
\end{equation}
Needless to say, the tensorial information is partially lost in the
usual measurements conducted at a single point. One of the main
stresses of the present review is that keeping as much of tensorial
information as possible can help significantly in disentangling
different scaling contributions to the statistical objects. Especially
when anisotropy implies different tensorial components with possible
different scaling exponents characterizing them, careful control of
the various contributions is called for.

To understand why irreducible representations of the symmetry group
may have an important role in determining the form of correlation
functions, we need to discuss the equations of motion which they
satisfy. We shall show that the isotropy of the Navier-Stokes equation
and the incompressibility condition implies the isotropy of the
hierarchical equations which the correlation functions satisfy. We
will use this symmetry to show that every component of the general
solution with a definite behavior under rotations (i.e., components of
a definite {\em irreducible representation} of the $SO(3)$ group) has
to satisfy these equations by itself - independently of components
with different behavior under rotations. This ``foliation'' of the
hierarchical equations may possibly lead to different scaling
exponents for each component of the general solution which belong to a
different $SO(3)$ irreducible representation.
\subsection{Tensorial Correlation Functions and $SO(3)$ Irreducible
Representations: General Theory}
\label{sec:basis}
The physical objects that we deal with are the moments of the velocity
field at different space-time locations. In this section we follow Ref.
\cite{ara99b}
which suggests a way of decomposing these objects into components with a
definite
behavior under rotations \cite{ara99b}. It will follow that components with
different behavior under rotation are subject to different dynamical
equations, and therefore, possibly, scale differently. Essentially, we
are about to describe the tensorial generalization of the well-known
procedure of decomposing a scalar function $\Psi ({\bf r})$ into
components of different irreducible representations using the
spherical harmonics:
\begin{equation}
\Psi ({\bf r})=\sum_{j,m}a_{jm}(r)Y_{jm}(\hat{{\bf r}})\ .
\label{scalar-case}
\end{equation}
\subsubsection{Formal Definition}
Consider the correlation function ${\bf F}^{(n)}$ of
Eq.~(\ref{defFtilde}).
This $n$-rank tensor is a
function of $2n$ spatial variables and $n$ temporal
variables. It transforms as a {\em tensor field}: if ${\bf F}^{(n)}$ is
measured in two frames $I$ and $\overline{I}$ which are connected by
the spatial transformation (say, a rotation)
$
\overline{x}^{\alpha }=\Lambda ^{\alpha \beta }x^{\beta }
$
then, the measured quantities in each frame will be connected by the
relation:
\begin{eqnarray}
&&\overline{{ F}}^{\alpha_{1}\ldots \alpha _{n}}(\overline{{\bf x}}
_{1},\overline{{\bf x}}_{1}^{\prime },\overline{t}_{1};\ldots ;\overline{
{\bf x}}_{n},\overline{{\bf x}}_{n}^{\prime },\overline{t}_{n})
=\Lambda ^{\alpha _{1}\beta _{1}}\cdots \Lambda ^{\alpha_n\beta _{n}}{
F}^{\beta_{1}\ldots \beta_{n}}({\bf x}_{1}, {\bf x}_{1}^{\prime
},t_{1};\ldots ;{\bf x}_{n},{\bf
x}_{n}^{\prime },t_{n})
\nonumber \\
&&=\Lambda ^{\alpha_{1}\beta_{1}}\cdots \Lambda ^{\alpha
_{n}\beta _{n}}{ F}^{\beta _{1}\ldots \beta _{n}}(\LL\overline{
{\bf x}}_{1},\LL\overline{{\bf x}}_{1}^{\prime },\overline{t}_{1};\ldots
;\LL
\overline{{\bf x}}_{n},\LL\overline{{\bf x}}_{n}^{\prime
},\overline{t}_{n}). \label{tensor-trans}
\end{eqnarray}
We see that as we move from one frame to another,
the {\em functional form}
of the tensor field changes. We want to classify the different tensor fields
according to the change in their functional form as we make that move. We
can omit the time variables from our discussion since under rotation they
merely serve as parameters. We thus define $\B T(\{\B x_i\})\equiv
\B F (\{\B x_i\},\{t_i=0\})$.
Consider coordinate transformations which are pure rotations. For such
transformations we may simplify the discussion further by separating the
dependence
on the amplitude of ${\bf x}_{i}$ from the dependence on the directionality
of ${\bf x}_{i}$: 
$$
T^{\alpha _{1}\ldots \alpha _{n}}({\bf x}_{1},\ldots ,{\bf x}_{p}) \\
=T^{\alpha _{1}\ldots \alpha _{n}}(x_{1},\ldots ,x_{p};\hat{{\bf x}}
_{1},\ldots ,\hat{{\bf x}}_{p}),
$$
where here we  have $p\le n$, i.e we  consider also
the possibility that $n-p$ spatial locations
in (\ref{tensor-trans}) coincide.
For pure rotations we may treat the amplitudes $x_{1},\ldots ,x_{p}$ as
parameters: the transformations properties of $T^{\alpha _{1}\ldots \alpha
_{n}}$ under rotation are determined only by the dependence of $T^{\alpha
_{1}\ldots \alpha _{n}}$ on the unit vectors $\hat{{\bf x}}_{1},\ldots
,\hat{{\bf x}} _{p}$. Accordingly it seems worthwhile to discuss tensor
fields
which are functions of the unit vectors {\em only}. Notice that in the
scalar case we follow the same procedure: by restricting our attention to
scalar functions that depend only on the unit vector $\hat{{\bf x}}$, we
construct the spherical harmonics. These functions are {\em defined} such
that each one of them has unique  transformation properties under rotations.
We then represent the most general scalar function as a linear combination
of the spherical harmonics with $x$-dependent coefficients, see
 Eq.~(\ref{scalar-case}).

The classification of the tensor fields $T^{\alpha _{1}\ldots \alpha _{n}}(
\hat{{\bf x}}_{1},\ldots ,\hat{{\bf x}}_{p})$ according to their functional
change under rotations follows immediately from group representation
theory \cite{Corn,Stern}.  But in order to demonstrate that, we must
first make some formal definitions. We define ${\mathcal S}_{p}^{n}$
to be the space of all smooth tensor fields of rank $n$ which depend
on $p$ unit vectors. This is obviously a linear space of infinite
dimension. With each rotation $\Lambda \in SO(3)$, we may now
associate a linear transformation ${\mathcal O}_{\Lambda }$ on that
space via the relation (\ref{tensor-trans}):
$$
\left[ {\mathcal O}_{\Lambda }T\right] ^{\alpha _{1}\ldots
\alpha_{n}}(\hat{{\bf x}}_{1},\ldots ,\hat{{\bf x}}_{p})
\equiv \Lambda ^{\alpha _{1}\beta _{1}}\cdots \Lambda ^{\alpha
_{n}\beta _{n}}T^{\beta _{1}\ldots \beta _{n}}(\LL\hat{{\bf x}}
_{1},\ldots ,\LL\hat{{\bf x}}_{p}).
$$
Using this definition, it is easy to see that the set of linear operators
$ {\mathcal O}_{\Lambda }$ furnishes a representation of the rotation group
$SO(3)$
since they satisfy the relations:
$$
{\mathcal O}_{\Lambda _{1}}{\mathcal O}_{\Lambda _{2}} ={\mathcal
O}_{\Lambda
_{1}\Lambda _{2}}, \qquad
{\mathcal O}_{\Lambda }^{-1} ={\mathcal O}_{\Lambda ^{-1}}.
$$
General group theoretical considerations imply that it is possible to
decompose ${\mathcal S}_{p}^{n}$ into subspaces which are invariant to the
action of all the group operators ${\mathcal O}_{\Lambda }$. Moreover, we
can
choose these subspaces to be {\em irreducible} in the sense that they will
not contain any invariant subspace themselves (excluding themselves and the
trivial subspace of the zero tensor field). For the $SO(3)$ group each of
these subspaces is conventionally characterized by an integer $
j=0,1,2,\ldots $ and is of dimension $2j+1$ \cite{Corn,Stern}. It should be
noted that unlike
the scalar case, in the general space ${\mathcal S}_{p}^{n}$, there might be
more than one subspace for each given value of $j$. We therefore use the
index $q$ to distinguish subspaces with the same $j$. For each irreducible
subspace $(q,j)$ we can now choose a basis with $2j+1$ components labeled by
the index $m$: 
$$
B_{q,jm}^{\alpha _{1},\ldots ,\alpha _{n}}(\hat{{\bf x}}_{1},\ldots ,\hat{
{\bf x}}_{p})\;;\;m=-j,\ldots ,+j.
$$
In each subspace $(q,j)$, the group operators ${\mathcal O}_{\Lambda }$
furnish
a $2j+1$ dimensional irreducible representation of $SO(3)$. Using the basis
$
B_{q,jm}^{\alpha _{1},\ldots ,\alpha _{n}}(\hat{{\bf x}}_{1},\ldots ,\hat{
{\bf x}}_{p})$, we can represent each operator ${\mathcal O}_{\Lambda }$ as
a $
(2j+1)\times (2j+1)$ matrix $D_{m^{\prime }m}^{(j)}(\Lambda )$ via the
relation: 
\begin{eqnarray}
\left[ {\mathcal O}_{\Lambda }B\right] _{q,jm}^{\alpha _{1},\ldots ,\alpha
_{n}}(\hat{{\bf x}}_{1},\ldots ,\hat{{\bf x}}_{p})
&=&\Lambda ^{\alpha _{1}\beta _{1}}\cdots \Lambda ^{\alpha
_{n}\beta _{n}}B _{q,jm}^{\beta _{1}\ldots \beta _{n}}(\LL\hat{{\bf x}}
_{1},\ldots ,\LL\hat{{\bf x}}_{p}) \nonumber\\&\equiv &\sum_{m^{\prime
}=-j}^{+j}D_{m^{\prime }m}^{(j)}(\Lambda
)B_{q,jm^{\prime }}^{\alpha _{1},\ldots ,\alpha _{n}}(\hat{{\bf x}}
_{1},\ldots ,\hat{{\bf x}}_{p}).
\nonumber \end{eqnarray}
It is conventional to choose the basis ${\bf B}_{q,jm}$ such that the
matrices $D_{m^{\prime }m}^{(j)}(\phi )$, that correspond to rotations of $
\phi $ radians around the 3 axis, will be diagonal, and given by:
$
D_{m^{\prime }m}^{(j)}(\phi )=\delta _{mm^{\prime }}e^{im\phi }$.
The ${\mathcal S}_{p}^{n}$ space possesses a natural inner-product:
$$
\left\langle {\bf T},{\bf U}\right\rangle \!\! \equiv \!\!\int \!\!d\hat{
{\bf x}}_{1}\dots d\hat{{\bf x}}_{p}
\cdot T^{\alpha _{1}\dots \alpha _{n}}(\hat{{\bf x}}_{1}\dots \hat{{\bf
x}}
_{p})g_{\alpha _{1}\beta _{1}}\ldots g_{\alpha _{n}\beta _{n}}U^{^{\beta
_{1}\ldots \beta _{n}}}(\hat{{\bf x}}_{1}\dots \hat{{\bf x}}_{p})^{\ast }\
$$
where $g_{\alpha \beta }$ is the 3-dimensional Euclidean metric
tensor.  By definition, the rotation matrices $\Lambda ^{\alpha
\beta}$ preserve this metric, and therefore it is easy to see that for
each $\Lambda
\in SO(3)$ we get: 
$$
\left\langle {\mathcal O}_{\Lambda }{\bf T},{\mathcal O}_{\Lambda }{\bf U}
\right\rangle =\left\langle {\bf T},{\bf U}\right\rangle
$$
so that, ${\mathcal O}_{\Lambda }$ are unitary operators. If we now choose
the
basis ${\bf B}_{q,jm}$ to be orthonormal with respect to the inner-product
defined above, then the matrices $D_{m^{\prime }m}^{(j)}(\Lambda )$ will be
unitary.

Finally, consider {\em isotropic tensor fields}. An isotropic tensor field
is a tensor field which preserves its functional form under any arbitrary
rotation of the coordinate system. In other words, it is a tensor field
which is invariant to the action of all operators ${\mathcal O}_{\Lambda }$.
The
one dimensional subspace spanned by this tensor-field is therefore invariant
under all operators ${\mathcal O} _{\Lambda }$ and therefore it must be a
$j=0$
subspace.

Once the basis ${\bf B}_{q,jm}$ has been selected, we may expand any
arbitrary tensor field $F^{\alpha _{1}\ldots \alpha _{n}}({\bf x}_{1},\ldots
,{\bf x}_{p})$ in this basis. As mentioned above, for each fixed set of
amplitudes $x_{1},\ldots ,x_{p}$, we can regard the tensor field $F^{\alpha
_{1}\ldots \alpha _{n}}({\bf x}_{1},\ldots ,{\bf x}_{p})$ as a tensor field
which depends only on the unit vectors $\hat{{\bf x}}_{1},\ldots ,\hat{{\bf
x }}_{p}$, and hence belongs to ${\mathcal S}_{p}^{n}$. We can therefore
expand
it in terms of the basis tensor fields ${\bf B}_{q,jm}$ with coefficients
that depend on $x_{1},\ldots ,x_{p}$:
\begin{equation}
F^{\alpha _{1}\ldots \alpha _{n}}({\bf x}_{1},\ldots ,{\bf x}_{p})
=\sum_{q,j,m}
 {\mathcal F}_{q,jm}(x_{1},\ldots ,x_{p})B_{q,jm}^{\alpha _{1},\ldots
,\alpha _{n}}(\hat{{\bf x}}_{1},\ldots ,\hat{{\bf x}}_{p}) \ .  \label{Texp}
\end{equation}
The goal of the following sections is to demonstrate the utility of such
expansions for the study of the scaling properties of the correlation
functions. For the important case of tensorial structure functions
(\ref{Sn}) the basis function depend on one spatial vector only $\B r$,
and we can expand 
\be
\B S^{(n)}(\B r) = \sum_{q,jm} S^{(n)}_{q,jm}(r) \B B^{(n)}_{q,jm}(\hat{ \B
r}).
\label{eq:so3sn}
\ee
\subsubsection{Construction of the Basis Tensors}
\label{sec:construction}
\paragraph{The Clebsch-Gordan machinery.}
A straightforward (although somewhat impractical) way to construct the basis
tensors ${\bf B}_{q,jm}$ is to use the well-known Clebsch-Gordan machinery.
In this approach we consider the ${\mathcal S}_{p}^{n}$ space as a {\em
direct
product space} of $n$ 3-dimensional Euclidean vector spaces with $p$
infinite dimensional spaces of single-variable continuous functions on the
unit sphere. In other words, we notice that ${\mathcal S}_{p}^{n}$ is given
by:
$$
{\mathcal S}_{p}^{n}=\left[ {\mathcal S}_{0}^{1}\right] ^{n}\otimes
\left[ {\mathcal S}
_{1}^{0}\right] ^{p},$$
and therefore every tensor $T^{\alpha_{1}\dots \alpha_{n}}(\hat{{\bf x}}
_{1}\dots \hat{{\bf x}}_{p})$ can be represented as a linear combination of
tensors of the form:$$
v_{1}^{\alpha_{1}} \ldots v_{n}^{\alpha_{n}} \varphi_{1} \left(
\hat{{\bf x}}_{1}\right) \cdot \ldots \cdot \varphi_{p}\left( \hat{{\bf x}}
_{p}\right) . 
$$
where 
$v_{i}^{\alpha _{i}}$ are constant Euclidean vectors and $\varphi _{i}(\hat{
{\bf x}}_{i})$ are continuous functions over the unit sphere. The
3-dimensional Euclidean vector space, ${\mathcal S}_{0}^{1}$, contains
exactly
one irreducible representation of $SO(3)$ - the $j=1$ representation - while
${\mathcal S}_{1}^{0}$, the space of continuous functions
$\otimes$ over the unit sphere,
contains every irreducible representation exactly once. The statement that $
{\mathcal S}_{p}^{n}$ is a direct  product space may now be written in a
group
representation notation as:
$$
{\mathcal S}_{p}^{n}= \stackrel{n}
{\overbrace{1 \otimes 1 \otimes
\ldots \otimes 1}} \otimes \stackrel{p}{\overbrace{\left(
0\oplus 1\oplus 2\dots \right) \otimes \ldots \left( 0\oplus 1\oplus 2\dots
\right) }}
$$
We can now choose an appropriate basis for each space in the product:
\begin{itemize}
\item  For the 3-dimensional Euclidean space we may choose:
$$
{\bf e}_{1}={\frac{1}{\sqrt{2}}}\left(
1,
i, 
0
\right) ,\quad {\bf e}_{0}={\frac{1}{\sqrt{2}}}\left(
0,
0, 
1
\right) ,\quad {\bf e}_{-1}={\frac{1}{\sqrt{2}}}\left(
1,
-i, 
0
\right) 
$$
\item  For the space of continuous functions over the unit sphere we may
choose the well-known spherical harmonic functions.
\end{itemize}

Once these bases have been chosen, we can construct a direct-product basis
for ${\mathcal S}_{p}^{n}$:
$$
E_{i_{1}\ldots i_{n}\left( l_{1}\mu _{1}\right) \ldots \left( l_{p}\mu
_{p}\right) }^{\alpha _{1}\ldots \alpha _{n}}\left( {\bf
\hat{x}}_{1},\ldots ,{\bf \hat{x}}_{p}\right) \\
\equiv e_{i_{1}}^{\alpha _{1}}\cdot \dots \cdot e_{i_{n}}^{\alpha_{n}}\cdot
Y_{l_{1},\mu _{1}}(\hat{{\bf x}}_{1})\cdot \dots \cdot
Y_{l_{p},\mu _{p}}(\hat{{\bf x}}_{p}).
$$ The unitary matrix that connects the ${\bf E}_{i_{1}\ldots i_{n}\left(
l_{1}\mu _{1}\right) \ldots \left( l_{p}\mu _{p}\right) }$ basis to
the ${\bf B}_{q,jm}$ basis can be calculated using the appropriate
Clebsch-Gordan coefficients. The calculation is straightforward but
very long and tedious.  However, the above analysis enables us to
count and classify the different irreducible representations of a
given $j$. By using the standard rules of angular-momentum addition:
$$
s\otimes l=\left| s-l\right| \oplus \ldots \oplus \left( s+l\right)
$$
we can count the number of irreducible representations of a given
$j$. For example, consider the space ${\mathcal S}_{1}^{2}$ of second-rank
tensors with one variable over the unit sphere. Using the
angular-momentum addition rules we see:
\begin{eqnarray}
{\mathcal S}_{1}^{2} &=&1\otimes 1\otimes \left( 0\oplus 1\oplus 2\oplus
3\oplus
\ldots \right)  \label{dps21} \\
&=&\left( 0\oplus 1\oplus 2\right) \otimes \left( 0\oplus 1\oplus
2\oplus 3\oplus \ldots \right) \nonumber \\ &=&\left( 3\times
0\right)\oplus \left( 7\times 1\right) \oplus \left( 9\times 2\right)
\oplus
\left( 9\times 3\right) \oplus \ldots \nonumber
\end{eqnarray}
We see that there are exactly three $j=0$ representations, seven $j=1$
representations and 9 representations for each $j>1$. It can be
further argued that the symmetry properties of the basis tensors with
respect to their indices come from the $1\otimes 1=0\oplus 1\oplus 2$
part of the direct product (\ref{dps21}). Therefore, out of the 9
irreducible representation of $j>1$, 5 will be symmetric and
traceless, 3 will be anti-symmetric and 1 will be trace-full and
diagonal. Similarly, the parity of the resulting tensors (with respect
to the single variable) can be calculated.

Once we know how many irreducible representations of each $j$ are
found in ${\mathcal S}_{p}^{n}$, we can construct them ``by hand'', in
some other, more practical method which will be demonstrated in the
following subsection.
\paragraph{Alternative derivation of the basis functions.}
\label{alternative}
The method found most useful in application  is based on the
simple idea that contractions with $r^{\alpha },\delta ^{\alpha \beta
},\epsilon ^{\alpha \beta \gamma }$ and differentiation with respect
to $r^{\alpha }$ are all {\em isotropic} operations \cite{ara99b}. Isotropic
in the
sense that the resulting expression will have the {\em same}
transformation properties under rotation as the expression we started
with. The proof of the last statement follows directly from the
transformation properties of $r^{\alpha },\delta ^{\alpha \beta
},\epsilon ^{\alpha \beta \gamma }$.

The construction of all ${\bf B}_{q,jm}$ that belongs to ${\mathcal
S}_{1}^{n}$
now becomes a rather trivial task. We begin by defining a scalar tensor
field with a definite $j,m$. An obvious choice will be the well-known
spherical harmonics $Y_{jm}(\hat{{\bf r}})$, but a better one will be:
$$
\Phi _{jm}({\bf r})\equiv r^{j}Y_{jm}(\hat{{\bf r}}).
$$
The reason that we prefer $\Phi _{jm}({\bf r})$ to $Y_{jm}(\hat{{\bf r}})$,
is that $\Phi _{jm}({\bf r})$ is polynomial in ${\bf r}$ (while
$Y_{jm}(\hat{{\bf r}})$ is polynomial in $\hat{{\bf r}}$) and therefore it
is easier to
differentiate it with respect to ${\bf r}$. Once we have defined $\Phi
_{jm}(
{\bf r})$, we can construct the ${\bf B}_{q,jm}$ by ``adding indices'' to $
\Phi _{jm}({\bf r})$ using the isotropic operations mentioned above. For
example, we may now construct:

\begin{itemize}
\item  $r^{-j}\delta ^{\alpha \beta }\Phi _{jm}({\bf r})$,

\item  $r^{-j+2}\delta ^{\alpha \beta }\partial ^{\tau }\partial ^{\gamma
}\Phi _{jm}({\bf r})$,

\item  $r^{-j-1}r^{\alpha }\Phi _{jm}({\bf r})$, \ etc...
\end{itemize}

Notice that we should always multiply the resulting expression with an
appropriate power of $r$, in order to make it $r$-independent, and thus a
function of $\hat{{\bf r}}$ only.

The crucial role of the Clebsch-Gordan analysis is to tell us how many
representations from each type we should come up with. First, it tells
us the highest power of $\hat{{\bf r}}$ in each representation, and
then it can also give us the symmetry properties of ${\bf B}_{q,jm}$
with respect to their indices. For example, consider the irreducible
representations of ${\mathcal S} _{1}^{2}$ - second rank tensors which
depend on one unit vector $\hat{{\bf r} }$. The Clebsch-Gordan
analysis shows us that this space contains the following irreducible
representations spelled out in (\ref{dps21}).
That is, for each $j>1$ we are going to have 9 irreducible representations.
The indices symmetry of the tensor comes from the ${\mathcal
S}_{0}^{1}\otimes
{\mathcal S}_{0}^{1}=1\otimes 1=0\oplus 1\oplus 2$ part of the direct
product.
This is a direct product of two Euclidean spaces, so its a second rank
constant
tensor. We can mark the representation number in this space with the letter
$s$, and the representation number of the ${\mathcal S}_{1}^{0}=0\oplus
1\oplus
2\oplus 3\oplus \ldots $ space with the letter $l$. This way each
representation in ${\mathcal S}_{1}^{2}$ of a given $j$ will have two
additional
numbers $(s,l)$, which actually serve as the index $q$ that distinguishes
irreducible representations of the same $j$. The $s$ index will determine
the indices symmetry of the tensor, while the $l$ index will determine the
highest power of $\hat{{\bf r}}$ in the tensor. If we now recall that in the
space of constant second-rank tensors, ${\mathcal S}_{0}^{1}\otimes
{\mathcal S}
_{0}^{1}=0\oplus 1\oplus 2$, the $s=0,2$ representations are symmetric while
the $s=1$ representation is anti-symmetric, we can easily construct the $
{ B}^{\alpha \beta}_{q,jm}$:
\begin{equation}
\begin{array}{ll}
\left( s,l\right) =\left( 0,j\right) & B^{\alpha \beta}_{1,jm}(\hat{{\bf
r}})\equiv
r^{-j}\delta ^{\alpha \beta }\Phi _{jm}({\bf r}), \\
\left( s,l\right) =\left( 1,j-1\right) & B^{\alpha \beta}_{2,jm}(\hat{{\bf
r}})\equiv
r^{-j+1}\epsilon ^{\alpha \beta \mu }\partial _{\mu }\Phi _{jm}({\bf r}), \\
\left( s,l\right) =\left( 1,j\right) & B^{\alpha \beta}_{3,jm}(\hat{{\bf
r}})\equiv r^{-j}
\left[ r^{\alpha }\partial ^{\beta }-r^{\beta }\partial ^{\alpha }\right]
\Phi _{jm}({\bf r}), \\
\left( s,l\right) =\left( 1,j+1\right) & B^{\alpha \beta}_{4,jm}(\hat{{\bf
r}})\equiv
r^{-j-1}\epsilon ^{\alpha \beta \mu }r_{\mu }\Phi _{jm}({\bf r}), \\
\left( s,l\right) =\left( 2,j-2\right) & B^{\alpha \beta}_{5,jm}(\hat{{\bf
r}})\equiv
r^{-j+2}\partial ^{\alpha }\partial ^{\beta }\Phi _{jm}({\bf r}), \\
\left( s,l\right) =\left( 2,j-1\right) & B^{\alpha \beta}_{6,jm}(\hat{{\bf
r}})\equiv
r^{-j+1}\left[ \epsilon ^{\alpha \mu \nu }r_{\mu }\partial _{\nu }\partial
^{\beta }+\epsilon ^{\beta \mu \nu }r_{\mu }\partial _{\nu }\partial
^{\alpha }\right] \Phi _{jm}({\bf r}), \\
\left( s,l\right) =\left( 2,j\right) & B^{\alpha \beta}_{7,jm}(\hat{{\bf
r}})\equiv r^{-j}
\left[ r^{\alpha }\partial ^{\beta }+r^{\beta }\partial ^{\alpha }\right]
\Phi _{jm}({\bf r}), \\
\left( s,l\right) =\left( 2,j+1\right) & B^{\alpha \beta}_{8,jm}(\hat{{\bf
r}})\equiv
r^{-j-1}\left[ r^{\beta }\epsilon ^{\alpha \mu \nu }r_{\mu }\partial _{\nu
}+r^{\alpha }\epsilon ^{\beta \mu \nu }r_{\mu }\partial _{\nu }\right] \Phi
_{jm}({\bf r}), \\ 
\left( s,l\right) =\left( 2,j+2\right) & B^{\alpha \beta}_{9,jm}(\hat{{\bf
r}})\equiv
r^{-j-2}r^{\alpha }r^{\beta }\Phi _{jm}({\bf r}).
\end{array}
\label{eq:second-rank-tensors}
\end{equation}
It should be stressed that these $B^{\alpha \beta}_{q,jm}$ are not exactly
the same
one we would have gotten from the Clebsch-Gordan machinery. For example,
they are not orthogonal among themselves for the same values of $j,m$
(although, they are orthogonal for different values of $j$ or $m$).
Nevertheless, they are linearly independent and thus span a given $(j,m)$
sector in the ${\mathcal S}_{1}^{2}$ space.
The  set of  eigenfunction, $B^{\alpha \beta}_{q,jm}$,  can be further
classified in terms of its properties under permutation of tensorial
indices,
$\alpha \beta$ and in terms of their parity properties,
i.e. how do they transform under the $\B r \rightarrow -\B r$ operation.
Taking in to account both  properties we may distinguish:
\begin{description}
\item{Subset I} Symmetric in $\alpha,\beta$ and with  parity $(-1)^j$:
$$B^{\alpha\beta}_{9,jm}(\hat{\B r}),
B^{\alpha\beta}_{7,jm}(\hat{\B r}),
B^{\alpha\beta}_{1,jm}(\hat{\B r}),
B^{\alpha\beta}_{5,jm}(\hat{ \B r})$$
\item{Subset II} Symmetric  to $\alpha,\beta$ exchange and with  parity
$(-1)^{j+1}$:
$$ B_{8,jm}^{\alpha \beta }(\hat{ \B r}),B_{6,jm}^{\alpha\beta}(\hat{ \B
r})$$
\item{Subset III} Antisymmetric  to $\alpha,\beta$ exchange and with  parity
$(-1)^{j+1}$:
$$ B_{4,jm}^{\alpha \beta }(\hat{ \B r}),B_{2,jm}^{\alpha\beta}(\hat{ \B
r})$$
\item{Subset IV} Antisymmetric  to $\alpha,\beta$ exchange and with  parity
$(-1)^{j}$:
$$B_{3,jm}^{\alpha \beta }(\hat{ \B r})$$
\end{description}

The reader may find  more details on the algebra of SO(3) decomposition of
second order tensor
 in the Appendix ~\ref{sec:general-form}.
\subsection{The Isotropy of the Hierarchy of Equations and
its Consequences}
\label{sec:hierachy}
In this section we follow Ref. \cite{ara99b} in deriving equations of motion
for the
statistical averages of the velocity and pressure fields
differences. We start from the Navier-Stokes equations, and show that
its isotropy implies the isotropy of the equations for the statistical
objects. Finally, we demonstrate the foliation of these equations to
different sectors of $j,m$.
Consider the  Navier-Stokes equations (\ref{NS}) in a bounded domain
$\Omega$.
In principle these equations can be the basis for deriving infinite linear
hierarchy
of equations for the Eulerian correlation functions and to study its
properties
under rotation.
Unfortunately the relevant dynamical time scales are revealed only when
the effect of sweeping is removed. Therefore we choose to work here
with the transformation
proposed in \cite{bel87} in which the flow is observed from the point of
view of one specific fluid particle which is located at ${\bf x}_{0}$ at
time $t_{0}$. Let ${\bf \rho }({\bf x}_{0},t_{0}|t)$ be the particle's
translation at time $t$:
$$
{\bf \rho }({\bf x}_{0},t_{0}|t)=\int\limits_{t_{0}}^{t}ds{\bf u}[{\bf x}
_{0}+{\bf \rho }({\bf x}_{0},t_{0}|s),s]\ . $$
We then redefine the velocity and pressure fields to be those seen from an
inertial frame whose origin sits at the current particle's position:
$$
{\bf v}({\bf x}_{0},t_{0}|{\bf x},t) \equiv {\bf u}[{\bf x}+{\bf \rho }(
{\bf x}_{0},t_{0}|t),t]\ , $$
$$\pi ({\bf x}_{0},t_{0}|{\bf x},t) \equiv p[{\bf x}+{\bf \rho }({\bf x}
_{0},t_{0}|t),t]\ .$$
Next, we define the differences of these fields:
$$
{\mathcal W}^{\alpha }({\bf x}_{0},t_{0}|{\bf x},{\bf x}^{\prime },t)
\equiv v^{\alpha }({\bf x}_{0},t_{0}|{\bf x},t)-v^{\alpha }({\bf
x}_{0},t_{0}|{\bf
x}^{\prime },t)\ , $$
$$
\Pi ({\bf r}_{0},t_{0}|{\bf x},{\bf x}^{\prime },t) \equiv \pi (
{\bf x}_{0},t_{0}|{\bf x},t)-\pi ({\bf x}_{0},t_{0}|{\bf x}^{\prime },t)\ .
$$
A straightforward calculation shows that the dynamical equations for $
{\bf {\mathcal W}}$ are:
\begin{eqnarray}
&&\partial _{t}{\mathcal W}^{\alpha }({\bf x},{\bf x}^{\prime },t) =-\left(
\partial _{\alpha }+\partial ^{\prime}_{ \alpha }\right) \Pi ({\bf x}
_{0},t_{0}|{\bf x},{\bf x}^{\prime },t)  \label{eq:WDyn}
+\nu \left( \partial ^{2}+\partial ^{\prime 2}\right) {\mathcal W}^{\alpha
}(
{\bf x}_{0},t_{0}|{\bf x,x}^{\prime },t)  \nonumber \\
&&-\partial _{\mu }{\mathcal W}^{\mu }({\bf x}_{0},t_{0}|{\bf
x,x}_{0},t){\mathcal
W}
^{\alpha }({\bf x}_{0},t_{0}|{\bf x,x}^{\prime },t)
-\partial _{\mu }^{\prime }{\mathcal W}^{\mu }({\bf x}_{0},t_{0}|{\bf x}
^{\prime }{\bf ,x}_{0},t){\mathcal W}^{\alpha }({\bf x}_{0},t_{0}|{\bf x,x}
^{\prime },t)\ ,  \nonumber \\
&&\partial _{\alpha }{\mathcal W}^{\alpha }({\bf x}_{0},t_{0}|{\bf x},{\bf
x}
^{\prime },t) =\partial _{\alpha }^{\prime }{\mathcal W}^{\alpha }({\bf x}
_{0},t_{0}|{\bf x},{\bf x}^{\prime },t)=0\ .
\end{eqnarray}
By inspection, $t_{0}$ merely serves as a parameter, and therefore we will
not denote it explicitly in the following discussion. Also, in order to make
the equations easier to understand, let us introduce some shorthand notation
for the variables $({\bf x}_{k},{\bf x}_{k}^{\prime },t_{k})$:
$$
{\bf X}_{k} \equiv ({\bf x}_{k},{\bf x}_{k}^{\prime },t_{k})\ , \quad
X_{k} \equiv (x_{k},x_{k}^{\prime },t_{k})\ , \quad
{\hat{{\bf X}}}_{k} \equiv (\hat{{\bf x}}_{k},\hat{{\bf x}}_{k}^{\prime
})\ .
$$
Using (\ref{eq:WDyn}), we can now derive the dynamical equations for the
statistical moments of ${\bf {\mathcal W}}$,$\Pi $: Let $\left\langle \cdot
\right\rangle $ denote a suitable ensemble averaging. We define two types of
statistical moments:
\begin{eqnarray*}
&&{\mathcal F}^{\alpha _{1}\ldots \alpha _{n}}({\bf x}_{0}|{\bf
X}_{1},\ldots ,{\bf
X}_{n}) 
\equiv \left\langle {\mathcal W}^{\alpha _{1}}({\bf x}_{0}|{\bf
X}_{1})\ldots
{\mathcal W}^{\alpha _{n}}({\bf x}_{0}|{\bf X}_{n})\right\rangle \ , \\
&&{\mathcal H}^{\alpha _{2}\ldots \alpha _{n}}({\bf x}_{0}|{\bf
X}_{1},\ldots ,{\bf
X}_{n}) 
\equiv \left\langle \Pi (x_{0}|{\bf X}_{1}){\mathcal W}^{\alpha _{2}}({\bf
x}
_{0}|{\bf X}_{2})\ldots {\mathcal W}^{\alpha _{n}}({\bf x}_{0}|{\bf X}
_{n})\right\rangle \ .
\end{eqnarray*}
Equation (\ref{eq:WDyn}) implies:
\begin{eqnarray}
&&\partial _{t_{1}}{\mathcal F}^{\alpha _{1}\ldots \alpha _{n}}({\bf
x}_{0}|{\bf X}
_{1},\ldots ,{\bf X}_{n})
=-\left( \partial _{(x_{1})}^{\alpha _{1}}+\partial _{(x_{1}^{\prime
})}^{\alpha _{1}}\right) {\mathcal H}^{\alpha _{2}\ldots \alpha _{n}}({\bf
x}_{0}|
{\bf X}_{1},\ldots ,{\bf X}_{n})  \nonumber \\
&&-\partial _{\mu }^{(x_{1})}{\mathcal F}^{\mu \alpha _{1}\ldots \alpha
_{n}}\left( {\bf x}_{0}|{\tilde{{\bf X}}},{\bf X}_{1},\ldots ,{\bf
X}_{n}\right)
-\partial _{\mu }^{(x_{1}^{\prime })}{\mathcal F}^{\mu \alpha _{1}\ldots
\alpha
_{n}}\left( {\bf x}_{0}|{\tilde{{\bf X}}}^{\prime }{\bf ,X}_{1},\ldots ,{\bf
X}_{n}\right)   \nonumber \\
&&+\nu \left( \partial _{(x_{1})}^{2}+\partial _{(x_{1}^{\prime
})}^{2}\right) {\mathcal F}^{\alpha _{1}\ldots \alpha _{n}}({\bf x}_{0}|{\bf
X}
_{1},\ldots ,{\bf X}_{n})\ ,  \label{eq:FDyn}
\end{eqnarray}
with ${\tilde{{\bf X}}} \equiv {(}{\bf x}_{0},{\bf x}^{\prime
},t)\;;\;{\tilde{
{\bf X}}}^{\prime }{\equiv (}{\bf x},{\bf x}_{0},t)\ ,$ with the
further constraint:
$$
\partial _{\alpha _{1}}^{(x_{1})}{\mathcal F}^{\alpha _{1}\ldots \alpha
_{n}}({\bf x
}_{0}|{\bf X}_{1},\ldots ,{\bf X}_{n}) =0\ ,  \qquad
\partial _{\alpha _{1}}^{(x_{1}^{\prime })}{\mathcal F}^{\alpha _{1}\ldots
\alpha
_{n}}({\bf x}_{0}|{\bf X}_{1},\ldots ,{\bf X}_{n}) =0\ . $$
Equations (\ref{eq:FDyn}),  are linear and homogeneous.
Therefore their solutions form a linear space. The most general solution to
these equations is given by a linear combination of a suitable basis of the
solutions space. To construct a specific solution, we must use the boundary
conditions in order to set the linear weights of the basis solutions. We
shall now show that the isotropy of these equations implies that our
basis of solutions can be constructed such that every solution will have a
definite behavior under rotations (that is, definite $j$ and $m$).
 But before we do that, note that in many aspects the situation
described here is similar to the well-known problem of Laplace equation in a
closed domain $\Omega $:
$$
\nabla ^{2}\Psi =0\, \qquad
\Psi |_{\partial \Omega } =\sigma \ .
$$
The Laplace equation is linear, homogeneous and isotropic. Therefore its
solutions form a linear space. One possible basis for this space is:
$$
\Psi_{jm}({\bf r})\equiv r^{j}Y_{jm}(\hat{{\bf r}})\ ,
$$
in which the solutions have a definite behavior under rotations (belong
to an irreducible representation of $SO(3)$ ). The general solution of the
problem is given as a linear combination of the $\Psi_{jm}({\bf r})$, cf.
Eq. (\ref{scalar-case}).
For a specific problem, we use the value of $\Psi ({\bf r})$ on the boundary
(i.e., we use $\sigma ({\bf r})$) in order to set the values of $a_{l,m}$.

To see that the same thing happens with the hierarchy equations
 (\ref{eq:FDyn}), we consider an arbitrary solution
$\{{\B {\mathcal F}}^{(n)},{\B {\mathcal H}}^{(n)}|\;n=2,3,\ldots \}$ of
these
equations. We may write the tensor fields
${\B{\mathcal F}}^{(n)},{\B {\mathcal H}}^{(n)}$ in terms of a basis ${\bf
B}_{q,jm}$:
\begin{equation}
 \label{def:jmExpansion}
{\mathcal F}^{\alpha _{1}\ldots \alpha _{n}}({\bf x}_{0}|{\bf X}_{1},\ldots
,{\bf 
X}_{n})  
\equiv \sum_{q,j,m} F_{q,jm}^{(n)}(x_{0},X_{1},\ldots ,X_{n})
\B B_{q,jm}^{(n)}(\hat{{\bf x}}_{0},{
\hat{{\bf X}}}_{1},\ldots {\hat{{\bf X}}}_{n})\ ,
\end{equation}
\begin{equation}
{\mathcal H}^{\alpha _{2}\ldots \alpha _{n}}({\bf x}_{0}|{\bf X}_{1},\ldots
,{\bf 
X}_{n}) 
\equiv \sum_{q,j,m} H_{q,jm}^{(n)}(x_{0},X_{1},\ldots ,X_{n})
\B  B_{q,jm}^{(n-1)}(\hat{{\bf r}}_{0},{
\hat{{\bf X}}}_{1},\ldots {\hat{{\bf X}}}_{n})\ ;
\end{equation}
where here and below we use the shorthand notation, $\B B_{q,jm}^{(n)}$
to denote the SO(3) basis of  $n$th order tensors,
 $ B_{q,jm}^{\alpha_1,\dots,\alpha_n}$.
Now all we have to show is that the pieces of
 ${\B {\mathcal F}}^{(n)},{\B {\mathcal
H}}^{(n)}$ with
definite $j,m$ solve the hierarchy equations {\em by themselves} -
independently of pieces with different $j,m$. The proof of the last
statement is straightforward though somewhat tedious. We therefore only
sketch it in general lines. The isotropy of the hierarchy equations implies
that pieces of ${\B {\mathcal F}}^{(n)},{\B {\mathcal H}}^{(n)}$ with
definite $j,m$,
maintain their
transformations properties under rotation {\em after} the linear and
isotropic operations of the equation have been performed. For example, if $
{\mathcal F}^{\alpha _{1}\ldots \alpha _{n}}({\bf x}_{0}|{\bf X}_{1},\ldots
,{\bf X}
_{n})$ belongs to the irreducible representation $(j,m)$, then so will the
tensor fields: 
$
\partial _{\alpha _{i}}^{(x_{k})}{\mathcal F}^{\alpha _{1}\ldots \alpha
_{n}},
\partial _{\alpha _{i}}^{(x_{k})}\partial_{\alpha _{i}}^{(x_{k})}{\mathcal
F}^{\alpha _{1}\ldots \alpha
_{n}},
$
although, they may belong to different ${\mathcal S}_{p}^{n}$ spaces
(i.e., have one less or one more indices). Therefore, if we choose the
bases $\left\{ {\bf B}_{q,jm}^{(n)}\right\} $ to be orthonormal, plug
the expansion (\ref {def:jmExpansion}) into the hierarchy equations
equations (\ref{eq:FDyn}), and take the inner product with ${\bf
B}_{q,jm}^{(n)}$, we
will obtain new equations for the scalar functions
$F_{q,jm}^{(n)},H_{q,jm}^{(n)}$:
\begin{eqnarray}
&&\partial _{t_{1}}F_{q,jm}^{(n)}(r_{0},X_{1},\ldots ,X_{n})
=\nonumber\\&&-\sum_{q^{\prime }}\left\langle \left( \partial
_{(x_{1})}^{\alpha
_{1}}+\partial _{(x_{1}^{\prime })}^{\alpha _{1}}\right) H_{q^{\prime
}jm}^{(n)}(r_{0},X_{1},\ldots ,X_{n}){\bf B}_{q^{\prime }jm}^{(n-1)},{\bf B}
_{q,jm}^{(n)}\right\rangle   \label{eq:ScalarDyn} \\
&&-\sum_{q^{\prime }}\left\langle \partial _{\mu }^{(x_{1})}F_{q^{\prime
}jm}^{(n+1)}(r_{0},\tilde{X},X_{1},\ldots ,X_{n}){\bf B}_{q^{\prime
}jm}^{(n+1)},{\bf B}_{q,jm}^{(n)}\right\rangle\nonumber\\&&
-\sum_{q^{\prime }}\left\langle \partial _{\mu }^{(x_{1}^{\prime
})}F_{q^{\prime }jm}^{(n+1)}(r_{0},\tilde{X}^{\prime },X_{1},\ldots ,X_{n})
{\bf B}_{q^{\prime }jm}^{(n+1)},{\bf B}_{q,jm}^{(n)}\right\rangle
\nonumber
\\
&&+\nu \sum_{q^{\prime }}\left\langle \left( \partial
_{(x_{1})}^{2}+\partial _{(x_{1}^{\prime })}^{2}\right) F_{q^{\prime
}jm}^{(n)}(r_{0},X_{1},\ldots ,X_{n}){\bf B}_{q^{\prime }jm}^{(n)},{\bf B}
_{q,jm}^{(n)}\right\rangle ,  \nonumber
\end{eqnarray}
\begin{eqnarray}
\sum_{q^{\prime }}\left\langle \partial _{\alpha _{1}}^{(x_{1})}F_{q^{\prime
}jm}^{(n)}(x_{0},X_{1},\ldots ,X_{n}){\bf B}_{q^{\prime }jm}^{(n)},{\bf B}
_{q,jm}^{(n-1)}\right\rangle  &=&0\ ,  \label{eq:ScalarIncomp} \\
\sum_{q^{\prime }}\left\langle \partial _{\alpha _{1}}^{(x_{1}^{\prime
})}F_{q^{\prime }jm}^{(n)}(r_{0},X_{1},\ldots ,X_{n}){\bf B}_{q^{\prime
}jm}^{(n)},{\bf B}_{q,jm}^{(n-1)}\right\rangle  &=&0\ .  \nonumber
\end{eqnarray}
Note that in the above equations, $\left\langle \cdot \right\rangle $ denote
the inner-product in the ${\mathcal S}_{p}^{n}$ spaces. Also, the sums over
$
q^{\prime },j^{\prime },m^{\prime }$ from (\ref{def:jmExpansion}) was
reduced to a sum over $q^{\prime }$ only - due to the isotropy. We thus see
explicitly from (\ref{eq:ScalarDyn},\ref{eq:ScalarIncomp}) the decoupling of
the equations for different $j,m$.\\
At this point we remind the reader that in the case of the most used
statistical objects
in the analysis of experimental and numerical data are the
longitudinal $n$th order structure functions:
$$
S^{(n)}(\B r) = \langle (\delta u_{\ell}(\B r))^n \rangle .
$$
For these objects the basis functions are simply the spherical harmonics
and the SO(3) decomposition reads:
\be
S^{(n)}(\B r) = \sum_{j,m} S^{(n)}_{jm}(r) Y_{jm}(\hat{\B r}).
\label{eq:fundamental}
\ee
A question of major interest for all that follows are the numerical values
of the scaling 
exponents which are defined by the power laws
$$
S^{(n)}_{jm}(r) \propto r^{\zeta_j^{(n)}} $$
\subsection{Dimensional  Analysis of Anisotropic Fluctuations}
\label{sec:dim_ana}
The actual calculation of  scaling exponents in the anisotropic
sectors is difficult, and  will be considered in the rest of this
review. It is worthwhile to have a phenomenological guess based on
dimensional analysis. Unfortunately, once anisotropies are considered,
dimensional considerations becomes tricky.  Historically, the first
successful attempt to introduce dimensional considerations in
anisotropic turbulence was the  approach discussed in
Sect.~\ref{sec:lumley}. There, the key role was played by the large-scale
mean-shear. However this work is limited to the analysis of second order
correlations, without discriminating among
 different $j$ sectors. In  light of the SO(3)
decomposition it should be considered as a prediction for
$\zeta_{2}^{(2)}$.
 Another  dimensional argument was presented in
\cite{bif02} extending the consideration of Sec.~\ref{sec:lumley}.
 This argument takes  into
account also the particular angular structure entering in the
interaction between small-scale fluctuations and large-scale shear.
 By decomposing the velocity field, $\B{u}$,
 in a small-scale component, $\B{v}$, and a large-scale
anisotropic component, $\B{U}$, one finds the following equation for
the time evolution of $\B{v}$:
$$
\partial_t v_{\alpha} + v_{\beta}\partial_{\beta} v_{\alpha} +
U_{\beta}\partial_{\beta} v_{\alpha} + v_{\beta}\partial_{\beta}
U_{\alpha} = -\partial_{\alpha} p + \nu\Delta v_{\alpha}.
$$
The major effect of the large-scale field is given by the
instantaneous shear $\partial_{\beta} U_{\alpha}$ which acts as an
anisotropic forcing term on small scales.
 We can write the balance
equation for two
point quantities $\la v_{\delta}(\Bx')v_{\alpha}(\Bx)\ra$ in the
stationary regime:
$$
\la v_{\delta}(\Bx') v_{\beta}(\Bx)
 \partial_{\beta} v_{\alpha}(\Bx) \ra \sim \la
\partial_{\mu} U_{\alpha}
v_{\delta}(\Bx') v_{\mu}(\Bx) \ra.
$$
The shear
term is a large-scale ``slow'' quantity and therefore, as far as
scaling properties are concerned, can be safely estimated as: $\la
\partial_{\mu} U_{\alpha}
 v_{\delta}(\Bx') v_{\mu}(\Bx) \ra \sim
D_{\alpha \mu} \la v_{\delta}(\Bx')v_{\mu}(\Bx)
\ra$. The tensor $D_{\alpha \beta}$ is associated to the joint
probability to have a given shear and a given small scale velocity
configuration. The $D_{\alpha \beta}$ being a constant tensor can possess at
most  angular momentum
up to $j=2$.
Similarly for three point  quantities
we may write: $\la v v v \partial v \ra \sim \la \partial U  vvv  \ra$,
which can
be easily generalized to velocity correlation of any order.  One may
therefore argue, by using simple composition
of angular momenta, ($j= 2 \oplus j-2$), the following dimensional
matching for structure functions in different anisotropic sectors:
\be {S}^{(n)}_{jm}(r) \;\sim\; r\, |D| \, \cdot \,
S_{j-2,m}^{(n-1)}(r),
\label{eq:gen}
\ee
where $ S_{j,m}^{(n)}(r)$ is a shorthand notation for the projection
on the $j$-{\it th} sector of the $n$-{\it th} order correlation
function introduced in the previous section, ${F}^{(n)}_{q jm}(r)$.
In (\ref{eq:gen}) with $|D|$ we denote the typical intensity of the
shear term $D_{\alpha \beta}$ in the $j=2$ sector.  From
Eq. (\ref{eq:gen}) one can obtain higher $j$ exponents of the higher
order structure functions from the lower order structure functions of
lower anisotropic sectors which appear on the RHS. For example, the
dimensional prediction for the third order scaling exponent in the
$j=2$ sector, $\zeta^{(3)}_2$ can be obtained by the
matching: $S_{2,m}^{(3)}(r) \sim r\, |D| S_{0,m}^{(2)}(r) \sim
r^{\zeta_{2}^{(3)}}$. By using the same argument and the known scaling of
the third order correlation for $j=0,2$, the scaling exponents of the
fourth order correlation for $j=2,4$ can be estimated. The following
expression is readily obtained for any order\,:
\begin{equation}
\zeta_{j}^{(n)}=\frac{(n+j)}{3} \quad \mbox{(dimensional prediction).}
\label{lumleygen}
\end{equation}
 This  formula coincides with the prediction (\ref{eq:Lumley})
for $n=2$ and $j=2$. We will see below that
both measurements and closure calculations exhibit
exponents which are anomalous, i.e.
different from these dimensional predictions.
\section{Exactly Solvable Models}
\label{chap:analytical}
In this section we review the work done on anomalous scaling in
the anisotropic sectors of exactly solvable models. The first of
these models is the Kraichnan model of passive scalar advection
in which the velocity field is rapidly varying in time
\cite{kra68,kra94,fal01}. This
model offers detailed understanding of the anomalous scaling in all the
anisotropic sectors both from the Lagrangian and the Eulerian
points of view. The scaling exponents can be calculated however
only in perturbation theory. The second model that we consider is
of passive advection of a magnetic field \cite{ver96}. In this case one can
compute non-perturbatively  the scaling exponents of the second order
correlation function in all the sectors of the symmetry group.
These two models show that the spectrum of scaling exponents is
discrete and strictly increasing as a function of $j$. If this
is true for systems with pressure, like the Navier-Stokes
equation, it may lead to problems of convergence of the integrals
induced by the existence of the pressure terms. To this aim we
review below a third exactly solvable model in which pressure is
used explicitly to keep an advected vector solenoidal.It was shown
that also here the spectrum is discrete and strictly increasing,
and it was explained how the putative divergences are avoided. The
mechanism discovered here is most likely also operating in the
Navier-Stokes case. The last model reviewed in this section
is the the second order structure function in the Navier-Stokes problem,
linearized for small anisotropies. Also in this case we find
a discrete spectrum of strictly increasing scaling exponents
as a function of $j$.
Most of the results here presented can also be reproduced within
the Renormalization Group approach. We do not enter here in this
subject which would deserve a whole review by itself.
 The interested reader can find the most important results
for passive scalar advection in \cite{anto01,anto01a}, for magnetic fields
 in \cite{anto00,anto99} and  for passive vectors in \cite{anto03}.\\
\subsection{Anomalous Scaling in the Anisotropic Sectors
of the Kraichnan Model  of Passive Scalar Advection}
\label{passcalar}
 Kraichnan's model of passive scalar advection in
which the driving (Gaussian) velocity field has fast temporal
decorrelation turned out to be a very important case model for
understanding the anomalous scaling behavior in turbulent
advection, including the anisotropic sectors of turbulent scalar fields.
We review here the derivation that shows that the solutions of the Kraichnan
equation for the
$n$ order correlation functions foliate into sectors that are
classified by the irreducible representations of the SO($d$)
symmetry group. A discrete spectrum of universal
anomalous exponents is found, with a different exponent characterizing the
scaling behavior in every sector. Generically the correlation
functions and structure functions appear as sums over all these
contributions, with non-universal amplitudes which are determined
by the anisotropic boundary conditions. The isotropic sector is
always characterized by the smallest exponent, and therefore for
sufficiently small scales local isotropy is always restored.
We start by
 presenting  the Eulerian calculation which results in actual values
of the scaling exponents (in perturbation theory) \cite{ara00}. The Eulerian
calculation of the anomalous exponents is done in two
complementary ways. In the first they are obtained from the
analysis of the correlation functions of {\em gradient fields}.
The theory of these functions involves the control of logarithmic
divergences which translate into anomalous scaling with the ratio
of the inner {\em and} the outer scales appearing in the final
result. In the second way one computes the exponents from the zero
modes of the Kraichnan equation for the correlation functions of
the scalar field itself. In this case the renormalization scale
is the outer scale. The two approaches lead to the same scaling
exponents for the same statistical objects, illuminating the
relative role of the outer and inner scales as renormalization
scales. To clarify this further,  Ref.~\cite{ara00} presented
an exact derivation of
fusion rules which govern the small scale asymptotic of the
correlation
functions in all the sectors of the symmetry group and in all
dimensions. The purpose of the Eulerian calculation is twofold. On the one
hand we are interested in the effects of anisotropy on the
universal aspects of scaling behavior in turbulent systems. On the other
hand
we are interested in clarifying the relationship between
ultraviolet and infrared anomalies in turbulent systems.
The two issues discussed in this subsection have an importance
that transcends the particular example that we treat here in
detail. Having below a theory of anomalous scaling in all the various
sectors of
the symmetry group allows us to explain clearly the relationship
between the two renormalization scales and the anomalous
exponents that are implied by their existence.
 Since we expect
that Kolmogorov type theories, which assume that no
renormalization scale appears in the theory, are generally
invalidated by the appearance of both the outer and the inner
scales as renormalization scales, the clarification of the
relation between the two is important also for other cases of
turbulent statistics.

The central quantitative result of the Eulerian calculation is the
expression for the scaling exponent $\xi^{(n)}_{j}$ which is
associated with the scaling behavior of the $n$-order correlation
function (or structure function) of the scalar field in the
$j$'th sector of the symmetry group. In other words, this is
the scaling exponent of the projection of the correlation
function on the $j$'th irreducible representation of the
SO($d$) symmetry group, with $n$ and $j$ taking on even values
only, $n=0,2, \dots$ and $j=0,2,\dots$:
\begin{equation}
\xi^{(n)}_{j}= n-\epsilon\Big[\frac{n(n+d)}{2(d+2)}
-\frac{(d+1)j(j+d-2)}{2(d+2)(d-1)}\Big] +O(\epsilon^2) \ .
\label{eq:perturbative}
\end{equation}
The result is valid for any even $j\le n$, and to $O(\epsilon)$
where $\epsilon$ is the scaling exponent of the eddy diffusivity
in the Kraichnan model (and see below for details). In the
isotropic sector ($j=0$) we recover the  result of
\cite{ber96}. It is noteworthy that for higher values of $j$
the discrete spectrum is a strictly increasing function of
$j$. This is important, since it shows that for diminishing
scales the higher order scaling exponents become irrelevant, and
for sufficiently small scales only the isotropic contribution
survives. As the scaling exponent appear in power laws of the
type $(r/\Lambda)^\xi$, with $\Lambda$ being some typical outer scale
and $r \ll \Lambda$, the larger is the exponent, the faster is the
decay of the contribution as the scale $r$ diminishes. This is
precisely how the isotropization of the small scales takes place,
and the higher order exponents describe the rate of
isotropization. Nevertheless for intermediate scales or for finite
values of the Reynolds and Peclet numbers the lower lying scaling
exponents will appear in measured quantities, and understanding
their role and disentangling the various contributions cannot be
avoided.
\subsubsection{Kraichnan's Model of Turbulent Advection and the
Statistical Objects} \label{s:Kra} The model of passive scalar
advection with rapidly decorrelating velocity field was introduced in
\cite{kra68}. In recent years
\cite{gaw95,che95,ber96,fai96,kra94,lvo94c,ben97} it was shown to be a
fruitful case model for understanding multi-scaling in the statistical
description of turbulent fields. The basic dynamical equation in this
model is for a scalar field $T({\B r},t)$ advected by a random
velocity field ${\B u}({\B x},t)$:
\begin{equation}
\label{advect}
        \big[\partial_t  - \kappa_0 \nabla^2  +
        {\B u}({\B x},t) \cdot {\bf \nabla}\big]
         T({\B x},t) = f({\B x},t)\ .
\end{equation}
In this equation $f({\B x},t)$ is the forcing.  In Kraichnan's
model the advecting field ${\B u}({\B x},t)$ as well as the
forcing field $f({\B x},t)$ are taken to be Gaussian, time and
space homogeneous, and delta-correlated in time:
$$
        \overline{  f({\B x},t)  f({\B x}',t') } =
           \Phi({\B x}-{\B x}')  \delta(t - t')\,,  \quad
 \langle u^\alpha({\B x},t) u^\beta({\B x}',t') \rangle
= \C W^{\alpha\beta}({\B x}-{\B x}')  \delta(t - t').
$$
Here the symbols ~$\overline{\cdots}$~ and ~$\langle \cdots
\rangle$~ stand for independent ensemble averages with respect to
the statistics of $f$ and ${\B u}$ which are given {\em a priori}. We
will study this model in the limit of large Peclet (Pe) number,
Pe$\equiv U_{\Lambda} \Lambda/\kappa_0$, where $U_\Lambda$ is the
typical size of the velocity fluctuations on the outer scale $\Lambda$
of the velocity field. We stress that the forcing is {\em not} assumed
isotropic, and actually the main goal of this section is to study the
statistic of the scalar field under anisotropic forcing.

The correlation function of the advecting velocity needs further
discussion.  It is customary to introduce $\C W^{\alpha \beta}(\B
r)$ via its $\B k$-representation: \be
\label{W1} \C
W^{\alpha\beta}(\B r)=\frac{\e\, D}{\Omega_d} \int \limits
_{\Lambda ^{-1}}^{\lambda^{-1}}     \frac{  d^d p} {p^{d+\e}} \,
P^{\alpha\beta}(\B p)\,
\exp (-i \B p\cdot \B r)\,, \quad
 P^{\alpha\beta}(\B p) = \Big[\delta_{\alpha\beta}
-\frac {p^\alpha p^\beta}{p^2} \Big]  \ee
 where
$P^{\alpha\beta}(\B p)$ is the transversal projector, $\Omega_d=(d-1)
\Omega(d)/d $ and $\Omega(d)$ is the volume of the sphere in $d$
dimensions. Equation (\ref{W1}) introduces the four important parameters
that determine the
statistics of the driving velocity field: $\Lambda$ and $\lambda$ are
the outer and inner scales of the driving velocity field
respectively. The scaling exponent $\e$ characterizes the correlation
functions of the velocity field, lying in the interval $[0,2]$. The
factor $D$ is related to the correlation function  as
follows:
\be\label{W3} \C W^{\alpha\beta}(0)=D\delta_{\a\b}(\Lambda^\e
-\lambda^\epsilon)\ . \ee The most important property of the driving
velocity field from the point of view of the scaling properties of the
passive scalar is the  ``eddy diffusivity" tensor \cite{kra68}
\be
\label{eddy-diff} K^{\alpha\beta}({\B r})
\equiv   2[\C W^{\alpha\beta}(0)-\C W^{\alpha\beta}(\B r)]\ . \ee
The scaling properties of the scalar depend sensitively on the
scaling exponent $\e$ that characterizes the $r$ dependence of
$K^{\alpha\beta}({\B r})
\propto   [\Lambda^\e
 -\lambda^\epsilon]\delta_{\alpha\beta}$, for
$ r \gg \Lambda$, namely:
\be
K^{\alpha\beta}({\B r})\propto  r^\e
 \Big[\delta_{\alpha\beta} -\frac{\epsilon}{d-1+\epsilon}{r^\alpha
 r^\beta
\over r^2}\Big], \quad \lambda\ll r\ll\Lambda \ .
\label{kappa1}
\ee
\subsubsection{The Statistical Objects}
In the statistical theory we are interested in the power laws
characterizing the $r$ dependence of the various correlation and
response functions of $T({\B x},t)$ and its gradients.  We will focus
on three types of quantities:

1) ``Unfused" structure functions of $T({\B x},t)$ are defined as
\begin{eqnarray}
\label{Sunfused}F_T^{(n)}(\B x_1,\B x'_1,\dots \B x_{n},\B x'_{n})
&\equiv &\langle [T(\B x_1,t)-T(\B x'_1,t)]\\ \nonumber \times [T(\B
x_2,t)-T(\B x'_2,t)]&\dots& [T(\B x_n,t)-T(\B x'_n,t)]\rangle \ ,
\end{eqnarray}
and in particular the standard ``fused" structure functions are
$$
       S_T^{(n)}({\B r}) \equiv \langle [ T({\B x}+{\B r}, t) - T({\B
        x}, t)]^{n} \rangle \ .
$$
In writing this equation we used the fact that the stationary and
space-homogeneous statistics of the velocity and the forcing fields
lead to a stationary and space homogeneous ensemble of the scalar
$T$. If the statistics is also isotropic, then $S^{(n)}_T(\B r)$ becomes a
function of $r$ only, independent of the direction of ${\B r}$.  The
``isotropic scaling exponents'' $\xi^{(n)}_0$ of the structure functions
$$
        S_T^{(n)}(r) \propto r^{\xi^{(n)}_0} ,
$$
characterize their $r$ dependence in the limit of large Pe, when
$r$ is in the ``inertial" interval of scales. This range is
$\lambda,\eta\ll r\ll\Lambda, \,L$ where
$\eta = \Lambda \left(\frac{\kappa_0}{D}\right) ^{1/\epsilon}$
is the dissipative
scale of the scalar field,.
When the ensemble is not isotropic we define the exponents
(\ref{eq:perturbative})
by   expanding $S^{(n)}_T(\B r)$  according to:
$$
S^{(n)}_T(\B r) = \sum_{jm} S^{(n)}_{T,jm}(r) Y_{jm}(\hat{\Br}) \ ; \qquad
S^{(n)}_{T,jm}(r) \propto  r^{\xi^{(n)}_j}
$$
2) In addition to structure functions we are also interested in
the simultaneous $n\,$th order correlation functions of the
temperature field which is time independent in stationary
statistics: 
\be 
\label{Fn} \C T^{(n)}(\{\B x_l \}) \equiv \la T(\B
x_1,t)\,T(\B x_2,t)\dots T(\B x_{n},t)\ra \,,
\ee 
where we used
the shorthand notation $\{\B x_l \}$ for the whole set of
arguments of $n$th order correlation function $\C T^{(n)}$, $\B
x_1,\B x_2\dots \B x_{n}$.

3)  Finally, we are interested in correlation functions of the
gradient field $\B \nabla T$. There can be a number of these, and
we denote 
$$ \C H^{\alpha_1\dots \alpha_n} (\{\B x_l\}) \equiv
\Big\langle \prod_{i=1}^{n} \Big[\nabla^{\a_i}T(\B x_i,t)\Big]\Big\rangle
 \,, $$
The tensor $\C H^{\alpha_1 \dots \alpha_n}$
 can be contracted in various ways. For example,
binary contractions $\alpha_1=\alpha_2, \alpha_3=\alpha_4$, {\em
etc.}  with $\B x_1=\B x_2, \B x_3=\B x_4$ {\em etc.}  produces
the correlation functions of dissipation field $|\B \nabla T|^2$.
Of particular interest is the coordinate independent tensor $\B H^{(n)}$
obtained by taking
all $\B x_i=\B x$:
\begin{equation}
H^{\alpha_1\ldots\alpha_n}={\C H}^{\alpha_1\dots \alpha_n} (\{\B x_i=\B
x\}) \ . \label{defHn}
\end{equation}

When the ensemble is not isotropic we need to take into account
the angular dependence of $\B x$, and the scaling behavior
consists of multiple contributions arising from anisotropic
effects. 
\subsubsection{The Eulerian calculation}
The correlation functions of the gradient field $\B H^{(n)}$ of Eq.
(\ref{defHn}) are tensors independent of the coordinates. Nevertheless
their calculation is somewhat heavy, and we do not reproduce it here;
we refer the reader to \cite{ara00} where the calculation is
presented in full detail. The final result of the calculation is for
the projection of $\B H^{(n)}$ onto the $j$ sector of the SO(3)
symmetry group reads
\begin{equation}
\B H^{(n)}_{j}\propto \left(\frac{\Lambda}{\eta}\right)^{n-\xi^{(n)}_j} \ ,
\label{Hscale}
\end{equation}
where the proportionality constant is a tensor in the limit $\eta \ll
\Lambda$.  The exponents $\xi^{(n)}_j$ are the same as those found
below for the correlation function in which all the scales are within
the inertial range. The appearance of both renormalization lengths
and the identity of the exponents in inertial and gradient objects is
a consequence of the fusion rules that were  explored in \cite{ara00}
with
some care.
The correlation functions ${\mathcal T}^{(n)}$ satisfy the Kraichnan's
equation \cite{kra68}
\begin{eqnarray}
        & & \Big[- \kappa_0 \sum_{i=1}^{n} \nabla^2_i+ \frac{1}{2}
        \sum_{i,k=1}^{n} K^{\a\b} (\B x_i-\B
        x_k)\nabla^\a_i\nabla^\b_k \Big] {\mathcal T}^{(n)}(\{{ \B x}_{l}
        \}) \nonumber \\ &&= \frac{1}{2} \sum_{\{i\ne k\}=1}^{n}
        \Phi({\B x}_i-{\B x}_k) {\mathcal T}^{(n-2)}(\{{\B x}_l\}_{l\ne
        i,k}) \,,
\label{bob1}
\end{eqnarray}
where $\{{\B x}_{l}\}_{l\ne i,k}$ is the set off all ${\B x}_{l}$
with $l$ from 1 to $n$, except of $l= i$ and $l= j$. Substituting
$ K^{\a\b} (\B x)$ from Eqs.~(\ref{W3},
\ref{eddy-diff})  one gets:
\begin{eqnarray}
        & & \Big[- \kappa \sum_{i=1}^{n} \nabla^2_i+ \sum_{\{i\ne
        k\}=1}^{n} \C W^{\a\b} (\B x_i-\B x_k)\nabla^\a_i\nabla^\b_k
        \Big] {\mathcal T}^{(n)}(\{{ \B x}_{l} \}) \nonumber \\
        &&=\frac{1}{2} \sum_{\{i\ne k\}=1}^{n} \Phi({\B x}_i-{\B x}_k)
        {\mathcal T}^{(n-2)}(\{{\B x}_l\}_{l\ne i,k}) \,,
\label{bob0}
\end{eqnarray}
where $
 \kappa =\kappa _0+
D[\Lambda^\e-\lambda^\epsilon]$.
 Here we used that in space
homogeneous case $ \sum_{i=1}^{n} \B \nabla_i=0$ and therefore
$$\Big|\sum_{i=1}^{n} \B \nabla_i\Big|^2=
\sum_{i=1}^{n}\nabla^2_i+\sum_{\{i\ne k\}=1}^{n}
\nabla^\a_i\nabla^\b_k=0\ .
$$

In this section we consider the zero-modes of Eq.~(\ref{bob1}). In
other words we seek solutions $Z^{(n)}(\{\B x_l\})$ which in the
inertial interval solve the homogeneous equation
\begin{equation}
\sum_{i\ne k=1}^{n}
       K^{\a\b} (\B x_i-\B x_k)\nabla^\a_i\nabla^\b_k
       Z^{(n)}(\{{ \B x}_{l} \}) =0 \ . \label{zeromodes}
\end{equation}
We allow anisotropy on the large scales. Since all the operators here
are isotropic and the equation is linear, the solution space foliate
into sectors ${j,m}$ corresponding the the irreducible
representations of the SO($d$) symmetry group.  Accordingly we write
the wanted solution in the form
$$
 Z^{(n)}(\{{ \B r}_{l} \})= \sum_{j,m}Z^{(n)}_{jm}(\{{ \B r}_{l}
\})
\ , $$
where $Z^{(n)}_{jm}$ are functions composed of irreducible
representations of SO($d$) with definite $j,m$. Each of these
components is now expanded in $\epsilon$. In other words, we write, in
the notation of Ref.\cite{ber96},
$$
Z^{(n)}_{jm} =E^{(n)}_{jm}+\epsilon
G^{(n)}_{jm}+O(\epsilon^2) \ .
$$
For $\epsilon=0$ Eq.~(\ref{zeromodes}) simplifies to
\begin{equation}
\sum_{i=1}^{n} \nabla_i^2 E^{(n)}_{jm}(\{{ \B r}_{l} \}) =0
\ , \label{EqE}
\end{equation}
for any value of $j,m$. Next we expand the operator in
Eq.~(\ref{zeromodes}) in $\epsilon$ and collect the terms of
$O(\epsilon)$:
\begin{equation}
\sum_{i=1}^{n} \nabla_i^2 G^{(n)}_{jm}(\{{ \B r}_{l}
\})=V_{n}E^{(n)}_{jm}(\{{ \B r}_{l} \}) \ , \label{firsto}
\end{equation}
where $\epsilon V_{n}$ is the first order term in the expansion
of the operator in (\ref{zeromodes}):
\begin{equation}
V_{n} \equiv \sum_{i\ne
k=1}^{n}\Big[\delta^{\alpha\beta}\log(r_{ik})-{r_{ik}^\alpha
r_{ik}^\beta\over
(d-1)r_{ik}^2}\Big]\nabla_i^\alpha\nabla_k^\beta \ , \label{defVn}
\end{equation}
where $\B r_{ik}\equiv \B x_i-\B x_k$.

In solving Eq.~(\ref{EqE}) we are led by the following
considerations: we want scale invariant solutions, which are
powers of $\B r_{ik}$. We want analytic solutions, and thus we
are limited to polynomials. Finally we want solutions that
involve all the $n$ coordinates for the function
$E^{(n)}_{jm}$; solutions with fewer coordinates do not
contribute to the structure functions (\ref{Sunfused}).  To see
this note that the structure function is a linear combination of
correlation functions. This linear combination can be represented
in terms of the difference operator $\delta_l(\B x,\B x')$
defined by:
\begin{equation}
\delta_l(\B x,\B x') {\mathcal T}^{(n)}(\{\B x_k\})\equiv {\mathcal
T}^{(n)}(\{\B
x_k\})|_{\B x_l=\B x} -{\mathcal T}^{(n)}(\{\B x_k\})|_{\B x_l=\B x'} \ .
\label{diffop}
\end{equation}
Then,
\begin{equation}
S_T^{(n)}(\B x_1,\B x'_1\dots \B x_{n}\B x'_{n}) = \prod_l \delta_l(\B
x_l,\B x'_l) {\mathcal T}^{(n)}(\{\B x_k\}) \ . \label{SinF}
\end{equation}

Accordingly, if ${\mathcal T}^{(n)}(\{\B x_k\})$ does not depend on $\B
x_i$, then $\delta_i (\B r_i,\B r'_i){\mathcal T}^{(n)}(\{\B r_k\})=0$
identically. Since the difference operators commute, we can have
no contribution to the structure functions from parts of ${\mathcal
T}^{(n)}$ that depend on less than $n$ coordinates. Finally we want the
minimal polynomial because higher order ones are negligible in
the limit $r_{ik}\ll \Lambda$. Accordingly, $E^{(n)}_{jm}$
with $j\le n$ is a polynomial of order $n$.  Following the procedure
outlined in Appendix~(\ref{app:d-dim})
 we can write the most general form of
$E^{(n)}_{jm}$, up to an arbitrary factor, as
\begin{equation}
E^{(n)}_{jm}=x_1^{\alpha_1}\dots x_{n}^{\alpha_{n}}
B^{\alpha_1\dots \alpha_{n}}_{n,jm} +[\dots] \ ,
\end{equation}
where $[\dots]$ stands for all the terms that contain less than
$n$ coordinates; these do not appear in the structure functions
but maintain the translational invariance of our quantities. Note that in
this case we carry
the index $n$ in the tensor basis functions since the theory mixes basis
functions
of different orders. The
appearance of the tensor $B^{\alpha_1\dots
\alpha_n}_{n,jm}$ 
 is justified by the fact
that $E^{(n)}_{jm}$ must be symmetric to permutations of any
pair of coordinates on the one hand, and it has to belong to the
$jm$ sector on the other hand. This requires the
appearance of the fully symmetric tensor (\ref{Birr}).
In light of Eqs.~(\ref{firsto}-\ref{defVn}) we seek solution for
$G^{(n)}_{j} (\{{ \B r}_{k}\})$ of the form
\begin{equation}
G^{(n)}_{jm}(\{{ \B r}_{k}\})=\sum_{i\ne
l}H^{il}_{jm}(\{{ \B r}_{k}\}) \log(r_{il})
+H_{jm}(\{{ \B r}_{k}\}) \ , \label{ansatzG}
\end{equation}
where $H^{il}_{jm}(\{{ \B r}_{k}\})$ and
$H_{jm}(\{{ \B r}_{k}\})$ are polynomials of degree $n$.
The latter is fully symmetric in the coordinates.  The former is
symmetric in $r_i$, $r_l$ and separately in all the other
$\{r_k\}_{k\ne l,i}$.
Substituting Eq.~(\ref{ansatzG}) into Eq.~(\ref{firsto}) and
collecting terms of the same type yields three equations:
\begin{eqnarray}
&&\sum_i \nabla_i^2 H^{lk}_{jm}=\nabla_l\cdot \nabla_k
E_{jm}^{(n)}\ ,
\label{zero1}\\
&&\big[d-2+\B r_{lk}\cdot (\B \nabla_l-\B
\nabla_k)\big]H^{lk}_{jm} +\frac{r_{lk}^\alpha
r_{lk}^\beta \nabla_l^\alpha \nabla_k ^\beta}{2d-2} E_{jm}^{(n)}
=-\frac{r_{lk}^2 K_{jm}^{lk}}{2}\ ,
\label{zero2}\\ &&\sum_i \nabla_i^2 H_{jm} =\sum_{l\ne
k}K_{jm}^{lk} \ . \nonumber
\end{eqnarray}
Here $K_{jm}^{lk}$ are polynomials of degree $n-2$ which
are separately symmetric in the $l,k$ coordinates and in all the
other coordinates except $l,k$. In Ref. \cite{ber96} it was proven
that for $j=0$ these equations possess a unique solution. The
proof follows through unchanged for any $j \ne 0$, and we thus
proceed to finding the solution.

By symmetry we can specialize the discussion to $l=1$, $k=2$. In
light of Eq.~(\ref{zero2}) we see that $H^{12}_{jm}$
must have at least a quadratic contribution in $r_{12}$. This
guarantees that (\ref{ansatzG}) is nonsingular in the limit
$r_{12}\to 0$.  The only part of $H^{12}_{jm}$ that will
contribute to structure functions must contain $\B r_3\dots \B
r_{n}$ at least once. Since $H^{12}_{jm}$ has to be a
polynomial of degree $n$ in the coordinates, it must be of the
form
\begin{equation}
H^{12}_{jm}=r_{12}^{\alpha_1}r_{12}^{\alpha_2}
r_3^{\alpha_3}\dots r_{n}^{\alpha_{n}} C^{\alpha_1\alpha_2\dots
\alpha_{n}} +[\dots]_{1,2} \ , \label{H}
\end{equation}
where $[\dots]_{1,2}$ contains terms with higher powers of
$r_{12}$ and therefore do not contain some of the other
coordinates $r_3\dots r_{n}$. Obviously such terms are
unimportant for the structure functions. Since
$H^{12}_{jm}$ has to be symmetric in $\B r_1,\B r_2$ and
$\B r_3\dots \B r_{n}$ separately, and it has to belong to an
$jm$ sector, we conclude that the constant tensor $\B C$
must have the same symmetry and must belong to the same sector.
Consulting Appendix~(\ref{app:d-dim}) the most general form of $\B C$ is
\begin{equation}
C^{\alpha_1\alpha_2\dots \alpha_{n}} =
aB_{n,jm}^{\alpha_1\alpha_2\dots
\alpha_{n}}+b\delta^{\alpha_1\alpha_2}
B_{n-2,jm}^{\alpha_3\alpha_4\dots
\alpha_{n}}+c\sum_{i\ne
l>2}\delta^{\alpha_1\alpha_i}\delta^{\alpha_2\alpha_l}
B_{n-4,jm}^{\alpha_3\alpha_4\dots \alpha_{n}} \ .
\end{equation}
Substituting in Eq.~(\ref{zero2}) one find
$$
(d+2)H^{12}_{jm}+\frac{r_{12}^{\alpha_1}r_{12}^{\alpha_2}r_3^{
\alpha_3} \dots r_{n}^{\alpha_{n}}}{2d-2} B_{n,jm}^{\alpha_1\dots
\alpha_{n}} +\frac{1}{2}r_{12}^{\alpha_1}r_{12}^{\alpha_2}
\delta^{\alpha_1\alpha_2}K^{1,2}_{jm}=[\dots]_{1,2} \ .
$$
Substituting Eq.~(\ref{H}) and demanding that coefficients of the
term $r_1^{\alpha_1}\dots r_{n}^{\alpha_{n}}$ will sum up to
zero, we obtain
$$
-2(d+2)a -\frac{2}{2d-2}=0 \ , \quad -2(d+2)c=0\ ;
\Longrightarrow c=0 \ . $$
The coefficient $b$ is not determined from this equation due to
possible contributions from the unknown last term. We determine
the coefficient $b$ from Eq.~(\ref{zero1}).  After substituting
the forms we find
$$
4\delta^{\alpha_1\alpha_2} r_3^{\alpha_3}\dots r_{n}^{\alpha_{n}}
[a B_{n,jm}^{\alpha_1\dots
\alpha_{n}}+b\delta^{\alpha_1\alpha_2}
B_{n-2,jm}^{\alpha_3\alpha_4\dots\alpha_{n}}]=\delta^{\alpha_1
\alpha_2}
r^{\alpha_3}\dots r^{\alpha_{n}} B_{n,jm}^{\alpha_1\dots
\alpha_{n}}+ [\dots]_{1,2} \ .
$$
Recalling the identity (\ref{iden1}) we obtain
$
b=\frac{z_{n,j}}{4d}[1-4a] \ . $
Finally we find that $a$ is $n,j$-independent,
$
a=-\frac{1}{2(d+2)(d-1)}\ , $
whereas $b$ does depend on $n$ and $j$, and we therefore
denote it as $b_{n,j}$
$$
b_{n,j}=\frac{(d+1)}{4(d+2)(d-1)} z_{n,j} \ . $$
In the next Subsect. we compute from these results the scaling
exponents in all the sectors of the SO($d$) symmetry group.
\subsubsection{The Scaling Exponents of the
Structure Functions}
We now wish to show that the solution for the zero modes of the
correlation functions $F^{(n)}_T$ (i.e $Z^{(n)}$) result in
homogeneous structure functions $S^{(n)}_T$. In every sector
$j,m$ we compute the scaling exponents, and show that
they are independent of $m$. Accordingly the scaling
exponents are denoted $\xi^{(n)}_{j}$, and we compute them to
first order in $\epsilon$.
Using (\ref{diffop}) and (\ref{SinF}), the structure function is
given by:
\begin{eqnarray} S^{(n)}_{T,jm}(\B r_1,\Obr_1;
\ldots ; \B r_{n},\Obr_{n} )= \Delta_1^{\alpha_1} \ldots
\Delta_{n}^{\alpha_{n}}  B_{n,jm}^{\alpha_1\dots
\alpha_{n}} + \nonumber \\
\epsilon \sum_{i \neq l}
\stackrel{\mbox{no } i,l} {\overbrace{ \Delta_1^{\alpha_1} \ldots
\Delta_{n}^{\alpha_{n}} } } f^{\alpha_i\alpha_l} (\B r_i,\Obr_i,
\B r_l,\Obr_l)   [a B_{n,jm}^{\alpha_1\dots \alpha_{n}} +b\Ctensor{i}{l}]
\ , \nonumber \end{eqnarray}
where $\Delta_i^{\alpha_i} \equiv
r_i^{\alpha_i}-\Or_i^{\alpha_i}$, and the function $f$ is defined
as:
\begin{eqnarray}
f^{\alpha_i\alpha_l} (\B r_i,\Obr_i,\B r_l,\Obr_l) & \equiv &
(r_i-r_l)^{\alpha_i}(r_i-r_l)^{\alpha_l} \ln|\B r_i -\B r_l| \nonumber\\&+&
(\Or_i - \Or_l)^{\alpha_i}(\Or_i - \Or_l)^{\alpha_l} \ln|\Obr_i -
\Obr_l| \\ & - & (r_i - \Or_l)^{\alpha_i}(r_i - \Or_l)^{\alpha_l}
\ln |\B r_i - \Obr_l| \nonumber\\&-& (\Or_i - r_l)^{\alpha_i}(\Or_i -
r_l)^{\alpha_l}\ln |\Obr_i-{\B r}_l| \ .
\end{eqnarray}
The scaling exponent of $S^{(n)}_{T,jm}$ can be found
by multiplying all its coordinates by $\mu$. A direct calculation
yields:
\begin{eqnarray*}
S^{(n)}_{T,jm}(\mu \B r_1, \mu \Obr_1 ; \ldots )  =
\mu^{n}S^{(n)}_{T,jm}(\B r_1,\Obr_1 ; \ldots) -
2\epsilon \mu^{n}\ln \mu \sum_{i\neq l}\stackrel{\mbox{no } i,l} {
\overbrace{\Delta_1^{\alpha_1} \ldots \Delta_{n}^{\alpha_{n}} } }
\Delta_i^{\alpha_i}\Delta_l^{\alpha_l} \\ \times  [a\Btensor +
b_{n,j}\Ctensor{i}{l}] + O(\epsilon^2), \\ = \mu^{n}S^{(n)}_{jm}(\B
r_1,\Obr_1 ; \ldots ) - 2\epsilon \mu^{n}
\ln \mu \Delta_1^{\alpha_1} \ldots \Delta_{n}^{\alpha_{n}}
\times \sum_{i\neq l} [a\Btensor + b{n,j}\Ctensor{i}{l} +
O(\epsilon^2) \ .
\end{eqnarray*}
Using (\ref{iden2}), we find that $ \sum\limits_{i\neq l}
[a\Btensor + b_{n,j}\Ctensor{i}{l}] = [n(n-1)a+b_{n,j}] \Btensor $, and
therefore, we finally obtain:
\begin{eqnarray*}
S^{(n)}_T(\mu \B r_1, \mu \Obr_1 ;\ldots ) = \mu^{n} \left\{
1 - 2\epsilon [n(n-1)a+b_{n,j}] \ln \mu \right\} S^{(n)}_T(\B r_1,
\Obr_1 ;\ldots ) + O(\epsilon^2) \\ = \mu^{\xi^{(n)}_{j}}
S^{(n)}_T(\B r_1, \Obr_1 ; \ldots ) + O(\epsilon^2)\ ,
\end{eqnarray*}
The result of the scaling exponent is:
$$
\xi_{j}^{(n)} = 
n-2\epsilon[-\frac{n(n-1)}{2(d+2)(d-1)}+\frac{(d+1)}{4(d+2)(d-1)}
z_{n,j}] +O(\epsilon^2)
$$
from which follows (\ref{eq:perturbative}).
 This is the
final result of this calculation.\\
It is noteworthy that this result is in full agreement with
(\ref{Hscale}), even though the scaling
exponents that appear in these result refer to different
quantities. The way to understand this is the fusion rules that
are discussed next.
\subsubsection{Fusion Rules}
The fusion rules address the asymptotic properties of the fully
unfused structure functions when two or more of the coordinates
are approaching each other, whereas the rest of the coordinates
remain separated by much larger scales. A full discussion of the
fusion rules for the Navier-Stokes and the Kraichnan model can be
found in \cite{lvo96,fai96,ben98}. In this section we quote the fusion rules
that
were
derived in Ref. \cite{ara00} directly from the zero modes that were computed
to $O(\epsilon)$, in all the sectors of the symmetry group. In
other words, we are after the dependence of the structure
function $ S^{(n)}_T({\bf r}_{1},\overline{{\bf r}}_{1};\ldots
)$ on its first $p$ pairs of coordinates ${\bf
r}_{1},\overline{{\bf r}}_{1};\ldots ;{\bf r}_{p}, \overline{{\bf
r}}_{p}$ in the case where these points are very close to each
other compared to their distance from the other $n-p$ pairs of
coordinates. Explicitly, we consider the case where ${\bf r}_{1},
\overline{{\bf r}}_{1};\ldots ;{\bf r}_{p},\overline{{\bf
r}}_{p}\ll {\bf r} _{p+1},\overline{{\bf r}}_{p+1};\ldots ;{\bf
r}_{n},\overline{{\bf r}}_{n}$. (We have used here the property
of translational invariance to put the center of mass of the
first $2p$ coordinates at the origin. The full calculation is
presented in \cite{ara00}, with the final result (to $O(\epsilon)$)
$$
S^{(n)}_{T,jm}({\bf r}_{1},\overline{{\bf
r}}_{1};\ldots ;{\bf r}_{n}, \overline{{\bf r}}_{n})
=\sum_{l=l_{\rm max}}^p\sum_{m^{\prime }}\psi_{l,m^{\prime
}}S^{(p)}_{T,lm^{\prime }}({\bf r}_{1},\overline{{\bf
r}}_{1};\ldots ;{\bf r}_{p},\overline{ {\bf r}}_{p}) \ .
$$
In this expression the quantity $ \psi_{l,m^{\prime }}$
is a tensor function of all the coordinates that remain separated
by large distances, and
$$
l_{\rm max}={\rm max}\{0,p+j-n\} \ , \quad j \le n.
$$
We have shown that the LHS has a homogeneity exponent
$\xi^{(n)}_j$. The RHS is a product of functions with
homogeneity exponents $\xi^{(p)}_l$ and the functions $
\psi_{l,m^{\prime}}$. Using the linear independence of the
functions $S^{(p)}_{T,lm'}$ we conclude that $
\psi_{l,m^{\prime}}$ must have homogeneity  exponent
$\xi^{(n)}_j-\xi^{(p)}_l$. This is precisely the
prediction of the fusion rules, but in each sector separately.
One should stress the intuitive meaning of the fusion rules. The
result shows that when $p$ coordinates approach each other, the
homogeneity exponent corresponding to these coordinates becomes
simply $\xi^{(p)}_l$ as if we were considering a $p$-order
correlation function. The meaning of this result is that $p$
field amplitudes measured at $p$ close-by coordinates in the
presence of $n-p$ field amplitudes determined far away behave
scaling-wise like $p$ field amplitudes in the presence of
anisotropic boundary conditions.
\subsubsection{The Lagrangian Approach to Anomalous Scaling}
An elegant approach to the correlation functions is furnished by
Lagrangian dynamics \cite{ap01,sra74,gat98,gat98b,fmnv99}. In this
formalism one recognizes that the actual value of the scalar at
position $\B x$ at time $t$ is determined by the action of the
forcing along the Lagrangian trajectory from $t=-\infty$ to $t$:
\begin{equation}
  T(\B x_0,t_0)=\int_{-\infty}^{t_0} dt
    \left< f(\B x(t),t) \right>_{\B \eta} \ ,
\label{path}
\end{equation}
with the trajectory $\B x(t)$ obeying
\be
    \B x(t_0) = \B x_0  \ ,\qquad
    \partial_t \B x(t) = \B u(\B x(t),t) +
        \sqrt{2 \kappa}{\B \eta}(t) \ ,
\label{traject1}
\ee
and $\B \eta$ is a vector of  zero-mean independent Gaussian
white random variables, $\left< \eta^\alpha (t) \eta^\beta (t')
\right> = \delta^{\alpha\beta} \delta(t-t')$. With this in mind,
we can rewrite $S_T^{(2n)}$ of Eq. (\ref{Sunfused}) by substituting each
factor
of $T(\B x_i)$  by its representation (\ref{path}). Performing the
averages  over the random forces, we end up with
\begin{eqnarray}
    S_T^{(2n)}(\B x_1, \ldots, \B x_{2n},t_0) =
        \Big< \int_{-\infty}^{t_0} dt_1 \cdots dt_n
        \Big[ \phi(\B x_1(t_1)-\B x_2(t_1)) \cdots \\
        \times \phi(\B x_{2n-1}(t_n)-\B x_{2n}(t_n))
        + \hbox{permutations} \Big] \Big>_{\B u,\{\B \eta_i\}} \ ,
\label{FnXi}
\end{eqnarray}
To understand the averaging procedure recall that each of the
trajectories $\B x_i$ obeys an equation of the form
(\ref{traject1}), where $\B u$ as well as $\{\B
\eta_i\}_{i=1}^{2n}$ are independent stochastic variables whose
correlations are given above. Alternatively, we refer the reader
to section II of \cite{fmnv99}, where the above analysis is carried
out in detail. Here we follow the derivation of Ref. \cite{ap01}.
In considering Lagrangian trajectories of {\em groups} of
particles, we should note that every initial configuration is
characterized by a {\em center of mass}, say $\B R$, a {\em
scale} $s$ (say the radius of gyration of the cluster of
particles) and a {\em shape} $\B Z$. In ``shape" we mean here all
the degrees of freedom other than the scale and $\B R$: as many
angles as are needed to fully determine a shape, in addition to
the Euler angles that fix the shape orientation with respect to a
chosen frame of coordinates. Thus a group of $2n$ positions $\{\B
x_i\}$ will be sometime denoted below as $\{\B R,s,\B Z\}$.

One component in the evolution of an initial configuration is a
rescaling of all the distances which increase on the average like
$t^{1/\xi_2}$; this rescaling is analogous to Richardson
diffusion. The exponent $\xi_2$ which determines the scale
increase is also the characteristic exponent of the second order
structure function \cite{kra68}. This has been related to the
exponent $\epsilon$ of (\ref{kappa1}) according to $\xi_2=2-\epsilon$.
After factoring  out this overall expansion we are left with a
normalized `shape'. It is the evolution of this shape that
determines the anomalous exponents.

Consider a final shape ${\B Z}_{0}$  with an overall scale $s_0$
which is realized at $t=0$. This shape has evolved during
negative times. We fix a scale $s>s_0$ and examine the shape when
the configuration reaches the scale $s$ for the last time before
reaching the scale $s_0$. Since the trajectories are random the
shape $\B Z$ which is realized at this time is taken from a
distribution $\gamma(\B Z;\B Z_0,s\to s_0)$. As long as the
advecting velocity field is scale invariant, this distribution
can depend only on the ratio
$s/s_0$. 

Next, we use the shape-to-shape transition probability to define
an operator $\hat \gamma (s/ s_0)$ on the space of functions
$\Psi(\B Z)$ according to
$$
    [\hat {\B \gamma} (s/ s_0) \Psi](Z_0) =
        \int d\B Z \gamma(\B Z;\B Z_0,s \to s_0) \Psi(\B Z).
$$
We will be interested in the eigenfunction and eigenvalues of
this operator. This operator has two important properties. First,
for an isotropic statistics of the velocity field the operator is
isotropic. This means that this operator commutes with all
rotation operators on the space of functions $\Psi(\B Z)$. In
other words, if ${\mathcal O}_\Lambda$ is the rotation operator that
takes the function $\Psi(\B Z)$ to the new function
$\Psi(\Lambda^{-1} \B Z)$, then
$    {\mathcal O}_\Lambda \hat{\B \gamma}
=\hat{\B \gamma} {\mathcal O}_\Lambda $.
This property follows from the obvious symmetry of the Kernel
$\gamma(\B Z;\B Z_0,s\to s_0)$ to rotating $\B Z$ and $\B Z_0$
simultaneously. Accordingly the eigenfunctions of $\hat {\B
\gamma}$ can be classified according to the irreducible
representations of SO($3$) symmetry group.
Because in this section we are not computing explicitly the exponents
we do not need to present the precise form of the eigenfunctions
and we will denote them for simplicity as
 $B_{qj m}(\B Z)$.
The second important property of $\hat {\B \gamma}$ follows from
the $\delta$-correlation in time of the velocity field. Physically
this means that the future trajectories of $n$ particles are
statistically independent of their trajectories in the past.
Mathematically, it implies for the kernel that
$$
  \gamma(\B Z;\B Z_0,s\to s_0)  = \int d\B Z_1 \gamma(\B Z;\B Z_1,s\to s_1)
    \gamma(\B Z_1;\B Z_0,s_1\to s_0) \ , \quad s>s_1>s_0
$$
and in turn, for the operator, that
$$
\hat {\B \gamma} (s/ s_0) =\hat {\B \gamma} (s/ s_1)\hat {\B
\gamma} (s_1/ s_0) \ .
$$
Accordingly, by a successive application of $\hat{\B
\gamma}(s/s_0)$ to an arbitrary eigenfunction, we get that the
eigenvalues of $\hat{\B \gamma}$ have to be of the form
$\alpha_{q,j}=(s/s_0)^{\xi^{(2n)}_{j}}$:
\begin{equation}
  ({s \over s_0})^{\xi^{(2n)}_{j}} B_{qj m}(\B Z_0) =
    \int d \B Z\gamma(\B Z; \B Z_0, s \to s_0) B_{qj m}(\B Z)
\label{eig}
\end{equation} From Schur's lemmas one can prove that the eigenvalues do not
depend on
$m$. On the other hand  they can still be a function of $q$
but for simplicity of notation we do not explicitly carry the $q$
index in $\xi$.\\
To proceed we want to introduce into the averaging process in
(\ref{FnXi}) by averaging over Lagrangian trajectories of the
$2n$ particles. This will allow us to connect the shape dynamics
to the statistical objects. To this aim consider any set of
Lagrangian trajectories that started at $t=-\infty$ and end up at
time $t=0$ in a configuration characterized by a scale $s_0$ and
center of mass $\B R_0=0$. A full measure of these have evolved
through the scale $L$ or larger. Accordingly they must have
passed, during their evolution from time $t=-\infty$ through a
configuration of scale $s>s_0$ at least once. Denote now
$
    \mu_{2n}(t,R,\B Z;s\to s_0,\B Z_0)dtd\B Rd\B Z
$
as the probability that this set of $2n$ trajectories crossed the
scale $s$ for the last time before reaching $s_0,\B Z_0$, between
$t$ and $t+dt$, with a center of mass between $\B R$ and $\B
R+d\B R$ and with a shape between $\B Z$ and $\B Z+d\B Z$.

In terms of this probability we can rewrite Eq.(\ref{FnXi})
(displaying, for clarity, $\B R_0=0$ and $t=0$) as
\begin{eqnarray}
 && S_T^{(2n)}(\B R_0=0,s_0,\B Z_0,t=0) = \int d\B Z\int_{-\infty}^0 dt \int
d\B
R
    \mu_{2n}(t,R,\B Z; s \to s_0, \B Z_0) \nonumber \\
 &&\times\left< \int_{-\infty}^{0}dt_1\cdots dt_n
    \Big[  \phi(\B x_1(t_1)-\B x_2(t_1)) \cdots
        \phi(\B x_{2n-1}(t_n) - \B x_{2n}(t_n)) +\hbox{perms}\Big]
    \Big \vert (s; \B R, \B Z, t) \right> _{\B u,\B \eta_i} \ .
\label{Fntraj} 
\end{eqnarray}
The meaning of the conditional averaging is an averaging over all
the realizations of the velocity field and the random $\B \eta_i$
for which Lagrangian trajectories that ended up at time $t=0$ in
$\B R=0,s_0, \B Z_0$ passed through $\B R ,s,\B Z$ at time $t$.

Next, the time integrations in the above equation are split to the
interval $[-\infty,t]$ and $[t,0]$ giving rise to $2^n$ different
contributions:
$$
    \int_{-\infty}^t dt_1 \cdots \int_{-\infty}^t dt_n +
    \int_{t}^0 dt_1 \int_{-\infty}^{t} dt_2 \cdots \int_{-\infty}^{t} dt_n
    + \dots
$$
Consider first the contribution with $n$ integrals in the domain
$[-\infty,t]$. It follows from the delta-correlation in time of
the velocity field, that we can write
\begin{eqnarray}
   && \left< \int_{-\infty}^t dt_1\cdots dt_n
    \Big[ \phi(\B x_1(t_1)-\B x_2(t_1)) \cdots
          \phi(\B x_{2n-1}(t_n)-\B x_{2n}(t_n)) + \hbox{perms} \Big]
    \Big \vert (s; \B R, \B Z,t) \right> _{\B u,\B \eta_i} \nonumber \\
   && = \left< \int_{-\infty}^t dt_1 \cdots dt_n
    \Big[\phi(\B x_1(t_1)-\B x_2(t_1)) \cdots \phi(\B x_{2n-1}(t_n) -
           \B x_{2n}(t_n)) + \hbox{perms} \Big]
   \right> _{\B u,\B \eta_i} \nonumber \\
   && = S^{(2n)}_T(\B R, s, \B Z, t)=S_T^{(2n)}(s, \B Z) \ .
\end{eqnarray}
The last equality follows from translational invariance in
space-time. Accordingly the contribution with $n$ integrals in
the domain $[-\infty,t]$ can be written as
$$
    \int d\B Z S_T^{(2n)}(s,\B Z) \int_{-\infty}^0 dt \int d\B R~~
    \mu_{2n}(t,R,\B Z;s\to s_0,\B Z_0) \ .
$$
We identify the shape-to-shape transition probability:
\begin{equation}
    \gamma(\B Z;\B Z_0,s \to s_0)=\int_{-\infty}^0 dt\int d\B R~~
    \mu_{2n}(t,R,\B Z;s \to s_0, \B Z_0) \ .
\end{equation}
Finally, putting all this added wisdom back in Eq.(\ref{Fntraj})
we end up with
\begin{equation}
    S_T^{(2n)}(s_0, \B Z_0) = I + \int d\B Z \gamma(\B Z; \B Z_0, s \to s_0)
        S_T^{(2n)}(s,\B Z)\ . \label{crucial}
\label{split}
\end{equation}
Here $I$ represents all the contributions with one or more time
integrals in the domain $[t,0]$.  The key point now is that only
the term with $n$ integrals in the domain $[-\infty,t]$ contains
information about the evolution of $2n$ Lagrangian trajectories
that probed the forcing scale $L$. Accordingly, the term denoted
by $I$ cannot contain information about the leading anomalous
scaling exponent belonging to $F_{2n}$, but only of lower order
exponents. The anomalous scaling dependence of the LHS of
Eq.(\ref{crucial}) has to cancel against the integral containing
$F_{2n}$ without the intervention of $I$.

Representing now
\begin{eqnarray}
  S_T^{(2n)}(s_0,\B Z_0) &=& \sum_{qj m} a_{q,j m}(s_0)
    B_{qj m}(\B Z_0)\ ,\nonumber \\
  S_T^{(2n)}(s,\B Z) &=& \sum_{qj m} a_{q, j m}(s)
    B_{qj m}(\B Z) \ ,\nonumber \\
I&=& \sum_{qj m} I_{q j m} B_{qj m}(\B Z_0)
\end{eqnarray}
and substituting on both sides of Eq.(\ref{crucial}) and using
Eq.(\ref{eig}) we find, due to the linear independence of the
eigenfunctions $B_{qj m}$
$$  a_{q, j m}(s_0) = I_{q j m} +
  \left(\frac{s}{s_0}\right)^{\xi^{(2n)}_{j}} a_{q, j m}(s).
$$ To leading order the contribution of $I_{q j m}$ is neglected,
leading to the conclusion that {\em the spectrum of anomalous
exponents of the correlation functions is determined by the
eigenvalues of the shape-to-shape transition probability
operator}. Calculations show that the leading exponent in the
isotropic sector is always smaller than the leading exponents in
all other sectors. This gap between the leading exponent in the
isotropic sector to the rest of the exponents determines the rate
of decay of anisotropy upon decreasing the scale of observation.

The derivation presented above has used explicitly the properties
of the advecting field, in particular the $\delta$-correlation in
time. Accordingly it cannot be immediately generalized to more
generic situations in which there exist time correlations.
Nevertheless we find it pleasing that at least in the present
case we can trace the physical origin of the exponents anomaly,
and connect it to the underlying dynamics. In more generic cases
the mechanisms may be more complicated, but one should still keep
the lesson in mind - higher order correlation functions depend on
many coordinates, and these define a configuration in space. The
scaling properties of such functions may very well depend on how
such configurations are reached by the dynamics. Focusing on
static objects like structure functions of one variable may be
insufficient for the understanding of the physics of anomalous
scaling. Important confirmation of this picture have been found recently
also for the case of passive scalars advected by a $2d$ turbulent flow
in the inverse cascade regime \cite{cel01a} and for the case of
shell models for passive scalars advection \cite{ara01a}.
\subsubsection{Summary and Discussion}
The main lesson from this subsection is that the scaling exponents
form a discrete and strictly increasing spectrum as a function of
$j$. This is the first example where this can be shown rigorously.
The meaning of this result is that for higher $j$ the anisotropic
contributions to the statistical objects decay faster upon decreasing
scales.  The rate of isotropization is determined by the difference
between the $j$ dependent scaling exponents, and is of course a power
law. The result shows that to first order in $\epsilon$ the
$j$-dependent part is independent of the order of the correlation
function. This means that the rate of isotropization of all the
moments of the distribution function of field differences across a
given scale is the same. This is a demonstration of the fact that,
to $O(\epsilon)$ the
distributions function itself tends toward a locally isotropic
distribution function. We note in passing that to
first order in $\epsilon$ the $j$ dependent part is also the same  for
$\xi^{(2)}$, a quantity whose isotropic value is {\em not}
anomalous. For all $j>1$ also $\xi^{(2)}_j$ is anomalous, and in
agreement with the $n=2$ value of Eq.~(\ref{eq:perturbative}).
 Significantly, for
$\xi_j^{(2)}$ we have a nonperturbative result that was derived in
\cite{fai96}, namely
$$
\xi^{(2)}_j=\frac{1}{2}\Big(2-d-\epsilon+\sqrt{(2-d-\epsilon)^2
+\frac{4(d+\epsilon-1)j(d+j-2)}{d-1}}\Big) \ , \quad
j\ge 2$$
valid for all values of $\epsilon$ in the interval (0,2) and for
all $j \ge 2$. This exact result agrees after expanding to
$O(\epsilon)$ with (\ref{eq:perturbative}) for $n=2$ and $j=2$.

The second lesson from this first exactly solvable example
was  the correspondence between
the scaling exponents of the zero modes in the inertial interval
and the corresponding scaling exponents of the gradient fields.
The latter do not depend on any inertial scales, and the exponent
appears in the combination
$\left(\Lambda/\eta\right)^{\xi^{(n)}_j}$ where $\eta$ is the
appropriate ultraviolet inner cutoff. We found exact agreement with
the exponents of the zero modes in all the sectors of the
symmetry group and for all values of $n$. The deep reason behind
this agreement is the linearity of the fundamental equation of
the passive scalar (\ref{advect}). This translates to the fact
that the viscous cutoff $\eta$  is $n$ and $j$
independent, and also does not depend on the inertial separations
in the unfused correlation functions. This point has been
discussed in detail in \cite{fai96,lvo96a}. In the case of
Navier-Stokes statistics we expect this ``trivial" correspondence
to fail. Nevertheless, many attempts have been done
to describe the matching between the inertial and dissipative
scaling properties \cite{lvo96,fri91,ben96b}
for the isotropic sector. Finally we note that in the present
case we have displayed the fusion rules in all the $j$
sectors, using the $O(\epsilon)$ explicit form of the zero modes.
We expect the fusion rules to have a nonperturbative validity for
any value of $\epsilon$.\\
An interesting modification of Kraichnan models has been recently
proposed in \cite{cel04} where the scaling properties of a passive
scalar advected by a Kraichnan-like shear flow are investigated.
The anisotropy introduced by the shear breaks the foliations of the
correlation functions equations. Nevertheless, the authors have been able
to explain the existence of a scaling range in the passive spectrum
with anomalous slope (i.e. different from the result obtained
in absence of shear), for scales larger than the typical  shear-length
in the system. This anomalous slope is due to the fast advection of passive
particles in the mean shear direction.
\subsection{Passively Advected Magnetic Field}
\label{chap:magnetic} Another exactly solvable system  of some interest
is the case of passively advected magnetic field. This model was first
proposed in \cite{ver96}. It describes the advection of a
magnetic field $\BB(\Bx,t)$ by the same Kraichnan stochastic velocity
field described in Eq.~(\ref{W1}).  The equation of motion for the
magnetic field is
\begin{eqnarray}
\partial_t \BB(\Bx,t) &&+ \Big[\Bu(\Bx,t)\cdot\B{\nabla}\Big]\BB(\Bx,t)
    - \Big[\BB(\Bx,t)\cdot\B{\nabla}\Big] \Bu(\Bx,t)
  \nonumber\\&&= \kappa \nabla^2\BB(\Bx,t) + \B{f}(\Bx,t) \ , \label{eq:B}
\end{eqnarray}
which has to be supplemented by the solenoidality condition
$\B{\nabla}\cdot\BB(\Bx,t)=0$. The source (``forcing'') term $\B{f}(\Bx,t)$
is a solenoidal vector field that is responsible for injecting the magnetic
field into the system at large scales.
 The second-order moment of the source field here is a
second-order solenoidal tensor
\begin{equation}
  \Big< f^\alpha(\Bx+\Br,t')f^\beta(\Bx,t) \Big> \equiv
     \delta(t-t')A^{\alpha\beta}\left(\frac{\Br}{L}\right) \ ,
\label{eq:ff}
\end{equation}
instead of a scalar. The tensor
$A^{\alpha\beta}(\By)$ is used to mimic large-scale anisotropic boundary
conditions and is therefore taken to be anisotropic, analytic in $\By$ and
vanishing rapidly for $y \gg 1$.
 Finally, the dissipative term
$\kappa \nabla^2\BB(\Bx,t)$ dissipates the magnetic field out of the system
at small scales.

Notice that in order to keep the magnetic field solenoidal, \Eq{eq:B}
contains a ``stretching'' term
$\Big[\BB(\Bx,t)\cdot\B{\nabla}\Big]\Bu(\Bx,t)$. This term may cause a
``dynamo effect'', which is what happens when the magnetic field amplifies
itself by extracting kinetic energy from the velocity field \cite{chi95}.
  Such effect
can destabilize the system, and prevents it from reaching a stationary
state.

Just as in the Kraichnan passive scalar case, we can use the fact that
both the velocity and source fields are white-noise Gaussian
processes, and derive a closed set of equations for the simultaneous
$n$th-order correlation-functions of the magnetic field.  For example,
the equation of motion for the second order magnetic correlation
function $$ C^{\alpha\beta}(\Br,t) \equiv \Big< B^\alpha(\Bx+\Br,t)
B^\beta(\Bx,t) \Big> \, $$ can be easily derived
\cite{ver96}:
\begin{eqnarray}
\partial_t C^{\alpha\beta}&=& K^{\mu\nu}\partial_\mu \partial_\nu
    C^{\alpha\beta} - [(\partial_\nu K^{\mu\beta})\partial_\mu
    C^{\alpha\nu} + (\partial_\nu K^{\alpha\mu}) \partial_\mu
    C^{\nu\beta}] \nonumber\\ &+&(\partial_\mu \partial_\nu
    K^{\alpha\beta}) C^{\mu\nu} + 2\kappa\nabla^2 C^{\alpha\beta} +
    A^{\alpha\beta} \equiv {\hat
    T}^{\alpha\beta}_{~~\sigma\rho}C^{\sigma\rho} + A^{\alpha\beta} \
    , \label{EqC} \end{eqnarray}
where one has to add also the solenoidal condition for the magnetic field,
$\partial_\alpha C^{\alpha\beta} = 0$
and the tensor $K^{\mu \nu}$
is the two-point velocity correlation (\ref{kappa1}).
  The solution of (\ref{EqC}) was found in
\cite{ver96}.  It was shown there that for $0<\epsilon <1$ no dynamo
occurs, while for $\epsilon >1$ a dynamo is developed. Consequently
for $0<\epsilon<1$ the system may reach a stationary state where the
correlation function of the magnetic field behaves like a power law in
the inertial range. In \cite{ver96} the zero modes of the
second-order correlation-function was calculated and its anomalous
scaling in the isotropic sector was found for any $0 \le \epsilon \le
1$.  Notice that for this passive vector model, the absence of any
conservation law for the magnetic energy allows for anomalous scaling
already for the second order correlation in the isotropic sector, at
difference from what happens in the passive scalar case discussed in
section (\ref{s:Kra}). This was the first case where a fully
non-perturbative analytical solutions was presented demonstrating the
possibility to have anomalous scaling in hydrodynamic problems.

In ref.~\cite{lan99,ara00a} this analysis was generalized to all
the sectors of the SO($3$) group using the SO($3$) decomposition.  Here we
review the results presented in ref.~\cite{ara00a} where a
systematic non-perturbative study of the solutions of (\ref{EqC}) was
given in all $(j,m)$ sectors of the SO($3$) group. As usual, it is
advantageous to decompose the covariance $C^{\alpha\beta}$ in terms of
basis functions that block-diagonalize the angular part of the
operator $\hat{ \B T}$,which  is invariant to all
rotations. In addition, $\hat{ \B T}$ is invariant to the
parity transformation $\B r \to -\B r$, and to the index permutation
$(\alpha,\mu)\Leftrightarrow (\beta,\nu)$. Accordingly, $\hat{ \B T}$
can be further block-diagonalized into blocks with definite parity and
symmetry under permutations.

In light of these consideration we seek solutions
in terms of the decomposition given in (\ref{Texp}):
\begin{equation}
C^{\alpha\beta}(\Br,t) =\sum_{q,j,m} C^{(2)}_{q,jm}(r,t) \,\,
    B^{\alpha\beta}_{q,jm}(\hat{\Br}).
\label{c-expand}\end{equation}
As discussed in sec. (\ref{sec:construction}) the nine basis functions
can be grouped in four sub-groups depending on their symmetries under
parity and index permutation (\ref{eq:second-rank-tensors}).
It should be noted that not all
subsets contribute for every value of $j$.
Space homogeneity implies the obvious symmetry of the covariance:
$
C^{\alpha\beta}(\B r,t)=C^{\beta\alpha}(-\B r,t) $.
Therefore representations symmetric to $\alpha,~\beta$ exchange must
also have even parity, while antisymmetric representations must have
odd parity.  Accordingly, even $j$'s are associated with subsets I and
III, and odd $j$'s are associated with subset II.  Subset IV cannot
contribute to this theory due to the solenoidal constraint.
%
%
\subsubsection{The Matrix Representation of the Operator $\hat {\bf T}$}
Having the angular basis functions we seek the representation of the
operator $\hat{ \B T}$ in this basis. In such a representation $\hat{ \B T}$
 is a
differential operator with respect to $r$ only. In Appendix A
of \cite{ara00a} it is shown
 how $\hat{ \B T}$ mixes basis functions within a given subset, but
not between the subsets - as is expected in the last section. In finding the
matrix representation of $\hat{ \B T}$ we are aided by the incompressibility
constraint. Consider first subset I made of the four
symmetric and with $(-)^j$ parity
 basis functions: $B^{\alpha\beta}_{q,jm}(\hat{\Br})$ with $q=1,5,7,9$
in a given $j,m$ sector. To simplify the notation we
denote the $a$'s coefficients according to
$a(r) \equiv C^{(2)}_{9jm}(r)$, $b(r) \equiv C^{(2)}_{7jm}(r)$,
$c(r) \equiv C^{(2)}_{1jm}(r) $ and $d(r) \equiv C^{(2)}_{5jm}(r)$.
Primes will denote
differentiation with respect to $r$. \\
In this basis the operator $\hat{ \B T}$ takes on the form
\begin{equation}
\hat{ \B T} \left [\left(
    \begin{array}{c} a\\ b \\ c \\ d \end{array} \right)\right]
=\B  T_1 \left( \begin{array}{c} a''\\ b'' \\ c'' \\ d'' \end{array} \right)
+ \B  T_2\left( \begin{array}{c} a'\\ b' \\ c' \\ d'\end{array} \right)
+ \B  T_3 \left(\begin{array}{c} a\\ b \\ c \\ d \end{array} \right) \ .
\label{T123}
\end{equation}
On the RHS we have matrix products. In addition, the solenoidal condition
implies the following two constrains on $a,~b,~c$ and $d$ (cf. the Appendix
of \cite{ara99b}):
\begin{eqnarray}
0 &=&a' + 2\frac{a}{r} + jb' - j^2\frac{b}{r} + c' - j\frac{c}{r}
\nonumber\\
0 &=&b' + 3\frac{b}{r} + \frac{c}{r} + (j-1)d' - (j-1)(j-2)\frac{d}{r}
\nonumber \end{eqnarray}
Using these conditions one can bring $\B T_1$ and $\B T_2$ to diagonal
forms,
$$
\B T_1=2(Dr^\epsilon+\kappa) \B 1 \; ;
\qquad
\B T_2=\frac{4}{r}[(Dr^\epsilon+\kappa)+\epsilon D r^\epsilon] \B 1
$$
where $\B 1$ is the unit matrix.
$\B T_3$ can be written in the form
$$
\B T_3 = Dr^{\epsilon-2} \B Q(j,\epsilon)+\kappa r^{-2}\B Q(j,0).
$$
The explicit expression for the  four columns of $\B Q(j,\epsilon)$
can be found in \cite{ara00a}
In Appendix B of (\cite{ara00a})
 the two remaining blocks (subsets II, III after the list
(\ref{eq:second-rank-tensors}) ), in the
matrix representation of $\hat{ \B T}$ as a function of $j$ have been also
investigated. The single basis
$B_{3,jm}$ (subset IV) cannot appear in the theory since $C^{(2)}_{3jm}=0$
by the
solenoidal condition (cf. Appendix of \cite{ara99b}).
Lastly, there are no solutions belonging to the $j=1$ sector. This is due
to the fact that such solutions correspond to subset II. In this subset the
$j=1$ solenoidal condition implies the equation:
$\frac{d}{dr} C^{(2)}_{81m}+\frac{3C^{(2)}_{81m}}{r}=0 $,
or $C^{(2)}_{81m} \propto r^{-3}$ which is not an admissible solution.
%
%

%
%
\subsubsection{Calculation of the Scaling Exponents}
Before turning to the computation of the exponents one should consider
 the existence of a stationary solution for
$t\to \infty$. In  \cite{ver96}
it was showed that there is not dynamo  in the isotropic
sector as long as $\epsilon<1$. In \cite{ara00a}
it has been  demonstrated that for the same values of $\epsilon$,
the dynamo effect is absent also in the anisotropic sectors.
The reader is referred to \cite{ara00a} for details
on this subject.
In the absence of a dynamo effect, we can consider a stationary state of the
system, maintained by the forcing term $\B f(\B r,t)$. The covariance in
such
a case will obey the following equation:
$$
{\hat T}^{\alpha\beta}_{~~\sigma\rho}C^{\sigma\rho}+A^{\alpha\beta}=0 \ .
$$
Deep in the inertial range we look for scale invariant solutions of
the above equation
neglecting the dissipative terms.
The most general scale invariant solution can be expressed  as a linear
superposition of  homogeneous (zero-modes)
and non-homogeneous solutions of the above equation:
$$C^{\sigma\rho}(\B r) =  C_{hom}^{\sigma\rho}(\B r) +
C_{non-h}^{\sigma\rho}(\B r).$$
In particular, only zero-modes can carry anomalous scaling, being the
scaling properties of the non-homogeneous solutions fixed by
 the dimensional matching $ {\hat T}^{\alpha\beta}_{~~\sigma\rho}
C^{\sigma\rho} \sim A^{\alpha\beta} $. Therefore,
the existence of a leading anomalous scaling contribution to small scales
magnetic fluctuations is connected to the existence of one, some,
zero-modes with scaling exponents smaller than the dimensional estimate.

The calculation of the scale-invariant solutions becomes rather immediate
once
we know the functional form of the operator $\hat{\B T}$ in the basis of the
angular tensors $\B B_{q,jm}$. Using the expansion (\ref{c-expand}), and the
fact that $\hat{\B T}$ is block diagonalized by such an expansion, we get a
set of 2nd order coupled ODE's for each block. To demonstrate this point,
consider the four dimensional block of $\hat{\B T}$, created by the four
basis
tensors $\B B_{q,jm}$ of subset I. According to the notation of the last
section, we denote the coefficients of these angular tensors in
(\ref{c-expand}), by the four functions $a(r),b(r),c(r),d(r)$:
$$
C^{\alpha\beta}(\B r) \equiv a(r)B^{\alpha\beta}_{9,jm} +
b(r)B^{\alpha\beta}_{7,jm} + c(r)B^{\alpha\beta}_{1,jm} +
d(r)B^{\alpha\beta}_{5,jm} + \dots \ ,
$$
where ($\dots$) stand for terms with other $j,m$ and other symmetries with
the same $j,m$. Let us first consider the case where $\xi>0$. According to
(\ref{T123}), well within the inertial range, these functions obey:
\begin{equation}
\B  T_1(\kappa=0)
\left(\begin{array}{c}a''\\ b'' \\ c'' \\ d'' \end{array} \right) +
\B T_2(\kappa=0)\left(\begin{array}{c} a'\\ b'\\ c'\\ d'\end{array}\right) +
\B T_3(\kappa=0) \left( \begin{array}{c} a\\ b\\ c\\ d \end{array}\right)=0
\ .
\label{zero-eq}
\end{equation}
Due to the scale-invariance of these equations, we look for scale-invariant
solutions in the form:
\begin{equation}
a(r)=ar^{\xi}, \quad b(r)=br^{\xi},\quad d(r)=cr^{\xi},
\quad d(r)=dr^{\xi} \ .
\label{si-form}
\end{equation}
Where $a,b,c,d$ are complex constants. Substituting (\ref{si-form}) into
(\ref{zero-eq}) results in a set of four
linear homogeneous equations for the unknowns $a,b,c,d$ :
$$
\left[
  \xi(\xi-1) \B T_1(\kappa=0) + \xi \B T_2(\kappa=0) + \B
T_3(\kappa=0)
\right]
\left(\begin{array}{c} a\\ b \\ c \\ d \end{array} \right) =0 \ .
$$
The last equation admits non-trivial solutions only when
$$
\det\left[
  \xi(\xi-1) \B T_1(\kappa=0) + \xi \B T_2(\kappa=0) + \B
T_3(\kappa=0)
\right]=0 \ .
$$
This solvability condition allows us to express $\xi$ as a function of $j$
and $\epsilon$. Using MATHEMATICA one finds eight possible values of $\xi$,
out-of-which, only four are in agreement with the solenoidal condition:
\begin{eqnarray}
\xi_{j}^{(2)}(i)&=&-\frac{1}{2}\epsilon -\frac{3}{2} \pm \frac{1}{2}
    \sqrt{H(\epsilon ,j) \pm 2 \sqrt{K(\epsilon ,j)}} \ , \quad{i=1,2,3,4}
\ ,\label{exponents-I} \\
K(\epsilon,j) &\equiv& \epsilon^4- 2\epsilon^3 + 2\epsilon^3j +
2\epsilon^3j^2 - 4\epsilon^2j -
    3\epsilon^2 - 4\epsilon^2 j^2 - 8 \epsilon j^2-8\epsilon j + 4\epsilon +
16j + 16j^2 + 4
    \nonumber \\
H(\epsilon ,j) &\equiv& -\epsilon^2 - 8\epsilon + 2\epsilon j^2 + 2\epsilon
j + 4j^2 + 4j + 5
\nonumber \ .
\end{eqnarray}
Not all of these solutions are physically acceptable because not all of them
can be matched to the zero mode solutions in the dissipative regime. To see
why this is so, consider the zero-mode equation for $\epsilon=0$:
\begin{equation}
    (2\kappa+2D) \nabla^2 \B C = 0 \ .
\label{zero-eq-zero-epsilon}
\end{equation}
The main difference between the $\epsilon=0$ case and the $\epsilon>0$ case
is that
in the
former the same scale-invariant equation holds {\em both} for the inertial
range and the dissipative range. As a result, for $\epsilon=0$, the zero
modes
scale
with the same exponents in the two regimes. These exponents are given
simply by
(\ref{exponents-I}) with $\epsilon=0$, because for $\epsilon=0$ the zero
modes
equation
with $\kappa=0$ is the same as (\ref{zero-eq-zero-epsilon}) up to the
overall
factor
$\frac{D}{D+\kappa}$ which does not change the exponent. For
$\epsilon=0$  th  solutions should be valid for the dissipative regime as
well as
for
the inertial regime, ruling out the two solutions with negative exponents in
(\ref{exponents-I}), for they will give a non-physical divergence as
$r \rightarrow 0$. Assuming now that the solutions (including the
exponents) are
continuous in $\epsilon$, (and not necessarily analytic!), one finds that
also
for
finite $\epsilon$ only the positive exponents appear in the inertial range
(an
exception to that is the $j=0$, to be discussed below). Finally there
are two branches of solutions corresponding to the ($-$) and ($+$) in the
square
root.
$$
\xi_{j}^{(2)}=-\frac{3}{2}-\frac{1}{2}\epsilon + \frac{1}{2}
\sqrt{H(\epsilon ,j)
    \pm 2 \sqrt{K(\epsilon ,j)}}\ , \quad \mbox {subset I}.
$$
 Note that for $j=0$,
only the branch with the $+$ sign under the square root exists
 since the other exponent is not admissible,
being negative for $\epsilon\to 0$, and therefore excluded by continuity.
$\xi_{0}^{(2)}$ however becomes negative as $\epsilon$ increases.
For $j\ge 2$ both solutions are admissible, and the
leading is that one with the minus  sign in the square root.\\
Let us also discuss
the behavior of the zero modes in the dissipative regime for
$\epsilon>0$. Here the dissipation terms become dominant and we can neglect
all
other
terms in $\hat{\B T}$. The zero mode equation in this regime becomes
$2\kappa
\nabla^2 C^{\alpha\beta} = 0$, which is again, up to an overall factor,
identical to the zero mode equation with $\kappa=0, \epsilon=0$. The
solutions in
this region are once again scale invariant with scaling exponents
$\xi_{j}^{(2)}|_{\epsilon=0}=j,j-2$. As expected, the correlation
function
$C^{\alpha\beta}(\B r)$ becomes smooth in the dissipative regime.

In \cite{ara00a}  the  computation of the exponents corresponding to
subsets II and III is also presented. The result is:
$$
\xi_{j}^{(2)}=-\frac{3}{2}-\frac{1}{2}\epsilon +\frac{1}{2}
    \sqrt{1-10\epsilon+ \epsilon ^{2}+2j^{2}\epsilon +2j\epsilon
+4j+4j^{2}}\ ,
    \quad \mbox {subset II}
$$
$$
\xi_{j}^{(2)}=-\frac{3}{2}-\frac{1}{2}\epsilon +\frac{1}{2}\sqrt{ \epsilon
^{2}
    + 2\epsilon + 1 + 4j^{2} + 2j^{2}\epsilon + 4j + 2\epsilon j}\ ,
    \quad \mbox {subset III}.
$$
For $j=0$ there is no contribution from this subset, as the exponent is
negative.
After matching the zero modes to the dissipative range, one has to
guarantee matching at the outer scale $L$. The condition to be fulfilled is
that
the sum of the zero-modes with the inhomogeneous solutions (whose exponents
are
2-$\epsilon$) must give $\B C(\B r)\to 0$ as $|\B r|\to L$. Obviously this
means
that
the forcing must have a projection on any sector $\B B_{q,jm}$ for which
$C^{(2)}_{q,jm}$ is nonzero.
%
%
\subsubsection{Summary and Conclusions}
The results of this section should be examined in the light of the
previous section on  passive scalars. That passive scalar case afforded
only perturbative calculation of anomalous exponents in all
anisotropic sectors. The present example offers exact,
non-perturbative calculations, of the whole spectrum of scaling
exponents that determines the covariance of a vector field in the
presence of anisotropy.  The main conclusions are: (i) scaling
exponents of the second order magnetic correlation functions are
anomalous; (ii) they are strictly increasing with the index of $j$ of
the sector, meaning that there is a tendency toward isotropization
upon decreasing the scales of observation.
 The equations for the magnetic covariance foliate into
independent closed equations for each set of irreducible
representations of the SO(3) group. Moreover, scaling properties of
the zero-modes do not show any dependence on the $q$ index labeling
projections on different irreducible representations of the SO(3)
groups for each fixed $(j,m)$. The consequence of the latter
property is that transversal and longitudinal correlation have
the same scaling exponents within each anisotropic sector.

%
%
\subsection{The Linear Pressure Model}
\label{chap:linearP}
In this subsection we discuss the scaling exponents
characterizing the power-law behavior of
  the anisotropic components of correlation functions in turbulent systems
  with pressure, exploring the fundamental question whether
also for such systems the scaling exponents increase as $j$
increases, or they are bounded from above.
  The equations of motion in systems with pressure contain nonlocal
integrals over all space. One could argue that the requirement of
convergence of these integrals bounds the exponents from above. It is
shown here on the basis of a solvable model (the ``Linear Pressure
Model"), that this is not necessarily the case. The model described  here
is of a passive vector advection by a rapidly varying velocity
field \cite{ara01}. The advected vector field is divergent free and the
equation contains a pressure term that maintains this condition.
The zero modes of the second order correlation function are found
in all the sectors of the symmetry group. We show that the
spectrum of scaling exponents can increase with $j$ without
  bounds, while preserving finite integrals. The conclusion is that
  contributions from higher and higher anisotropic sectors can disappear
  faster and faster upon decreasing the scales also in systems with
  pressure.
To demonstrate that, consider a
typical integral term of the form,
\begin{equation}
\label{eq:ex}
  \int\!\! d\B y\, G(\B r-\B y) C(\B y) \ .
\end{equation}
Here $G(\B r) = -1/(4\pi r)$ is the infinite domain Green
function of the Laplacian operator, and $C(\B r)$ is some
statistical object which is expected to be scale invariant in the
inertial range. If $C(\B r)$ has an infrared cross over at scale
$L$ (or equivalently, the integral has an infrared cutoff at
scale $L$), then the above expression will not be a pure power
law of $r$, not even inside the inertial range. Then how is it
possible that such an expression will cancel out a local term of
$C(\B r)$, as is required by the typical equations of motion?
This puzzle has led in the past to the introduction of the
concept of ``window of locality'' \cite{lvo95c,fuk00}. The window
of locality is the range for the scaling exponents in which no
divergence occurs, even if the cross over length $L$ is taken to
infinity. For these exponents integrals of type (\ref{eq:ex}) are
dominated by the range of integration $y\approx r$ and are
therefore termed ``local". In a ``local" theory no infrared
cutoff is called for.

In this subsection we present solutions for the scaling exponents
in the anisotropic sectors of a linear model of turbulence with
pressure. 
This model reveals two
mechanisms that allow an unbounded spectrum of scaling exponents.
First, a careful analysis of the window of locality in the
anisotropic sectors shows that it widens as $j$ increases. We
always have a leading scaling exponent within the window of
locality. Secondly, there is a more subtle mechanism that comes
to play when sub-leading exponents exist outside the window of
locality. In these cases we show that there exist counter-terms
in the exact solution (not the zero modes!) which maintain the
locality of the integrals. The bottom line is that in these
models the anisotropic exponents are unbounded from above leading
to a fast decay of the anisotropic contributions in the inertial
range.
 The Linear Pressure model captures some of the
 aspects of the pressure term in Navier-Stokes turbulence, while being
 a linear and therefore much simpler problem. The non linearity of the
 Navier-Stokes equation is replaced by an advecting Kraichnan field
 $\B u(\B x,t)$ and an advected field $\B v(\B x,t)$. The advecting
 field $\B u(\B x, t)$ is taken, as before, the Kraichnan field
 (\ref{W1}).  Both fields are assumed incompressible. The equation
 of motion for the vector field $v^\alpha(\B x, t)$ is:
\begin{eqnarray}
    \partial_t v^\alpha + u^\mu \partial_\mu v^\alpha +
        \partial^\alpha p - \kappa \partial^2 v^\alpha &=& f^\alpha \ ,
        \label{eq:v-p} \\
    \partial_\alpha v^\alpha = \partial_\alpha u^\alpha &=& 0 \ .
\end{eqnarray}
In  this equation, $\B f(\B x,t)$
is the same as the one in Eq.~(\ref{eq:B}).
 Analyticity of $\B f(\B x,t)$ is an
important requirement. It means that $A^{\alpha\beta}(\B x)$ can
be expanded for small $|\B x|$ as a power series in $x^\alpha$;
as a result its leading contribution in the $j$-sector is
proportional to $x^{j-2}$, given by
$\partial^\alpha\partial^\beta x^j Y_{j m}(\hat{\B x})$. To
see that this is the leading contribution the reader can consult
the general discussion of the construction of the irreducible
representations in Ref.\cite{ara99b}. All other analytic
contributions contain less derivatives and are therefore of
higher order in $x$.

In order to derive the statistical equations of the correlation
function of $v^\alpha(\B x,t)$, we need a version of (\ref{eq:v-p})
without the pressure term. Following the standard treatment of the
pressure term in Navier-Stokes equation, we take the divergence of
(\ref{eq:v-p}) and arrive at,
$$
    \partial_\nu \partial_\mu u^\mu v^\nu +
        \partial^2 p = 0 \ .
$$
The Laplace equation is now inverted using the Green function of
infinite domain with zero-at-infinity boundary conditions:
$$
  p(\B x) = -\int\!\! d\B y\, G (\B x - \B y)
     \partial_\nu \partial_\mu u^\mu(\B y) v^\nu(\B y) \ ,$$
with 
  $ G(\B x) \equiv -1/4\pi  x $.
With this expression for $p(\B x)$, Eq.~(\ref{eq:v-p}) can be
rewritten as:
\begin{eqnarray}
 \partial_t v^\alpha(\B x, t)
     &+& u^\mu(\B x,t) \partial_\mu v^\alpha(\B x,t)
- \partial^\alpha_{(\B x)} \int\!\! d\B y\,
    G(\B x-\B y)\partial_\nu\partial_\mu u^\mu(\B y) v^\nu(\B y)
 \nonumber\\&-& \kappa \partial^2 v^\alpha(\B x,t) = f^\alpha(\B x,t) \ .
\nonumber 
\end{eqnarray}
In \cite{ara01} the
equation of motion for  the 2-point
correlation function,$
    C^{\alpha\beta}(\B r) \equiv
       \left<v^\alpha(\B x+\B r) v^\beta(\B x) \right>$
was found:
\begin{eqnarray}
  && \partial_t C^{\alpha\beta}(\B r) - T^{\alpha\beta}(\B r)
    -T^{\beta\alpha}(-\B r)
+ \int\!\!d\B y\, G(\B r - \B y) \partial^\beta\partial_\nu
     T^{\alpha\nu}(\B y)\nonumber\\
  &&+ \int\!\!d\B y\, G(-\B r - \B y) \partial^\alpha\partial_\nu
        T^{\beta\nu}(\B y)  -2\kappa\partial^2 C^{\alpha\beta}(\B
r)\nonumber\\
  && = \left<v^\alpha(\B x+\B r) f^\beta(\B x)\right> +
       \left<v^\beta(\B x) f^\alpha(\B x+\B r)\right> \ . \nonumber
\end{eqnarray}
where
to simplify the equations we have  defined an auxiliary function
$T^{\alpha\beta}(\B r)$:
    $$
  T^{\alpha\beta}(\B r) \equiv
    \partial_\mu^{(r)} \left< v^\alpha(\B x+\B r) u^\mu(\B x) v^\beta(\B x)
      \right> \ .
$$
This equation is identical to the equation for the second
order correlation function in the usual Navier-Stokes turbulence,
provided that $u^\mu$ is replaced with $v^\mu$ in
the expression above. Indeed, the vexing problem that we face is
being made very clear: if the triple correlation function has a
power law dependence on $\B r$ with an arbitrarily large
exponent, how can the integral converge in the infrared? One
possibility is that the scaling exponent of $T^{\alpha\beta}(\B
r)$ is sufficiently low, making the integral convergent. The
other possibility is that the correlation function is scale
invariant only in the inertial range and vanishes quickly after
that, which is equivalent to the introduction of an infrared
cutoff. However the integral terms in the equation probe the
correlation function throughout the entire space.  Therefore, a
cross over behavior of the correlation function at the outer
scale $L$, seems to contradict a pure scaling behavior of the
correlation function in the inertial range itself. This in turn
implies the saturation of the anisotropic scaling exponents.

To proceed, we use the fact that the field $\B u(\B x,t)$, as well as
the forcing, are Gaussian white noises with correlation given by
Eq.~(\ref{kappa1}). This enables us to express
$T^{\alpha\beta}(\B r)$ and the correlation of the force in terms of
$C^{\alpha\beta}(\B r)$ and $A^{\alpha\beta}(\B r)$. One can use
 the well known method of Gaussian integration
by parts \cite{fri95} which leads to the final equations (see also
appendix of (\cite{ara01}):
\begin{eqnarray} 
  \partial_t C^{\alpha\beta}(\B r) &=& T^{\alpha\beta}(\B r) +
  T^{\beta\alpha}(-\B r) - \int\!\!d\B y\, G(\B r - \B y)
  \partial^\beta\partial_\nu T^{\alpha\nu}(\B y) \nonumber \\ &-&
  \int\!\!d\B y\, G(-\B r - \B y) \partial^\alpha\partial_\nu
  T^{\beta\nu}(\B y) + 2\kappa\partial^2 C^{\alpha\beta}(\B r) +
A^{\alpha\beta}(\B r) \ , \label{eq:dtC} \\ T^{\alpha\beta}(\B r)
  &=& -\frac{1}{2}K^{\mu\nu}\partial_\mu\partial_\nu
  C^{\alpha\beta}(\B r) + \frac{1}{2}\partial^\alpha_{(\B r)} \int\!\!
  d\B y\, G(\B r-\B y) \partial_\tau\Big[ K^{\mu\nu}(\B y)
  \partial_\mu \partial_\nu C^{\tau\beta}(\B y) \Big] \nonumber \\ &-&
  \frac{1}{2} \int \!\! d\B y\, G(\B y) \partial^\beta\partial_\tau
  \Big[ K^{\mu\nu}(\B y)\partial_\mu\partial_\nu C^{\alpha\tau}(\B
  r-\B y)\Big] \ . \label{eq:Tab}
\end{eqnarray}
These equations have to be supplemented with two more equations that
follow directly from the definition of $C^{\alpha\beta}(\B r)$:
$$
 \partial_\alpha C^{\alpha\beta}(\B r) = 0 \ ,
 \quad
  C^{\alpha\beta}(\B r) = C^{\beta\alpha}(-\B r) \ .
$$
Finally we note that Eqs.~(\ref{eq:dtC},\ref{eq:Tab}) can be
interpreted in a transparent way, utilizing two projection
operators which maintain the RHS of \Eq{eq:dtC} divergence free
in both indices. To define them, let us consider a tensor field
$X^{\alpha\beta}(\B r)$ which vanishes sufficiently fast at
infinity. Then the two projection operators $\PLO$ and $\PRO$ are
defined by:
\begin{eqnarray}
  \PLO X^{\alpha\beta}(\B r) &\equiv& X^{\alpha\beta}(\B r) -
     \partial^\alpha_{(r)}\int d\B y\, G(\B r-\B y)
        \partial_\mu  X^{\mu\beta}(\B y) \ , \nonumber \\
  \PRO X^{\alpha\beta}(\B r) &\equiv& X^{\alpha\beta}(\B r) -
     \partial^\beta_{(r)}\int d\B y\, G(\B r-\B y)
         \partial_\mu  X^{\alpha\mu}(\B y) \ .\nonumber
\end{eqnarray}
We observe that $\PLO X^{\alpha\beta}$ and $\PRO X^{\alpha\beta}$
are divergence free in the left and right indices respectively.
Using these operators we can rewrite
Eqs.(\ref{eq:dtC}-\ref{eq:Tab}) in the form
\be
  \partial_t C^{\alpha\beta}(\B r) =
    \PRO T^{\alpha\beta}(\B r) + \PRO T^{\beta\alpha}(-\B r)
   + 2 \kappa\partial^2 C^{\alpha\beta}(\B r)
    + A^{\alpha\beta}(\B r) \ , \label{eq:dtC1}  \ee
\be
  T^{\alpha\beta}(\B r) = -\frac{1}{2} \PLO
K^{\mu\nu}\partial_\mu\partial_\nu
    C^{\alpha\beta}(\B r)
    - \frac{1}{2} \int \!\! d\B y\, G(\B y) \partial^\beta\partial_\tau
    \Big[ K^{\mu\nu}(\B y)\partial_\mu\partial_\nu
        C^{\alpha\tau}(\B r-\B y)\Big] \ . \label{eq:Tab1}
\ee
The projection in \Eq{eq:Tab1} guarantees that
$T^{\alpha\beta}(\B r)$ is divergence free in its left index,
while the projection in \Eq{eq:dtC1} guarantees divergence
freedom in the right index.

Not all the terms in these equations are of the same nature. The
integrals due to the projection operator are easy to deal with by
applying a Laplacian on them. For example,
$
  \partial^2 \PRO T^{\alpha\beta}(\B r) = \partial^2 T^{\alpha\beta}(\B r)
     - \partial^\beta \partial_\nu T^{\alpha\nu}(\B r) \ $.
On the other hand, there seems to be no way to eliminate the last
integral in \Eq{eq:Tab1}, and therefore we shall refer to it as
the ``non-trivial integral''.  Only when the velocity
scaling exponents in (\ref{kappa1}) are  $\epsilon=0$ and $\epsilon=2$ it
trivializes: the integral vanishes when $\epsilon=0$ and is
proportional to $C^{\alpha\beta}(\B r)$ when $\epsilon=2$.
Unfortunately, in these extreme cases also the projection
operator trivializes, and the effect of the pressure cannot be
adequately assessed. We prefer to study the problem for a generic
value $\epsilon$ for which the incompressibility constraint and the
pressure terms are non-trivial.

We deal with the this problem head-on in
Sect.\ref{sec:zeromodes}. Due to the non-trivial integral, we
will not be able to provide a full solution of
$C^{\alpha\beta}(\B r)$, but only of the zero modes.  However
before doing so we would like to study a model that affords an
exact solution in order to understand in detail the issues at
hand. In the next section we therefore consider a simplified
model of the Linear Pressure model, yet posing much of the same
riddle.
\subsubsection{An exactly Solvable Toy model} \label{sec:Toy}

We construct a toy model which is inspired by equations
(\ref{eq:dtC}, \ref{eq:Tab}) for the correlation function in the
Linear Pressure model. Within this model we demonstrate the
strategy of dealing with the non-local pressure term. Since it is
a simplification of the {\em statistical} equation of the Linear
Pressure model, the toy model has no obvious underlying dynamical
equation.

In the toy model, we are looking for a ``correlation function''
$C^\alpha(\B r)$, whose equations of motion are:
\begin{eqnarray}
&&  \partial_t C^{\alpha}(\B r) =
     -K^{\mu\nu}(\B r)\partial_\mu\partial_\nu C^\alpha(\B r) \label{eq:C}
 -\partial^\alpha_{(r)} \int \!\! d\B x G(\B r-\B x)
\partial_\tau
      K^{\mu\nu}(\B x)\partial_\mu\partial_\nu C^\tau(\B x)\nonumber\\&&
+\ \kappa \partial^2 C^\alpha(\B r)+A^\alpha(\B r/L)  \ , \nonumber
\\
&& \partial_\alpha C^\alpha(\B r) = 0 \label{eq:toy-incomp} \ .
\end{eqnarray}
Here $A^\alpha(\B x)$ is a one-index analog of the correlation
function of the original forces $A^{\alpha\beta}(\B x)$.
Accordingly, we take it anisotropic, analytic in $x^\alpha$ and
rapidly vanishing for $ x \gg 1$. As in the previous model,
also here analyticity requires that the leading contribution for
small $ x$ is proportional to $\partial^\alpha x^j Y_{j
m}(\hat{\B x})$ in the $j$-sector. Accordingly it is of order
$x^{j-1}$.

The toy model is simpler than the Linear Pressure model in two
aspects: First, the ``correlation function'', $C^\alpha(\B r)$
has one index instead of two and therefore can be represented by
a smaller number of scalar functions. Second, the unpleasant
non-trivial term of the Linear Pressure model is absent. This
will allow us to solve the model exactly for every value of
$\epsilon$. Nevertheless, the toy model confronts us with the same
conceptual problems that exist in the Linear Pressure model and
in NS: can a scale invariant solution in the inertial range with
a cross over to a decaying solution at scale $L$, be consistent
with the integral term? If not, is there a saturation of the
anisotropic exponents?

\Eq{eq:C} can be rewritten in terms of a new projection operator
$\PO$, which projects a vector $X^\alpha(\B r)$ on its divergence
free part:
$$
  \partial_t C^{\alpha} =
    -\PO \Big[ K^{\mu\nu}\partial_\mu\partial_\nu C^\alpha \Big]
     + \kappa \partial^2 C^\alpha +A^\alpha  \ ,
$$
where
$$
  \PO X^\alpha(\B r) \equiv X^\alpha(\B r) -
    \partial^\alpha
 \int\!\! d\B y\, G(\B r-\B y)\partial_\mu X^\mu(\B y)
$$
We shall solve this integro-differential equation by first
turning it into a PDE using the Laplacian operator, and then
turning it into a set of decoupled ODE's using the SO($3$)
decomposition.
As in the Linear Pressure model, the non locality of the
projection operator can be removed by considering a differential
version of the operator:
$$
  \partial^2 \PO T^\alpha(\B r) = \partial^2 T^\alpha(\B r) -
     \partial^\alpha \partial_\mu T^\mu(\B r) \ .
$$
In stationary condition $\partial_t C^\alpha=0$, and therefore the
differential form of the toy model is given by:
\begin{eqnarray}
 && \partial^2 \PO
  \Big[K^{\mu\nu}(\B r)\partial_\mu\partial_\nu C^\alpha(\B r)\Big] =
\partial^2 K^{\mu\nu}(\B r)
      \partial_\mu\partial_\nu C^\alpha(\B r) -
    \partial^\alpha \partial_\tau
          K^{\mu\nu}(\B r)\partial_\mu\partial_\nu C^\tau(\B r)\nonumber\\
 &&= \kappa \partial^2\partial^2 C^\alpha(\B r) +\partial^2 A^\alpha(\B r)
\ ,
         \label{eq:diff-C}  \\
   &&\partial_\alpha C^\alpha(\B r) = 0 \ \nonumber .
\end{eqnarray}
We have reached a linear PDE of order 4. This PDE will be solved
by exploiting its symmetries, i.e., isotropy and parity
conservation, as demonstrated in the next subsection.

\Eq{eq:diff-C} and the incompressibility condition of
$C^\alpha(\B r)$ are both isotropic and parity conserving.
Therefore, if we expand $C^\alpha(\B r)$ in terms of spherical
vectors with a definite behavior under rotations and under
reflections, we would get a set of decoupled ODE's for their
coefficients.

For each sector $(j,m), j>0$ of SO($3$) we have three
spherical vectors:
\begin{eqnarray}
  B_{1jm}^\alpha(\hat{\B r}) &\equiv& r^{-j-1}r^\alpha \Phi_{j m}(\B r) \
,
    \nonumber \\
  B_{2jm}^\alpha(\hat{\B r}) &\equiv&
      r^{-j+1}\partial^\alpha \Phi_{j m}(\B r) \ ,
    \nonumber \\
  B_{3jm}^\alpha(\hat{\B r}) &\equiv& r^{-j}
    \epsilon^{\alpha\mu\nu}r_\mu\partial_\nu \Phi_{j m}(\B r) \ .
\nonumber
\end{eqnarray}
Here $\Phi_{j m}(\B r)=r^j Y_{j m}(\hat{\B r})$, and see
\cite{ara99b} for further details. The first two spherical vectors
have a different parity than the third vector, hence the
equations for their coefficients are decoupled from the equation
for the third coefficient. In the following, we shall consider
the equations for the first two coefficients only, as they have a
richer structure and larger resemblance to the Linear Pressure
model. Finally note that the isotropic sector, i.e., $j=0$, is
identically zero. To see why, notice that in this special sector
there is only one spherical vector, $B_{100}^\alpha(\hat{\B r})
\equiv r^{-1}r^\alpha$. Hence the isotropic part of $C^\alpha(\B
r)$ is given by $c(r)r^{-1}r^\alpha$, $c(r)$ being some scalar
function of $r$. But then the incompressibility condition
(\ref{eq:toy-incomp}) implies that $c(r) \sim r^{-2}$, which has
a UV divergence. We therefore conclude that $c(r)=0$, and
restrict the  calculation to $j>0$.

By expanding $C^\alpha(\B r)$ in terms of the spherical vectors
$\B B_{1jm}, \B B_{2jm}$, we obtain a set of ODEs (decoupled in the
$(j,m)$ labels) for the scalar functions that are the
coefficients of these vectors in the expansion. The equations for
these coefficients can thus be written in terms of matrices and
column vectors. To simplify the calculations, we find the matrix
forms of the Kraichnan operator and of the Laplacian of the
projection operator separately, and only then combine the two
results to one. \vskip 0.3 cm
\subsubsection{The Matrix Form of the  Operators and the Solution of the Toy
Model}
\begin{figure}
\includegraphics[scale=0.65]{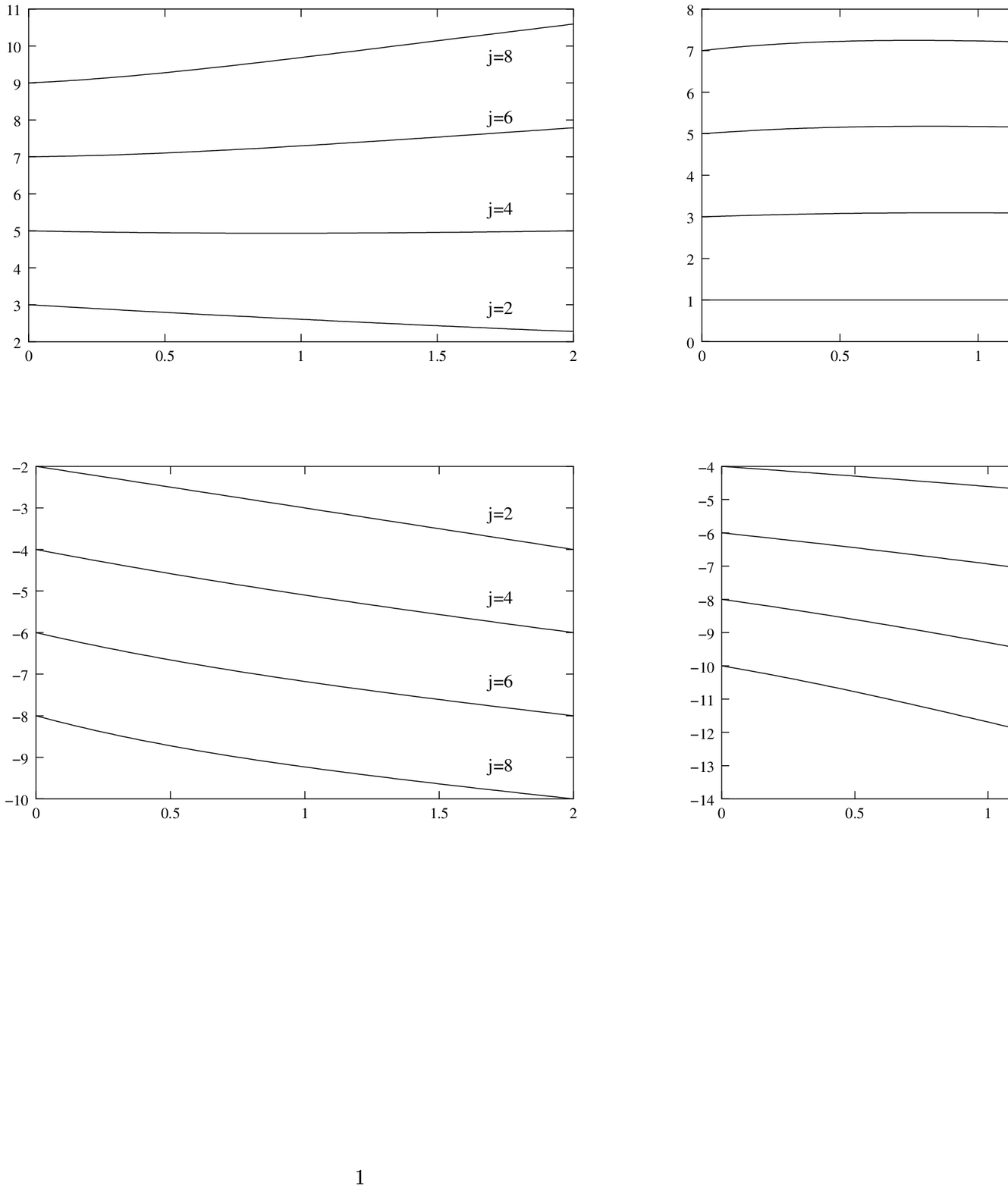}
\caption{Scaling exponents of the first few $j$s  as a function of
$\epsilon$. Top panels show:  set 1 (left);  set 2 (right). Bottom panel:
 set 3 (left);  set 4 (right)}\label{fig:toy1}
\end{figure}

In this subsection we derive the matrix form of the Kraichnan operator
and of the Laplacian of the Projection operator in each $j$ sector.
To obtain the matrix of the Kraichnan operator in the basis of
$\B B_{1jm}, \B B_{2jm}$, we expand $C^\alpha(\B r)$:
$$
  C^\alpha(\B r) =
    c_1(r)B_{1jm}^\alpha(\hat{\B r}) + c_2(r)B_{2jm}^\alpha(\hat{\B r}) \ .
$$ in appendix (\ref{app:lp1}) we show how to find the operator
on $ C^\alpha(\B r) $ in a matrix form
which results in the final equation for $c_1(r)$ and $c_2(r)$:
\begin{eqnarray}
  && r^\epsilon \MM_4 \VecII{c^{(4)}_1}{c^{(4)}_2} +
    r^{\epsilon-1} \MM_3 \VecII{c^{(3)}_1}{c^{(3)}_2} +
    r^{\epsilon-2} \MM_2 \VecII{c^{(2)}_1}{c^{(2)}_2}  \nonumber \\
  && \quad + r^{\epsilon-3} \MM_1 \VecII{c^{(1)}_1}{c^{(1)}_2} +
    r^{\epsilon-4} \MM_0 \VecII{c_1}{c_2} =   \VecII{\rho_1}{\rho_2} \ .
\label{eq:C-mat}
\end{eqnarray}
In addition  also the incompressibility constraint $\partial_\alpha
C^\alpha(\B r)=0$, can be expressed as a relation between
$c_1(r)$ and $c_2(r)$:
\begin{equation}
\label{eq:incomp}
  c'_1 + 2\frac{c_1}{r} + j c'_2 - j(j-1)\frac{c_2}{r} = 0 \ .
\end{equation}
This constraint has to be taken into account when solving
\Eq{eq:C-mat}.
The solution of \Eq{eq:C-mat} is somewhat tricky due to the
additional constraint (\ref{eq:incomp}). Seemingly the two
unknowns $c_1(r), c_2(r)$ are over determined by the three
equations (\ref{eq:C-mat}, \ref{eq:incomp}), yet this is not the
case for the two equations (\ref{eq:C-mat}) are not independent.
To see that this is the case and find the solution, it is
advantageous to work in the new basis
$$
  d_1 = c_1 + j c_2 \ , \qquad
  d_2 = -2c_1 + j(j-1)c_2 \ .$$
In this basis the incompressibility constraint becomes very
simple: $d_2 = rd_1' $,
allowing us to express $d_2$ and its derivatives in terms of
$d_1$. To do that in the framework of the matrix notation, we
define the transformation matrix $\UM$:
$$
  \UM \equiv \MatII{1}{j}{-2}{j(j-1)} \ ,
\qquad
  \UM^{-1} = \frac{1}{j(j+1)}\MatII{j(j-1)}{-j}{2}{1} \ ,
$$
so that, $ \VecII{d_1}{d_2} = \UM \VecII{c_1}{c_2} \ .$
The equations of $d_i(r)$ are the same as the equations for
$c_i(r)$, with the matrices $\MM_i$ replaced by
$
  \NM_i \equiv \UM \MM_i \UM^{-1} \ ,
$
and the sources $\rho_i$ replaced by
$$
  \VecII{\rho^*_1}{\rho^*_2} = \UM \VecII{\rho_1}{\rho_2} \ .
$$
Notice that a divergence free forcing $A^\alpha(\B r)$ will cause
$\rho^*_1(r), \rho^*_2(r)$ to be related to each other in the
same way that $d_1(r), d_2(r)$ are related to each other, i.e.,
$
  \rho^*_2 = r(\rho^*_1)' \ .
$
Next, we perform the following replacements:
$$
  d_2       = rd^{(1)}_1, \quad
  d_2^{(1)} = rd^{(2)}_1 + d^{(1)}_1, \quad
  d_2^{(2)} = rd^{(3)}_1 + 2d^{(2)}_1, $$
$$  d_2^{(3)} = rd^{(4)}_1 + 3d^{(3)}_1, \quad
  d_2^{(4)} = rd^{(5)}_1 + 4d^{(3)}_1.
$$
We get an equation written entirely in terms of the function
$d_1(r)$ and its derivatives:
\be
r^{\epsilon}(r V_5 d^{(5)}_1 +  V_4 d^{(4)}_1 +
r^{-1}V_3 d^{(3)}_1
  + r^{-2}V_2 d^{(2)}_1  + r^{-3}V_1 d^{(1)}_1 +
r^{-4}V_0 d_1) =
    \VecII{\rho^*_1}{\rho^*_2}, \label{V5}
\ee
where $V_i$ are two dimensional vectors given by:
\begin{eqnarray}
  V_5 \equiv \NM_4 \VecII{0}{1}, \nonumber  \qquad
  V_4 \equiv \NM_4 \VecII{1}{4} + \NM_3\VecII{0}{1}, \qquad
  V_3 \equiv \NM_3 \VecII{1}{3} + \NM_2\VecII{0}{1}, \\
  V_2 \equiv \NM_2 \VecII{1}{2} + \NM_1\VecII{0}{1}, \qquad
  V_1 \equiv \NM_1 \VecII{1}{1} + \NM_0\VecII{0}{1}, \qquad
  V_0 \equiv \NM_0 \VecII{1}{0}. \nonumber
\end{eqnarray}
Their explicit values can be found in \cite{ara01}.
The Eq.~(\ref{V5})  are for a column vector, and can be regarded as
two scalar differential equations that we refer to as the
``upper" and the ``lower". The upper ODE is of the fourth order,
while the lower ODE is of fifth order. Non surprisingly, the lower
equation is the first derivative of the upper equation, provided
that $A^\alpha(\B r)$ is divergence free. Hence the two equations
are dependent, and we restrict the  attention to the upper
equation.  To simplify it, we divide both sides by $Dr^\epsilon$,
replace $d_1(r)$ by $\psi(r)$ and define the RHS to be the
function $S(r)$:
\begin{equation}
  S(r)\equiv D^{-1}r^{-\epsilon}\rho^*_1(r) \ .
\label{eq:S(r)}
\end{equation}
After doing so, we reach the following equation:
\begin{equation}
  \psi^{(4)} + a_3\frac{\psi^{(3)}}{r} + a_2\frac{\psi^{(2)}}{r^2} +
    a_1 \frac{\psi^{(1)}}{r^3} + a_0 \frac{\psi}{r^4} =S(r) \ .
\label{eq:psi}
\end{equation}
Its homogeneous solution is easily found once we substitute,
$
  \psi(r) = \psi_0 r^\xi$.
The scaling exponents are the roots of the polynomial,
\begin{eqnarray}
\label{def:Polynomial}
   P(\xi) &=& \xi(\xi-1)(\xi-2)(\xi-3)
   + a_3\xi(\xi-1)(\xi-2) + a_2\xi(\xi-1) + a_1\xi + a_0 \ .
\nonumber
\end{eqnarray}
The polynomial roots are found to be real and non-degenerate. Two
of them are positive while the other two are negative. They are
given  by:
\begin{equation}
\label{def:exponents}
  \xi_j(i) = -\frac{1}{2} - \frac{1}{2}\epsilon \pm \frac{1}{2}
     \sqrt{A(j,\epsilon) \pm \sqrt{B(j,\epsilon)}} \
 \qquad {\mbox i=1,2,3,4} ,
\end{equation}
where,
$$
  A(j,\epsilon) \equiv \epsilon^2 +\epsilon j^2 + \epsilon j - 2\epsilon +
5 + 4j
    + 4j^2 \ ,$$ and $$
  B(j,\epsilon) \equiv  -8\epsilon^2j - 7\epsilon^2j^2 + 16\epsilon^2 +
2\epsilon^2j^3
    + \epsilon^2j^4  - 8\epsilon j^2 - 8\epsilon j - 32\epsilon + 16 + 64j
+64j^2.
$$
In the limit $\epsilon \to 0$ the roots become, in decreasing order:
$$
   \xi_j(1) = j+1, \qquad
 \xi_j(2) = j-1, \qquad
 \xi_j(3) = -j, \qquad
 \xi_j(4) = -j-2, 
 $$
Fig.~(\ref{fig:toy1})
 displays the first few exponents as a function of $\epsilon$.
We note that the spectrum has no sign of saturation as $j$
increases. Before we discuss the meaning of this observation we
will make sure that these solutions are physically relevant and
participate in the full (exact) solution including boundary
conditions.
The general solution of \Eq{eq:psi} is traditionally given as the
sum of a special solution of the non-homogeneous equation plus a
linear combination of the zero modes. However when attempting to
match the solution to the boundary conditions it is convenient to
represent it as:
\begin{equation}
\label{eq:psi-general-solution}
 \psi(r) = \sum_{i=1}^4
  \frac{r^{\xi_j(i)}}{\underbrace{ (\xi_j(i)-\xi_j(1)) \ldots
       (\xi_j(i)-\xi_j(4))}_{\scriptsize\mbox{all {\em different} roots}}}
   \int_{m_i}^r \!\! dx \, x^{3-\xi_j(i)}\, S(x) \ ,
\end{equation}
where the free parameters of the solution are the four constants
$m_i$. Indeed a change in $m_i$ is equivalent to adding to the
solution a term proportional to $ r^{\xi_j(i)}$. In the next subsection we
find the values of $m_i$ to match the boundary conditions, and
discuss the properties of the solution. \vskip 0.3 cm
 \subsubsection{Boundary Conditions and Inertial-range Behavior}
From Eq.(\ref{eq:psi-general-solution}) it is clear that the only
values of $m_i$ that guarantee that the solution remains finite
as $r\to 0$ and that it decays as $r\to\infty$ are
$m_1=m_2=+\infty$, $m_3=m_4 = 0$:
\begin{eqnarray}
\psi(r) =  
-\frac{r^{\xi_j(1)}}{(\xi_j(1)-\xi_j(2))(\xi_j(1)-\xi_j(3))(\xi_j(1)-\xi_j(4
))}
  \int_r^\infty \!\! dx \, x^{3-\xi_j(1)}\, S(x) \nonumber \\
-\frac{r^{\xi_j(2)}}{(\xi_j(2)-\xi_j(1))(\xi_j(2)-\xi_j(3))(\xi_j(2)-\xi_j(4
))}
  \int_r^\infty \!\! dx \, x^{3-\xi_j(2) }\, S(x) \nonumber\\
+\frac{r^{\xi_j(3)}}{(\xi_j(3)-\xi_j(1))(\xi_j(3)-\xi_j(2))(\xi_j(3)-\xi_j(4
))}
  \int_0^r \!\! dx \, x^{3-\xi_j(3) }\, S(x) \nonumber \\
+\frac{r^{\xi_j(4)}}{(\xi_j(4)-\xi_j(1))(\xi_j(4)-\xi_j(2))(\xi_j(4)-\xi_j(3
))}
  \int_0^r \!\! dx \, x^{3-\xi_j(4)}\, S(x) \ .
\label{eq:fullsol}
\end{eqnarray}
To understand the asymptotic of this solution we find from
\Eq{eq:S(r)} that for $x \ll L$, $S(x)$ has a leading term which
goes like $x^{j-1-\epsilon}$, whereas for for $x \gg L$, $S(x)$
decays rapidly. It is now straightforward to prove that for $r
\ll L$, the $\xi_j(3), \xi_j(4)$ terms scale like $r^{j+3-\epsilon}$,
the $\xi_j(2)$ term scales like $r^{\xi_j(2)}$ and the $\xi_j(1)$
term scales like $r^{\xi_j(1)}$ for values of $\epsilon$ for which
$\xi_j(1) < j+3-\epsilon$ and like $r^{j+3-\epsilon}$ otherwise. In
addition it is easy to see that for $r \gg L$, $\psi(r)$ exhibits
an algebraic decay: the $\xi_j(1), \xi_j(2)$ terms decay rapidly due
to the decay of $S(x)$ whereas the $\xi_j(3),\xi_j(4)$ terms decay
algebraically like $r^{\xi_i}$ respectively. The asymptotic of
the full solution are thus given by
\begin{equation}
  \psi(r) \sim \left\{ \begin{array}{lcr}
    r^{\xi_j(2)} &,& r \ll L \\
    r^{\xi_j(3)} &,& r \gg L \end{array} \right. \ .
\end{equation}
The obvious conclusion is that there is no saturation in the
anisotropic scaling exponents as $j$ increases. The lack of
contradiction with the existence of an integral over all space
has two aspects. The main one is simple and obvious. The
Integro-differential equation (\ref{eq:C}) for $C^\alpha$ has a
differential version (\ref{eq:diff-C}). Solving the differential
version we are unaffected by any considerations of convergence of
integrals and therefore the solution may contain exponents that
increase with $j$ without limit. Nevertheless the full solution
(\ref{eq:fullsol}) exhibits a cross over at $L$: it increases in
the inertial range $r\ll L$ and decays for $r\gg L$. Thus
plugging it back to the Integro-differential equation we are
guaranteed that no divergence occurs.

The question why the cross-over length $L$ does not spoil the
scale invariance in the inertial range still remains. The answer
is found in differential form of the equation of motion, given by
\Eq{eq:diff-C}. From this equation we find that the integrand is
a Green's function times a Laplacian of a tensor. By definition
such an integral localizes, i.e. it is fully determined by the
value of the tensor at the external vector $\B r$. In the
language of Eq.(\ref{eq:ex}) $A(\B y)=\nabla^2 B(\B y)$!

The second and less obvious aspect is that the window of locality
widens up with $j$. This is due to the cancellations in the
angular integration of the anisotropic solutions that are due to
the orthogonality of the $Y_{j
  m}(\hat{\B r})$ and their generalizations $B^{\alpha}_{q j
  m}(\hat{\B r})$. To demonstrate this consider again the simple integral
(\ref{eq:ex}), and assume that $C(\B y)$ belongs to $(j, m)$
sector, i.e.
$
  C(\B y) =a(y) Y_{j m}(\hat{\B y})
$.
For $y \gg r$, we may expand the Green function in $r/y$:
\begin{eqnarray}
  G(\B r-\B y) &=& -\frac{1}{4\pi|\B r-\B y|} = -\frac{1}{4\pi
y}\sum_{n=0}^\infty
     a_n \left[\left(\frac{r}{y}\right)^2
       - 2\frac{\B r \cdot \hat{\B y}}{y}\right]^n \ . \nonumber
\end{eqnarray}
Here $a_n$ are Taylor coefficients. Obviously the dangerous terms
for the infrared convergence are those with low values of $n$.
However all these terms will vanish for $n<j$ due to the
angular integration against $Y_{j m}(\hat{\B y})$. The reason
is that all these terms are of the form $r^{n_1} y^{n_2} (\B
r\cdot \hat{\B y})^{n_3}$ with $n_3< j$. The angular part here
has projections only $Y_{j' m'}$ with $j'\le k_3<j$. The
first term to contribute comes when $n=j$, and is proportional
to the amplitude integral $\int_r^\infty \!\! dy \, y^2 a_{j
m}(y) y^{-j -1}$.  For a power law $a_{j m}(y)\sim
y^\lambda$ this implies locality for
$
  \lambda< j-2,
$,
instead of $\lambda<-2$ in the isotropic sector. The lower bound
of the window of locality is also extended, and a similar
analysis for $y\ll r$ leads to $\lambda>-j-3$. For the toy
model this translates to the window of locality
$$
  -j-\epsilon < \xi_j(i) < j +1 -\epsilon \ .
$$ From the previous analysis we find that the leading power law of
the full solution in the inertial range is $r^{\xi_j(2)}$, which
is inside this ``extended'' window of locality. Nevertheless, the
subleading power $r^{\xi_j(1)}$ originating from the first term in
\Eq{eq:fullsol} is above this window, and its presence in the
solution can be explained only using the first mechanism.

We will see when we turn back to the Linear Pressure model that
both these mechanisms operate there as well, leading again to a
lack of saturation in the exponents.
\subsubsection{Solving the Linear Pressure Model}
\label{sec:zeromodes}
We now return to the Linear Pressure model.
The methods used to solve it follow very close those
developed for the toy model and therefore will not be described
in the full way. 
Contrary to the toy model where we can have the full solution
in the present case we can solve only for the zero modes.
 These are scale invariant solutions which solve an
equation containing an integral. Their exponent must therefore
lie within the ``extended'' ($j$ dependent) window of
locality. Finally one can argue that these zero modes are a part
of the full solution that decays for $r \gg L$, and therefore
solve the original equation as well.
We start from Eqs (\ref{eq:dtC1}) and (\ref{eq:Tab1}). In the Appendix
of (\cite{ara01})  \Eq{eq:Tab1}
was brought to the form:
\begin{equation}
\label{eq:simpleT}
   T^{\alpha\beta}(\B r) = -\frac{1}{2}\PLO
     K^{\mu\nu}\partial_\mu\partial_\nu C^{\alpha\beta}(\B r)  - \frac{1}{2}
\frac{12\epsilon
D}{(\epsilon-3)(\epsilon-5)}
      \int\!\! d\B y \, G(\B y) y^{\epsilon-2}
         \partial^2 C^{\alpha\beta}(\B r-\B y),
\end{equation}
which is true for every $\epsilon\neq 1$. The $\epsilon=1$ case will not be
treated here explicitly. Nevertheless, in \cite{ara01} it was  argued that
that the results for $\epsilon=1$ can be deduced from the $\epsilon\neq 1$
results by continuity.

Looking at \Eq{eq:simpleT}, we note that when $\epsilon=2$, the
integral on the RHS of the above equation trivializes to a local
term $C^{\alpha\beta}(\B r)$. In this limiting case the model can
be fully solved utilizing the same machinery used in the previous
section. The solution can then be used to check the zero modes
computed below for arbitrary values of $\epsilon$.

\begin{figure}
 \epsfbox{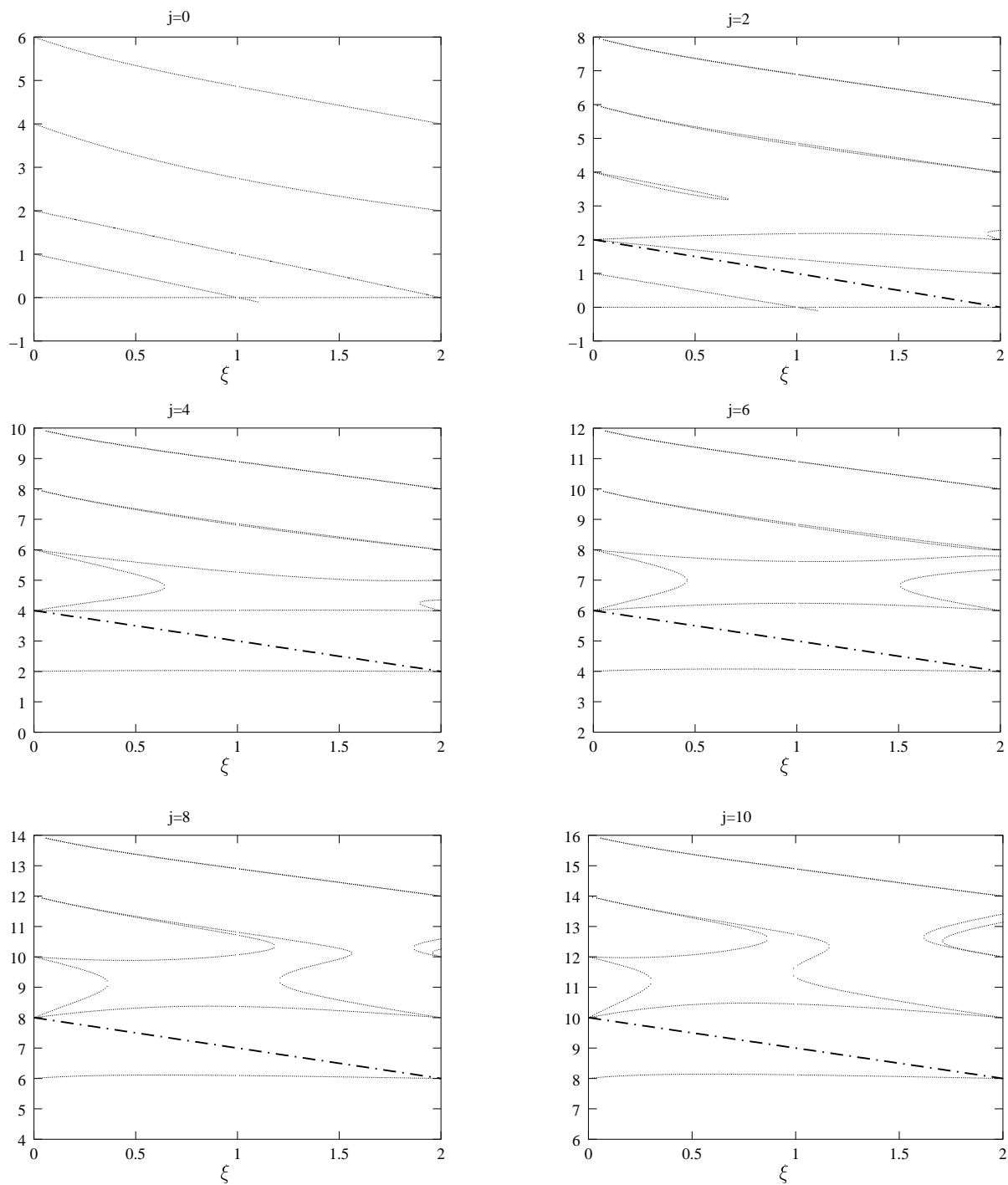}
\caption{Leading scaling exponents for the first
 few $j$s. The dashed line indicates the upper bound of the window of
locality}
\label{fig:toy2}
\end{figure}

To proceed, we substitute \Eq{eq:simpleT} into \Eq{eq:dtC1},
noting that the projector $\PRO$ leaves the non-trivial integral
in (\ref{eq:simpleT}) invariant since it is divergence-free in
both indices. Setting $\partial_t C^{\alpha\beta}(\B r,t) = 0$ in
the stationary case, we arrive to following equation
\begin{eqnarray}
\label{eq:fulleq}
&&0 = -\Big[\PRO\PLO K^{\mu\nu}\partial_\mu\partial_\nu
       C^{\alpha\beta}\Big](\B r) \\&& -\frac{12\epsilon
D}{(\epsilon-3)(\epsilon-5)}
      \int\!\! d\B y\, G(\B y) y^{\epsilon-2} \partial^2
         C^{\alpha\beta}(\B r-\B y) + 2\kappa\partial^2 C^{\alpha\beta}(\B
r)+A^{\alpha\beta}(\B r) \ . \nonumber
\end{eqnarray}
As in the toy model, we apply two Laplacians to the above
equation in order to get rid of the integrals of the projection
operators, and obtain
\begin{eqnarray}
\label{eq:full-diff-eq}
&&0 = -\partial^4\Big[\PRO\PLO K^{\mu\nu}\partial_\mu\partial_\nu
       C^{\alpha\beta}\Big](\B r)\\&&-\frac{12\epsilon
D}{(\epsilon-3)(\epsilon-5)}
      \int\!\! d\B y\, G(\B y) y^{\epsilon-2} \partial^6
         C^{\alpha\beta}(\B r-\B y)  + 2\kappa\partial^6 C^{\alpha\beta}(\B
r)+\partial^4
A^{\alpha\beta}(\B r) \ . \nonumber
\end{eqnarray}
Here and in the sequel, the operator $\partial^{2n}$ should be
interpreted as $(\partial^2)^n$. We now seek the homogeneous
stationary solutions of $C^{\alpha\beta}(\B r)$ in the inertial
range (zero modes). These satisfy the equations obtained by
neglecting the dissipation, and setting the forcing and time
derivative to zero:
\begin{eqnarray}
\label{eq:zeroCab} && 0 = \partial^4
       K^{\mu\nu}\partial_\mu\partial_\nu C^{\alpha\beta}(\B r)
   + \partial^\alpha\partial^\beta\partial_\tau\partial_\sigma
       K^{\mu\nu}\partial_\mu\partial_\nu C^{\tau\sigma}(\B r)
\\&&-\
\partial^\alpha\partial_\tau\partial^2
       K^{\mu\nu}\partial_\mu\partial_\nu C^{\tau\beta}(\B r)
       - \partial^\beta\partial_\tau\partial^2
       K^{\mu\nu}\partial_\mu\partial_\nu C^{\alpha\tau}(\B r) \nonumber \\
&&\ +\ \frac{12\epsilon D}{(\epsilon-3)(\epsilon-5)}
   \int\!\! d\B y \, G(\B y) y^{\epsilon-2}
     \partial^6 C^{\alpha\beta}(\B r-\B y) \ , \nonumber
\end{eqnarray}
Let us now define the RHS of the above equation as the ``zero
modes operator'' $\OO(\epsilon)$, and write the zero modes equation
compactly as
$$
  0 = \Big[ \OO(\epsilon) C^{\alpha\beta} \Big] (\B r) \ .
$$
The solutions of this problem is obtained as before
by expanding  $C^{\alpha\beta}(\B r)$ in a
basis that diagonalizes $\OO(\epsilon)$. Full detail of this
procedure are available in \cite{ara01}.
We turn now to discuss the results.
In Fig. (\ref{fig:toy2})
we show the leading scaling exponents of the Linear Pressure
model for $j=0,2,4,6,8,10$. The results are shown for
From Fig. (\ref{fig:toy2}),
 we see that in the isotropic sector and in the
$j=2$ sector, the leading exponent is $\xi_j^{(2)} = 0$,
corresponding to the trivial $C^{\alpha\beta}(\B r) = const$
solution. These zero modes will not contribute to the second
order structure function, which is given by
$$
  S^{\alpha\beta}(\B r) =
     2\Big[C^{\alpha\beta}(\B r) - C^{\alpha\beta}(\B 0)\Big] \ ,
$$
and so we have to consider the zero mode with the consecutive
exponent. In the isotropic sector this exponent is exactly
$\xi_0^{(2)}=2-\epsilon$, as can be proven by passing to Fourier space. This
special solution is a finger-print of the existence of a constant
energy flux in this model.
Returning to the main question of this subsection, we see that no
saturation of the anisotropic exponents occurs since the leading
exponent in every $j>2$ sector is $\xi_j^{(2)} \simeq
j-2$. These exponents are within the window of locality of
\Eq{eq:fulleq} which is given by $-j-3 < \xi_j^{(2)} < j - \epsilon$.
However the next-to-leading exponents (which are the leading ones
in the structure function for $j=0,2$), are already out of
this window, and their relevance has to be discussed. in \cite{ara01} it was
proposed
that the same mechanism that works in the toy model
 also operates here, and that all these higher
exponents can be found in the full solution. To understand this,
let us write a model equation for the correlation function in the
spirit of Eq.(\ref{eq:ex}):
\begin{equation}
  \hat{\mathcal D}C(\B r) + \int d\B y\, K(\B r-\B y) C(\B y) = F(r) \ ,
\label{eq:model}
\end{equation}
with $K$ being some kernel, and $\hat{\mathcal D}$ being some local
differential operator. In view of \Eq{eq:fulleq}, the
differential operator $\hat{\mathcal D}$ should be regarded as the
Kraichnan operator, and the integral term should be taken for all
integral terms in the equation, including integrals due to the
projection operators. These integrals create a window of locality
that we denote by $\lambda_{\mbox{\small low}} < \lambda <
\lambda_{\mbox{\small hi}}$. Any pure scaling solution $C(\B r)
\sim r^\lambda$ with $\lambda$ outside the window of locality
will diverge and hence will not solve the homogeneous part of
\Eq{eq:model}. Nevertheless, we will now demonstrate how this
zero mode can be a part of a full solution without breaking scale
invariance. For this we act with a Laplacian on both sides of
\Eq{eq:model}, in order to get rid of the projection operators
integrals.  Of course, like in the Linear Pressure model, this
will not eliminate all integral terms, and thus we can write the
resultant equation as
\begin{equation}
  \partial^2\hat{\mathcal D}C(\B r)
     + \int d\B y \, K(\B r-\B y)\partial^2 C(\B y) = \partial^2 F(r) \ .
\label{eq:lap-model}
\end{equation}
The main assumption, which was proven analytically in the simple
case of the toy model, is that the above equation has a solution
which is finite for all $r$, and decays for $r \gg L$. Let us now
consider the zero modes of \Eq{eq:lap-model}; their exponents
have to be within the ``shifted'' window of locality
$\lambda_{\mbox{\small low}}+2 < \lambda < \lambda_{\mbox{\small
hi}}+2$. Suppose now that $r^\lambda$ with $\lambda_{\mbox{\small
hi}} < \lambda < \lambda_{\mbox{\small hi}}+2$ is such a
solution, which is therefore part of the full solution of
\Eq{eq:lap-model}. We now claim that this solution also solves
the original equation \Eq{eq:model}, hence allowing the existence
of scaling exponents outside of its window of locality. To see
that, we first notice that since the full solution decays for
$r\gg L$, then all integrals in \Eq{eq:model} converge, and are
therefore well defined. All that is left to show is that the
equation is indeed solved by $C(\B r)$. But this is a trivial
consequence of the uniqueness of the solution for Laplace
equation with zero at infinity boundary conditions. Indeed, if we
denote the integral term in \Eq{eq:model} by
$$  I(\B r) = \int d\B y\, K(\B r-\B y) C(\B y) \ ,$$
then from \Eq{eq:lap-model} we have
$$
  \partial^2 I(\B r) = \partial^2 [F(r) - \hat{\mathcal D}C(\B r)] \ ,
$$
and since both $I(\B r)$ and $F(r) - \hat{\mathcal D}C(\B r)$ decay as
$r\to\infty$, then they must be equal. Of course no breaking of
scale invariance occurs because the equation is satisfied and
$F(r) - \hat{\mathcal
  D}C(\B r)$ is a sum of an inhomogeneous solution and power laws.

Returning to the Linear Pressure model, we have shown that not
only the first, leading exponents in every sector are legitimate,
but also the next few exponents. These exponents are inside the
shifted window of locality of the ``Laplaced'' equation
(\ref{eq:zeroCab}), which is given by $-j+1 < \lambda <
j+4-\epsilon$.
At this point, we may ask whether this is also the case for the
other exponents, which are outside this shifted window of
locality. In light of the above discussion, it is clear that all
of them may be part of the full solution, for we can always
differentiate \Eq{eq:fulleq} sufficient number of times, thus
shifting the window of locality to include any of these
exponents. However this procedure is unnecessary once we have
written the prefactor $A(\lambda; j, \epsilon)$ as an infinite sum
of poles in $\lambda$. In that case the equation is defined for
all values of $\lambda$ except for a discrete set of poles,
enabling us to look for exponents as high as we wish.
\subsubsection{Summary and Conclusions}
\label{sec:summary}
The main question raised and answered in this subsection is whether the
existence of the pressure terms necessarily leads to a saturation
of the scaling exponents associated with the anisotropic sectors.
Such terms involve integrals over all space, and seem to rule out
the existence of an unbounded spectrum. We have discussed a
mechanism that allows an unbounded spectrum without spoiling the
convergence of the pressure integrals. The mechanism is
demonstrated fully in the context of the simple toy model, and it is
proposed that it also operates in the case of the Linear Pressure
model. The mechanism is based on two fundamental observations.
The first one is that the window of locality widens up linearly
in $j$ due to the angular integration. The second, and more
important, is that a scaling solution with an unbounded spectrum
can exist {\em as a part of a full solution,
  which decays at infinity}. Indeed pure scaling
solutions cannot solve
themselves the zero modes equation if their scaling exponent is
out of the window of locality. However the zero modes are always
part of the full solution which decays to zero once $r \gg L$,
and we have shown that if such a solution solves a differential
version of the full equation, it must also solve the original
equation. Therefore by differentiating the full equation
sufficiently many times, we can always reach a differential
equation with a window of locality as high as we wish.  In that
equation we can find zero mode solutions with arbitrarily high
exponents (notice that in the toy model, it was sufficient to
differentiate once to get rid of all integrals, thus obtaining an
``infinitely wide'' window of locality). But since these zero
modes are part of a full solution that decays at infinity, then
this solution is also valid for the original equation, hence
showing that in the full solution there can be power laws with
arbitrarily high exponents.
Finally we want to comment about the relevance of these calculations
to Navier-Stokes turbulence. If we substitute blindly $\epsilon=4/3$ in our
results, we predict the exponents 2/3, 1.25226, 2.01922, 4.04843,
6.06860 and 8.08337 for $j=0,2,4,6,8$ and $10$ respectively. It would
be tempting to propose that similar numbers may be expected for
Navier-Stokes flows with weak anisotropy, and indeed for $j=0$ and $2$
this is not too far from the truth.  We return to this issue after
analyzing the Navier-Stokes case in next section.  The closeness of
the Linear Pressure Model with Navier-Stokes equations has also
been used in \cite{ben01a} to propose a closure for the non-linear
turbulent problem. 
\subsection{A Closure  Calculation of Anisotropic Exponents for
Navier-Stokes Turbulence}
In this subsection we start from the Navier-Stokes equations, and
write down an approximate equation satisfied by the second order
correlation function, in a closure approximation (renormalized
perturbation theory in 1-loop order) \cite{yos01,lvo03}.
 This equation is nonlinear.
For a weakly anisotropic system we follow \cite{lvo03} in linearizing the
equation, to
define a linear operator over the space of the anisotropic
components of the second order correlation function. The solution
is then a combination of forced solutions and ``zero modes" which
are eigenfunctions of eigenvalue zero of the linear operator.
\subsubsection{Model Equations for Weak Anisotropy in the Closure
Approximation} 
\label{modeleqs}
It is customary to discuss the closure equations in $\B k,t$
representation. The Fourier transform of the velocity field $\B
u(\B r,t)$ is defined by
$$
{\B u}({\B k},t)\equiv \int d{\B r}\, \exp[-i({\B r}\cdot{\B
k}    )] {\B u}({\B x},t).
$$
The Navier-Stokes equations for an incompressible fluid then read
$$
\Big[ {\partial \over \partial t}+\nu k^2\Big]u^\alpha  (\B k,t) =
\frac{i} {2}\Gamma ^{\alpha \beta \gamma }(\B k)\int {d^3 q d^3 p\over
(2\pi)^3}
\delta (\B k+\B q+\B p){u^*}^\beta (\B q,t)
{u^*}^\gamma  (\B p,t).$$
The interaction amplitude $\Gamma ^{\alpha \beta \gamma }(\B k)$
is defined by
$ \Gamma ^{\alpha \beta \gamma }(\B k) =-\left[P^{\alpha\gamma}(\B
k) k^\beta +P^{\alpha\beta}(\B k)k^\gamma \right] \ ,
$ with the transverse projection operator $P^{\alpha\beta}$ defined
as
$ P^{\alpha\beta} \equiv \delta^{\alpha\beta} -\frac{k^\alpha
k^\beta}{k^2}.$
The statistical object that is the concern of this subsection is the
second order (tensor) correlation function $\B F(\B k,t)$,
$$
(2\pi)^3 \FT^{\alpha\beta}(\B k,t)\delta (\B k-\B q) \equiv \langle
u^\alpha (\B k,t){u^*}^\beta(\B q,t)\rangle \ .$$ In stationary conditions
this object is time independent. Our aim
is to find its $k$-dependence, especially in the anisotropic
sectors.

It is well known that there is no close-form theory for the
second order simultaneous correlation function. We therefore need
to resort to standard closure approximations that lead to model
equations.  Such a closure leads to
approximate equations of motion of the form
\begin{equation}
\frac{\partial \FT^{\alpha\beta}(\B k,t)}{2\partial t} =
I^{\alpha\beta} (\B k,t)-\nu k^2 \FT^{\alpha\beta}(\B k,t)  \ ,
\end{equation}
where
\begin{equation}
I^{\alpha\beta}(\B k) = \int\frac {d^3q d^3p}{(2\pi)^3} \delta(\B
k+\B p+\B q) \Phi^{\alpha\beta} (\B k,\B q,\B p) \ .
\label{Integral}
\end{equation}
In this equation $
\Phi^{\alpha\beta} (\B k,\B q,\B p)
=\frac{1}{2}[\Psi^{\alpha\beta} (\B k,\B q,\B p)
+\Psi^{\beta\alpha} (\B k,\B q,\B p) ]$,
and
\begin{eqnarray}
&&\Psi^{\alpha\beta} (\B k,\B q,\B p) = \Theta(\B k,\B q,\B p)
\Gamma^{\alpha\gamma\delta}(\B k)[\Gamma^{\delta\beta'\gamma'}(\B q)
 \FT^{\gamma\gamma'}(\B p) ^{\beta'\beta}(\B k) +\nonumber\\&&
\Gamma^{\gamma\beta'\delta'}(\B p)
 \FT^{\delta\delta'}(\B q) A^{\beta'\beta}(\B k)
+ \Gamma^{\beta\delta'\gamma'}(\B k)
 \FT^{\delta\delta'}(\B q) A^{\gamma\gamma'}(\B p)] \ . \label{Phi}
\end{eqnarray}
In stationary conditions and for $k$ in the inertial range we need
to solve the integral equation $I^{\alpha\beta}(\B k) = 0$.

The process leading to these equations is long; one starts with
the Dyson-Wyld perturbation theory, and truncates (without
justification) at the first loop order. In addition one asserts
that the time dependence of the response function and the
correlation functions are the same. Finally one assumes that the
time correlation functions decay in time in a prescribed manner.
This is the origin of the ``triad interaction time" $\Theta(\Bk,\B q,\B p)$.
If one assumes that all the correlation functions
involved decay exponentially (i.e. like $\exp(-\gamma_{\B k}|t|)$,
then
\begin{equation}
\Theta(\B k,\B q,\B p) =\frac{1}{\gamma_{\B k}+\gamma_{\B
q}+\gamma_{\B p}} \ . \label{expdecay}
\end{equation}
For Gaussian decay, i.e. like $\exp[-(\gamma_{\B k} t)^2/2]$,
\begin{equation}
\Theta(\B k,\B q,\B p) =\frac{1}{\sqrt{\gamma^2_{\B
k}+\gamma^2_{\B q}+\gamma^2_{\B p}}} \ . \label{gaussdecay}
\end{equation}
All these approximations are uncontrolled. Nevertheless this type
of closure is known to give roughly correct estimates of scaling
exponents and even of coefficients in the isotropic sector.

Eq. (\ref{Integral}) poses a nonlinear integral equation which is
closed once  $\gamma_{\B k}$ is modeled. One may use the estimate
$\gamma_{\B k}\sim k U_k$ where $U_k$ is the typical velocity
amplitude on the inverse scale of $k$, which is evaluated as
$U^2_k\sim k^3 \FT^{\alpha\alpha}(\B k)$.
\begin{equation}
\gamma_{\B k} =  C_\gamma k^{5/2}\sqrt {\FT^{\alpha\alpha}(\B k)} \
. \label{gamma}
\end{equation}
In isotropic turbulence Eqs. (\ref{Integral}) and (\ref{gamma})
have an exact solution with K41 scaling exponents,
\be
\FT^{\alpha\beta}_0(\B k) = P^{\alpha\beta}(\B k) F(k), \quad
F(k) = C\epsilon^{2/3} k^{-11/3}, \quad \gamma_k= \tilde
C_\gamma \epsilon^{1/3} k^{2/3}. \label{j0}
\ee
Note that the scaling exponents in $\B k$-representation, denoted as $\tilde
\zeta$, have a
$d$-dependent difference from their numerical value in $\B r$-
representation. In 3-dimensions $\tilde \zeta^{(2)}=\zeta^{(2)}+3$, and
the exponent 2/3 turns to 11/3 in Eq.(\ref{j0}).
For weak anisotropic turbulence Eq.(\ref{Integral}) will pose a
{\em linear} problem for the anisotropic components which depends
on this isotropic solution.
\subsubsection{Closure with Weak Anisotropy}
In weakly anisotropic turbulence one has to  consider a small anisotropic
correction $f^{\alpha\beta} (\B k)$ to the fundamental isotropic
background
$$
\FT^{\alpha\beta}(\B k)=\FT^{\alpha\beta}_0(\B k) + f^{\alpha\beta}
(\B k).$$
The first term vanishes with the solution (\ref{j0}). Linearizing
the  integral equation with respect to the anisotropic correction
leads to:
\begin{eqnarray}
&&I^{\alpha\beta}(\B k)\!=\!\! \int\!\!\frac {d^3q d^3p}{(2\pi)^3}
\delta(\B k+\B p+\B q)
[S^{\alpha\beta\gamma\delta} (\B k,\B q,\B p) f^{\gamma\delta}(\B k)
+2T^{\alpha\beta\gamma\delta} (\B k,\B q,\B p)
f^{\gamma\delta}(\B q)]=0 \ , \nonumber\\
&&S^{\alpha\beta\gamma\delta} (\B k,\B q,\B p)\equiv
\frac{\delta\Phi^{\alpha\beta} (\B k,\B q,\B p)}{\delta
\FT^{\gamma\delta}(\B k)}\ ,\quad
T^{\alpha\beta\gamma\delta} (\B k,\B q,\B p)\equiv
\frac{\delta\Phi^{\alpha\beta} (\B k,\B q,\B p)}{\delta
\FT^{\gamma\delta}(\B q)} \ . \label{allj}
\end{eqnarray}
We reiterate that the functional derivatives in Eq.(\ref{allj})
are calculated in the isotropic ensemble. In computing these
derivatives we should account also for the implicit dependence
of  $\Theta(\B k,\B q,\B p)$ on the correlation function through
Eq. (\ref{gamma}). We can rewrite Eq. (\ref{allj}) in a way that
brings out explicitly the linear integral operator $\hat L$,
\begin{equation}
\hat L |\B f\rangle\equiv \int\!\!\frac {d^3q}{(2\pi)^3} {\mathcal
L}^{\alpha\beta\gamma\delta}(\B k,\B q) f^{\gamma\delta}(\B q) = 0
\ , \label{opereq}
\end{equation}
where the kernel of the operator is
\begin{equation}
{\mathcal L}^{\alpha\beta\gamma\delta}(\B k,\B q)\equiv \delta(\B
k-\B q)\int\frac {d^3p}{(2\pi)^3}S^{\alpha\beta\gamma\delta} (\B
k,\B p,-\B k-\B p) +2T^{\alpha\beta\gamma\delta}
(\B k,\B q,-\B k-\B q) \ . \label{oper}
\end{equation}

\subsubsection{Symmetry Properties of the Linear Operator}
The first observation to make is that the linear operator is
invariant under all rotations. Accordingly we can block
diagonalize it by expanding the anisotropic perturbation in the
irreducible representation of the SO(3) symmetry group. These
have principal indices $j$ with an integer $j$ going from 0
to $\infty$. The zeroth component is the isotropic sector.
Correspondingly the  integral equation takes the form
\begin{equation}
I^{\alpha\beta}(\B k) = I^{\alpha\beta}_0(\B
k)+\sum_{j=1}^\infty I^{\alpha\beta}_j(\B k) =0 \ .
\label{intj}
\end{equation}
The block diagonalization implies that each $j$-block provides
an independent set of equations (for every value of $\B k$):
$
I^{\alpha\beta}_j(\B k) =0$.
The first term of (\ref{intj}) vanishes with the solution
(\ref{j0}). For all higher values of $j$ we need to solve the
corresponding equation
\begin{equation}
\hat L |\,\B f_j\rangle = 0 \ . \label{operell}
\end{equation}
We can block diagonalize further by exploiting additional
symmetries of the linear operator. In all  discussion we
assume that the  turbulent flow has zero helicity. Correspondingly
all the correlation functions are invariant under the inversion
of $\B k$.
Consequently there are no odd $j$ components, and we can write
$$
f^{\alpha\beta} (\B k)= \sum_{j=2,4,...}^\infty
f^{\alpha\beta}_j (\B k).$$
We also note that in general $\B u(-\B k)=\B u^*(\B k)$.
Accordingly, the correlation functions are real. From this fact
and the definition it follows that the correlation functions are
symmetric to index permutation,
$ \FT^{\alpha\beta}_0(\B k) =\FT^{\beta\alpha}_0(\B k)$ and
$f^{\alpha\beta}_j (\B k)=f^{\beta\alpha}_j (\B k)$.
As a result the  linear operator is invariant to permuting the
first ($\alpha,\beta$) and separately the second
($\gamma,\delta$) pairs of indices. In addition, the operator is
symmetric to $\B k\to -\B k$ together with $\B q\to -\B q$. This
follows from the inversion symmetry  and from the appearance
of products of two interaction amplitudes (which are
antisymmetric under the inversion of all wave-vectors by
themselves).
Finally, the  kernel is a homogeneous function of the wavevectors,
meaning that in every block we can expand in terms of basis
functions that have a definite scaling behavior, being
proportional to $k^{-\tilde\zeta}$.
\subsubsection{SO(3) Decomposition} \label{SO3} As a result of the
symmetry properties the operator $\hat L$ is block diagonalized
by tensors that have the following properties:
\begin{itemize}
\item They belong to a definite sector $(j, m)$ of the SO($3$) group.
\item They have a definite scaling behavior.
\item They are either symmetric or antisymmetric under permutations of
  indices.
\item They are either even or odd in $\B k$.
\end{itemize}
We have already explicitly presented the tensors involved
 for the case of passive vector advection. Here
we only quote the final results translated into $\B k$ space.
In every sector $(j, m)$ of
the rotation group with $j > 1$, one can find 9 independent
tensors $X^{\alpha\beta}(\B k)$ that scale like
$k^{-x}$. They are given by $k^{-x} \tilde
B_{j,j m}^{\alpha\beta}(\hat{\B k})$, where the index $j$ runs
from 1 to 9, enumerating the different spherical tensors. The
unit vector $\hat{\B k}\equiv \B k/k$. These nine tensors can be
further subdivided into four subsets exactly like the real-space
decomposition of
Sect.~\ref{alternative}:
\begin{itemize}
\item \textbf{Subset I} of 4 symmetric tensors with $(-)^j$ parity.
\item \textbf{Subset II} of 2 symmetric tensors with $(-)^{j+1}$ parity.
\item \textbf{Subset III} of 2 antisymmetric tensors with $(-)^{j+1}$
parity.
\item \textbf{Subset IV} of 1 antisymmetric tensor with $(-)^j$ parity.
\end{itemize}
Due to the diagonalization of $\hat L$ by these subsets, the
equation for the zero modes foliates, and we can compute the zero
modes in each subset separately. In this subsection, we choose to
focus on subset I, which has the richest structure. The four
tensors in this subset are given  here by
\begin{eqnarray}
  \tilde B_{1,j m}^{\alpha\beta}(\hat{\B k}) &=&
        k^{-j-2}k^\alpha k^\beta \phi_{j m}(\B k) \nonumber , \\
 \tilde B_{2,j m}^{\alpha\beta}(\hat{\B k}) &=&
        k^{-j}[k^\alpha \partial^\beta + k^\beta\partial^\alpha]
          \phi_{j m}(\B k) \ , \nonumber \\
 \tilde B_{3,j m}^{\alpha\beta}(\hat{\B k}) &=&
        k^{-j}\delta^{\alpha\beta} \phi_{j m}(\B k) \ , \nonumber \\
 \tilde B_{4,j m}^{\alpha\beta}(\hat{\B k}) &=&
        k^{-j+2}\partial^\alpha \partial^\beta \phi_{j m}(\B k) \ ,
       \label{basis}
\end{eqnarray}
where $\phi_{j m}(\B k)$ are the standard spherical harmonics.

The last property to employ is the incompressibility of the target
function $f^{\alpha\beta} (\B k)$. Examining the basis
(\ref{basis}) we note that we can find two linear combinations
that are transverse to $\B k$ and two linear combinations that are
longitudinal in $\B k$. We need only the former, which have the
form
\begin{eqnarray}
\label{basis1}
 B_{1,j m}^{\alpha\beta}(\hat{\B k}) &=&
        k^{-j}P^{\alpha\beta}(\B k) \phi_{j m}(\B k) ,
\\
 B_{2,j m}^{\alpha\beta}(\hat{\B k}) &=&
        k^{-j}[k^2 \partial^\alpha \partial^\beta -(j-1)( k^\beta
\partial^\alpha
+k^\alpha\partial^\beta) +j(j-1)
\delta^{\alpha\beta} ]
          \phi_{j m}(\B k) \ .\nonumber
\end{eqnarray}
Using this basis we can now expand the  target function as
\begin{equation}
\label{expand} f^{\alpha\beta}_j (\B k) = k^{-\tilde\zeta_j^{(2)}}
   \Big[ c_1 B_{1,j m}^{\alpha\beta}(\hat{\B k}) +
         c_2 B_{2,j m}^{\alpha\beta}(\hat{\B k})
    \Big] \ .
\end{equation}
\subsubsection{Calculation of the Scaling Exponents}
\label{calculation}
Substituting Eq.(\ref{expand}) into Eq.(\ref{operell}) we find
\begin{equation}
\hat L q^{-\tilde\zeta_j^{(2)}}|\B B_{1,j m}\rangle c_1+\hat L
q^{-\tilde\zeta_j^{(2)}}|\B B_{2,j m}\rangle c_2 =0 \ . \label{LonB}
\end{equation}
Projecting this equation on  the two function of the basis
(\ref{basis1}) we obtain for the  matrix $L_{i,l}(j, \tilde\zeta_j^{(2)})
\equiv \langle \B B_{i,j m}|\hat
L q^{-\tilde\zeta_j^{(2)}}|
\B B_{l,j m}\rangle$ the form:
\begin{equation}
\label{mateq}
L_{i,l}(j, \tilde\zeta_j^{(2)}) = \int\!\!\frac {d^3q}{(2\pi)^3}\,d\hat {\B
k}\,B^{\alpha\beta}_{i,j m}(\hat{\B k}) {\mathcal
L}^{\alpha\beta\gamma\delta}(\B k,\B
q)q^{-\tilde\zeta_j^{(2)}}B^{\gamma\delta}_{l,j m}(\hat{\B q}).
\ee
Here we have full integration with respect to $\B q$, but only
angular integration with respect to $\B k$. Thus the matrix
depends on $k$ as a power, but we are not interested in this
dependence since we demand the solvability condition
\begin{equation}
\det L_{i,l}(j, \tilde\zeta_j^{(2)}) = 0 \ . \label{solvability}
\end{equation}
It is important to stress that in spite of the explicit $m$
dependence of the basis functions, the matrix obtained in this way
has no $m$ dependence. In the calculation below we can therefore
put, without loss of generality, $m=0$. This is like having
cylindrical symmetry with a symmetry axis in the direction of the
unit vector $\hat{\B  n}$. In this case we can write the matrix
$\B  B_{i, j}(\hat{\B  k})$ (in the vector space $\alpha,\,
\beta= x,\,y,\,z$) as
\begin{equation}\label{B-as-operator}
B^{\alpha\beta}_{i, j}(\hat{\B  k}) = k^{-j} \hat {\C
B}^{\alpha\beta}_{i, j,\B  k} (k^{j} P_{j} (\hat{\B  k}
\cdot \hat{\B  n}))\,,
\end{equation}
where $\hat {\C  B}^{\alpha\beta}_{i, j,\B  k}$ are matrix
operators, acting on wave vector $\B  k$:
\begin{eqnarray}\label{def-B-operator}
\hat {\C  B}^{\alpha\beta}_{1, j,\B  k} & \equiv  &
\delta^{\alpha\beta} - \frac{k^\alpha k^\beta}{k^2}, \\  \nonumber
\hat {\C  B}^{\alpha\beta}_{2, j,\B  k} & \equiv  & \frac{
k^2\, \partial^2}{ \partial k^\alpha \partial k^\beta} - (j -
1) \Big ( \frac{   k^\alpha \partial}{\partial k^\beta}
 +\frac{   k^\beta \partial}{\partial k^\alpha }
- j \, \delta^{\alpha\beta}\Big ) \, ,
\end{eqnarray}
and $P_j(x)$ denote $j$-th order Legendre polynomials.
The technical details of the calculations
were presented in \cite{lvo03}. Here we
present and discuss the results.
\subsubsection{Results and Concluding Remarks}
\label{conclusions}
The determinants $\det[L_{i,l}(j,\tilde\zeta^{(2)}_j]$ were computed as
functions of the scaling exponents $\tilde \zeta^{(2)}_j$ in every
$j$-sector
separately, and the scaling exponent was determined from the zero
crossing. The procedure is exemplified in Fig. \ref{Fig.1v} for
the isotropic sector $j=0$. We expect for this sector
$\tilde\zeta^{(2)}_0=11/3$, in accordance with $\zeta_0^{(2)}=2/3$. Indeed,
for
both decay models, i.e the exponential decay (\ref{expdecay}),
shown in dark line, and the Gaussian decay (\ref{gaussdecay})
shown in light line, the zero crossing occurs at the same point,
which in the inset can be read as 3.6667.
\begin{figure}
\includegraphics[scale=0.45]{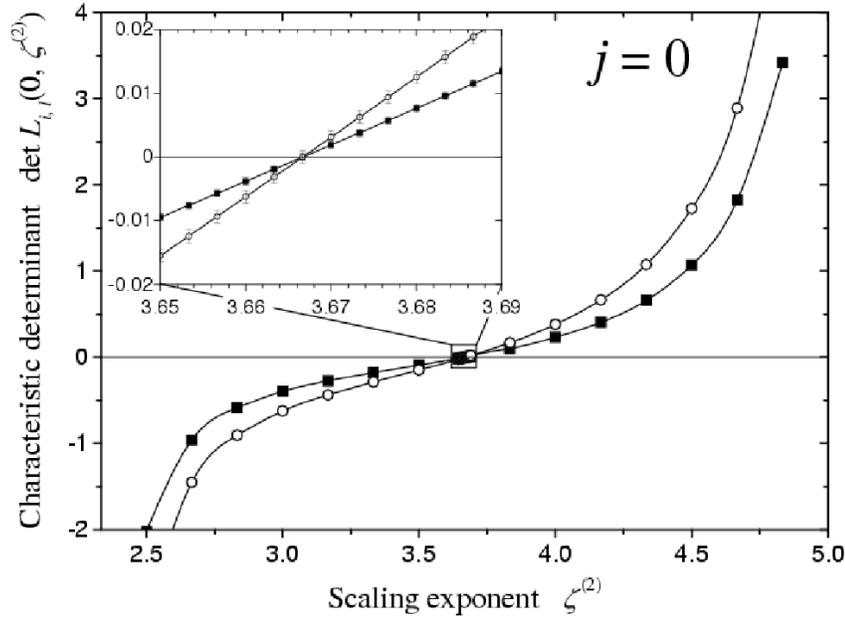}
\caption{determinant and zero crossing for the sector $j=0$. The scaling
exponent
computed from the zero crossing is $\zeta^{(2)}_0\approx
0.667$.} \label{Fig.1v}
\end{figure}
For the higher $j$-sectors the agreement between the
exponential and Gaussian models is not as perfect, indicating
that the  procedure is not exact. In Fig. \ref{Fig.2} we present
the determinant and zero crossings for $j=2$.
\begin{figure}
\includegraphics[scale=0.45]{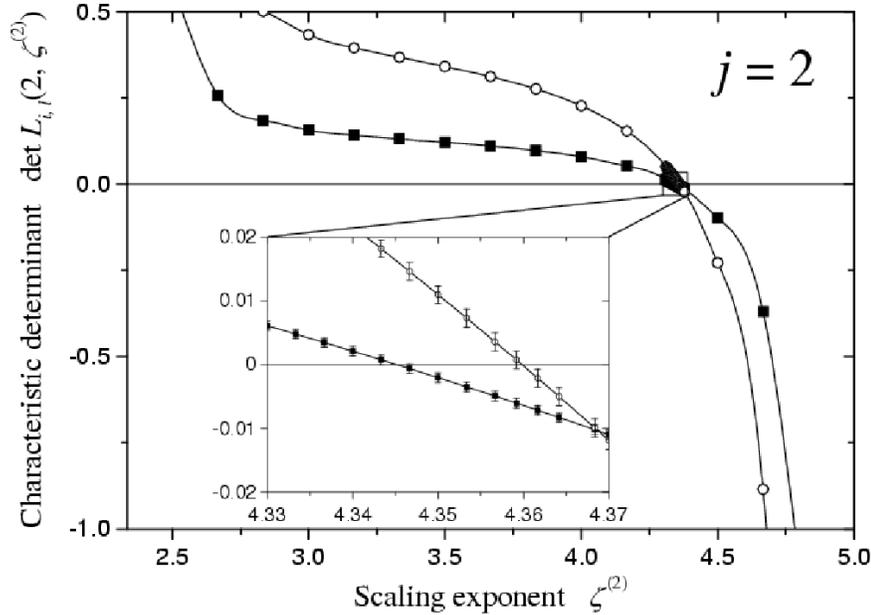}
\caption{determinant and zero crossing for the sector $j=2$. The scaling
exponent
computed from the zero crossing is $\zeta^{(2)}_2\approx
1.36-1.37$.} \label{Fig.2}
\end{figure} From the inset we can read the exponents
$\tilde\zeta_2^{(2)}=4.351$ and
4.366 for the exponential and Gaussian models respectively. This
is in correspondence with $\zeta_2^{(2)}=1.351$ and $1.366$
respectively. These numbers are in excellent correspondence with
the experimental measurements reported in
\cite{ara98,kur00a}, cf. the next
Section.
The results for $j=4$ are presented in Fig. \ref{Fig.3}. Here
the zero crossing, as seen in the inset, yields very close
results for $\tilde\zeta_4^{(2)}$ between the exponential and Gaussian decay
models, i.e. $\tilde\zeta_4^{(2)}\approx 4.99$. Note that this result is
very close to the boundary of locality as discussed in \cite{lvo03}.
Nevertheless the zero crossing is still easily
resolved by the numerics, with the prediction that
$\zeta_4^{(2)}\approx 1.99$. The simulation estimate of this number
in \cite{bif01a} was $1.7\pm 0.1$. We note that while the  result
$\zeta_4^{(2)}\approx 1.99$ is not within the error bars of the
simulational estimate, it is very possible that the closeness of
the exponent to the boundary of the window of locality gives rise
to very slow convergence to asymptotic scaling. We therefore have
to reserve judgment about the agreement with simulations until
larger scaling ranges were available.

Similar results are obtained for $j=6$, see Fig. \ref{Fig.4}.
Also this case exhibits zero crossing close to the boundary of
locality, with $\tilde\zeta_6^{(2)}\approx 6.98$. Again we find close
correspondence between the exponential and Gaussian models. In
terms of $\zeta^{(2)}$ this means $\zeta_6^{(2)}\approx 3.98$. This
number appears higher than the simulational result from
\cite{bif01a}, which estimated $\zeta_6^{(2)}\approx 3.3\pm 0.3$. We
note however that for $j=6$ the log-log plots measured in DNS
 \cite{bif01a}
possess a short scaling range.
\begin{center}
\begin{figure}
\includegraphics[scale=0.45]{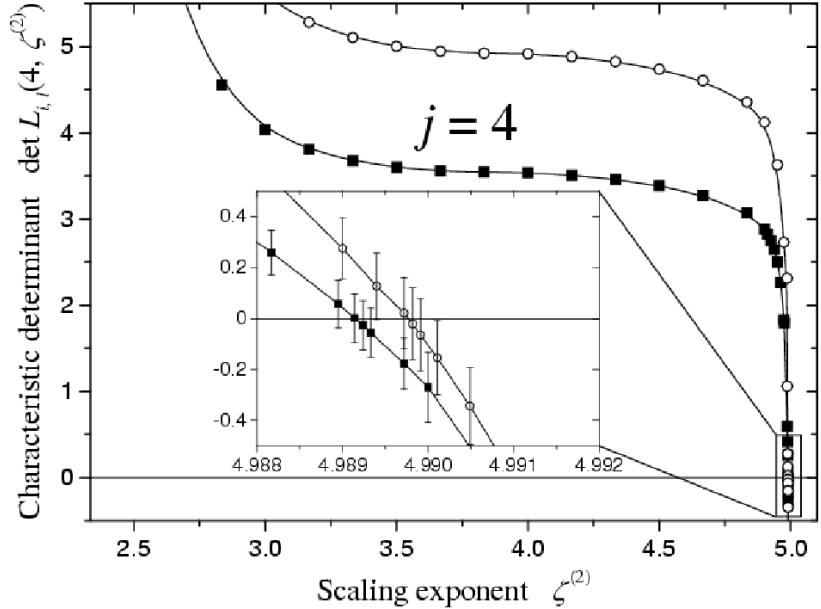}
\caption{determinant and zero crossing for the sector $j=4$. The scaling
exponent
computed from the zero crossing is $\zeta^{(2)}_4\approx 1.99$.}
\label{Fig.3}
\end{figure}
\end{center}
\begin{figure}
\includegraphics[scale=0.45]{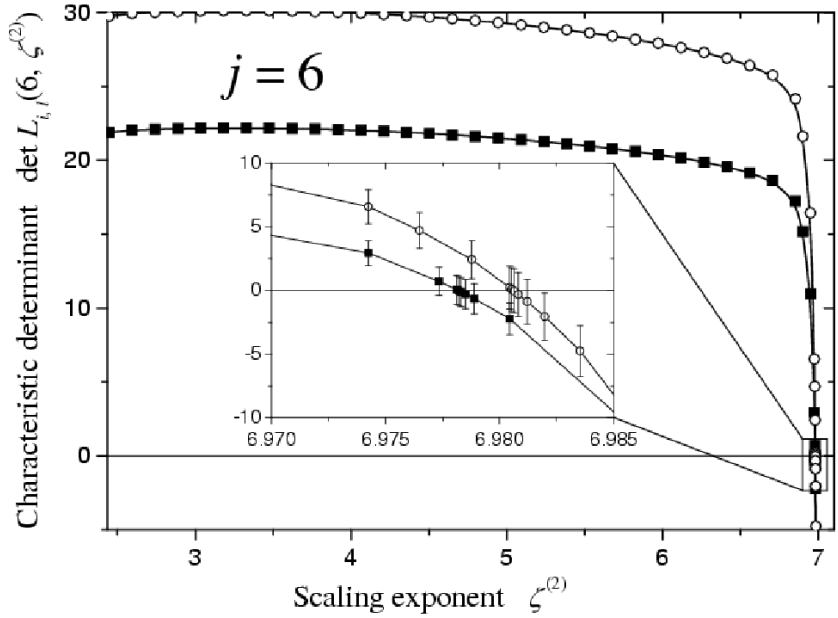}
\caption{determinant and zero crossing for the sector $j=6$. The scaling
exponent
computed from the zero crossing is $\zeta^{(2)}_6\approx 3.98$.}
\label{Fig.4}
\end{figure}
Interestingly enough, the set of exponents $\zeta_j^{(2)}$=2/3,
1.36, 1.99 and 3.98 for $j=$0, 2, 4 and 6 respectively are in
close agreement with the numbers obtained for the linear pressure
model, $\xi_j^{(2)}$=2/3, 1.25226, 2.01922, 4.04843, for
$j=0,2,4$ and 6 respectively. We reiterate at this point that
the latter set is exact for the linear pressure model, whereas
the former set is obtained within the closure approximation. In fact, the
close correspondence
is not so surprising since he linearization of Navier-Stokes equations for
small anisotropy
results in a linear operator which is very close to the one that exists
naturally in the Linear
Pressure Model.
Numerical results \cite{bif01a,bif02,bif03a} obtained at moderate Re
and  with strong anisotropies
show a small disagreement with the numbers calculated  in the closure
approximation.
We do not expect a much more precise
theoretical evaluation of these exponents before numerical and experimental
data at higher Re are obtained and the
intermittency problem in the isotropic sector is fully settled.
\section{Analysis of Experimental Data}
\label{chap:experiment}
The major difficulty in applying the SO(3) decomposition to
experimental data lies in the fact that one never has the whole
field $\B u(\B x)$. We thus cannot project the statistical objects
onto chosen basis functions $\B B_{qj m}$ and simply integrate out
all other contributions. Rather, we need to extract the wanted information
laboriously by fitting  partially resolved data,
or to measure quantities that do not have projections on the isotropic
sector, to see right away the anisotropic contributions. We begin with the
first
approach.
\subsection{Anisotropic Contribution to the Statistics of the
Atmospheric Boundary Layer}
\label{atmospheric}
The atmospheric boundary layer offers a natural laboratory of
turbulence that is unique in offering extremely high Re
number. Students of turbulence interested in the scaling
properties that are expected to be universal in the limit $Re\to
\infty$ are thus attracted to atmospheric measurements. On the
other hand the boundary layer suffers from strong inhomogeneity
(explicit dependence of the mean velocity on the height) which
leads to strong anisotropies such that the vertical and the
horizontal directions are quite distinguishable. In addition, one
may expect the boundary layer to exhibit large-scale quasi
2-dimensional eddys whose typical decay times and statistics may
differ significantly from the generic 3-dimensional case. The aim
of this section is to review systematic methods of data analysis
that attempt to resolve such difficulties, leading to a useful
extraction of the universal, 3-dimensional aspects of turbulence.

Obviously, to isolate tensorial components belonging to other than
isotropic sectors one needs to collect data from more than one vector
component of the velocity field.  Having {\em two} probes is actually
sufficient to read surprisingly rich information about anisotropic
turbulence. In the experiments discussed in this subsection two types
of geometry were employed, one consisting of two probes at the same
height above the ground and the other with the two probes separated
vertically. In both cases the inter-probe separation is orthogonal to
the mean wind. 
\subsubsection{Experiments, Data Sets and the Extraction of
Structure Functions}
\label{expsetup}
The results presented in this subsection are based on two experimental
setups \cite{ara98,kur00,kur00a}, which are denoted
throughout as {\rm I} and {\rm II}
respectively. In both setups the data were acquired over the salt
flats in Utah with a long fetch. In set {\rm I} the data were
acquired simultaneously from two single hot wire probes at a
height of 6 m above the ground, with a horizontal separation of
55 cm, nominally orthogonal to the mean wind. The Taylor
microscale Reynolds number was about 10,000. Set {\rm II} was
acquired from an array of three cross-wires, arranged {\em above}
each other at heights 11 cm, 27 cm and 54 cm respectively. The
Taylor microscale Reynolds numbers in this set were 900, 1400 and
2100 respectively.
Table~1 lists a few relevant facts about the data records
analyzed here. The various symbols have the following meanings:
$\overline U$ = local mean velocity, $u^{\prime}$ =
root-mean-square velocity, $\eb$ = energy
dissipation rate obtained by the assumption of local isotropy and
Taylor's hypothesis, $\eta$ and $\lambda$ are the Kolmogorov and
Taylor length scales, respectively, the microscale Reynolds number
$R_{\lambda} \equiv u^{\prime} \lambda/\nu$, and $f_s$ is the
sampling frequency.

For set I it is important to  test whether the separation between the two
probes is indeed orthogonal to the mean wind. (We do not need to
worry about this point in set {\rm II}, since the probes are
above each other). To do so one computes the cross-correlation
function $\langle u_1(t+\tau)u_2(t)\rangle$. Here, $u_1$ and
$u_2$ refer to velocity fluctuations in the direction of the mean
wind, for probes 1 and 2 respectively. If the separation were
precisely orthogonal to the mean wind, this quantity should be
maximum for $\tau=0$. Instead, for set I, it was  found that the maximum
shifted slightly to $\tau=0.022$ s, implying that the separation
was not precisely orthogonal to the mean wind. To correct for
this effect, the data from the second probe were time-shifted by
0.022 s. This amounts to a change in the actual value of the
orthogonal distance. The  effective distance is
$\Delta \approx 54$ cm (instead of the 55 cm that was set
physically).
\begin{table}
\begin{tabular} {|c|c|c|c|c|c|c|c|c|}
\hline
Height&$\overline U$ & $u^\prime$ &$10^2 \langle \eb
\rangle $,& $\eta$ & $\lambda$ & $R_{\lambda}$ & $f_s,$ per & \#
of \\meters& ms$^{-1}$ & ms $^{-1}$
& m $^2$ s$^{-3}$ & mm & cm & & channel, Hz & samples\\
\hline 6 & 4.1 & 1.08 & $1.1$ & 0.75 & 15 & 10,500 & 10,000 & $4 \times
10^7$\\
\hline\hline 0.11 & 2.7
& 0.47 & $6.6 $ & 0.47 & 2.8 & 900 & 5,000 & $ 8 \times 10^6$\\
0.27 &3.1 & 0.48 & 2.8& 0.6 & 4.4 & 1400& 5,000 &$8 \times 10^6$\\
0.54 &3.51 &0.5& 1.5& 0.7& 6.2&2100& 5,000& $8 \times
10^6$\\
\hline
\end{tabular}
\caption{Data sets I (first line) and II (second-fourth lines).}
\end{table}
The  coordinates were chosen such that the mean wind direction is along
the 3-axis, the vertical is  along the 1-axis and the third
direction orthogonal to these is the 2-axis. We denote these
directions by the three unit vectors $\hat {\B n}$, $\hat {\B m}$,
and $\hat {\B p}$ respectively. The raw data available from set I
is $u^{(3)}(t)$ measured at the positions of the two probes. In
set II each probe reads a linear combination of $u^{(3)}(t)$ and
$u^{(1)}(t)$ from which each component is extractable. From this
raw data we would like to compute the scale-dependent structure
functions, using the Taylor hypothesis to surrogate space for
time. This needs a careful discussion.
\subsubsection{Theoretical constructs: the Taylor Hypothesis, Inner
and Outer Scales} \label{Taylor}
Decades of research on the statistical aspects of thermodynamic
turbulence are based on the Taylor Hypothesis \cite{tay38}, which
asserts that the fluctuating velocity field measured by a given
probe as a function of time, $\B u(t)$ is the same as the velocity
$\B u(r/\overline{U})$ where $\overline{U}$ is the mean velocity
and $r$ is the distance to a position ``upstream'' where the
velocity is measured at $t=0$. The natural limitation on the
Taylor hypothesis is provided by the typical decay time of
fluctuations of scale $r$. Within a K41 scaling theory this time
scale is the turn-over time $r/\sqrt{S(r)}$ where $S(r)\equiv
S^{\alpha\alpha}(r)$. With this estimate the Taylor Hypothesis is
expected to be valid when $\sqrt{S(r)}/\overline{U} \to 0$. Since
$S(r)\to 0$ when $r\to 0$, the Taylor hypothesis becomes exact in
this limit. We will use this to calibrate the units when we
employ two different probes and read a distance from a
combination of space and time intervals.

The Taylor hypothesis has also been employed when the mean
velocity vanishes, and instead of $\overline{U}$ one uses the
root-mean-square $u'$. Ref.\cite{lpp99} has presented a
detailed analysis of the consequences of the Taylor hypothesis on
the basis of an exactly soluble model. In particular
 ways were proposed there  to minimize the systematic errors introduced by
the
use of the Taylor hypothesis. In light of that analysis we will
use here an ``effective" wind $U_{eff}$ which for surrogating the
time data of a single probe is made of a combination of the mean
wind $\overline{U}$ and the root-mean-square $u'$,
\begin{equation} U_{eff}
\equiv \sqrt{\overline{U}^2+(\tilde b u')^2} \ , \label{defUeff}
\end{equation}
where $\tilde b$ is a dimensionless parameter. Evidently, when we employ
the Taylor hypothesis in log-log plots of structure functions
using time series measured in a {\em single} probe, the value of
the parameter $\tilde b$ is irrelevant, changing just the (arbitrary)
units of length (i.e the arbitrary intercept). When we used data
collected from two probes, we mix read distance and surrogated
distance, and the parameter $\tilde b$ becomes a unit fixer. The
numerical value of this  parameter is found in \cite{lpp99} by
the requirement that the surrogated and directly measured
structure functions coincide in the limit $r\to 0$. When we do
not have the necessary data we will use values of $\tilde b$ suggested
by the exactly soluble model treated in \cite{lpp99}. The choice
of these values can be justified a posteriori by the quality of the
fit of to the predicted scaling functions.

When we have two probes placed at different heights the mean
velocity and $u'$ as measured by each probe do not coincide. In
applying the Taylor hypothesis one needs to decide which value of
$U_{eff}$ is most appropriate. This question has been addressed
in detail in ref.\cite{lpp99}
 with the final conclusion that the
choice depends on the velocity profile between the probe. In the
case of {\em linear} shear the answer is the precise average
between the two probes, \begin{equation} U_{eff} \equiv
\sqrt{{\overline{U_1}^2+\overline{U_2}\over 2}^2+\tilde b{{u_1'}^2
+{u_1'}^2\over2}} \ , \label{defUeff2}
\end{equation}
where the subscripts 1,2 refer to the two probes respectively.
In all the subsequent expressions we will therefore denote separations by
$r$, and invariably this will mean Taylor-surrogated time differences. The
effective velocity will be (\ref{defUeff}) or (\ref{defUeff2}) depending
on having probes at the same height or at different heights. The value of
$\tilde b$ will be $\tilde b=3$ following ref.\cite{lpp99}. It can be shown
 that the
computed scaling exponents are not sensitive to the changing $b$.
(They change by a couple of percents upon changing $b$ by 30\%.)
In seeking scaling behavior one needs to find the inner and outer
scales. Below the inner scale all structure functions have an
analytic dependence on the separation, $S(r)\sim r^2$, and above
the outer scale the structure functions should tend to a constant
value. We look at the longitudinal structure functions
$$
S^{33}(r) = \langle(u^3(x + r) - u^3(x))^2\rangle $$
computed from a single probe in set I and from the probe at
$0.54$m in set II , see Fig.\ref{susanpaper.scrange_long.eps}.
\begin{figure}
\includegraphics[scale=0.7]{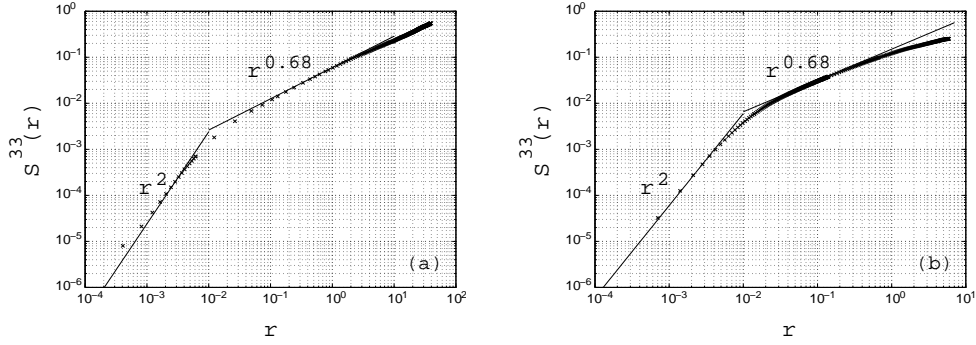}
\caption{Raw log-log plot of the longitudinal component of the
2nd order structure function.} \label{susanpaper.scrange_long.eps}
\end{figure}
We simultaneously consider the transverse structure function
$$
S^{11}(r) = \langle(u^1(x + r) - u^1(x))^2\rangle $$
computed from the probe at $0.54$m in set II, see Fig.
\ref{susanpaper.scrange_trans.eps}.
\begin{figure}
\includegraphics[scale=0.7]{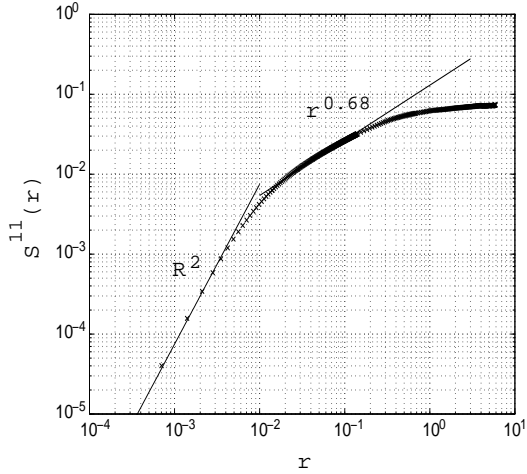}
\caption{Raw log-log plot of the transverse component of the 2nd
order structure function.} \label{susanpaper.scrange_trans.eps}
\end{figure}
The spatial scales are computed using the local mean wind in both
cases since  the scaling exponent for the
single-probe structure function are not expected
to be affected by the choice of
convection velocity. This choice does determine the value of $r$
corresponding to a particular time scale however. One may expect
that any correction to the numerical value of $r$ is small for a
different choice of convection velocity, and not crucial for the
qualitative statements that follow. In
Fig.~\ref{susanpaper.scrange_long.eps} we clearly see the $r^2$
behavior characterizing the transition from the dissipative to
the inertial range. As is usual, this behavior persists about a
half-decade above the ``nominal" Kolmogorov length scale $\eta$.
There is a region of cross-over and then the isotropic scaling
$\sim r^{0.68}$ expected for small scales in the inertial range
begins. We thus have no difficulty at all in identifying the
inner scale, it is simply revealed as a natural crossover length
in this highly resolved data.
\noindent
We understand by now that we cannot expect to be
able to fit with this single exponent for larger scales and must
include scaling contributions due to anisotropy. We expect that
the contributions due to anisotropy will account for scaling
behavior up to the outer scale of a 3-dimensional flow patterns.
The question therefore is how to identify what this large scale
is.
One approach would be to simply use the scale where the structure
function tends to a constant, which corresponds to the scale
across which the velocity signal has decorrelated. It becomes
immediately apparent that this is not a reasonable estimate of
the relevant large scale. Fig.~\ref{susanpaper.scrange_long.eps}
shows that the structure function stays correlated up to scales
that are at least an order of magnitude larger than the height at
which the measurement is made. On the other hand, if we look at
the transverse structure function computed from the probe at
$0.54$m, Fig.~\ref{susanpaper.scrange_trans.eps} we see that it
ceases to exhibit scaling behavior at a scale that is of the
order of the height of the probe.
\noindent 
It appears that we are observing extremely flat eddys that are
correlated over very long distances in the horizontal direction
but have a comparatively small correlation lengths in the
direction perpendicular to the boundary. Since we know that the
presence of the boundary must limit the size of the largest
3-dimensional structures, the height of the probe should be
something of an upper bound on the largest 3-dimensional flow
patterns that we can detect in experiments. Thus we arrive at a
qualitative understanding of the kind of flow that is observed in
these atmospheric measurements. The size of the largest
3-dimensional structures is determined by the decorrelation
length of the transverse structure function. This is because the
transverse components of the velocity are unaffected by the
extended, persistent, 2-dimensional eddys that govern the
behavior of the longitudinal components. The theory of scaling
behavior in 3-dimensional turbulence can usefully be applied to
only those flow patterns that are truly 3-dimensional. The
extended flat eddys must be described in terms of a separate
theory, including maybe notions of 2-dimensional turbulence which
has very different scaling properties \cite{kra67}. Such considerations
are outside the scope of this review.
Rather, in the following
analysis,  the outer-scale was chosen to be of the order of the
decorrelation length of the {\em transverse} structure function
(where available) or of the height of the probe. We will see
below that up to a factor of 2 these are the same; taking $L$ to
be as twice the height of the probe is consistent with all
data. We use this estimate in the  study of both transverse and
longitudinal objects.
\subsubsection{Extracting the universal exponents of higher $j$
sectors}
\label{extract}
We consider the second order structure function
\begin{equation}\label{Sab} S^{\alpha\beta}({\bf r}) =
\la (u^\alpha({\bf x} + {\bf r}) - u^\alpha({\bf x})) (u^\beta({\bf
x} + {\bf r}) - u^\beta({\bf x})) \ra. \end{equation} The lowest
order anisotropic contribution to the symmetric (in indices),
even parity (in ${\bf r}$ due to homogeneity), second-order
structure function is the $j=2$ component of the SO(3) symmetry
group. Ref.~\cite{ara98} presents a derivation of the $m=0$
axisymmetric (invariant under rotation about the 3-axis) part of
the $j=2$ contribution to this structure function in homogeneous
turbulence.  The derivation of the full $j=2$ contribution to
the symmetric, even parity structure function appears in Appendix
\ref{app:fullj2}.
 Fig.~\ref{Fig.1}~b shows the
fit to the structure function computed from a single probe in set
I
\begin{equation}\label{tta0}
S^{33}(r,\theta = 0) = \langle (u_1^{(3)}(x+r) -
u_1^{(3)}(x))^2\rangle, \end{equation} where the subscript $1$
denotes one of the two probes, with just the $j=0$ contribution.
The best-fit exponent for the range $0<r/\Delta<4.5$ is
$\zeta^{(2)}_0=0.68\pm0.01$ (Fig.~\ref{Fig.1}~a). Above this range, was
impossible  to obtain a good fit to the data with just the
isotropic exponent and Fig.~\ref{Fig.1}~b shows the peel-off from
isotropic behavior above $r/\Delta = 4$.
\begin{center}
\begin{figure}
\includegraphics[scale=0.7]{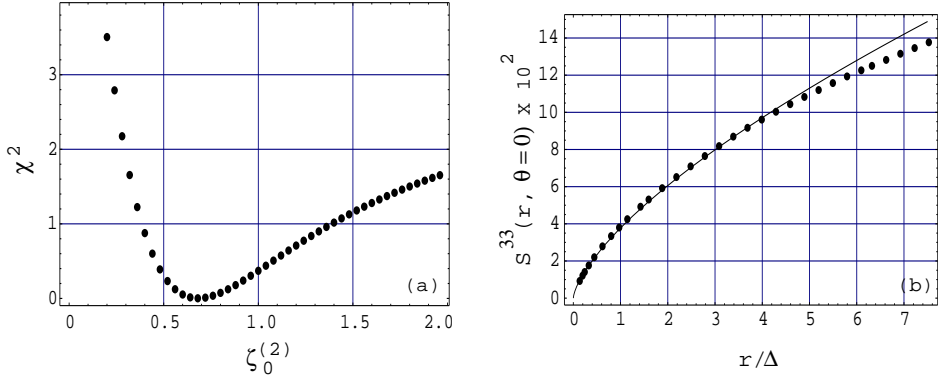}
\caption{The single-probe structure
function computed from data set I. (a) shows the $\chi^2$
minimization by the best-fit value of the exponent in the
isotropic sector $\zeta_0^{(2)}\approx 0.68$ for the single-probe
structure function in the range $0 < r/\Delta <4.5$. (b) shows
the fit using the best value of $\zeta_0^{(2)}$ obtained in (a),
indicating the peel-off from isotropic  behavior at the end of
the fitted range.}\label{Fig.1}
\end{figure}
\end{center}
To find the $j=2$ anisotropic exponent one needs to use data taken
from the two probes. To clarify the procedure we show in
Fig.\ref{Figexpsetup} the geometry of set I.
\begin{figure}
\includegraphics[scale=0.3]{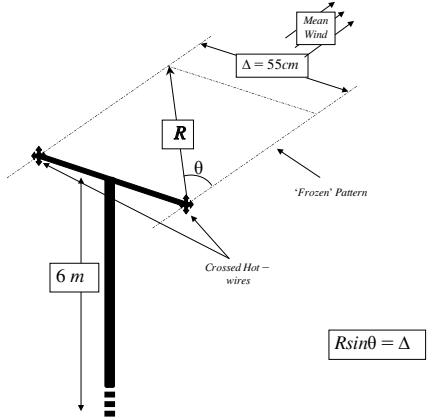}
\caption{Diagrammatic illustration of the experimental set-up.
Shown is the positioning of the probes with respect to the mean
wind and how the Taylor hypothesis is employed}
\label{Figexpsetup} \end{figure}
What was computed is actually
$$
S^{33}(r,\theta)=\langle [u^{(3)}_1( U_{eff} t + U_{eff}t_{\tilde
r})- u^{(3)}_2( U_{eff}t)]^2\rangle $$ Here
$\theta=\arctan(\Delta/ U_{eff}t_{\tilde r})$, $t_{\tilde
r}=\tilde r/ U_{eff}$, and $r=\sqrt{\Delta^2+(\bar
U_{eff}t_{\tilde r})^2}$. $U_{eff}$ was as in Eq.(\ref{defUeff})
with $b=3$. We will refer from now on to such quantities as
\begin{equation}\label{ttadep}
S^{33}(r,\theta) = \langle (u_1^{(3)}(x+r) -
u_2^{(3)}(x))^2\rangle \ . \end{equation}

Next, one may  fix the scaling exponent of the isotropic sector as
$0.68$ and find the $j=2$ anisotropic exponent that results from
fitting to the full $j=2$ tensor contribution. Finally, one needs
to  fit the objects
in Eqs.~(\ref{tta0}) and (\ref{ttadep}) to the sum of the $j=0$ (with
scaling exponent $\zeta^{(2)}_0 = 0.68$) and the $j=2$ contributions
(see Appendix \ref{app:fullj2})
\begin{eqnarray}
&&S^{33}(r,\theta)=S^{33}_{j=0}(r,\theta)+ S^{33}_{j=2}(r,\theta)
=c_0\left({r\over \Delta}\right)^{\zeta^{(2)}_0} \Big[ 2
+\zeta^{(2)}_0-\zeta^{(2)}_0 \cos^2\theta\Big] \nonumber\\
&+&a\left({r\over \Delta}\right)^{\zeta_2^{(2)}} \Big[
(\zeta_2^{(2)}+2)^2 -\zeta_2^{(2)}
(3\zeta_2^{(2)}+2)\cos^2\theta+2\zeta_2^{(2)}(\zeta_2^{(2)}-2)\cos^4\theta\B
ig] \nonumber\\
&+&b\left({r\over \Delta}\right)^{\zeta_2^{(2)}} \Big[
(\zeta_2^{(2)}+2) (\zeta_2^{(2)}+3)-
\zeta_2^{(2)}(3\zeta_2^{(2)}+4)\cos^2\theta+(2\zeta_2^{( 2)}+1)
(\zeta_2^{(2)}-2)\cos^4\theta\Big] \nonumber \\ &+&a_{9,2,1} \left({r\over
\Delta}\right)^{\zeta_2^{(2)}}
\Big[-2\zeta_2^{(2)} (\zeta_2^{(2)}+2) \sin\theta\cos\theta
+ 2\zeta_2^{(2)}(\zeta_2^{(2)}-2)\cos^3\theta\sin\theta \Big]\nonumber\\
&+&a_{9,2,2} \left({r\over
\Delta}\right)^{\zeta_2^{(2)}}\Big[-2\zeta_2^{(2)}
(\zeta_2^{(2)}-2) \cos^2\theta\sin^2\theta\Big]
\nonumber\\
&+&a_{1,2,2}\left({r\over \Delta}\right)^{\zeta_2^{(2)}}
\Big[-2\zeta_2^{(2)} (\zeta_2^{(2)}-2)\sin^2\theta\Big].
\label{fulltens}
\end{eqnarray}
The above fit was performed using values of $\zeta_2^{(2)}$ ranging
from $0.5$ to $3$. The best value of this exponent is the one that
minimizes the $\chi^2$ for the fits.  From Fig.\ref{chj2} one may read
the best value to to be $1.38\pm0.15$. The fits with this choice of
exponent are displayed in Fig.\ref{sfj2full}.  The corresponding
values of the 5 fitted coefficients can be found in the paper
\cite{ara98}.  The range of scales that are fitted to this expression
is $1 < r/\Delta < 25$.  We thus conclude that that structure
functions which is symmetric in $\B r$ exhibits scaling behavior
over the whole scaling range, but this important fact is missed if one
does not consider a superposition of the $j=0$ and $j=2$
contributions.
\begin{figure}
\includegraphics[scale=0.7]{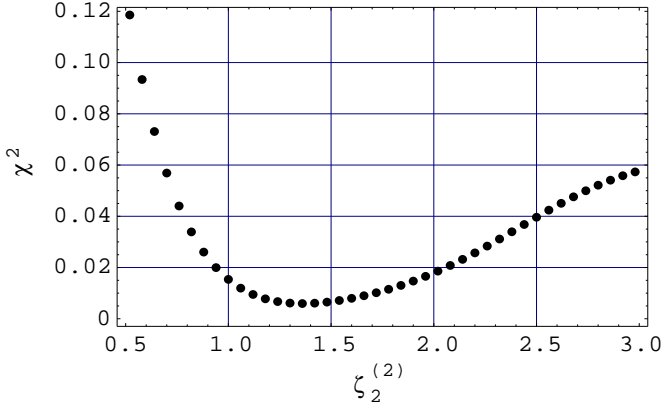}
\caption{The $\chi^2$ minimization
by the best-fit value of the exponent in the $j=2$ anisotropic
sector from the fit to both the $\theta=0$ and the
$\theta$-dependent structure function in the range $0 < r/\Delta
<25$. }\label{chj2} \end{figure}
\begin{figure}
\includegraphics[scale=0.7]{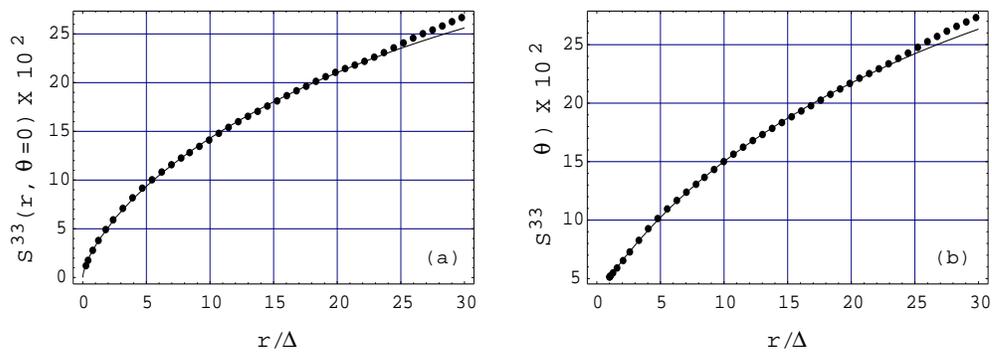}
\caption{The structure functions
computed from data set I and fit with the $j=0$ and full $j=2$
tensor contributions using the best fit values of exponents
$\zeta^{(2)}_0=0.68$ and $\zeta_2^{(2)}=1.38$ in the range $0 <
r/\Delta <25$. Panel (a) shows the fit to the single-probe
($\theta =0$) structure function and panel (b) shows the fit to
the $\theta$-dependent structure function.}\label{sfj2full}
\end{figure}
 Finally, let us note that the value of the exponent is
perfectly in agreement with the analysis of numerical simulations
\cite{ara99} in which one can comfortably integrate the
structure function against the basis functions, eliminating all
contributions except $j=2$ (see next section).
\subsubsection{Extracting the j=1 component}
\label{j1com}
In homogeneous flows, it follows from symmetry and parity that the often
computed and 
widely analyzed structure
function as defined in Eq. (\ref{Sab}) does not possess any contribution
from the
$j=1$ sector. The lowest order anisotropic contribution belongs to the
$j=2$ sector. In order to isolate the scaling
behavior of the $j=1$ contribution in atmospheric shear flows
we must either explicitly construct a new tensor object which will
allow for such a contribution, or see if it can be extracted
from  the structure function itself computed in the case of {\em
inhomogeneity}. We have pursued both avenues. In the former, we
construct the tensor
\begin{equation}\label{Tab}
T^{\alpha\beta}({\B  r}) = \langle u^\alpha({\B  x} + {\B  r}) -
u^\alpha({\B  x}))(u^\beta({\B  x} + {\B  r}) + u^\beta({\B
x}))\rangle. \end{equation} It is easily seen that the function
vanishes both in the case of $\alpha=\beta$ and when ${\B  r}$ is
in the direction of homogeneity. From data set II we can
calculate this function for non-homogeneous (in the shear
direction) scale-separations. In general, this will exhibit mixed
parity and symmetry and therefore, to minimize as far as possible
the final number of fitting parameters we look at only the
antisymmetric contribution. We derive the tensor contributions in
the $j=1$ sector for the antisymmetric case in Appendix \ref{app:fullj1}
and
use this to fit for the unknown $j=1$ exponent.
Below we describe the results of this analysis. Next, we computed
the $\theta$-dependent structure function from set II. We expect
that this could exhibit the $j=1$ component as inhomogeneity does
not allow us to apply incompressibility in the different symmetry
and parity sectors to eliminate this contribution as in the case
of the homogeneous structure function. This structure function is
symmetric but of mixed parity. We derive the tensor contributions
in the $j=1$ sector for the symmetric case in Appendix \ref{app:fullj1}
and use this to fit for the $j=1$ exponent.
\subsubsection{Antisymmetric Contribution}
We consider the tensor object in Eq.~(\ref{Tab}). In order to have
as few parameters as possible in the  fitting procedure, we take
the antisymmetric part
$$
{\widetilde T}^{\alpha\beta}({\B  r}) = {T^{\alpha\beta}({\B  r})
- T^{\beta\alpha}({\B  r}) \over 2} = \langle u^\alpha({\B
x})u^\beta({\B  x} + {\B  r})\rangle - \langle u^\beta({\B
x})u^\alpha({\B  x} + {\B  r})\rangle
$$ which will only have contributions from the antisymmetric
$j=1$
basis tensors. An additional useful property of this object is
that it does not have any contribution from the isotropic
helicity-free $j=0$ sector due to its antisymmetry. This allows
us to isolate the $j=1$ contribution and determine its scaling
exponent $\zeta_1^{(2)}$ starting from the smallest scales
available. Using data from the probes at $0.27$m (probe 1) and at
$0.11$m (probe 2) we calculate
$$
{\widetilde T}^{31}({\B  r}) = \langle u_2^{(3)}(\B
x)u_1^{(1)}(\B x + \B r)\rangle - \langle u_1^{(3)}(\B x + \B
r)u_2^{(1)}(\B x)\rangle \ ,$$  where again
super-scripts denote the velocity component and sub-scripts
denote the probe at which this component is measured. We want to
fit this object to the tensor form derived in Appendix (\ref{app:fullj1}),
 namely:
$$ {\widetilde
T}^{31}(r,\theta,\phi=0) = - a_{3,1,0}r^{\zeta_1^{(2)}}\sin\theta
+ a_{2,1,1}r^{\zeta_1^{(2)}} +
a_{3,1,-1}r^{\zeta_1^{(2)}}\cos\theta.
$$
Fig.~\ref{chj1T} gives the $\chi^2$ minimization of the fit as a
function of $\zeta_1^{(2)}$ and we use the best value of $1 \pm
0.15$ for the final fit. This is shown in the left panel.
\begin{figure*}
\includegraphics[scale=0.55]{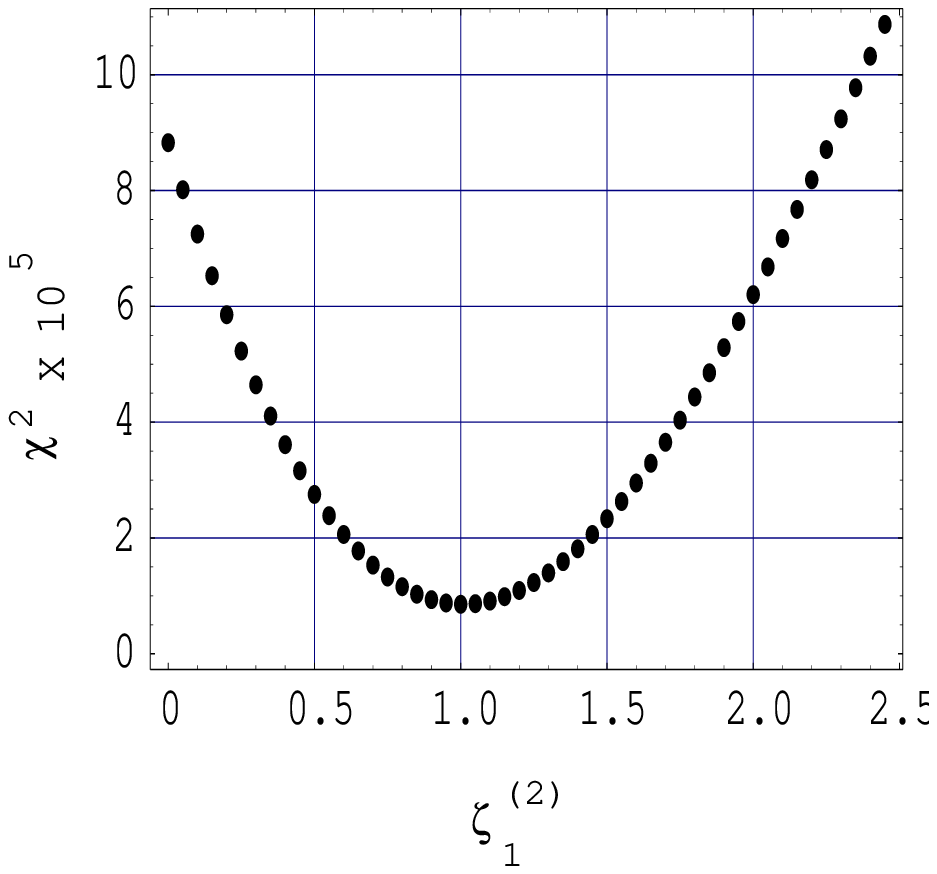}
\includegraphics[scale=0.8]{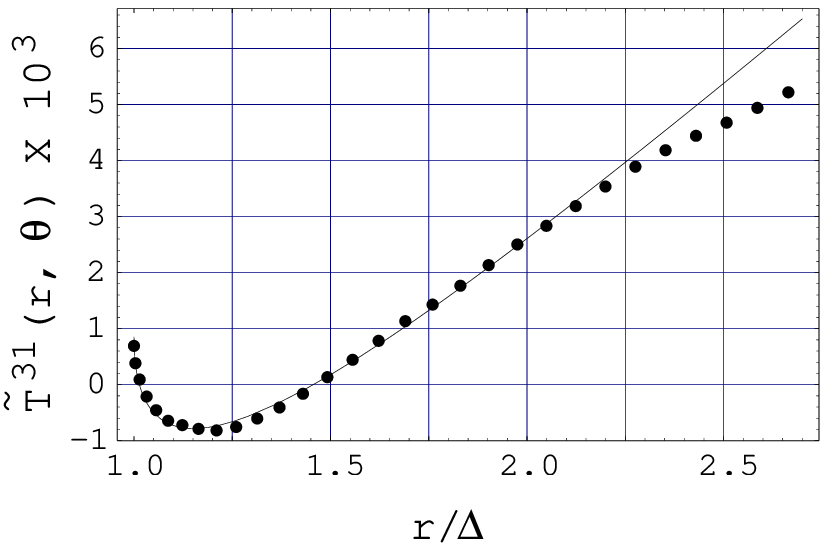}
\caption{Right: the $\chi^2$ minimization
by the best-fit value of the exponent $\zeta_1^{(2)}$ of the
$j=1$ anisotropic sector from the fit to $\theta$-dependent
${\widetilde T}^{31}(r,\theta)$ function in the range $0 <
r/\Delta < 2.2$. Left: The fitted ${\widetilde
T}^{31}(r,\theta)$ function. The dots indicate the data and the
line is the fit. }\label{chj1T}
 \end{figure*}
The fit in Fig.~\ref{chj1T} peels off at the end of the fitted
range. 
The maximum range over which one can fit is of the order of the
height of the probes and again, this is consistent with the
considerations presented above.
\subsubsection{Symmetric Contribution}
Finally, we compute the structure function Eq.~(\ref{ttadep}) where
the subscripts denote probe 1 at $0.27$m and probe 2 at $0.11$cm.
As discussed in Appendix (\ref{app:fullj1}),
 since the scale separation
has an inhomogeneous component, we expect a contribution from the
$j=1$ anisotropic sector and we would like to extract what the
scaling exponent in this sector is. Note that the $j=0$ sector
contributes {\em two}  independent tensor forms with coefficients
we will denote by $c_1$ and $c_2$, since incompressibility does
not provide a constraint to relate them. This fact combined with
Eq.~(\ref{b15}) gives us the the tensor form to which we must fit our
function
\begin{eqnarray}\label{sj1tens}
&&S^{33}(r,\theta) = c_1r^{\zeta_0^{(2)}} + c_2r^{\zeta_0^{(2)}}\cos^2\theta
+a_{1,1,0}r^{\zeta_1^{(2)}}\cos\theta
+a_{7,1,0}r^{\zeta_1^{(2)}}2\cos\theta \\&&
+a_{9,1,0}r^{\zeta_1^{(2)}} \cos^3\theta +
a_{8,1,1}r^{\zeta_1^{(2)}} (-2\cos\theta\sin\theta)
+ a_{1,1,-1}r^{\zeta_1^{(2)}} \sin\theta +
a_{9,1,-1} r^{\zeta_1^{(2)}}\cos^2\theta\sin\theta \nonumber \end{eqnarray}
We fix the exponent $\zeta^{(2)}_0$ to be $0.68$ and perform fits with
varying values of $\zeta_1^{(2)}$ for 8 unknown coefficients. The
best value of $\zeta_1^{(2)}$ is obtained for the range
$0<r/\Delta<4.2$ and is $1.05 \pm 0.15$ as is shown in
Fig.~\ref{chj1S6}. In the left panel we show the fit to the
data using this value of the exponent.
-8.2
The fit peels off at the end of the fitted range at the scale on
the order of twice the height of the probe, consistent with the
earlier discussion. There does not exist a well-defined
$\zeta_1^{(2)}$ as given by the standard $\chi^2$ minimization
procedure for ranges smaller or larger than that fitted for in
Fig.~\ref{chj1S6}. The quality of the fit  is
good although, as was expected from the large number of
parameters in the fitting function Eq.~(\ref{sj1tens}), $\chi^2$ as
a function of the $\zeta_1^{(2)}$ is not as smooth as for all
previous fits and its minimum is a relatively weak one.
Therefore, we present this result mainly as it provides support
to that of the antisymmetric case in the previous section.
\begin{figure*}
\includegraphics[scale=0.6]{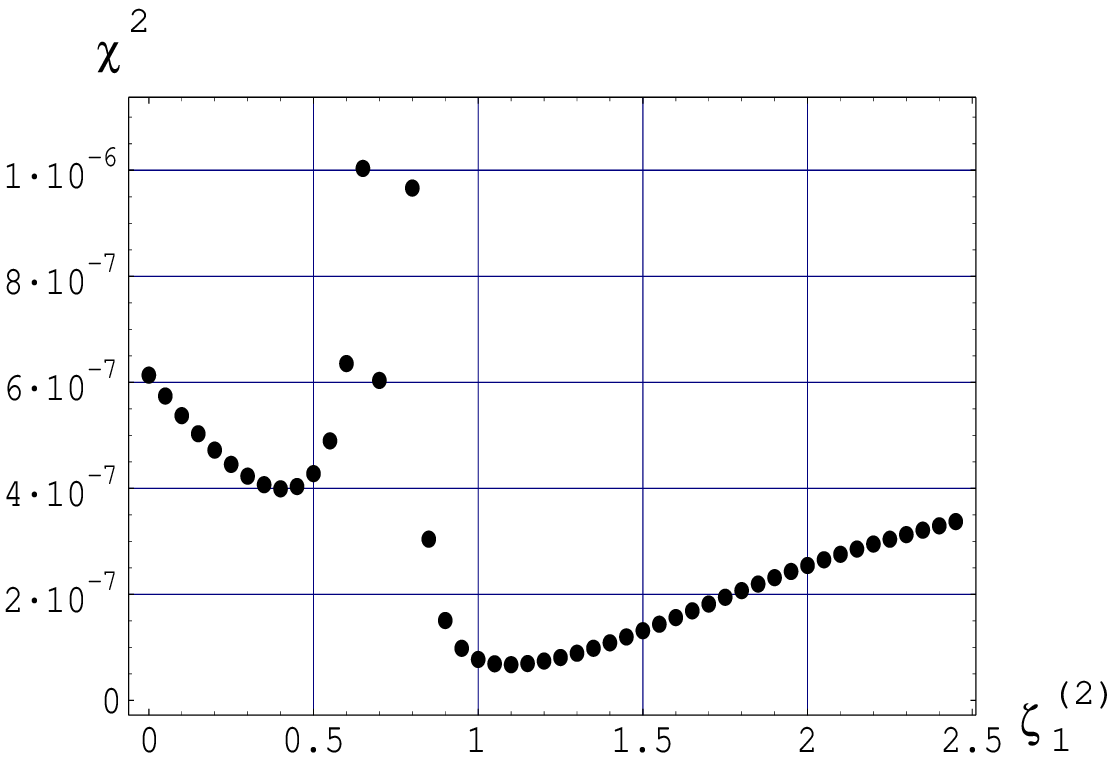}
\includegraphics[scale=0.55]{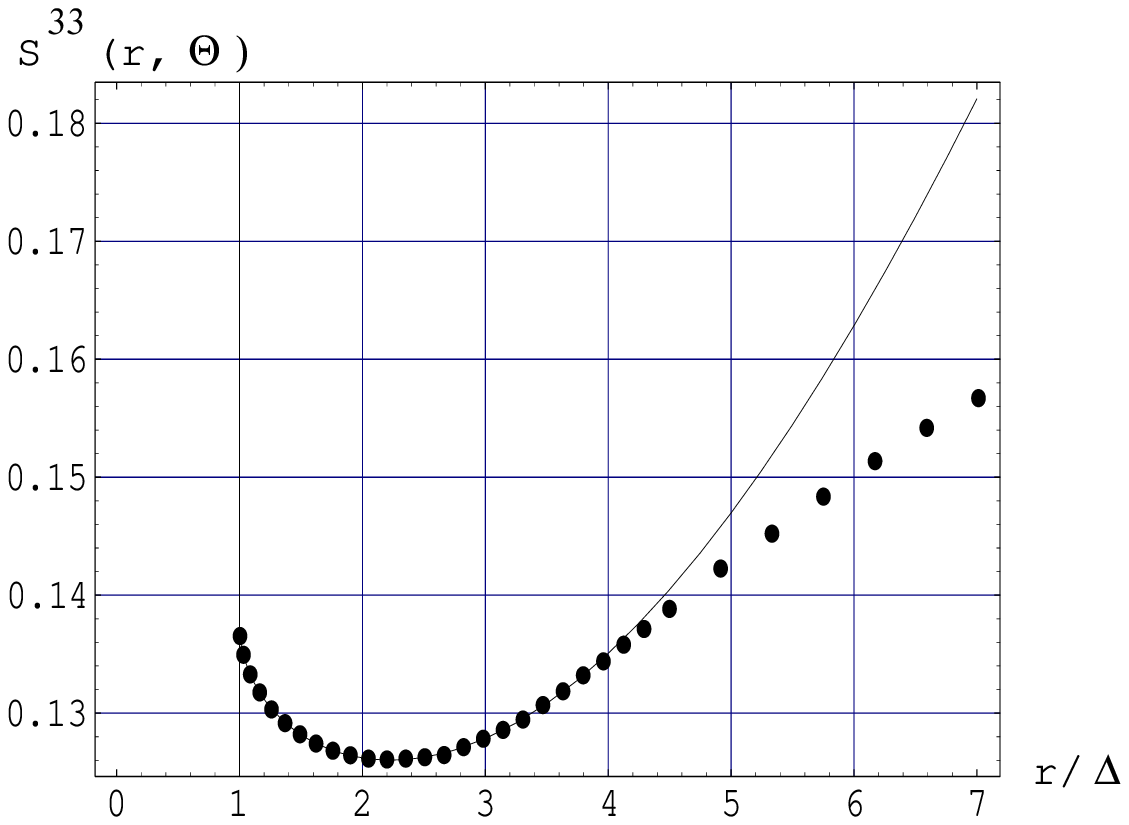}
\caption{Right: the $\chi^2$ minimization
by the best-fit value of the exponent $\zeta_1^{(2)}$ of the
$j=1$ anisotropic sector from the fit to $\theta$-dependent
inhomogeneous structure function in the range $0 < r/\Delta <
2.2$. Left: The fit to the
inhomogeneous structure function computed as in Eq.(\ref{sj1tens}). The dots
indicate the data and the line is the fit.}\label{chj1S6}
\end{figure*}
dots
\subsubsection{Summary and conclusions}
In summary, we considered the 2nd order tensor structure
functions of velocity differences in the atmospheric boundary
layers. The following conclusions appear important:
\begin{enumerate}
\item The atmospheric boundary layer exhibits 3-dimensional statistical
turbulence
intermingles with activities whose statistics are quite
different. The latter are eddys with quasi-two dimensional
nature, correlated for hundreds of meters, having little to do
with the three-dimensional fluctuations discussed above. We found
that the

\item We found that the ``outer scale of turbulence" as measured by the
three-dimensional
statistics is of the order of twice the height of the probe.

\item The inner scale is the the usual dissipative crossover, which is
clearly seen
as the scale connecting two different slopes in log-log plots.

\item Between the inner and the outer scales
the sum of the components up to $j=2$
 appears
to offer
an excellent representation of the structure function.
\item The scaling exponents $\zeta_j^{(2)}$ are measured as $0.68\pm 0.01,~
1\pm 0.15, ~1.38\pm0.10$ for $j=0,1,2$ respectively.
\end{enumerate}

We note that as far as the low order $j$ sectors are concerned, the picture
that emerges for Navier-Stokes turbulence is not different from the linear
advection
problems that were treated in the previous section. If the trends seen here
continue for higher
$j$ values, we can rationalize the apparent tendency toward isotropy with
decreasing scales. If indeed every anisotropic contribution
introduced by the large scale forcing (or boundary conditions)
decays as $(r/L)^{\zeta_j^{(2)}}$ with increasing $\zeta_j^{(2)}$
as a function of $j$, then obviously when $r/L\to 0$ only the
isotropic contribution survives. This is a pleasing notion that
justifies the modeling of turbulence as isotropic at the small
scales.
\subsection{Homogeneous Shear}
In this subsection we discuss recent experiments in which anisotropy is
created without 
inhomogeneity  \cite{she00,she02,she02a}; such experiments are
particularly appealing for our purposes.  {\it Homogeneous}-shear
flow can be realized in a wind tunnel by
 using a variable
solidity screen followed by flow straighteners.
Such a set-up results in a shear flow that remains approximately
constant for the length of the tunnel. To produce high Reynolds
numbers one places an active grid before the shear generating screen
\cite{she00}.  In this way $Re_{\lambda}$ can be as high as 1000.\\ To
assess directly the effects of anisotropy it is useful to measure
statistical objects that vanish identically in the isotropic sector. A
possible choice is the set of skewness and hyper-skewness
\cite{fer00,sch00,sch01} as explained in Sect.~(\ref{sec:persis}).
Other purely anisotropic inertial range observables can be defined by
mixing longitudinal and transversal increments with an odd number of
transversal components: \be S^{(p,2q+1)}(\Br) = \left< \delta
u_\ell^{p}(\Br) \delta u_t^{2q+1}(\Br) \right>.
\label{eq:sn_aniso} \ee
Systematic measurements of these anisotropic mixed correlation functions was
reported 
in \cite{she02,she02a,kur00}.
 From the experimental data it is not possible to exactly disentangle
  different anisotropic projections in different sectors. This is because
SO(3) projection requires the knowledge of the whole velocity field
in a 3d sub-volume, something clearly out of reach in any experimental
apparatus.  The simplest
  working hypotheses one can make is that, due to the hierarchical
  organization of anisotropic scaling exponents, the statistical
  behavior of quantities as (\ref{eq:sn_aniso}) is dominated, at
  scales small enough, by the leading $j=2$ sector. In other words,
  the experimental measurements of the scaling properties of
  (\ref{eq:sn_aniso}) is the best estimate of the
  exponent $\zeta_{2}^{(n=p+2q+1)}$.
In \cite{she02,she02a} the plots of purely anisotropic quantities like
(\ref{eq:sn_aniso}) up to order $n=8$ with $n=p+2q+1$ were shown.  The
data clearly shows that these purely anisotropic structure functions
have quite good power-laws behavior with exponents that are
sub-leading with respect to the exponents of the isotropic structure
functions of the same order, $n$.  For example $ S^{(1,3)}(r) \sim
r^{1.56}$ while the fourth-order longitudinal structure function in
isotropic ensembles is known to scale as $ S^{(4,0)}(r) \sim
r^{1.27}$.  Similar qualitative and quantitative results were obtained
by analyzing data from an atmospheric boundary layer in \cite{kur00}
and in the boundary layer close to a wall \cite{iacob04}.  In the
latter two works, a phenomenological fitting procedure to the large
scale behavior allowed the authors to find a power law for the
anisotropic structure functions which pertain to a much larger range
of scales.  We draw the reader's attention to the discrepancy in the
best fit for the scaling exponents founds for $ S^{(1,3)}(r)$ and
$S^{(3,1)}(r)$ in \cite{she02,she02a}. Similar discrepancies are also
reported for higher order structure functions.  In our view, this
cannot be taken as evidence that there is a $q$-dependence of the
scaling exponents of the SO(3) projections. First, the anisotropic
exponents are relatively inaccurate due to statistical errors; the
amplitudes of the anisotropic fluctuations are relatively small.
Second, as already said, the experimental data cannot disentangle
exactly the contribution of the $j=2$ sector. Therefore, it may well
be that contributions from the $j=4$ (and higher) sectors affect
differently the correlation functions with different tensorial
structure. Similarly, other experimental investigation focused on the
SO(3) decomposition \cite{sta02,sta03,wander} have found results
depending on the geometric set-up of the analyzing probes. The
experimental analysis of anisotropic turbulence via the SO(3)
decomposition is at its infancy; more refined experimental techniques
are needed before a firm conclusion can be reached on these issues.

\subsubsection{Explanation of Persistence of Anisotropies}
\label{pers_ani}
As discussed in Sect.~(\ref{sec:persis}) there are
  numerical and experimental evidences of the persistence of small-scale
 anisotropic fluctuations in various instances
\cite{gar98,she00,kur00,pum96,bif01}.
The issue has many important consequences. We would like to
refer to the
violation of the {\it return-to-isotropy} in  different
meanings \cite{bif01}. A {\it strong} violation would be implied if the
following set of inequalities between different anisotropic exponents of the
same correlation function were broken:
\begin{equation}
\zeta^{(n)}_0 < \zeta^{(n)}_1 < \cdots < \zeta^{(n)}_j \ ,
\label{eq:SnDecomp}
\end{equation}
 i.e. if one, or more, anisotropic sector becomes leading with respect
 to the isotropic one. This would destroy the phenomenology of turbulence as
developed
 since Kolmogorov's theory in 1941. Turbulence would become more and
 more anisotropic at smaller and smaller scales. As a result,
 strong non-universalities in small-scales statistics would show up
 depending on which anisotropic sector is switched on/off by the
 large-scale forcing. A {\it strong} violation of the {\it
 return-to-isotropy} postulate has never been observed in
 Navier-Stokes turbulence. On the other hand, when the hierarchy
 (\ref{eq:SnDecomp}) holds, any dimensionless anisotropic observables
 made of ratios between anisotropic and isotropic projections of the
 {\it same} correlation function vanishes in the small-scales limit.
For example, focusing on the decomposition of longitudinal structure
functions
(\ref{eq:fundamental}) we may write:
\be
 \lim_{r\rightarrow 0}
\frac{S_{jm}^{(n)}(r)}{(S_{00}^{(n)}(r))}\;\sim
r^{\zeta^{(n)}_j-\zeta^{(n)}_0}
\rightarrow 0.
\label{eq:ret-b}
\ee
  A new phenomenon occurs when anisotropic fluctuations are assessed
by using dimensionless observables made of {\it different}
correlation functions.
 For instance, by
 using again the  SO($3$) decomposition of longitudinal structure
 function (\ref{eq:fundamental})
 one may build up anisotropic observables defined as:
 \be R_{jm}^{(n)}(r) =
\frac{S_{jm}^{(n)}(r)}{(S_{00}^{(2)}(r))^{n/2}}\;\sim r^{\chi_j^{(n)}}
\;\;\;
\mbox{ with } \;\; {\chi_j^{(n)} = \zeta_j^{(n)}-{ n \over 2} \zeta_0^{(2)}}
\ .
\label{eq:ret-a}
\ee 
This is the $n$th order moment of the velocity
probability density function, normalized by its isotropic
second order moments. The quantities defined in (\ref{eq:ret-a}) must
be exactly zero in isotopic ensembles, and should
 go to zero as power laws, $R_{jm}^{(n)}(r) \sim r^{j/3}$,
in an  anisotropic ensemble in which
the dimensional scaling (\ref{lumleygen}) is satisfied. On the
other hand, results from experiments and numerics show
a  much slower decay, and, in some cases, no decay at all
\cite{she00,bif01}. We refer to this phenomenon as {\it weak} violation
of the {\it return-to-isotropy}. Such a weak violation is not in
contradiction
with the inequalities (\ref{eq:SnDecomp}); there the
 relative importance of anisotropic fluctuations with respect to
isotropic fluctuation of the {\it same} correlation function are
implied. The violation of the dimensional recovery-of-isotropy is
simply due to the existence of anomalous scaling in the anisotropic
sectors. Indeed, in  this  case, the exponents, $\chi_j^{(n)}$, governing
the LHS of (\ref{eq:ret-a}) can assume values much smaller than the
dimensional estimate (including negative values!).
 This is exactly what is observed in the experiments and numerics. From
Table~3 one realizes that due to the
presence of anomalous scaling in the anisotropic sectors  we have a
slow recovery-of-isotropy,
in agreement with what was explained before.

The anisotropic observables built in terms of the generalized flatness
or skewness discussed in section (\ref{sec:persis}) are nothing but
Eq.~(\ref{eq:ret-a}) evaluated at the dissipative length scale,
$r=\eta$. Therefore, the ``persistence-of-anisotropies" discussed in
\cite{pum96,she00}  can be
explained invoking the very same reasoning.
\subsubsection{Summary of experimental results: universality of the
anisotropic sectors} \label{expuniver}

Comparing the results obtained in
\cite{she02,she02a,kur00} the following picture emerges.
First, all the correlation functions up to $n=10$, show  {\it
anomalous} scaling behavior, where
anomalous is meant with respect to the dimensional
Lumley-like prediction discussed in Sect.~(\ref{sec:dim_ana}).
Second, the values of scaling exponents extracted from the two
different experiments  \cite{she02a} and  \cite{kur00} are in good
qualitative  agreement
 (see Table~2).
\begin{table}[ht]
\label{tab:scaling_exp}
\begin{center}
\begin{tabular}{|c|c|c|c|c|c|c|c|c|c|}
\hline 
$(p,2q+1)$ & (1,1) & (1,3)     & (3,1) & (5,1) & (3,3)    &  (1,5) & (7,1) &
(5,3) &  (3,5) \\ 
\hline 
WS &  1.05      & 1.56      & 1.42  & 2.02  & 1.89     &  1.71  & 2.33  &
2.22 & 1.99  \\
KS &  1.22,1.12 & 1.58,1.54 &  -    & -     & 2.14,2.00&  -     & -     &
-  & - \\
 \hline
\end{tabular}
 \caption{Measured scaling exponents for $S^{(p,2q+1)}(\B r)$
 of various orders in two experiments. WS corresponds
to \cite{she02a} and KS to \cite{kur00}}
 \end{center}
\end{table}
 This is an important first confirmation of the {\it universality} of
 scaling exponents in the $j=2$ anisotropic sector.  Finally, there
 exists a clear hierarchy between isotropic and anisotropic scaling
 exponents, the latter being always larger for any given order, $n$ of
 the correlation function. This hierarchical organization is the
 necessary and sufficient requirement for the {\it
 return-to-isotropy} to hold, i.e. the small scales statistics
 of any correlation function is dominated by the isotropic
 fluctuations. Nevertheless the gap between isotropic and anisotropic
 exponents, $\zeta_0^{(n)} - \zeta_{2}^{(n)}$, tends to shrink when $n$
increases, implying that anisotropic contributions may exhibit
 important sub-leading effects also at very high Re.
%
%
\section{Analysis of DNS Data}
\label{chap:numerics}
Direct numerical simulations  of turbulence are natural grounds
where the utility of the SO($3$) decomposition can be exploited to its
maximum benefit. The reason
for this is that numerical simulations, in contrast to current
experiments, provide access to the full velocity field at all points
of the turbulent domain. Therefore, the full SO($3$) decomposition can
be realized, without the constraints of best-fits to partial data. Given a
tensor structure function $\B S^{(n)}(\Br)$, cf. Eq.(\ref{Sn}),
we can integrate it against the spherical tensors,
$\B B^{(n)}_{q j m}(\hat{\Br})$ [e.g., (\ref{eq:so3sn})],
 on a sphere of radius $r$. These integrations yield the
projection of the structure function on the different sectors of the
SO($3$) group, by virtue of the orthogonality of the basis tensors. On the
other hand DNS suffer from limited Reynolds numbers;
consequently they have relatively short inertial ranges.

Prior to the introduction of the SO(3) decomposition the numerical
investigations of anisotropic flows were focused on either single
point or two-points correlations, limited, often, to the analysis of
the Fourier transforms in wavevector space. The most recent,
highly-resolved, numerical investigation of this kind was reported in
\cite{ish02}; there the full tensorial properties of the Fourier
transform of the two-point velocity correlation, $Q^{\alpha
\beta}(\Bk) \EqDef \int d\Br e^{i
\Bk \Br} \la u^\alpha(\Bx+\Br)u^\beta(\Bx) \ra $, were calculated in a {\it
homogeneous} shear \cite{hin75,rog82}. The main result is a
confirmation of Lumley's prediction for the scaling exponent of the purely
anisotropic
co-spectrum:
\begin{equation}
E^{\alpha \beta}(k) \sim k^{-7/3}\;\;\; \mbox
{where}\;\;\; E^{\alpha \beta}(k) = \int_{{k \over 2}<|\Bp|<2k} d\Bp
Q^{\alpha \beta}(\Bp) \ , \label{cospec}
\end{equation}
where $\alpha \neq \beta$ to eliminate  the
isotropic contribution. Only recently DNS were performed to probe
the anisotropic component in a systematic way by
exploiting the SO($3$) decomposition
\cite{ara99,bif01a,bif02a,bif02,bif03a}.
Here we review the main findings, showing that
\begin{enumerate} 
\item The scaling laws (log-log plots) at moderate Reynolds numbers are
significantly improved by projecting the raw correlation functions onto
each $j$-sectors. The improvement is particular noticeable whenever strong
anisotropies are present in
the system, as in the case of channel flows \cite{ara99,bif02a};
\item Anisotropic sectors with $j \ge 2$ (inaccessible in present
experimental data) 
possess good scaling laws \cite{bif01a,bif02};
\item The scaling exponents  are discrete and increasing as a
function of $j$.
\item The exponents are {\it anomalous}; i.e.  they differ from the
dimensional
prediction (\ref{lumleygen}).
\item  There exists
 preliminary evidence that also for $j > 2$, the anomalous  exponents
 are {\it universal}, i.e. the scaling properties are independent of the
 external forcing mechanism \cite{bif03a}.
\end{enumerate}
DNS were performed both in wall-bounded flows and in homogeneous (but
anisotropic) 
turbulence. In wall-bounded flows the anisotropies are accompanied by {\it
inhomogeneous} effects. The presence of such effects may
 spoil the very meaning of scaling, and the SO($3$) decomposition
 should be supplemented by some tool to project on the homogeneous
 components.  Otherwise, the SO(3) decomposition must be used
 carefully, and locally, only in those regions of the tested flow
 where inhomogeneous effects are confined mostly to
 large-scales \cite{ara99,bif02a}.  In the second part of
 this section, we discuss numerical experiments
 built such as to have a perfectly {\it homogeneous} and {\it
 anisotropic} statistics at all scales.  One such example is {\it
 homogeneous shear flows} \cite{sch00,cas00}. More recently,
 other homogeneous anisotropic flows have been invented and simulated,
 in particular the random-Kolmogorov-flow
 \cite{bif01a,bif02,bif03b} and  a convective cell with an imposed linear
mean profile of
 temperature \cite{bif03a}.
\subsection{Anisotropic and inhomogeneous statistics: Channel flows}
\label{sec:num_channel}
In this section, we discuss the analysis of a DNS of a channel flow
using the SO($3$) decomposition.  The coordinates are chosen
 such that $\hat{x}$, $\hat{y}$ and $\hat{z}$ are the stream-wise,
span-wise,
and wall-normal direction respectively. The simulation was done on a
grid with $256$ points in the stream-wise direction and $128\times128$
points in the two other directions. The boundary conditions were
periodic in the span-wise and stream-wise directions and no-slip on the
walls. The Reynolds-number based on the Taylor micro-scale was quite
moderate,
$R_{\lambda}\approx 70$ at the center of the channel $(z=64)$. The
simulation was fully symmetric with respect to the central plane. For
more details on the averaged quantities and on the numerical code, see
refs.~\cite{ara99,ama97,ama97a}.

The analysis focused on longitudinal second-, forth- and sixth-order
structure functions:
$$
  S^{(n)}(\Br^c,\Br) \equiv \left<\big[\delta
  u_\ell(\Br^c,\Br)\big]^n\right>, \quad \delta u_\ell(\Br^c,\Br) \equiv
  \hat{\Br}\cdot \big[\Bu(\Br^c+\Br,t)-\Bu(\Br^c-\Br,t) \big] \ .
$$
The $\Br^c$ coordinate specifies the location of the measurement
(i.e., the center of mass of the two measurement points), and $2\Br$
is the separation vector.  Previous analysis of the same data-base
\cite{ama97} as well as of other DNS \cite{bor96} and
experimental data \cite{sad94,gau98} in anisotropic flows
found that the scaling exponents of energy spectra, energy co-spectra
and of longitudinal structure functions exhibit strong dependence on
the position $r_c$.  For example, in \cite{tos99}
the authors studied the longitudinal structure functions at fixed
distances from the walls: $$S^{(n)}(r,z)\equiv \la
(u_x(x+r,y,z)-u_x(x,y,z))^n\ra_z$$ where $\la \cdots \ra_z$ denotes a
spatial average on a plane at a fixed height $z$, $1<z<64$. For this
set of observables they found that: (i) These structure functions did
not exhibit clear scaling behavior as a function of the distance
$r$. Consequently, one needed to resort to Extended-Self-Similarity
(ESS) \cite{ben93b} in order to extract a set of {\rm relative}
scaling exponents $ \hat{\zeta}^{(n)}(z) \equiv
\zeta^{(n)}(z)/\zeta^{(3)}(z)$;
(ii) the relative exponents, $ \hat{\zeta}^{(n)}(z)$ depended strongly on
the height $z$. Moreover, only at the center of the channel and very
close to the walls the error bars on the relative scaling exponents
extracted by using ESS were small enough to claim the very existence
of scaling behavior in any sense.  Similarly, an experimental analysis
of a turbulent flow behind a cylinder \cite{gau98} showed a
strong dependence of the relative scaling exponents on the position
behind the cylinder for not too big distances from the obstacle,
i.e. where anisotropic effects may still be relevant in a wide range
of scales.  In the following we present an interpretation of the
variations in the scaling exponents observed in non-isotropic and
non-homogeneous flows upon changing the position in which the analysis
is performed. In particular, we will show that decomposing the
statistical objects into their different $(j,m)$ sectors rationalizes
the findings, i.e.  scaling exponents in given $(j,m)$ sector appear
quite independent of the spatial location; only the {\em amplitudes}
of the SO(3) decomposition depend strongly on the spatial location.
The analysis showed three major results. The first was the vast
improvement in scaling behavior of the structure functions as a result
of the decomposition. A typical example is found in
Fig.(\ref{fig:channel.fig1} )
\begin{figure}[!h]
\includegraphics[scale=0.9]{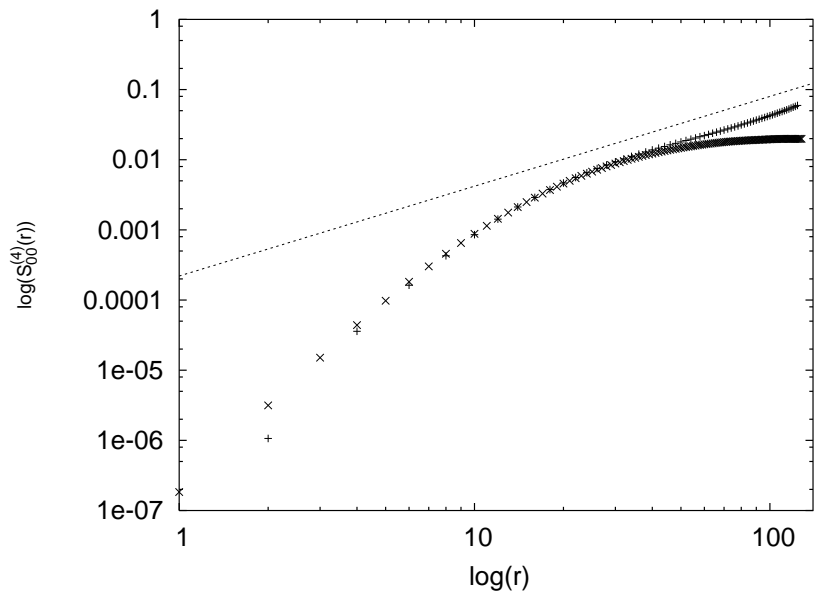}
\caption{Log-log plot of the isotropic sector of the 4th order structure
function $S^{(4)}_{0,0}(r)$, vs. $r$ at the center of the channel
$r^c_z=64$ (+). The data represented by ($\times$) correspond to the
raw longitudinal structure function, $S^{(4)}(r^c_z=64,r\hat x)$
averaged over the central plane only. The dashed line corresponds to
the intermittent isotropic high-Reynolds numbers exponents
$\zeta^{(4)}_0 = 1.28$.}
\label{fig:channel.fig1}
\end{figure}
 where the raw fourth-order
structure function, evaluated on the central plane, is compared to its
$j=0$ component. Without the SO($3$) decomposition there is no
scaling behavior at all and one needs ESS
to estimate the scaling exponents. On the other hand, the
$j=0$ component of the structure function shows a clear scaling
behavior with the expected exponent, $\zeta^{(4)}_{0} = 1.28$. This
strengthens the foliation hypothesis, according to which, the raw
structure function is a superposition of power laws from different
sectors of the SO($3$) group.  Such a sum looses its scale invariance
once the weights of the different exponents are of the same order and
the inertial range is small. In such cases, one needs the SO($3$)
decomposition to isolate the different sectors and retain scale
invariance.
\begin{figure}[!h]
\includegraphics[scale=0.9]{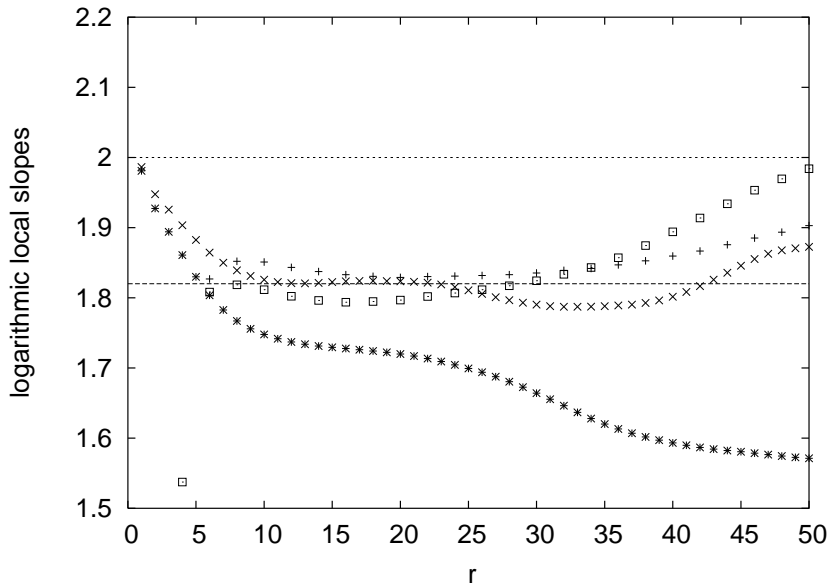}
\caption{
Logarithmic local slopes
  of the ESS plot of raw structure function,$\frac{d log(S^{(4)}(r,r_c)}{d
log(S^{(2)}(r,r_c)}$,
of order 4 versus raw structure function of order 2
at $r_c^z=64$ ($\times$), at $r_c^z=32$ ($\star$) and of the $j=0$
projection, $\frac{d log(S^{(4)}_{00}(r,r_c)}{d log(S^{(2)}_{00}(r,r_c)}$,
centered at $r_c^z=64$ ($+$), and at $r_c^z=32 $($\Box$).
Also two horizontal lines
corresponding to the high-Reynolds number limit, $1.82$, and to the
K41 non-intermittent value, $2$, are shown.}
\label{fig:channel.fig3}
\end{figure}

A second prominent result is the apparent universality of the
isotropic exponents. To show this in \cite{ara99} the local slopes,
$\frac{d log(S^{(4)}_{00}(r)}{d log(S^{(2)}_{00}(r)}$,
of the ESS curves  of the isotropic forth-order structure-function versus
the
isotropic second-order structure function were calculated at varying
the
distance from the wall.
 Despite their different locations, all curves  show the same ESS slope
1.82, which is the expected (anomalous) value. In
Fig.(\ref{fig:channel.fig3}) one  picture is presented for the
logarithmic local slopes at two different distances from the
channel boundary. To appreciate the improvements in scaling
and universality, also the slopes of the ESS on the raw structure
functions are presented
Finally, the analysis provided another evidence that the $j=2$
scaling exponent of the second order structure function is about
$4/3$, which is the dimensional theoretical prediction given in
(\ref{eq:Lumley})
(see also \cite{lum67,yak94,fal95,gro94,bif02}).
 Considering the relatively low
Reynolds-number and the fact that the prefactors $a_{j,m}$ in the
SO($3$) decomposition (\ref{eq:SnDecomp}) are non-universal, together
with the experimental result reported in
\cite{ara98,kur00,she02,she02a}, these findings give
strong support to the view that the scaling exponent in the $j=2$
sector is universal.
Before concluding this section we cite that
SO(3) and SO(2) decomposition have also been exploited in the analysis
of channel flow data to highlights the importance of structures as streaks
and 
hairpin filaments typical of many wall bounded flow \cite{bif02a}.
Preliminary investigation of the importance of SO(3) decomposition
to evaluate the performance of sub grid models used in Large Eddy
Simulations
\cite{mene00}
have also been reported in \cite{bif03b}.
\subsection{Anisotropic-homogeneous flows}
\label{sec:anis-hom}
Direct numerical simulations offer the unique opportunity to
study the physics of anisotropy in {\it ideal} situations, that is in
perfectly homogeneous flows. Recently, considerable effort has been spent
on simulating a Random-Kolmogorov-Flow (RKF)
\cite{bif01a,bif02,bif03b}.  The RKF  is
fully periodic, incompressible and with anisotropic large-scale energy
injection. A convenient choice for the forcing is ${\f}=(0,0,f_z(x))$
with $f_z(x)=F_1 \cos[2\pi x/L_x +\phi_1(t)] + F_2
\cos[4 \pi x/ L_x +\phi_2(t)]$, with constant amplitudes $F_{1,2}$ and
independent, uniformly distributed, $\delta$-correlated in time and with
 random
phases $\phi_{1,2}(t)$. The random phases
lead to a homogeneous statistics.  To give a first validation of the
statistical properties of the RKF flow we plot in
fig.~\ref{fig:RKF.spectra}
\begin{figure}[h]
\includegraphics[scale=0.9]{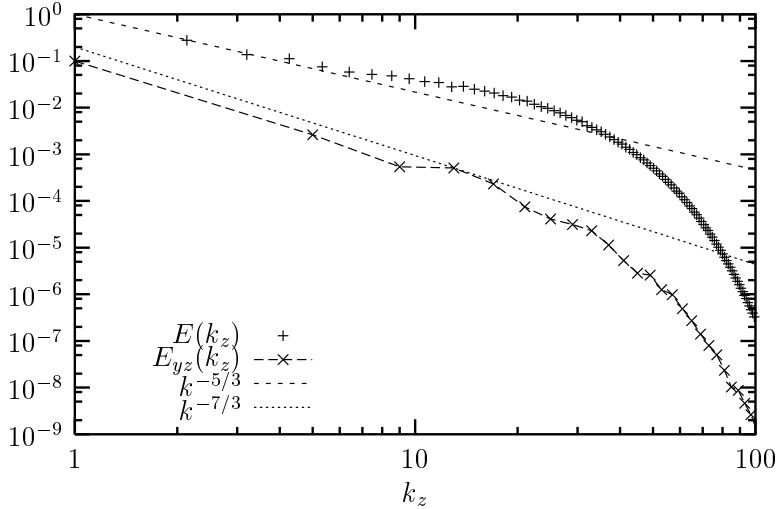}
\caption{Log-log plot of instantaneous energy spectrum in the
isotropic sector $E(k)$ (top). The straight line is the reference isotropic
$k^{-5/3}$ power law. Instantaneous co-spectrum $E_{yz}(k_z)$
(bottom). Here the straight line gives the reference $k_z^{-7/3}$
anisotropic Lumley prediction. The two spectra have been shifted along
the vertical direction for the sake of presentation.}
\label{fig:RKF.spectra}
\end{figure}
the instantaneous energy spectrum,
$$E(k)=\int_{|\Bq|=k} \la \Bu(\Bq)\cdot \Bu^*(\Bq) \ra d\Bq.$$  It
exhibits a scaling law in close agreement with the K41 isotropic behavior
$k^{-5/3}$. Also purely anisotropic quantities as the co-spectra
(\ref{cospec}),
show a good agreement with the Lumley  $k^{-7/3}$.
DNS of the RKF were reported in \cite{bif01a,bif02,bif03b}. The
resolution was $256^3$ reaching  $Re_{\lambda}\sim 100$,
 collecting up to $70$ eddy
 turn over times.  A long time  average is necessary because of the
formation of persistent
 large-scale structures inducing strong oscillations of the mean
 energy evolution. This is typical to many strongly anisotropic flows.
The viscous term was replaced by a
 second-order hyper-viscosity, $-\nu \Delta^2{\Bu}$.
  Thanks to both the high degree of homogeneity and to
 the high number of independent samples, a quantitative analysis of
 scaling laws of longitudinal structure functions up to the  anisotropic
 sector $j=6$ and up to order $n=6$ was possible.  In other words, the
longitudinal structure functions could be decomposed according to
$$
S^{(n)}(\Br) = \sum_{j=0}^{6}\sum_{m=-j}^{j} S^{(n)}_{jm}(r)
Y_{jm}(\unitr ) \quad \rm{for }~n\le 6 .
$$
In fig.~\ref{fig:RKF.fig1} we present the results for the
  isotropic sector. Here we compare the raw structure
  functions in the three directions with the projection
  $S^{(2)}_{00}(r)$, and their logarithmic local slopes (inset).
\begin{figure}[!h]
\includegraphics[scale=0.9]{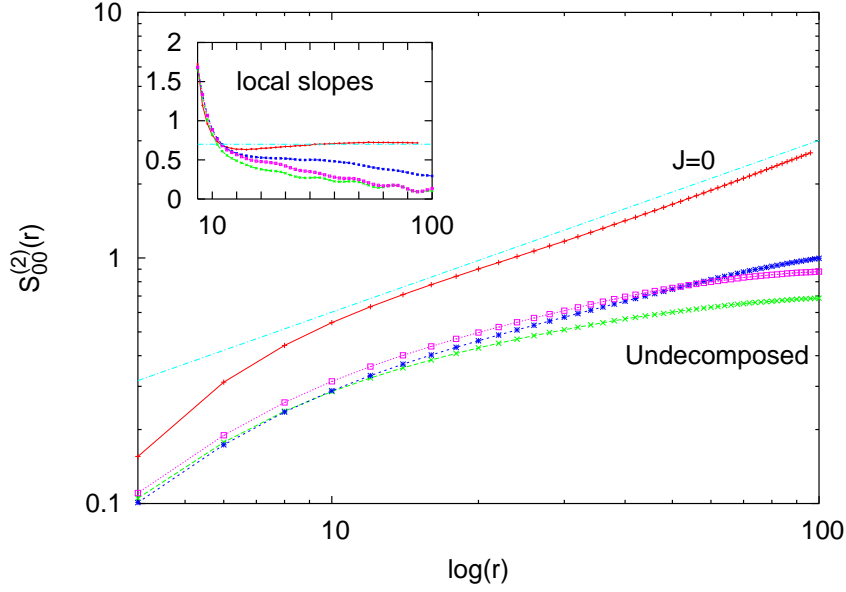}
\caption{ Analysis on
the real space. Log-log plot of $S^{(2)}_{00}(r)$ versus $r$ (top
curve), and of the
three undecomposed longitudinal structure functions in the three
directions $x,y,z$ (three bottom curves).  The
straight line gives the best fit slope $\zeta_{0}^{(2)}=0.7$. Inset:
logarithmic 
local slopes of the same curves in the main body of the figure (same
symbols). Notice that only the projected curve shows a nice plateau }
\label{fig:RKF.fig1}
\end{figure}
Only for the projected correlation it is possible to measure (with
$5\%$ of accuracy) the scaling exponent by a direct log-log fit versus
the scale separation. The best fit gives $\zeta_{0}^{(2)} = 0.70 \pm
0.03$.  On the contrary, the undecomposed structure functions are
overwhelmed by the anisotropic effects present at all scales, and the
scaling law is completely spoiled. We stress the accuracy
of these results; already at these modest Reynolds numbers it is
possible to ascertain the isotropic scaling laws if the anisotropic
fluctuations are disentangled properly.

In figure \ref{fig:RKF.fig3} there is an overview for the second order
structure functions in all the sectors, isotropic and anisotropic, for which
the signal-to-noise ratio is high enough to ensure statistically
stable results. Sectors with odd $j$ are absent due to the
parity symmetry of the longitudinal structure function. We conclude from
figure \ref{fig:RKF.fig3} a clear foliation in terms of the $j$
index\,: sectors with the same $j$ but different $m$ exhibit very
close scaling exponents.
In Table~3 the measured exponents are compiled,
showing the best power law fits for
structure functions of orders $n=2,4,6$.
\begin{figure}[!h]
\includegraphics[scale=0.9]{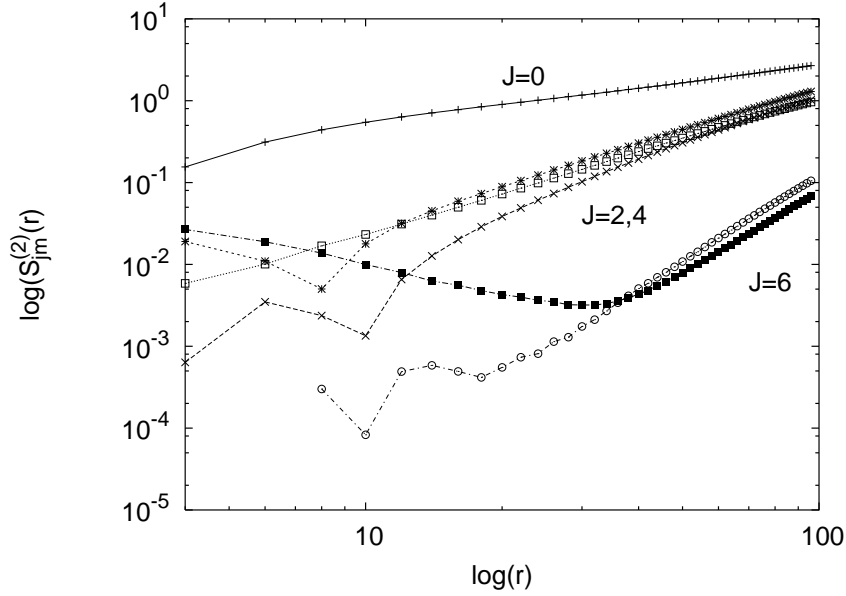}
\caption{Log-log plot of the second order structure function in all
sectors with a strong signal. Symbols refer to sectors (j,m) as follows:
  $(0,0)$, ($+$); $(2,2)$,
($\times$); $(4,0)$, ($\Box$); $(4,2)$, ($\star$); $(6,0)$, ($\circ$);
$(6,2)$, ($\blacksquare$). The statistical and numerical noise affecting
the SO(3) projection is estimated as the threshold where the $j=6$
sector starts to deviate from the monotonic decreasing behavior,
i.e. ${\mathcal O}(10^{-3})$}
\label{fig:RKF.fig3}
\end{figure}
We stress again the discreteness and monotonicity of the scaling
exponents as assumed in Eq.~(\ref{eq:SnDecomp});
there is no saturation of the exponents as a function of
$j$. Second, the measured exponents in the sectors $j=4$
and $j=6$ are anomalous, i.e. they differ from the dimensional
estimate given in \ref{lumleygen}.
\begin{table}[!h]
\label{tab:compile}
\begin{center}
\begin{tabular}{|c|c|c|c|c|}
\hline
$n$  & $j=0$  & $j=2$ & $j=4$ & $j=6  $\,\\
\hline
& $\zeta_{0}^{(n)}$ --- $n/3  $ \,& $\zeta_{2}^{(n)}$ ---
$ (n+2)/3   $\, & $\zeta_{4}^{(n)}$ --- $ (n+4)/3  $\, &
$\zeta_{6}^{(n)}$ --- $  (n+6)/3  $\,\\
\hline
2 & 0.70 (2) --- 0.66 & 1.1 (1) --- 1.33 & 1.65 (5) --- 2.00 & 3.2 (2)
--- 2.66 \\ 
4 & 1.28 (4) --- 1.33 & 1.6 (1) --- 2.00 & 2.25 (10) --- 2.66
& 3.1 (2) --- 3.33 \\
6 & 1.81 (6) --- 2.00 & 2.1 (1) --- 2.33 & 2.50
(10) --- 3.33 & 3.3 (2) --- 4.00 \\
\hline \end{tabular}
\caption{Summary of  the  numerical and experimental
findings for the scaling exponents in the isotropic and anisotropic
sectors. The values for the anisotropic sector $j=2$ are taken from
the experiments {\protect \cite{kur00,she02a}}. For the values
extracted from the numerical simulation (columns $j=0,4,6$), error
bars are estimated from the oscillation of the local slopes {\protect
\cite{bif01a,bif03b}}.  For the experimental data the error is
given as the mismatch between the two experiments.  For all sectors we
also give the dimensional estimate $\zeta_{j}^{(n)} = (n+j)/3$
{\protect \cite{bif02}}.}
\end{center}
\end{table}
Unfortunately, from the RKF data it was not possible to obtain clean
results for  the $j=2$ sector. This is because of the presence of
an annoying change of sign in the projections $S^{(n)}_{2m}(r)$ for
any $m$ (and any order $n$).  Still, the overall consistency of the
foliation and hierarchical organization of scaling exponents can be
checked by collecting  the scaling exponents in the $j=2$ sector from
the two sets of   experiments
\cite{kur00,she02a} previously discussed.
 In Fig.~\ref{fig:RKF+EXP} we show both numerical data and the
 experimental values as extracted from
\cite{kur00,she02a}.
 The resulting  picture
is fully coherent\,: experimental data coming from the $j=2$ sector
fit well in the global trend.  As one can see from table~3
all the anisotropic sectors show {\it anomalous} scaling laws.
\begin{figure}[!h]
\includegraphics[scale=0.9]{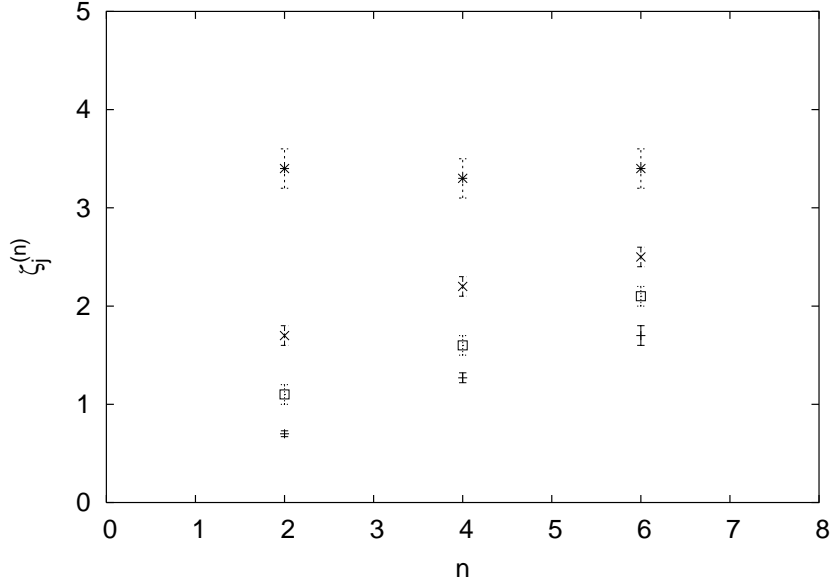}
\caption{Scaling exponents, $\zeta_{j}^{(n)}$,  of
  structure functions of order $n=2,4,6$ for isotropic and anisotropic
  sectors. From the DNS of RKF we have\,: isotropic sector, $j=0$
  ($+$); anisotropic sectors, $j=4$ ($\times$) and $j=6$
  ($\star$). From the experimental data {\protect \cite{kur00,she02a}},
  we have $j=2$ ($\Box$). For an estimate
 of error bars see Table 3.
}
\label{fig:RKF+EXP}
\end{figure}
\subsubsection{Universality of Anisotropic Fluctuations}
The
third numerical experiment that  we discuss here is devoted to study  {\it
universal}
properties of anisotropic scaling. We have already commented that
there is a nice qualitative and quantitative agreement between the
values extracted for the $j=2$ sector (the only one available from
experimental data) from different experiments. In order to check
whether this universality holds also in higher anisotropic sectors one
has to rely on DNS. In \cite{bif03a} a first direct comparison
between anisotropic scaling of longitudinal structure functions from
two different homogeneous systems, the RKF and a {\it homogeneous \rb}
convective flow was reported.
 
A  Homogeneous \rb system is a convective cell with fixed
linear mean temperature profile along the vertical direction.  The
flow is obtained by decomposing the temperature field as the sum of a
linear profile plus a fluctuating part, $T(x,y,z;t)= T'(x,y,z;t) + (\Delta
T/2 - z \Delta T/H) $, where $H$ is the cell height and $\Delta T$
the background temperature difference.  The evolution of the system can
be described by a modified version \cite{gro02} of the Boussinesq system
\cite{kad01}:
\bea
 \partial_t \Bu &+& \lp\Bu\cdot\nabla\rp \Bu =
 -\nabla p + \nu \nabla^2 \Bu + \alpha g T' {\unitz} \nonumber \\
 \partial_t T' &+& \lp\Bu\cdot\nabla\rp T' =
  \kappa \nabla^2 T' - {{\Delta T} \over H} v_z. \nonumber
\eea
where $\alpha$ is the thermal expansion constant,
$\nu$ and $\kappa$ the kinematic
viscosity and the thermal diffusivity coefficients,
and $g$ is the acceleration due to  gravity.
In   \cite{bif03a}
fully periodic boundary conditions were  used for the velocity field,
$\Bu$, and temperature, $T'$, fields.

Anisotropic effects in the \rb system were analyzed in \cite{bif03a}
starting from the  stationary equation for the second order
velocity structure functions; the extension of K\'arm\'an-Howarth
equation in the presence of a buoyancy term \cite{yak92}. The result
is, neglecting for simplicity tensorial symbols:
\bea
\label{general2}
\la \delta u(\r)^3 \ra &\sim& \ \ \eb \ r \ + \ \alpha g\unitz r\
\cdot \ \la \delta T(\r)\ \delta u(\r)\ra\\ _{j = 0,1,\dots} &\ & \ \
_{j=0} \quad \ \ _{j=1 \quad \otimes \quad j=1,2,\dots} \nonumber \eea
where  $\bar \epsilon$  denotes the energy dissipation, $\la
\delta u(\r)^3 \ra$ and $\la \delta T(\r)\ \delta u(\r)\ra$, the
general third-order velocity correlation and temperature-velocity
correlation, respectively.  In Eq.~(\ref{general2})  for each term the
value of its total angular momentum, $j$, is indicated.  Notice that
the
energy dissipation term in (\ref{general2}) has a non-vanishing limit,
for high Re, only in the isotropic sector, $j=0$. On the
other hand, the buoyancy coupling, $\alpha g\unitz$, brings only
angular momentum $j=1$. Due to the usual rule of composition of angular
momenta we have that the buoyancy term, $\alpha g \unitz \cdot \la
\delta T(\r)\ \delta u(\r)\ra$, has a {\it total} angular momentum
given by: $j_{tot}= 1 \otimes j = \{j-1,j,j+1\}$.  Using the
angular momenta summation rule for $j$, one can decompose the previous
equation obtaining the following dimensional matching, in the
isotropic sector: $$
\la \delta u(r)^3 \ra_{j=0} \sim  \epsilon\ r +
\alpha g \hat{{\bf z}} r \la \delta u(r) \ \delta
T(r)\ra_{j=1} + \dots $$
and in the anisotropic sectors, $j >0$:
\be
\la \delta u(r)^3 \ra_{j} \sim  \ \ \alpha g\ \hat{{\bf z}}
r\ \la \delta u(r) \ \delta T(r) \ra_{(j-1)} + \dots
\label{aniso}
\ee
where sub-dominant contributions coming from the $j$ and $j+1$ sectors
of $\la \delta v(r) \ \delta T(r) \ra$ are neglected.

In the isotropic sector the buoyancy term is sub-dominant with respect
to the dissipation term at scales smaller than the Bolgiano length,
$L_B=(\bar \epsilon)^{5/4}N^{-3/4}(\alpha g)^{-3/2}$ where $N$ is the rate
of temperature dissipation.
 This is the case for the numerical simulation presented in
\cite{bif03a}, where velocity fluctuations are closer to the typical
Kolmogorov scaling, $\delta u(r) \sim r^{1/3}$, rather than to the
Bolgiano-Obukhov scaling \cite{my75},
$\delta u(r) \sim r^{3/5}$.\\ Regarding the
anisotropic sectors, Eq.(\ref{aniso})
is the  {\it dimensional prediction} for the system,
consistent with the anisotropic properties of the buoyancy term,
sector by sector.   \\ In \cite{bif03a} the SO(3)
decomposition was applied in this system
to  velocity structure functions (\ref{eq:fundamental}) and to objects
 $$G^{(q,1)}(\r) =
\la\left[\lp\Bu\lp\r\rp-\Bu\lp 0 \rp\rp\cdot \unitr\right]^{q}\lp T\lp\r
\rp-T\lp 0 \rp\rp \ra\, = \sum_{jm}  G^{(q,1)}_{jm}(r)$$

The dimensional matching of Eq.~(\ref{aniso}) can be extended to any
order, giving:
$$
   S^{(p)}_{jm}(r) \sim r { G}^{(p-2,1)}_{j-1,m}(r).
$$
Denoting with $\chi_j^{(q,1)}$ the anisotropic scaling exponents of
the buoyancy terms, ${G}^{(q,1)}_{jm}(r) \sim r^{\chi_j^{(q,1)}}$
we
get the dimensional estimate:
\be
\zeta_j^{(p)} = 1 + \chi_{j-1}^{(p-2,1)}, \quad \mbox{ (dimensional
prediction)}
\label{dim_exp}
\ee
In \cite{bif03a} it was shown that
this dimensional prediction is not obeyed; the exponents $\zeta_j^{(n)}$
appear to be systematically smaller than the prediction (\ref{dim_exp}).
 Interestingly, enough the log-log plots computed in the sectors
$j=4,6$ show a good
qualitative agreement with those calculated in the RKF
of ref \cite{bif01a} as can be seen in
fig.~\ref{fig:RKF+HRB.1} where we  compare  the
projection on the $j=4$ sector of  structure
functions of different orders.
\begin{figure}[h]
\includegraphics[scale=0.9]{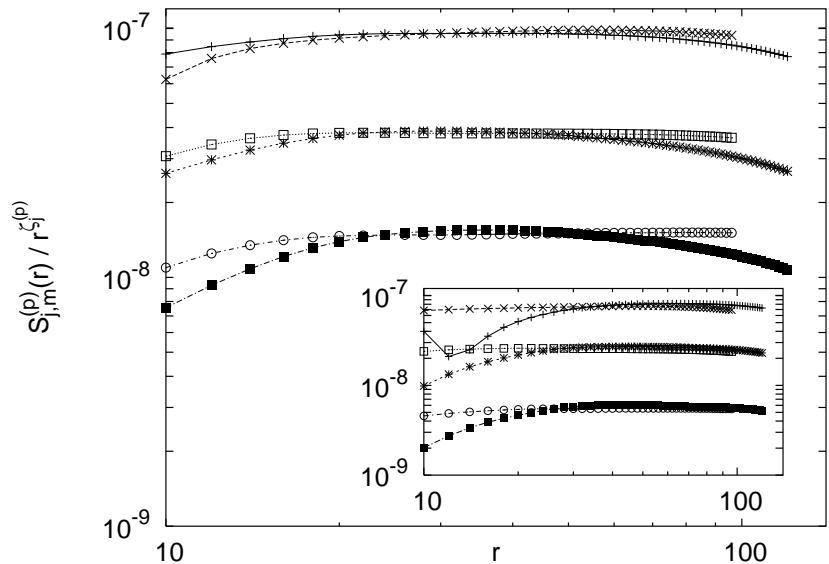}
\caption{ Log-log plot of compensated anisotropic $j=4,m=0$ projections
  $S^{(p)}_{4,0}(r)/r^{\zeta_4^{(p)}}$ {\it vs} $r$, for both HRB and RKF
  flows. Top curves refer to $p=2$: the best fit exponents which
  compensate HRB and RKF curves are $\zeta_4^{(2)}=1.7$ and
  $\zeta_4^{(2)}=1.66$, respectively. Curves in the middle refer to the
  same quantities but for $p=4$: compensation has been obtained with
  $\zeta_4^{(4)}=2.05$ for HRB, and $\zeta_4^{(4)}=2.2$ for RKF. Bottom
curves
  refer to $p=6$: here $\zeta_4^{(6)}=2.3$ for HRB, and $\zeta_4^{(6)}=2.5$
for
  RKF. Notice that the curves of the two flows are compensated with
  very similar values of the exponents (within $10\%$). Inset: the
  same but for $j=4,m=2$, compensation has been done with the same
  values used for $j=4,m=0$, to show the independence of the scaling
  exponents from the choice of the reference axis labeled with $m$ .
}
\label{fig:RKF+HRB.1}
\end{figure} 
Similar results are obtained for $j=6$
sector. These preliminary findings, if confirmed by other
independent measurements, would support  {\it universality}
for anisotropic scaling exponents in three dimensional turbulence.
\subsection{Scaling of Longitudinal and Transversal structure functions}
As discussed in
Sect.~\ref{sec:long-tran}
 there  exists  experimental and numerical data  suggesting that
{\it longitudinal} and {\it transversal } structure functions
in supposedly isotropic flows show
different scaling exponents \cite{she02,che97,dhr97,got02,wat99}.
  One needs to clearly
distinguish between experimental and numerical data.  The former can
never be considered fully isotropic; the best one can do is to try to
perform a multi-fit procedure to clean out sub-leading anisotropic
contributions as already explained in details in Sect.
\ref{chap:experiment}. This fitting procedure is, of
course, affected by experimental errors which cannot be eliminated.
Therefore it
is quite dangerous to make any firm conclusion about
supposed different scaling exponents of longitudinal and transversal
isotropic scaling on the basis of only experimental data .
Numerical data are not much safer. Here anisotropy can be much better
controlled.
With isotropic forcing the only source of
anisotropy is the
 3-dimensional grid whose effect is usually too small
to explain
possible discrepancies between longitudinal and transversal
scalings. Indeed some state-of-the-art isotropic
DNS indicate the possibility
of different scaling exponents both for inertial range structure functions
\cite{got02} and for coarse grained energy and enstrophy measures
\cite{che97,che97b}.   In Table~4 we
summarize the best-fit values of the scaling exponents measured in
\cite{got02}.
The small scale fluctuations were probed \cite{che97b}
by comparing the 
scaling of the coarse grained energy dissipation
 over a box of size $r$, $\tilde \epsilon (r)$ (Eq.~\ref{eq:coarse}),
and of the coarse grained enstrophy dissipation:
$\omega(r,\Bx) = {1\over r^3}
\int_{|y| < r} d\By ~\omega(\Bx +\By)$
where $\omega(\Bx)$ is the local enstrophy dissipation.  Different
scaling exponents were measured for the averaged quantities, $\la
(\tilde \epsilon)^p(r)\ra$, $\la\omega^p(r)\ra$.  Being scalar
quantities one expects that in isotropic ensembles they would not have
different exponents. From the theoretical point of view, different scaling
exponents of
longitudinal and transverse structure functions in isotropic ensembles
are unlikely.  In the language of the SO(3) decomposition it amounts
to the scaling exponents depending on the $q$-index which labels
different basis functions with the same rotation properties. In the
exactly solvable models examined before, this had never happened. In
general one would need a different symmetry to lift the degeneracy of
different $q$ dependent basis functions. At this point this problem
remains somehow unsettled. New numerical tests on larger grids and/or
with a better resolved viscous behavior are needed before a firm
statement can be made.
\begin{table}[h]
\begin{tabular}{|c|c|c|c|c|c|}
\hline
n&2&4&6&8&10 \\
\hline
$\zeta^{(n)}_0$ &0.701 (14)& 1.29 (3)&  1.77 (4)& 2.17 (7)& 2.53 (9) \\
\hline
$\zeta^{(n)}_0$ & 0.709 (13) &1.27 (2)& 1.67 (4) & 1.93 (9) & 2.08 (18)\\
\hline
\end{tabular}
\label{table:gotoh}
\caption{Measured values of
the longitudinal (first raw) and transverse (second raw)
 scaling exponents
at $Re_{\lambda}=460$ taken from \cite{got02}. One should note that
the scaling range displayed by the scaling plots in \cite{got02} are
relatively short, indicating that finite Re effects may still
be rather important}
\end{table}
\subsection{Anisotropies in decaying turbulence}
Decaying turbulence has attracted the attention of various communities
and is often considered in experimental, numerical and theoretical
investigations \cite{bat53,fri95,my75}. It is in fact quite common
that even experiments aimed at studying stationary properties of
turbulence involve processes of decay. Important examples are provided
by a turbulent flow behind a grid (see \cite{skr00} and references therein)
or the turbulent flow created
at the sudden stop of a grid periodically oscillating within a bounded
box \cite{des94}.  In the former case, turbulence is slowly decaying
going farther and farther away from the grid and its characteristic
scale becomes larger and larger (see \cite{skr00} for a thorough
experimental investigation). Whenever there is sufficient separation
between the grid-size $L_{in}$ and the scale of the tunnel or the tank
$L_{0} \gg L_{in}$, a series of interesting phenomenological
predictions can be derived.  For example, the decay of the two-point
velocity correlation function, for both isotropic and anisotropic
flows, can be obtained under the so-called self-preservation
hypothesis (see \cite{my75} chapter XVI). That posits the existence of
rescaling functions allowing to relate correlation functions at
different spatial and temporal scales. By inserting the
rescaling function
into the equations of motion, asymptotic results can be obtained both
for the final viscosity-dominated regime and for the intermediate
asymptotic when nonlinear effects still play an important role.

Here, we review some recent attempts to investigate the decay of
three-dimensional homogeneous and anisotropic turbulence by direct
numerical simulations of the Navier-Stokes equations in a periodic box
\cite{bif03c} for both short and large times.  The initial
conditions are taken from the stationary ensemble of the Random
Kolmogorov Flow discussed in the previous subsection.  Here the
correlation length-scale of the initial velocity field $L_{in}$ is of
the order of the size of the box $L_{0}\approx L_{in}$.

 On the one hand, one is interested  in the
long time decay regime where the typical interesting questions are: (i) how
do global quantities, such as single-point velocity and vorticity
correlations, decay? (ii) What is the effect of the outer boundary on
the decay laws~?  (iii) Do those quantities keep track of the initial
anisotropy~? (iv) As for the statistics of velocity differences within
the inertial range of scales, is there a recovery of isotropy at large
times~? (v) If so, do strong fluctuations get isotropic at a
faster/slower rate with respect to those of average intensity~? (vi)
Do anisotropic -and isotropic- fluctuations decay self-similarly~?
(vii)If not, do strong fluctuations decay slower or faster than
typical ones~? \
On the other hand, the
interest in the early stages of the decay is led by a hope of
establishing a link between the small-scale velocity statistics in
this phase and in forced turbulence. If such links existed, they would
shed additional light on the universality of forced
turbulence. 
As turbulence decays, the effective Re
 decreases, while the viscous characteristic scale and time
increase.  In \cite{bif03c} an offline analysis at fixed
multiples $\{0,1,10,10^{2},10^{3},10^4,10^5,10^6\}\,\tau_0$ of the
initial large-scale eddy turnover time $\tau_0 = L_{0}/u_{rms}^{t=0}$
was performed. \\ A first hint on the restoration of isotropy at large
times can be extracted from the analysis of single point quantities as
$$ E_{il} = {\overline{u_i(\Bx, t)u_l(\Bx, t)}},\quad
\Omega_{il} = {\overline{\omega_i(\Bx, t) \omega_l(\Bx, t)}}.  $$
Here with $\overline{\cdots}$ we denote the average over spatial
coordinates only, whereas $\la \cdots \ra$  indicates the
average over both initial conditions and space. The symmetric
matrices $E_{il}(t)$ and $\Omega_{il}(t)$ can be diagonalized at
each time-step and the eigenvalues $E_1(t),E_2(t),E_3(t)$ and
$\Omega_1(t),\Omega_2(t),\Omega_3(t)$ can be extracted.  The typical
decay of $E_{i}(t)$ and $\Omega_{i}(t)$ for $i=1,\dots,3$ is
shown in Fig.~\ref{fig:2}.
\begin{figure}[h] 
\epsfbox{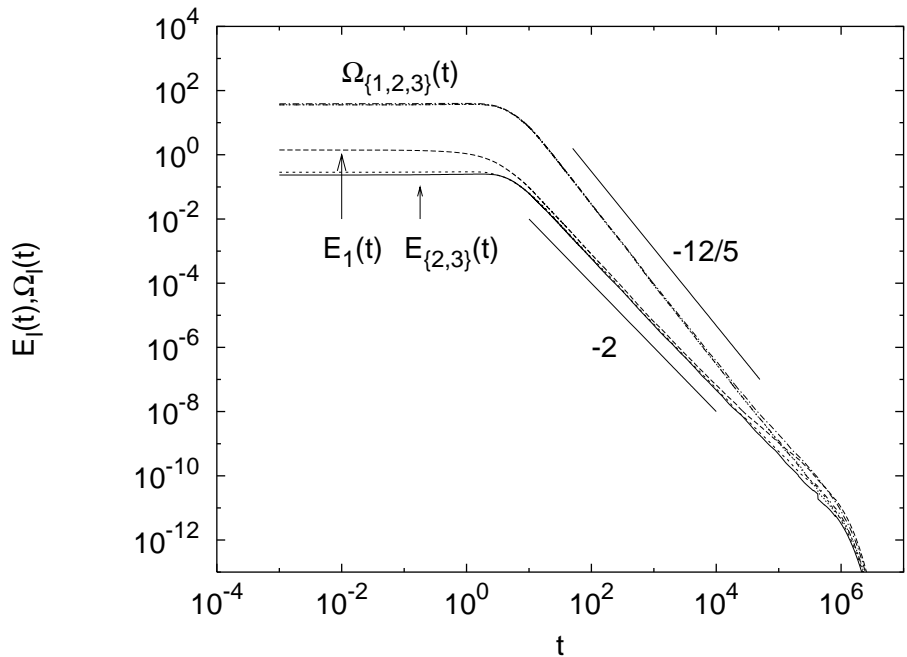}
\vspace{4pt}
\caption{Log-log plot of the eigenvalues of energy and vorticity matrices
{\it vs.} time, expressed in $\tau_0$ unit.}
\label{fig:2} 
\end{figure} 
 During the self-similar stage, $t\in
[10,10^6]$, the energy eigenvalues fall off as $E_{\{1,2,3\}}\sim
t^{-2}$, as expected for the decay in a bounded domain
\cite{skr00,bor95}. The enstrophy eigenvalues,
$\Omega_{\{1,2,3\}}$ decay as $t^{-12/5}$ as predicted from a
simple dimensional argument \cite{bif03c}.
  To  focus on the process of recovery of isotropy
in terms of global quantities one may track the behavior of
 two sets of observables
$$ \Delta_{il} E(t) =\frac{\la E_i(t)-E_l(t) \ra}{\la E_i(t) +
E_l(t)}
\ra\,,\quad
\Delta_{il} {\Omega}(t) = \frac{ \la \Omega_i(t)-\Omega_l(t) \ra}
{\la \Omega_i(t)  + \Omega_l(t)\ra } \,, $$
which vanish for isotropic statistics.  Their rate of decay is therefore
a direct measurement of the return to isotropy. The energy matrix $E_{il}$
is particularly sensitive to the large scales while small-scale
fluctuations are sampled by $\Omega_{il}$.  As  seen from
Fig.~\ref{fig:3},
\begin{figure}[h] 
\epsfbox{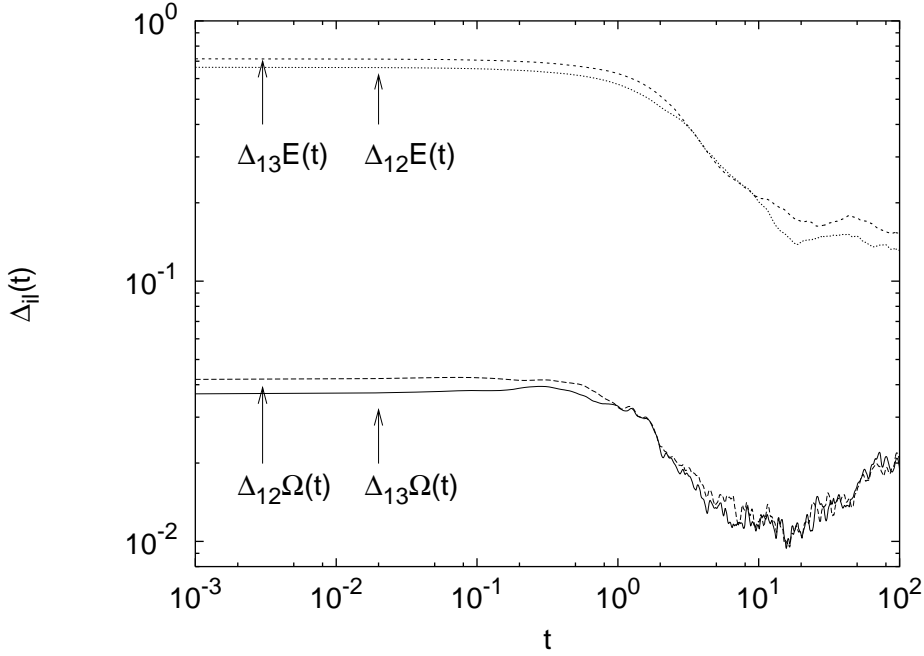}
\caption{Log-log plot of the anisotropy content at the large scales
($\Delta_{12}E(t)$, $\Delta_{13}E(t)$, top curves) and the small scales
($\Delta_{12}\Omega(t)$, $\Delta_{13}\Omega(t)$, bottom curves) as a
function
of time, expressed in $\tau_0$ unit. The large-scale (small-scale)
anisotropy 
content is defined as the mismatch between the eigenvalues of the
single-point 
velocity (vorticity) correlation.}
\label{fig:3} 
\end{figure}
 both large and small scales begin to isotropize
after roughly one eddy turnover time and become fully isotropic
(within statistical fluctuations) after $100$ eddy turnover
times. However, small scales show an overall degree of anisotropy much
smaller than the large scales.

Concerning small scales properties, in \cite{bif03c}
a  simple anisotropic generalization of the
self-preservation hypothesis (see, e.g.~Ref.~\cite{fri95}) was proposed:
$$
 S^{(n)}_{jm}\lp r,t\rp =
V^{(n)}_{jm}\lp t \rp f^{(n)}_{jm} \lp r/L_{jm}(t) \rp .
$$
Here  with  $V^{(n)}_{jm}\lp t \rp$ we take explicitly into account
the fact that large-scale velocity  properties may depend in a nontrivial
way
on both $(j,m)$ and the order $n$.  Furthermore, $L_{jm}(t)$ accounts
for the possibility that the characteristic length scale depend on the
SO(3) sector.  In analogy with the observations made in the
stationary case \cite{gar98,ara98,ara99,kur00,bif01a,bif02,she02,bif01}
a scaling law  was postulated:
\be S^{(n)}_{jm}\lp r,t \rp \sim a^{(n)}_{jm}(t) \lp \frac{r}
{L_{jm}(t)}\rp^{\zeta_{j}^{(n)}}\,.
\label{anyt}
\ee
The time behavior is encoded in both the decay of the overall
intensity, accounted by the prefactors $a^{(n)}_{jm}(t)$, and the
variation of the integral scales $L_{jm}(t)$. The representation
Eq.~(\ref{anyt}) is the simplest one fitting the initial time statistics
for $t=0$ and agreeing with the evolution given by the self
preservation hypothesis in the isotropic case. The power law behavior
for $f^{(n)}_{jm} \lp r/L_{jm}(t) \rp$ can be expected only in a
time-dependent inertial range of scales $ \eta(t) \ll r \ll L(t)$.  As
for the exponents appearing in (\ref{anyt}), their values are
expectedly the same as in the stationary case.  Concerning the time
evolution, it seems difficult to disentangle the dependence due to the
decay of $a^{(n)}_{jm}(t)$ from the one due to the growth of the
integral scale $L_{jm}(t)$.  The existence of a running reference
scale, $L_{jm}(t)$ introduces some non-trivial relations between the
spatial anomalous scaling and the decaying time properties, and those
relations might be subject to experimental verification.  In the case
discussed in \cite{bif03c}, the fact that the initial condition
has a characteristic length-scale comparable with the box size,
simplifies the matter.  Indeed we expect that $L_{jm}(t) \approx
L_{0}$, and the decay is due only to the fall-off of
 $a^{(n)}_{jm}(t)$.  An obvious shortcoming is that the width
of the inertial range $L_{0}/\eta(t)$ shrinks monotonically in time,
thereby limiting the possibility of precise quantitative statements.
\subsubsection{Long time decay}
A quantitative way to define the temporal rate of recovery of isotropy
at a fixed scale in the inertial range is given by the
dimensionless ratio:
\be \Pi^{(n)}_{jm}\lp r,t \rp \equiv
\frac{S^{(n)}_{jm}\lp r,t \rp}{S^{(n)}_{0,0}\lp r,t \rp} \sim
t^{-\Xi_{j}^{(n)}}\,.
\label{ratio_n}
\ee
In Fig.~\ref{fig:9},
\begin{figure}[h] 
\epsfbox{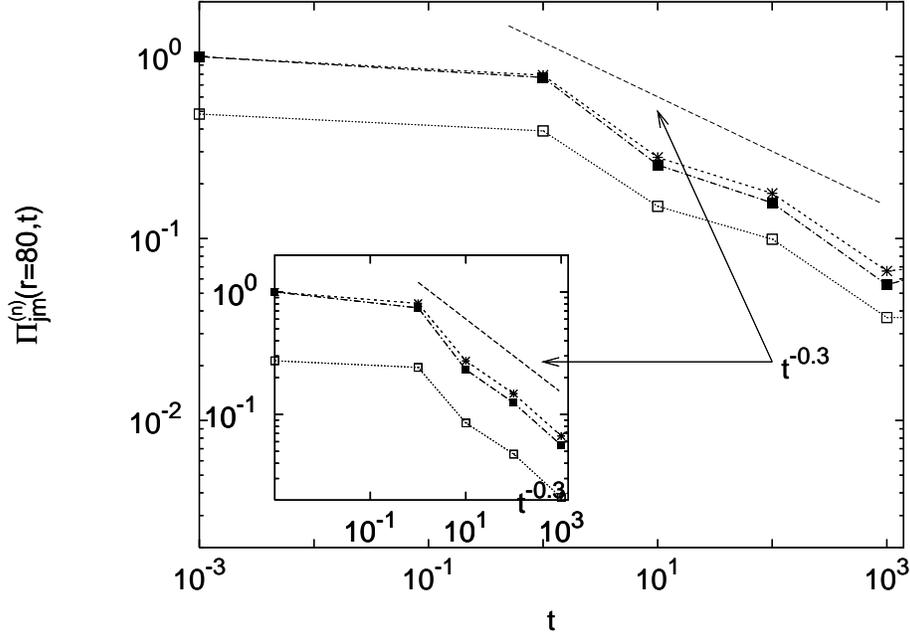}
\caption{Hierarchical organization of anisotropic fluctuations at long
times. Log-log plot of the anisotropic projections normalized by the
corresponding isotropic projection (see text), at two fixed scales $r=80$
and
$r=40$ (inset) for $n=2, 4, 6$ in the anisotropic sector $j=4,m=0$.
Symbols read as follows\,: $\Pi^{(2)}_{40}$ (full box);
$\Pi^{(4)}_{40}$ (star);
 $\Pi^{(6)}_{40}$ (empty box). The straight line
is  $t^{-\Xi}$ with $\Xi \sim 0.3$.
 Same symbols in the inset.}
\label{fig:9} 
\end{figure}
 we  plot  $\Pi^{(n)}_{jm}\lp r,t \rp$ at $r=80$
for structure functions of order $n=2,4,6$ and for the most intense
anisotropic sector, $(j,m)=(4,0)$.  All anisotropic sectors,
for all measured structure functions, decay faster than the isotropic
one. The measured slope in the decay is about $\Xi_j^{(n)} \sim 0.3$
for all $n$, within the statistical errors.
Note that these results agree with the simple picture that the
time-dependence in (\ref{anyt}) is entirely carried by the prefactors
$a_{jm}^{(n)}(t)$ and the value of the integral scales $L_{jm}(t)$ is
saturated at the size of the box. Indeed, by assuming that large-scale
fluctuations are almost Gaussian we have that the leading
time-dependence of $a_{jm}^{(2n)}$ is given by $a_{jm}^{(2)}
a_{00}^{(2n-2)}$.  For the isotropic sector, $a_{00}^{(2n)} \sim
(a_{00}^{(2)})^n$, and plugging that in (\ref{ratio_n}), one get:
$\Pi^{(n)}_{jm}\lp r,t \rp \sim a^{(2)}_{jm}(t)/a^{(2)}_{00}(t) \sim
t^{-\Xi}$ with $\Xi \sim 0.3 (\pm 0.1)$ independent of $n$. The
quality of data is insufficient to detect possible residual effects
due to $L_{jm}(t)$, which would make $\Xi_j^{(n)}$ depend on $n$ and
$j$ because of spatial intermittency.\\

 Let us denote with $\cP(\Delta,\r;t)$ the probability to observe a
 given longitudinal fluctuation, $\delta u_{\ell}(\B r,t)=\Delta$
 in the direction $\r$ at a given time,
 $t$. For any given fixed value $\Delta$ and for any given time, $t$,
 we can project $\cP(\Delta,\r;t)$ on the SO(3) basis functions:
\be 
\cP(\Delta,\r;t) =  \sum_{j=0}^{\infty}\sum_{m=-j}^{j}
\cP_{jm}\lp r,\Delta;t\rp Y_{jm}(\unitr),
\label{so3_pdf} 
\ee 
where now the projection, $\cP_{jm}\lp r,\Delta;t\rp$ play the role
of an {\it effective PDF} for each  SO(3) sector. The projection
of any longitudinal structure function,
$ S^{(n)}(\r,t) $ on any sector, $(j,m)$ can be reconstructed from the
corresponding  
projection of the  PDF on the same sector,
 $\cP_{jm}\lp r,\Delta;t\rp$, by  averaging over all possible $\Delta$:
$$ 
S^{(n)}_{jm}\lp r,t\rp = \int d\Delta \Delta^n \cP_{jm}\lp r,\Delta;t\rp
$$
which establish  the link between decomposition (\ref{eq:fundamental}) and
(\ref{so3_pdf}).\\ 

The interesting fact that the
decay properties of the anisotropic sectors are almost independent of
$n$ indicates that a non-trivial time
dependence in the shape of the PDF's $\cP_{jm}\lp r ,\Delta;t\rp$ for
$j>0$ must be expected. The most accurate way to probe the rescaling
properties of $\cP_{jm}\lp r,\Delta;t\rp$ in time is to compute the
generalized flatness:
$$
K_{jm}^{(n)}(r,t)\equiv \frac{S^{(n)}_{jm}\lp r,t \rp}{\lp
S^{(2)}_{jm}\lp r,t \rp \rp^{\frac{n}{2}}} \sim t^{\alpha_{j}^{(n)}}
$$
Were the PDF projection in the $(j,m)$ sector self-similar for $t \gg
\tau_0$, then $K_{jm}^{(n)}(r,t)$ would tend to constant values. This
is not the case for anisotropic fluctuations, as it is shown in
Fig. (\ref{fig:10}).
\begin{figure}[h] 
\epsfbox{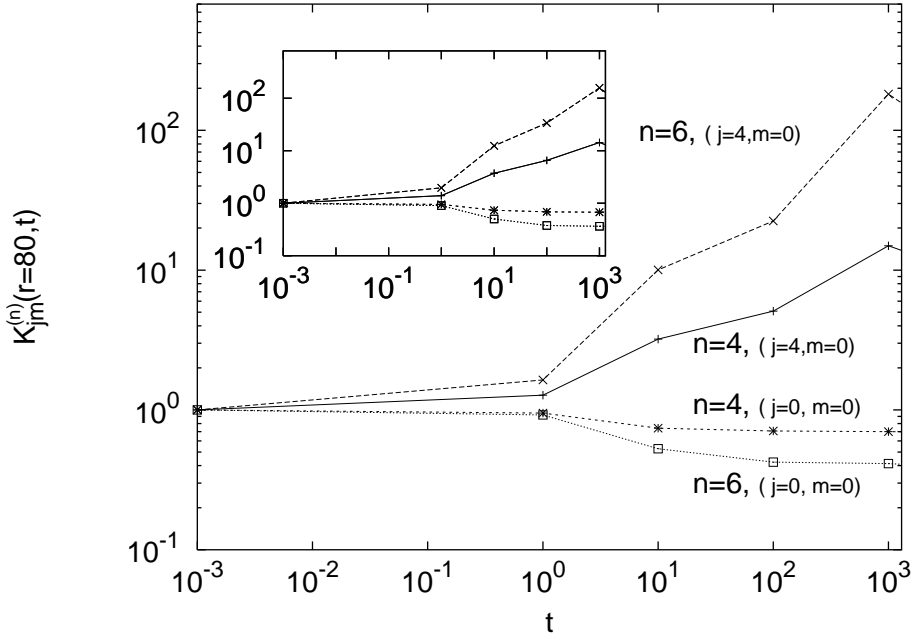}
\caption{Log-log plot of the generalized flatness, $K_{jm}^{(n)}(r,t)$
of order $n=4,6$ for both the isotropic (two bottom curves), and
the anisotropic sector $(j=4,m=0)$ (two top curves) at $r=80$, and as
a function of time. In the inset we plot the same quantities, in the same
order, at a different inertial range scale, $r=40$.}
\label{fig:10} 
\end{figure} 
The curves $K_{jm}^{(n)}(r,t)$ are collected for
two fixed inertial range separations, $r=80$ and $r=40$ (inset), for
two different orders, $n=4,6$ and for both the isotropic and one of
the most intense anisotropic sectors $(j=4,m=0)$ . The isotropic
flatness tends toward a constant value for large $t$. Conversely, its
anisotropic counterparts are monotonically increasing with $t$,
indicating a tendency for the anisotropic fluctuations to become more
and more intermittent as time elapses. Also the behavior in
Fig.~\ref{fig:10} is in qualitative agreement with the observation
previously made that all the time dependence can be accounted for  by the
prefactors $a_{jm}^{(n)}(t)$. Indeed, assuming that the length scales
$L_{jm}(t)$ have saturated and that the large scale PDF is close to
Gaussian, it is easy to work out the prediction $K_{jm}^{(n)}(r,t)
\sim t^{-\Xi(1-n/2)}$, i.e.  $\alpha_{j}^{(n)}=\Xi(n/2-1)$.
 We conclude this
section by a brief summary of the results.  It was found that
isotropic fluctuations persist longer than anisotropic ones,
i.e. there is a time-recovery, albeit slower than predicted by dimensional
arguments, of isotropy during the decay process.  It was  also found
that isotropic fluctuations decay in an almost self-similar way while
the anisotropic ones become more and more intermittent. Qualitatively,
velocity configurations get more isotropic but anisotropic
fluctuations become, in relative terms, more ``spiky'' than the
isotropic ones as time elapses.
\subsubsection{Short-time decay}
\label{shorttimes}
It is interesting to notice that it is possible to relate the
small-scale universal properties of forced turbulent statistics to
those of short-time decay for an ensemble of initial configurations
\cite{bif03c}. As already remarked, one cannot expect an
universal behavior for all statistical observables, as the very
existence of anomalous scaling is the signature of the memory of the
boundaries and/or the external forcing throughout all the scales.
Indeed, the main message we want to convey here is that only the
scaling exponents of both isotropic and anisotropic small-scale
fluctuations are universal, at least for forcing concentrated at large
scales.  The prefactors are not expected to be so.  There is therefore
no reason to expect that quantities such as the skewness, the kurtosis
and in fact the whole PDF of velocity increments or gradients be
universal. \\ 
This is the same situation that we discussed in great details in previous
sections
 for the passive transport
of scalar and vector fields.
 However, carrying over the analytical
knowledge developed for linear hydrodynamic problems involve some
nontrivial, yet missing, steps. For the Navier-Stokes dynamics, linear
equations of motion appear when we consider
the whole set of correlation functions as discussed  in
Sect.~\ref{sec:hierachy}.
 These equations
can be rewritten in a  schematic form:
\be
\partial_t C^{(n)} = \Gamma^{(n+1)}C^{(n+1)} +\nu D^{(n)} C^{(n)} + F^{(n)},
\label{compact}
\ee
where $\Gamma^{(n+1)}$ is the integro-differential linear operator
coming from the inertial and pressure terms,  $C^{(n)}$ is a
shorthand notation for a generic $(n)$-th order correlator and
  $D^{(n)}$ is the linear
operator describing dissipative effects. Finally, $F^{(n)}$ is the
correlator involving increments of the large-scale forcing $\f$ and of
the velocity field.  The balance between inertial and injection
terms cannot lead to anomalous scaling.  A natural possibility is that
a mechanism similar to the one identified in linear transport problems
be at work in the Navier-Stokes case as well. The anomalous
contributions to the correlators would then be associated with
statistically stationary solutions of the unforced equations
(\ref{compact}). The scaling exponents would {\it a fortiori} be
independent of the forcing and thus universal. As for the prefactors,
the anomalous scaling exponents are positive and thus the anomalous
contributions grow at infinity. They should then be matched at the
large scales with the contributions coming from the forcing to ensure
that the resulting combination vanish at infinity, as required for
correlation functions. The  aim here is not to prove the previous
points but rather to test whether they fail: the
Navier-Stokes equations, being integro-differential and  non-local,
 might directly couple inertial and injection scales and
spoil the argument. This effect might be particularly relevant for
anisotropic fluctuations where infrared divergences may appear in the
pressure integrals (see Sect.~\ref{chap:linearP}).
In order to investigate the previous point, we performed two sets of
numerical experiments in decay.

 The first set, A, is of the same kind
as in the previous section, i.e. we integrated the unforced
Navier-Stokes equations
with initial conditions picked from an ensemble obtained from a forced
anisotropic stationary run.  Statistical observables are measured as
an {\it ensemble} average over the different initial conditions.
 The ensemble at the initial time of
the decay process therefore coincides with  the stationary
state in forced runs. If correlation functions are indeed dominated
at small scales by statistically stationary solutions of the unforced
equations then the field should not decay. Specifically, the field
should not vary for times smaller than the large-scale eddy turnover
time $\tau_0$. Those are the times when the effects
of the forcing terms start to be felt. Note that this should hold
at all scales, including the small ones whose turnover times are much
faster than $\tau_0$.

The second set of numerical simulations (set B) takes the same initial
conditions but for the random scrambling of the phases\,:
$\Bu_i(\k)
\rightarrow P_{il}(\k)  \B u_l(\k) \exp(i \theta_l(\k)) $, with
$\theta_l(\k)$ i.i.d. random variables.  In this way, the spectrum and
its scaling exponent are preserved but the wrong organization of the
phases is expected to spoil the statistical stationarity of the
initial ensemble. As a consequence, two different decays are expected
for the two sets of initial conditions.  In particular, contrary to
set A, set B should vary at small scales on times of the order of the
eddy turnover times $\tau_r \sim r^{2/3}$.  This is exactly what has
been found in the numerical simulations for both isotopic and
anisotropic statistics as can be seen for the anisotropic case in
Fig.~\ref{fig:12},
\begin{figure}[h] 
\epsfbox{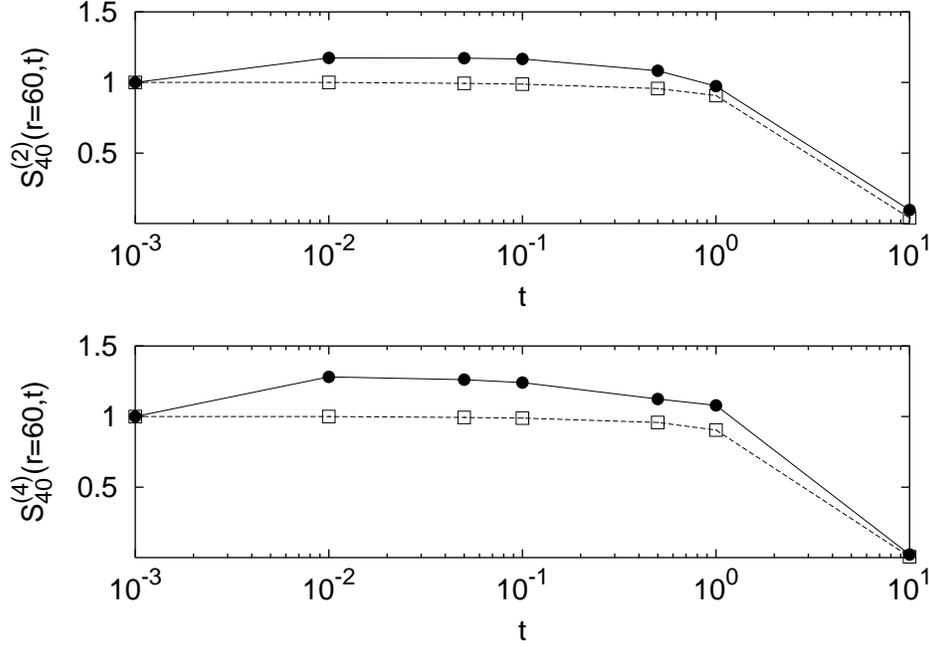}
\caption{
Top: Temporal decay of the second-order anisotropic structure
function $S^{(2)}_{40}(r,t)$, rescaled by its value at $t=0$.
Here $r=30$, inside the inertial
range. The two curves refer to the time evolution of the structure
function starting from the forced-stationary velocity
fields (squares, set A) and from the randomly dephased velocity
fields (circles, set B). Time is normalized by the integral eddy turnover
time.
Notice that for set B we observe changes on a time
scale faster than the integral eddy turnover time. That is to be
contrasted with the case A, where structure functions are strictly constant
in time up to an integral eddy turnover
time. Bottom: The same curves but for the fourth-order structure function.}
\label{fig:12} 
\end{figure} 
 where the temporal behavior of
longitudinal structure functions of order 2 and 4 is shown. The
scaling exponents of the contributions responsible for the observed behavior
at
small scales are thus forcing independent.\\
To conclude, the data presented here support the conclusion that
nonlocal effects peculiar to the Navier-Stokes dynamics do not spoil
arguments on universality based on analogies with passive turbulent
transport. The picture of the anomalous contributions to the
correlation functions having universal scaling exponents and
non-universal prefactors follows.

\newpage
%
%

\section{Concluding Discussion, }
\label{chap:conclusions}

In this review we presented a mathematical framework in which
anisotropy in turbulence can be studied, and we have tested its
utility in the context of experimental analysis, numerical simulations
and analytical models. The basic idea is to express the various
statistical quantities of turbulence (e.g., structure functions,
correlation functions) in terms of their projections on the different
sectors of the SO($3$) group.

The utility of the SO($3$) decomposition should be assessed in two
main aspects. The first aspect is its functionality as a tool for
characterizing anisotropy, whereas the second, and deeper aspect, is its
physical relevance and the theoretical and analytical advantages that
are gained by using it.

As a tool for describing anisotropy, the SO($3$) decomposition is
probably the most natural and general method. It is of \emph{high
resolution} - it subdivides the observed anisotropy into different
sectors - the $(j,m)$ sectors. The weights of the various sectors
give us a fine resolution of the anisotropy structure. Instead of
having one measure for anisotropy (e.g., the overall percentage of
anisotropy), we have an infinite set of numbers that compose a
detailed profile of the anisotropy structure.

The SO($3$) decomposition is also very \emph{general}. It is
applicable to any physical observable that has a well-defined
transformation under rotations. These can be, for example, correlation
functions, structure functions or Green's functions (response
functions). The observables themselves may depend on any number of
space coordinates or even be space independent. They may also be
scalars, vectors or tensors. Any such quantity can be presented as
a sum of parts that belong to the different $(j,m)$ sectors of the
rotation group. Additionally, since the SO($3$) decomposition is
invariant to isotropic operations, it is invariant to the most common
operations that we use. For example, to obtain the $n$th order
longitudinal structure function we can take the full $n$th order
structure function (which is a tensor) and contract it with $n$ unit
vectors in the direction of the separation distance. Since this
operation is linear and isotropic, it will preserve the $(j,m)$
sectors of the full structure function. That is, the $(j,m)$ sector
of the full structure function will be transformed into the same
$(j,m)$ sector of the longitudinal structure functions. The same
thing happens for operations such as differentiations (for example
when we look at moments of the gradient fields), space averaging, time
averaging, coordinate fusion etc... From the pure theoretical point of view
the  first and most obvious advantage of using the
SO($3$) decomposition is its elegance and the overwhelming
simplification that it offers in analytical calculations.
The SO($3$) decomposition may also have a deeper physical
justification if it produces universal quantities such as distinct
scaling behavior in the anisotropic sectors. There are several
different evidences that suggest that this is indeed the
case. Experimental results clearly show that a better scaling is
achieved if we take higher $j$ components into account.
Additionally, different experimental setups
seem to lead to the same numerical values of the anisotropic exponent.
 This is
 a strong support for the hypothesis that the
anisotropic sectors of the structure functions have  universal
exponents. 
Another support for the idea that the SO($3$) decomposition exposes
universal quantities, comes from numerical simulations.
In DNS, the SO($3$) decomposition
can be performed directly (since the velocity field is accessible in
every point in space and time) which makes the results much less
ambiguous. We clearly see that even in the very moderate Reynolds
numbers of the simulations, a scaling behavior is detected once we use
the SO($3$) decomposition. In some cases, without the SO($3$)
decomposition no scaling behavior is seen at all. Furthermore, the
resulting exponents in the isotropic sector are remarkably similar to
the experimental values which are measured at very high Reynolds
numbers. This is a strong indication that at least the isotropic
sector has a universal profile, and therefore by disentangling it from
the anisotropic sectors we get universal results.  In other sectors of
the rotation group, the scaling behavior is not as good, and in some
sectors there is no scaling at all. However, in those sectors where
scaling was detected, the scaling exponents seem to agree with the
theoretical and experimental predictions.
 Sectors with same $j$ and different $m$s had
the same scaling exponents (when scale-invariance was observed). And
finally, all exponents increased as a function of $n$ (order of the
structure-function) as well as $j$. It is still not clear whether
the ``bad'', non scale-invariant behavior that was detected in some
sectors, is a result of the poor Reynolds numbers of the simulations,
or is a genuine effect that tells us that the foliation picture is
incomplete. A further research with higher resolution is probably
needed to settle this issue.
In  the Navier-Stokes case
one can prove a  ``weak foliation''. Weak
foliation is an approximate foliation that happens in the case of weak
anisotropy, when we linearize the anisotropic part of the theory
around its isotropic part.
In such case, the linearized anisotropic part of the
theory is subject to a linear and isotropic equation whose kernel
contains the isotropic solution, and hence foliation occurs. This is a
very robust approximation since it holds for virtually any nonlinear
and isotropic theory in the case of weak anisotropy.
 Additionally, we know that as the Reynolds number
increases, the statistics becomes more and more isotropic and
therefore the linear approximation becomes better and better.

The SO($3$) decomposition has a physical relevance also in the
presence of strong anisotropy. If we merely consider the direct
application of the Navier-Stokes and continuity equations on the
various structure functions, we get a family of linear and isotropic
constraints. These constraints are valid independently of the
amount of anisotropy in the system. Their linearity and isotropy lead
to foliation and hence one can discuss them in every sector
independently. In some sectors, they are sufficient to determine the
full solution, whereas in others they can reveal some general
properties of the solution. For example, the isotropic sector of the
third-order structure function is completely determined by the
different constraints, and is given by the well-known 4/5 law of
Kolmogorov. Notice that because of foliation, this is true also in the
presence of anisotropy, which means that the 4/5 law holds also in an
anisotropic turbulence. In the $j=2$ sectors, on the other hand,
the third-order structure-function is given by two undetermined scalar
functions, whereas in the $j=4,6,\ldots$ it is given by
three. Another example is the $j=1$ sectors in the second-order
correlation function, which
 must all vanish. 

To conclude, the framework of the SO($3$) decomposition provides an
elegant and efficient way to describe anisotropy in turbulence. It
also greatly simplifies many analytical calculations that involve
anisotropic quantities, mainly through the mechanism of
foliation. This mechanism is present in simplified models of
turbulence, and may also be valid approximately in Navier-Stokes
turbulence. It predicts that the anisotropic sectors of the statistics
have universal properties such as scaling exponents. Further research
is needed to accurately measure these anisotropic exponents in
experiments as well as in numerical simulations.

A quantitative computation of the anisotropic exponents in Navier-Stokes
turbulence from first principles may very well be an illusive goal.
Nevertheless, we believe that the general principles of the physics
behind these exponents and anisotropy in general can be
understood.
\section{Acknowledgements}
We benefited from many discussion with many colleagues, without which we
could not reach the picture presented in this review. We thank in particular
Itai Arad, Roberto Benzi, Guido Boffetta, Carlo Casciola, Antonio Celani,
Isabelle Daumont, Siegfried Grossmann, Boris Jacob, Susan
Kurien, Alessandra  Lanotte,  Detlef Lohse, Victor S. Lvov, Irene
Mazzitelli,
Evgeny Podivilov, K.R. Sreenivasan, Federico Toschi and Massimo Vergassola.
Much
of the work reviewed here was
supported in part by the European Commission through a TMR grant (No.
HPRN-CT 2000-00162),
 the Italian Ministry of Education (MURST), the Instituto Nazionale di
Fisica
della Materia,  the German Israeli
Foundation, the Minerva
Foundation, Munich, Germany, the Israel Science Foundation, and the US
Israel Bi-National
Science Foundation.
\appendix
\section{The General Form of the 2nd Rank Tensor}
\label{sec:general-form}
In this appendix we discuss the general structure of the second rank
correlation functions
\begin{equation}
F^{\alpha \beta }({\bf r})\equiv\langle u^{\alpha }({\bf x}+{\bf r})u^{\beta
}({\bf 
x})\rangle \ ,  \label{cf}
\end{equation}
In (\ref{cf}) {\em homogeneity} of the flow is assumed, but not isotropy.
Note that this object is more general than the structure function $S^{\alpha
\beta }$ in being nonsymmetric in the indices, and having no definite
parity.  We wish to find the basis functions $B_{q,jm}^{\alpha \beta }(
{\bf \hat{r}})$, with which we can represent $F^{\alpha \beta }({\bf r})$ in
the form: 
\begin{equation}
F^{\alpha \beta }({\bf r})=\sum_{q,jm}a_{q,jm}(r)B_{q,jm}^{\alpha \beta
}({\bf
\hat{r}})  \label{trial-tensor}
\end{equation}
and derive some constraints among the functions $\ a_{q,jm}(r)$ that result
from incompressibility. We shall see, that due to the isotropy of the
incompressibility conditions, the constraints are among $a_{q,jm}(r)$ with
the {\em same} $j,m$ only.

We begin by analyzing the incompressibility condition: An incompressible
flow with constant density is characterized by the relation:
$
\partial _{\alpha }u^{\alpha }({\bf x},t)=0
$
as a result, one immediately gets the following constraints on $F^{\alpha
\beta }({\bf r})$: 
$$
\partial _{\alpha }F^{\alpha \beta }({\bf r}) =0, \qquad
\partial _{\beta }F^{\alpha \beta }({\bf r}) =0.
$$
Plugging the trial tensor (\ref{trial-tensor}) into the last two equations
we obtain 2 equations connecting the different $a_{q,jm}$:
\be
\partial _{\alpha }\sum_{q,jm}a_{q,jm}(r)B_{q,jm}^{\alpha \beta }({\bf
\hat{r}}
) =0  \label{incomp-1} \qquad
\partial _{\beta }\sum_{q,jm}a_{q,jm}(r)B_{q,jm}^{\alpha \beta }({\bf
\hat{r}})
=0  
\ee
We first notice that the differentiation action is isotropic. As a result,
if $T^{\alpha \beta }\left( {\bf r}\right) $ is some arbitrary tensor with a
definite $j,m$ transformation properties, then the tensor $\partial _{\alpha
}T^{\alpha \beta }({\bf r})$ will have {\em the same} $j,m$ transformation
properties. Components with different $j,m$ are linearly independent.
Therefore equations (\ref{incomp-1}) should hold for each $j,m$ separately.

Next, we observe that (\ref{incomp-1}) are invariant under the
transformation $
F^{\alpha \beta }\longrightarrow F^{\beta \alpha }$. As a result, the
symmetric and anti-symmetric parts of $F^{\alpha \beta }$ should satisfy
(\ref{incomp-1}) independently. To see that, let us write $F^{\alpha \beta
}$
as a sum of a symmetric term and an anti-symmetric term: $F^{\alpha \beta
}=F_{S}^{\alpha \beta }+F_{A}^{\alpha \beta }$, we then get:
\begin{eqnarray*}
\partial _{\alpha }F^{\alpha \beta } &=&\partial _{\alpha }F_{S}^{\alpha
\beta }+\partial _{\alpha }F_{A}^{\alpha \beta }=\partial _{\alpha
}F_{S}^{\beta \alpha }-\partial _{\alpha }F_{A}^{\beta \alpha }=0 \\
\partial _{\beta }F^{\alpha \beta } &=&\partial _{\beta }F_{S}^{\alpha \beta
}+\partial _{\beta }F_{A}^{\alpha \beta }=0
\end{eqnarray*}
from which we conclude:$
\partial _{\alpha }F_{S}^{\alpha \beta }=\partial _{\alpha }F_{A}^{\alpha
\beta }=0 .
$
Finally, (\ref{incomp-1}) is invariant under the transformation $F^{\alpha
\beta }({\bf r)}\longrightarrow F^{\alpha \beta }(-{\bf r)}$ and as a result
the odd parity and the even parity parts of $F^{\alpha \beta }$ should
fulfill (\ref{incomp-1}) independently. We conclude that a necessary and
sufficient condition for (\ref{incomp-1}) to hold is that it holds
separately for parts with definite $j,m$, definite symmetry in the $\alpha
,\beta $ indices and a definite parity in ${\bf r}$:
\[
\partial _{\alpha }\sum_{q}a_{q,jm}({ r})B_{q,jm}^{\alpha \beta }({\bf
\hat{r}})=0\mbox{ \
\begin{tabular}{l}
~~{\small summation is over} \\
~~$B_{q,jm}^{\alpha \beta }$ {\small with definite symmetries}
\end{tabular}
}
\]
where the summation is over $q$ such that $B_{q,jm}^{\alpha \beta }$ has a
definite indices symmetry and a definite parity.

According to (\ref{eq:second-rank-tensors}) we can write these
$B_{q,jm}^{\alpha \beta }$ as:
\begin{enumerate}
\item  $(-)^{j}$ parity, symmetric tensors:

\begin{itemize}
\item  $B_{1,jm}^{\alpha \beta }(\hat{{\bf r}})\equiv r^{-j}\delta ^{\alpha
\beta }\Phi _{jm}(${\bf $r$}$)$,
\item  $B_{7,jm}^{\alpha \beta }(\hat{{\bf r}})\equiv r^{-j}\left[ r^{\alpha
}\partial ^{\beta }+r^{\beta }\partial ^{\alpha }\right] \Phi _{jm}(${\bf
$r$
}$)$,
\item  $B_{9,jm}^{\alpha \beta }(\hat{{\bf r}})\equiv r^{-j-2}r^{\alpha
}r^{\beta }\Phi _{jm}(${\bf $r$}$)$,
\item  $B_{5,jm}^{\alpha \beta }(\hat{{\bf r}})\equiv r^{-j+2}\partial
^{\alpha }\partial ^{\beta }\Phi _{jm}(${\bf $r$}$)$.
\end{itemize}
\item  $(-)^{j}$ parity, anti-symmetric tensors:
\begin{itemize}
\item  $B_{3,jm}^{\alpha \beta }(\hat{{\bf r}})\equiv r^{-j}\left[ r^{\alpha
}\partial ^{\beta }-r^{\beta }\partial ^{\alpha }\right] \Phi _{jm}(${\bf
$r$
}$)$.
\end{itemize}
\item  $(-)^{j+1}$ parity, symmetric tensors
\begin{itemize}
\item  $B_{8,jm}^{\alpha \beta }(\hat{{\bf r}})\equiv r^{-j-1}\left[
r^{\alpha }\epsilon ^{\beta \mu \nu }r_{\mu }\partial _{\nu }+r^{\beta
}\epsilon ^{\alpha \mu \nu }r_{\mu }\partial _{\nu }\right] \Phi _{jm}(${\bf
$r$}$)$,
\item  $B_{6,jm}^{\alpha \beta }(\hat{{\bf r}})\equiv r^{-j+1}\left[
\epsilon ^{\beta \mu \nu }r_{\mu }\partial _{\nu }\partial ^{\alpha
}+\epsilon ^{\alpha \mu \nu }r_{\mu }\partial _{\nu }\partial ^{\beta }
\right] \Phi _{jm}(${\bf $r$}$)$.
\end{itemize}
\item  $(-)^{j+1}$ parity, anti-symmetric tensors:
\begin{itemize}
\item  $B_{4,jm}^{\alpha \beta }(\hat{{\bf r}})\equiv r^{-j-1}\epsilon
^{\alpha \beta \mu }r_{\mu }\Phi _{jm}(${\bf $r$}$)$,
\item  $B_{2,jm}^{\alpha \beta }(\hat{{\bf r}})\equiv r^{-j+1}\epsilon
^{\alpha \beta \mu }\partial _{\mu }\Phi _{jm}(${\bf $r$}$)$.
\end{itemize}
\end{enumerate}
In order to differentiate these expressions we can use the following
identities: 
\begin{eqnarray*}
r^{\alpha }\partial _{\alpha }r^{\zeta }Y_{jm}(\hat{{\bf r}}) &=&\zeta
r^{\zeta }Y_{jm}(\hat{{\bf r}}) \ ,\\
\partial ^{\alpha }\partial _{\alpha }r^{\zeta }Y_{jm}(\hat{{\bf r}}) &=&
\left[ \zeta \left( \zeta +1\right) -j(j+1)\right] r^{\zeta -2}Y_{jm}(
\hat{x})
\end{eqnarray*}
which give rise to:
\begin{eqnarray*}
r^{\alpha }\partial _{\alpha }\Phi _{jm}({\bf r}) &=&j\Phi _{jm}({\bf r}) \
.\\
\partial ^{\alpha }\partial _{\alpha }\Phi _{jm}({\bf r}) &=&0 \ .
\end{eqnarray*}
From this point, it is a matter of simple (though somewhat lengthy)
algebra to derive the differential constraints among $a_{q,jm}(r)$. The
results are as follows:
\begin{enumerate}
\item  $q\in \{1,7,9,5\}$
\begin{eqnarray}
&a&_{1,jm}^{\prime }(r)-jr^{-1}a_{1,jm}+ja_{7,jm}^{\prime
}-j^{2}r^{-1}a_{7,jm}+a_{9,jm}^{\prime }+2r^{-1}a_{9,jm} =0 \ ,
\label{eq:in1795} \\
&r&^{-1}a_{1,jm}+a_{7,jm}^{\prime }+3r^{-1}a_{7,jm}+\left( j-1\right)
a_{5,jm}^{\prime }-\left( j^{2}-3j+2\right) r^{-1}a_{5,jm} =0  \ .
\nonumber
\end{eqnarray}
\item  $q\in \{3\}$
\begin{eqnarray}
a_{3,jm}^{\prime }-jr^{-1}a_{3,jm} &=&0  \label{eq:in3} \ , \\
a_{3,jm}^{\prime }+r^{-1}a_{3,jm} &=&0 \ . \nonumber
\end{eqnarray}
 These equations have no solutions other than: $a_{3,jm}(r)=0$.
\item  $q\in \{8,6\}$
\begin{equation}
a_{8,jm}^{\prime }+3r^{-1}a_{8,jm}+(j-1)a_{6,jm}^{\prime }-\left(
j^{2}-2j+1\right) r^{-1}a_{6,jm}=0  \ . \label{eq:in86}
\end{equation}
\item  $q\in \{4,2\}$
\begin{equation}
r^{-1}a_{4,jm}-a_{2,jm}^{\prime }+(j-1)r^{-1}a_{2,jm}=0 \ . \label{eq:in42}
\end{equation}
\end{enumerate}
There are obviously more unknowns than equations, since we merely exploited
the incompressibility conditions. Nevertheless, we believe
that the missing equations that arise from the dynamical hierarchy of
equations will preserve the distinction between $a_{q,jm}$ of different
$j,m$
(again, due to the isotropy of these equations).
Note also, that the above analysis holds also for the second-order structure
function 
$$
S^{\alpha \beta }({\bf r})\equiv \left\langle \left[ u^{\alpha }({\bf x}+
{\bf r})-u^{\alpha }({\bf x})\right] \left[ u^{\beta }({\bf x}+{\bf r}
)-u^{\beta }({\bf x})\right] \right\rangle .
$$
Only that in this case we should only consider the representations $
q=1,7,9,5 $ for even $j$ and the representations $q=8,6$ for odd $j$. This
follows from the fact that $S^{\alpha \beta }({\bf r})$ is symmetric with
respect to its indices and it has an even parity in \ ${\bf r}$. Also, in
that case, it is possible to go one step further by assuming a specific
functional form for the $a_{q,jm}(r)$. We know that the $S^{\alpha \beta }(
{\bf r})$ is expected scale in the inertial range, and we therefore may {\em
assume}: 
$$
a_{q,jm}(r)\equiv c_{q,jm}r^{\zeta _{2}^{(j)}}.
$$
where $c_{q,jm}$ are just numerical constants. If we now substitute this
definition into the equations (\ref{eq:in1795},\ref{eq:in86}), we get a set
of linear equations among the $c_{q,jm}$. These relations can be easily
solved and give us two possible tensors for even $j$ ($q=1,7,9,5$) and one
tensor form for odd $j$ (from $q=8,6$). This kind of approach was taken in
the two-probes experiment which is described in Sect.~\ref{chap:experiment}.
\section{Anisotropy in d-dimensions}
\label{app:d-dim}

To deal with anisotropy in d-dimensions we need classify the
irreducible representations of the group of all $d$-dimensional
rotations, SO(d) \cite{62Ham}, and then to find a proper basis
for these representations. The main linear space that we work in
(the carrier space) is the space of constant tensors with $n$
indices. This space possesses a natural representation of SO(d),
given by the well known transformation of tensors under
$d$-dimensional rotation.

The traditional method to find a basis for the irreducible
representations of SO(d) in this space, is using the Young
tableaux machinery on the subspace of traceless tensors
\cite{62Ham,horvai}. It turns out that in the context of the present
work, we do not need the explicit structure of these tensors.
Instead, all that matters are some relations among them. A
convenient way to derive these relations is to construct the
basis tensors from functions on the unit $d$-dimensional sphere
which belong to a specific irreducible representation. Here also,
the explicit form of these functions in unimportant. All that
matters for the  calculations is the action of the Laplacian
operator on these functions.

Let us therefore consider first the space $\mathcal{S}_d$ of
functions over the
 unit
$d$-dimensional sphere. The representation of SO(d) over this
space is naturally defined by:
\begin{equation}
    \mathcal{O}_{\mathcal R} \Psi(\hat{u}) \equiv \Psi({\mathcal
R}^{-1}\hat{u}) \  ,
    \label{d-Rot}
\end{equation}
where $\Psi(\hat{u})$ is any function on the $d$-dimensional
sphere, and ${\mathcal R}$ is a $d$-dimensional rotation.

${\mathcal S}_d$ can be spanned by polynomials of the unit vector
$\hat{u}$. Obviously (\ref{d-Rot}) does not change the degree of a
polynomial, and therefore each irreducible representation in this
space can be characterized by an integer $j=0,1,2,\ldots$,
specifying the degree of the polynomials that span this
representation. At this point, we cannot rule out the possibility
that some other integers are needed to fully specify all
irreducible representations in ${\mathcal S}_d$ and therefore we will
need below another set of indices to complete the specification.

We can now choose a basis of polynomials $\{ Y_{j,\sigma}(\hat
u) \}$ that span all the irreducible representations of SO(d)
over ${\mathcal S}_d$. The index $\sigma$ counts all integers other
than $j$ needed to fully specify all irreducible
representations, and in addition, it labels the different
functions within each irreducible representation.

Let us demonstrate this construction in two and three dimensions.
In two dimensions $\sigma$ is unneeded since all the irreducible
representation are one-dimensional and are spanned by
$Y_j(\hat u)= e^{ij\phi}$ with $\phi$ being the angle
between $\hat u$ and the the vector $\hat e_1\equiv (1,0)$.  Any
rotation of the coordinates in an angle $\phi_0$ results in a
multiplicative factor $e^{i\phi_0}$. It is clear that
$Y_j(\hat u)$ is a polynomial in $\hat u$ since $Y_j(\hat
u)=[\hat u\cdot \hat p]^j$ where $\hat p\equiv (1,i)$.  In
three dimensions $\sigma=m$ where $m$ takes on $2j+1$ values
$m=-j, -j+1,\dots,j$. Here $Y_{j,m}\propto e^{im\phi}
P^m_j(\cos\theta)$ where $\phi$ and $\theta$ are the usual
spherical coordinates, and $P^m_j$ is the associated Legendre
polynomial of degree $j-m$. Obviously we again have a
polynomial in $\hat u$ of degree $j$.

We now wish to calculate the action  of the Laplacian operator
with respect to $u$ on the $Y_{j,\sigma}(\hat u)$. We prove
the following identity:
\begin{equation}
u^2\partial^\alpha\partial_\alpha Y_{j,\sigma}(\hat u) =
-j(j+d-2)Y_{j,\sigma}(\hat u) \ . \label{LapY}
\end{equation}
One can easily check that for $d=3$ (\ref{LapY}) gives the factor
$j (j + 1)$, well known from the theory of angular-momentum
in Quantum Mechanics. To prove this identity for any $d$, note
that
\begin{equation}
|u|^{2-j}\partial^2 |u|^j Y_{j,\sigma}(\hat u)) =0 \ .
\label{firststep}
\end{equation}
This follows from the fact that the Laplacian is an isotropic
operator, and therefore is diagonal in the $Y_{j,\sigma}$. The
same is true for the operator $|u|^{2-j}\partial^2 |u|^j$.
But this operator results in a polynomial in $\hat u$ of degree
$j-2$, which is spanned by $Y_{j',\sigma'}$ such that
$j'\le j -2$. Therefore the RHS of (\ref{firststep}) must
vanish. Accordingly we write
\begin{equation}
\partial^2|u^j|Y_{j,\sigma}(\hat
u)+2\partial^\alpha|u^j|\partial^\alpha
Y_{j,\sigma}+|u^j|\partial^2 Y_{j,\sigma}(\hat u) = 0 \ .
\end{equation}
The second term vanishes since it contains a radial derivative
$u^\alpha\partial_\alpha$ operating on $Y_{j,\sigma}(\hat u)$
which depends on $\hat u$ only. The first and third terms, upon
elementary manipulations, lead to (\ref{LapY}).

Having the $Y_{j,\sigma}(\hat u)$ we can now construct the
irreducible representations in the space of constant tensors. The
method is based on acting on the $Y_{j,\sigma}(\hat u)$ with
the {\em isotropic} operators $u^\alpha,~\partial^\alpha$ and
$\delta^{\alpha\beta}$. Due to the isotropy of the above
operators, the behavior of the resulting expressions under
rotations is similar to the behavior of the scalar function we
started with. For example, the tensor fields
$\delta^{\alpha\beta}Y_{j,\sigma}(\hat u),
\partial^{\alpha}\partial^{\beta}Y_{j,\sigma}(\hat u) $
transform under rotations according to the $(j, \sigma)$
sector of SO(d).

Next, we wish to find the basis for the irreducible
representations of the space of constant and fully symmetric
tensors with $n$ indices. We form the basis
\begin{equation}
B^{\alpha_1,\dots,\alpha_{n}}_{n,j,\sigma}\equiv
\partial^{\alpha_1} \dots \partial^{\alpha_{n}} u^{n}
Y_{j,\sigma}(\hat u) , \quad j\le n \ . \label{Birr}
\end{equation}
Note that when $j$ {\em and} $n$ are even
$B^{\alpha_1,\dots,\alpha_{n}}_{j,\sigma,n}$ no longer depends
on $\hat u$, and is indeed fully symmetric by construction.
Simple arguments can also prove that this basis is indeed
complete, and spans {\em all} fully symmetric tensors with $n$
indices. Other examples of this procedure for the other spaces
are presented directly in the text.

Finally let us introduce two identities involving the
$B_{n,j,\sigma}$. The first one is
\begin{eqnarray}
\delta_{\alpha_1\alpha_2}B^{\alpha_1,\dots,\alpha_{n}}_{n,j,\sigma}
&=&z_{n,j}B^{\alpha_3,\dots,\alpha_{n}}_{n-2,j,\sigma} \ ,
\label{iden1}\\ z_{n,j}&=&[n(n+d-2)-j(j+d-2)] \ .
\label{znl}
\end{eqnarray}
It is straightforward to derive this identity using (\ref{LapY}).
The second identity is
\begin{equation}
\sum_{i\ne j} \delta^{\alpha_i\alpha_j}B^{\{\alpha_m\},m\ne
i,j}_{n-2,j,\sigma}
=B^{\alpha_1,\dots,\alpha_{n}}_{n,j,\sigma} \ , \quad j\le
n-2 \ . \label{iden2}
\end{equation}
This identity is proven by writing $u^{n}$ in (\ref{Birr}) as
$u^2 u^{n-2}$, and operating with the derivative on $u^2$. The
term obtained as $u^2\partial^{\alpha_1} \dots
\partial^{\alpha_{n}} u^{n-2} Y_{j,\sigma}(\hat u) $ vanishes
because we have $n$ derivatives on a polynomial of degree $n-2$.
It is worthwhile noticing that these identities connect tensors
from two different spaces. The space of tensors with $n$ indices
and the space of tensors with $n-2$ indices. Nevertheless, in
both spaces, the tensors belong to the same $(j, \sigma)$
sector of the SO(d) group. This is due to the isotropy of the
contraction with $\delta^{\alpha_1\alpha_2}$ in the first
identity, and the contraction with $\delta^{\alpha_i\alpha_j}$ in
the second identity.
\section{Full Form for the $j=2$ Contribution for the Homogeneous Case}
\label{app:fullj2}
In this appendix we focus on the decomposition of second order
tensorial structure functions up to $j=2$. For this purpose we define:
$$
 S^{\alpha\beta}({\bf r}) =  S_{j=0}^{\alpha\beta}({\B r})+
 S_{j=2}^{\alpha\beta}({\B r})
$$
The $j=0$ is well-known and given explicitly by:
 \begin{equation} S_{j=0}^{\alpha\beta}({\bf
r})=c_0r^{\zeta_0^{(2)}} \left[(2+\zeta_0^{(2)}) \delta^
{\alpha\beta}-\zeta_0^{(2)}{r^\alpha r^\beta\over r^2}\right]\ ,
\label{Siso}
\end{equation} where $\zeta_0^{(2)} \approx 0.68$
is the known universal scaling
exponent for the isotropic contribution and $c_0$ is
an unknown coefficient that depends on the boundary conditions of
the flow. For the $j=2$ sector which is the lowest contribution
to anisotropy to the homogeneous structure function, the $m=0$
(axisymmetric) terms were derived from constraints of symmetry,
even parity (because of homogeneity) and incompressibility on the
second order structure function \cite{ara98}
\bea
&S&^{\alpha\beta}_{j=2,m=0} ({\bf r}) =
ar^{\zeta_2^{(2)}}\Big[(\zeta_2^{(2)} -2)\delta^{\alpha\beta} -
\zeta_2^{(2)}(\zeta_2^{(2)}+6)
\delta^{\alpha\beta}
{(\B n\cdot \B r)^2 \over r^2}+2\zeta_2^{(2)}(\zeta_2^{(2)}-2){r^\alpha
r^\beta(\B n\cdot \B r)^2 \over r^4} \nonumber \\\label{m0}
&+&([\zeta_2^{(2)}]^2+3\zeta_2^{(2)}+6)n^\alpha n^\beta
-{\zeta_2^{(2)}(\zeta_2^{(2)}-2)\over r^2}(r^\alpha n^\beta +
r^\beta n^\alpha)(\B n\cdot \B r)\Big]\\ &+& \nonumber
br^{\zeta_2^{(2)}}\Big [-(\zeta_2^{(2)}
+3)(\zeta_2^{(2)}+2)\delta^ {\alpha\beta}(\B n\cdot \B r)^2 +
{r^\alpha r^\beta \over r^2} + (\zeta_2^{(2)} +3)
(\zeta_2^{(2)}+2)n^\alpha n^\beta\\& + &(2\zeta_2^ {(2)}+1)(\zeta_2^{(2)}-2)
{{r^\alpha}{r^\beta}{(\B n\cdot \B r)^2} \over r^4}-
([\zeta_2^{(2)}]^2 - 4)(r^\alpha n^\beta + r^\beta n^\alpha)(\B
n\cdot \B r)\Big] \ . \nonumber
\end{eqnarray}
where $\zeta_2^{(2)}$ is the universal scaling exponent for the
$j=2$ anisotropic sector and $a$ and $b$ are independent unknown
coefficients to be determined by the boundary conditions. We would
now like to derive the remaining $m=\pm1$, and $m=\pm2$ components
$$
S_{2m}^{\alpha\beta}=
    \sum_q {a_{q,2, m}r^{\zeta_2^{(2)}}B_{q,2, m}^{\alpha\beta} (\bf {\hat
r}
)}
$$
As usual the $q$
label denotes the different possible ways of arriving at the
same $j$ and runs over all such terms with the same parity and
symmetry (a consequence of homogeneity and hence the constraint
of incompressibility).
In all that follows, we work
closely with the procedure outlined in \cite{ara99b}. Following the
convention in \cite{ara99b} the $q$'s to sum over are
$q=\{1,7,9,5\}$. The incompressibility condition $\partial_\alpha
u^\alpha = 0$ coupled with homogeneity can be used to give
relations between the $a_{q,jm}$ for a given $(j,m)$. That is,
for $j=2$, $m=-2\dots 2$ \begin{eqnarray} (\zeta_2^{(2)} -
2)a_{1,2, m} + 2(\zeta_2^{(2)} - 2)a_{7,2 m} + (\zeta_2^{(2)} +2)a_{9,2, m}
&=& 0 \\ \nonumber a_{1,2, m} + (\zeta_2^{(2)} +
3)a_{7,2, m} + \zeta_2^{(2)}a_{5,2,m} &=& 0.
 \end{eqnarray}
We solve the above equations in order to obtain $a_{5,2,m}$ and
$a_{7,2m}$ in terms of linear combinations of $a_{1,2,m}$ and
$a_{9,2m}$. \begin{eqnarray} a_{5,2,m} &=&
{a_{1,2m}([\zeta_2^{(2)}]^2 - \zeta_2^{(2)} - 2) + a_{9,2,m}
([\zeta_2^{(2)}]^2 + 5 \zeta_2^{(2)} + 6) \over
2\zeta_2^{(2)}(\zeta_2^{(2)} - 2)} \\ \nonumber a_{7,2,m} &=&
{a_{1,2m}(2-\zeta_2^{(2)}) - a_{9,2,m}(2+\zeta_2^{(2)}) \over
2(\zeta_2^{(2)} - 2)}.
\end{eqnarray}
Using the above constraints on the coefficients, we are now left
with a linear combination of just two linearly independent tensor
forms {\em for each m}
\begin{eqnarray}\label{genl-s2m}
S^{\alpha\beta}_{2m} &=&
a_{9,2,m}r^{\zeta_2^{(2)}}[-\zeta_2^{(2)}(2+\zeta_2^{(2)})
B_{7,2,m}^{\alpha\beta}({\B  {\hat r}}) +
2\zeta_2^{(2)}(\zeta_2^{(2)} - 2)
B_{9,2,m}^{\alpha\beta}({\B  {\hat r}}) \nonumber \\
&+&([\zeta_2^{(2)}]^2+5\zeta_2^{(2)}+6)B_{5,2,m}^{\alpha\beta}({\B  {\hat
r}})] \nonumber\\
&+&a_{1,2,m}r^{\zeta_2^{(2)}}[2\zeta_2^{(2)}(\zeta_2^{(2)} - 2)
B_{1,2,m}^{\alpha\beta}({\B  {\hat r}}) -
\zeta_2^{(2)}(\zeta_2^{(2)}-2)B_{7,2,m}^{\alpha\beta}({\B  {\hat
r}})
\nonumber\\
&+&([\zeta_2^{(2)}]^2-\zeta_2^{(2)}-2)B_{5,2,m}^{\alpha\beta}({\B
{\hat r}})]. \end{eqnarray}

The task remains to find the explicit form of the basis tensor
functions $B_{q,2,m}^{\alpha\beta}({\B  {\hat r}})$,
$q\in\{1,7,9,5\}$, $m\in\{\pm1,\pm2\}$
\\
$\bullet$\,$ B_{1,2,m}^{\alpha\beta}({\B  {\hat r}}) \equiv
r^{-2}\delta^{\alpha\beta}r^j Y_{2m}({\B  {\hat r}})$ \\
$\bullet$\,$ B_{7,2,m}^{\alpha\beta}({\B  {\hat r}}) \equiv
r^{-2}[r^\alpha \partial^\beta +
r^\beta \partial^\alpha]r^2 Y_{2m}({\B  {\hat r}})$ \\
$\bullet$\,$ B_{9,2,m}^{\alpha\beta}({\B  {\hat r}}) \equiv r^{-4}r^\alpha
r^\beta r^2 Y_{2m}({\B  {\hat r}})$ \\
$\bullet$\,$ B_{5,2,m}^{\alpha\beta}({\B  {\hat r}}) \equiv
\partial^\alpha \partial^\beta r^2 Y_{jm}({\B  {\hat r}})$

We obtain the $m=\{\pm1,\pm2\}$ basis functions in the following
derivation. We first note that it is more convenient to form a
real basis from the $ r^2 Y_{2m}({\B  {\hat r}})$ since we
ultimately wish to fit to real quantities and extract real
best-fit parameters. We therefore form the $r^2 {\widetilde
Y}_{2k}({\B  {\hat r}})$ ($k= -1,0,1$) as follows:
\begin{eqnarray}
r^2 {\widetilde Y}_{2\; 0}(\hat {\B  r})&=&r^2 Y_{2\;0}(\hat {\B
r}) =r^2 \cos^2 \theta = r{_3}^2\nonumber \\ r^2 {\widetilde
Y}_{2\;-1}(\hat {\B  r}) &=&r^2 {Y_{2\;-1}(\hat {\B  r}) -
Y_{2\;+1}(\hat {\B  r}) \over 2}\nonumber\\ &=&r^2 {(\cos\phi -
i\sin\phi)\cos\theta\sin\theta + (\cos\phi +
i\sin\phi)\cos\theta\sin\theta \over 2} \nonumber\\ &=&r^2
\cos\theta\sin\theta\cos\phi = r_3 r_1 \nonumber\\ r^2 {\widetilde
Y}_{2\;+1}(\hat {\B  r}) &=&r^2 {Y_{2\; -1}(\hat {\B  r}) +
Y_{2\; +1}(\hat {\B  r}) \over -2i} \nonumber\\ &=&r^2 {(\cos\phi
- i\sin\phi)\cos\theta\sin\theta - (\cos\phi +
i\sin\phi)\cos\theta\sin\theta \over -2i}\nonumber\\ &=&r^2
\cos\theta\sin\theta\sin\phi = r_3 r_2\nonumber\\ r^2 {\widetilde
Y}_{2\;-2}(\hat {\B  r}) &=& r^2 {Y_{2\; 2}(\hat {\B  r}) -
Y_{2\; -2}(\hat {\B  r}) \over 2i} \nonumber\\ &=& r^2
{(\cos2\phi + i\sin2\phi)\sin^2\theta - (\cos2\phi -
i\sin2\phi)\sin^2\theta \over 2i} \nonumber \\ &=& r^2
\sin2\phi\sin^2\theta = 2r_1 r_2 \nonumber \\ r^2 {\widetilde
Y}_{2\;+2}(\hat {\B  r}) &=& r^2 {Y_{2\; 2}(\hat {\B  r}) +
Y_{2\; -2}(\hat {\B  r}) \over 2} \nonumber\\ &=& r^2 {(\cos2\phi
+ i\sin2\phi)\sin^2\theta + (\cos2\phi - i\sin2\phi)\sin^2\theta
\over 2} \nonumber \\ &=& r^2 \cos2\phi\sin^2\theta = r_1^2 -
r_2^2 \end{eqnarray} This new basis of $r^2 {\tilde Y}_{2k}{(\B
r)}$ is equivalent to using the $r^2 Y_{jm}{(\B  r)}$ themselves
as they form a complete, orthogonal (in the new $k$'s) set. We omit
the normalization constants for the spherical harmonics for
notational convenience. The subscripts on $r$ denote its
components along the 1 ($m$), 2 ($p$) and 3 ($n$) directions.
${\B  m}$ denotes the shear direction, ${\B  p}$ the horizontal
direction parallel to the boundary and orthogonal to the mean
wind direction and ${\B  n}$ the direction of the mean wind. This
notation makes it simple to take the derivatives when we form the
different basis tensors and the only thing to remember is that
\begin{eqnarray}
&&\partial^\alpha r_1 = \partial^\alpha (\B  {r \cdot m}) = m^\alpha
\nonumber\\ &&\partial^\alpha r_2 = \partial^\alpha (\B  {r \cdot
p})= p^\alpha \nonumber\\ &&\partial^\alpha r_3 = \partial^\alpha
(\B  {r \cdot \B n}) = n^\alpha \end{eqnarray}

We use the above identities to proceed to derive the basis tensor
functions
\begin{eqnarray}
B_{1,2, -1}^{\alpha\beta}(\B{  {\hat r}}) &=&
r^{-2}\delta^{\alpha\beta} (\B { r \cdot  n})(\B{ r \cdot m})
\nonumber\\ B_{7,2, -1}^{\alpha\beta} (\B{ {\hat r}}) &=&
r^{-2}[(r^\alpha m^\beta + r^\beta m^\alpha)(\B{  r \cdot  n}) +
(r^\alpha n^\beta + r^\beta n^\alpha)(\B{  r \cdot m})]
\nonumber\\ B_{9,2, -1}^{\alpha\beta}(\B{  {\hat r}}) &=&r^{-2}
r^\alpha r^\beta (\B{  r \cdot n})(\B{  r \cdot m})\nonumber\\
B_{5,2, -1}^{\alpha\beta}(\B{  {\hat r}}) &=&n^\alpha m^\beta +
n^\beta m^\alpha \nonumber\\ B_{1,2, 1}^{\alpha\beta}(\B{  {\hat
r}}) &=& r^{-2}\delta^{\alpha\beta}(\B{  r \cdot n})(\B{  r \cdot
p})\nonumber\\ B_{7,2, 1}^{\alpha\beta}(\B{  {\hat r}}) &=&
r^{-2}[(r^\alpha p^\beta + r^\beta p^\alpha)(\B{  r \cdot n}) +
(r^\alpha n^\beta + r^\beta n^\alpha)(\B{  r \cdot p})]
\nonumber\\ B_{9,2, 1}^{\alpha\beta}(\B{  {\hat r}}) &=&r^{-2}
r^\alpha r^\beta (\B{  r \cdot n})(\B{  r \cdot p})\nonumber\\
B_{5,2,1}^{\alpha\beta}(\B{  {\hat r}}) &=& n^\alpha p^\beta +
n^\beta p^\alpha \nonumber\\ B_{1,2, -2}^{\alpha\beta}(\B{  {\hat
r}}) &=&2 r^{-2}\delta^{\alpha\beta}(\B{  r \cdot m})(\B{  r
\cdot p})\nonumber\\ B_{7,2, -2}^{\alpha\beta}(\B{  {\hat r}})
&=&2 r^{-2}[(r^\alpha p^\beta + r^\beta p^\alpha)(\B{  r \cdot
m}) + (r^\alpha m^\beta + r^\beta m^\alpha)(\B{  r \cdot
p})]\nonumber \\ B_{9,2, -2}^{\alpha\beta}(\B{  {\hat r}}) &=& 2
r^{-2} r^\alpha r^\beta (\B{  r \cdot m})(\B{  r \cdot
p})\nonumber\\ B_{5,2,-2}^{\alpha\beta}(\B{  {\hat r}}) &=& 2
(m^\alpha p^\beta + m^\beta p^\alpha) \nonumber\\ B_{1,2,
2}^{\alpha\beta}(\B{  {\hat r}}) &=&
r^{-2}\delta^{\alpha\beta}[(\B{  r \cdot m})^2 - (\B{  r \cdot
p})^2] \nonumber\\ B_{7,2, 2}^{\alpha\beta}(\B{  {\hat r}}) &=&2
r^{-2}[(r^\alpha m^\beta + r^\beta m^\alpha)(\B{  r \cdot m}) -
(r^\alpha p^\beta + r^\beta p^\alpha)(\B{  r \cdot p})]\nonumber
\\ B_{9,2, 2}^{\alpha\beta}(\B{  {\hat r}}) &=& r^{-2} r^\alpha
r^\beta [(\B{  r \cdot m})^2 - (\B{  r \cdot p})^2] \nonumber\\
B_{5,2,2}^{\alpha\beta}(\B{  {\hat r}}) &=& 2 (m^\alpha m^\beta -
p^\alpha p^\beta) \end{eqnarray}

Note that for each dimension $k$ the tensor is bilinear in some
combination of two basis vectors from the set $\B{  m}$, $\B{ p}$
and $\B{ n}$.
 Substituting these tensors forms into Eq.
\ref{genl-s2m} we obtain the full tensor forms for the $j=2$
non-axisymmetric terms, with two independent coefficients for
each $k$.
\begin{eqnarray}\label{ktens}
S^{\alpha\beta}_{j=2,k=-1}(\B{  r})
&=&a_{9,2,-1}r^{\zeta_2^{(2)}}{\Big
[}-\zeta_2^{(2)}(2+\zeta_2^{(2)}) r^{-2}[(r^\alpha m^\beta +
r^\beta m^\alpha)(\B{  r \cdot n})\nonumber\\ &+& (r^\alpha
n^\beta + r^\beta n^\alpha)(\B{  r \cdot m})] +
2\zeta_2^{(2)}(\zeta_2^{(2)} - 2) r^{-4} r^\alpha r^\beta (\B{  r
\cdot n})(\B{  r \cdot m})\nonumber\\
&+&([\zeta_2^{(2)}]^2+5\zeta_2^{(2)}+6)(n^\alpha m^\beta +
n^\beta m^\alpha) {\Big ]}\nonumber\\&+&
a_{1,2,-1}r^{\zeta_2^{(2)}}{\Big [}2\zeta_2^{(2)}(\zeta_2^{(2)} -
2) r^{-2}\delta^{\alpha\beta}(\B{  r \cdot n})(\B{  r \cdot m})
\nonumber\\ &-& \zeta_2^{(2)}(\zeta_2^{(2)}-2) r^{-2}[(r^\alpha
m^\beta + r^\beta m^\alpha)(\B{  r \cdot n})
+ (r^\alpha n^\beta + r^\beta n^\alpha)(\B{  r \cdot m})] \nonumber\\
&+&([\zeta_2^{(2)}]^2-\zeta_2^{(2)}-2)(n^\alpha m^\beta + n^\beta m^\alpha)
{\Big ]} \nonumber\\
S^{\alpha\beta}_{j=2,k=1}(\B{  r})
&=&a_{9,2,1}r^{\zeta_2^{(2)}}{\Big
[}-\zeta_2^{(2)}(2+\zeta_2^{(2)}) r^{-2}[(r^\alpha p^\beta +
r^\beta p^\alpha)(\B{  r \cdot n})\nonumber\\ &+& (r^\alpha
n^\beta + r^\beta n^\alpha)(\B{  r \cdot p})] +
2\zeta_2^{(2)}(\zeta_2^{(2)} - 2)
r^{-4} r^\alpha r^\beta (\B{  r \cdot n})(\B{  r \cdot p})\nonumber\\
&+&([\zeta_2^{(2)}]^2+5\zeta_2^{(2)}+6)(n^\alpha p^\beta + n^\beta p^\alpha)
{\Big ]}\nonumber\\
&+&a_{1,2,1}r^{\zeta_2^{(2)}}{\Big [}2\zeta_2^{(2)}(\zeta_2^{(2)}
- 2)
r^{-2}\delta^{\alpha\beta}(\B{  r \cdot n})(\B{  r \cdot p}) \nonumber\\ &-&
\zeta_2^{(2)}(\zeta_2^{(2)}-2) r^{-2}[(r^\alpha p^\beta + r^\beta p^\alpha)
(\B{  r \cdot n}) + (r^\alpha n^\beta + r^\beta n^\alpha)(\B{  r \cdot p})]
\nonumber\\
&+&([\zeta_2^{(2)}]^2-\zeta_2^{(2)}-2)(n^\alpha p^\beta + n^\beta p^\alpha)
{\Big ]}\nonumber\\
S^{\alpha\beta}_{j=2,k=-2}(\B{  r})
&=&a_{9,2,-2}r^{\zeta_2^{(2)}}{\Big
[}-2\zeta_2^{(2)}(2+\zeta_2^{(2)}) r^{-2}[(r^\alpha p^\beta +
r^\beta p^\alpha)(\B{  r \cdot m})\nonumber\\ &+& (r^\alpha
m^\beta + r^\beta m^\alpha)(\B{  r \cdot p})] +
2\zeta_2^{(2)}(\zeta_2^{(2)} - 2) r^{-4} r^\alpha r^\beta (\B{  r
\cdot p})(\B{  r \cdot m}) \nonumber\\ &+&
([\zeta_2^{(2)}]^2+5\zeta_2^{(2)}+6)(m^\alpha p^\beta + m^\beta
p^\alpha) {\Big ]}\nonumber\\&+& a_{1,2,-2}r^{\zeta_2^{(2)}}{\Big
[}2\zeta_2^{(2)}(\zeta_2^{(2)} - 2)
r^{-2}\delta^{\alpha\beta}(\B{  r \cdot m})(\B{  r \cdot p}) \nonumber\\
&-&2\zeta_2^{(2)}(\zeta_2^{(2)}-2) r^{-2}[(r^\alpha p^\beta + r^\beta
p^\alpha) (\B{  r \cdot m}) + (r^\alpha m^\beta + r^\beta m^\alpha)(\B{  r
\cdot p})] \nonumber\\
&+&2([\zeta_2^{(2)}]^2-\zeta_2^{(2)}-2)(m^\alpha p^\beta + m^\beta p^\alpha)
{\Big ]}\nonumber\\
S^{\alpha\beta}_{j=2,k=2}(\B{  r})
&=&a_{9,2,2}r^{\zeta_2^{(2)}}{\Big
[}-2\zeta_2^{(2)}(2+\zeta_2^{(2)}) r^{-2}[(r^\alpha m^\beta +
r^\beta m^\alpha)(\B{  r \cdot m})\nonumber\\ &-& (r^\alpha
p^\beta + r^\beta p^\alpha)(\B{  r \cdot p})] +
2\zeta_2^{(2)}(\zeta_2^{(2)} - 2) r^{-4} r^\alpha r^\beta [(\B{
r \cdot m})^2-(\B{  r \cdot p})^2]\nonumber\\
&+&2([\zeta_2^{(2)}]^2+5\zeta_2^{(2)}+6)(m^\alpha m^\beta -
p^\beta p^\alpha) {\Big ]}\nonumber\\&+&
a_{1,2,2}r^{\zeta_2^{(2)}}{\Big [}2\zeta_2^{(2)}(\zeta_2^{(2)} -
2)
r^{-2}\delta^{\alpha\beta}[(\B{  r \cdot m})^2-(\B{  r \cdot
p})^2]\nonumber\\ &-&2\zeta_2^{(2)}(\zeta_2^{(2)}-2) r^{-2}[(r^\alpha
m^\beta + r^\beta m^\alpha) (\B{  r \cdot m}) - (r^\alpha p^\beta + r^\beta
p^\alpha)(\B{  r \cdot p})] \nonumber\\
&+&2([\zeta_2^{(2)}]^2-\zeta_2^{(2)}-2)(m^\alpha m^\beta -
p^\beta p^\alpha) {\Big ]}
\end{eqnarray}
Now we want to use this form to fit for the scaling exponent
$\zeta_2^{(2)}$ in the structure function $S^{33}(\B  r)$ from
data set I where $\alpha=\beta=3$ and the azimuthal angle of $\B
r$ in the geometry is $\phi = \pi/2$.
\begin{eqnarray}
&S&^{33}_{j=2,k=-1}(r,\theta,\pi/2)=0\nonumber\\
&S&^{33}_{j=2,k=1}(r,\theta,\pi/2)
=a_{9,2,1}r^{\zeta_2^{(2)}}[-2\zeta_2^{(2)}(\zeta_2^{(2)}+2)
\sin\theta\cos\theta \nonumber +
2\zeta_2^{(2)}(\zeta_2^{(2)}-2)\cos^3\theta\sin\theta]
\\ &S&^{33}_{j=2,k=-2}(r,\theta,\pi/2)=0\nonumber \\
&S&^{33}_{j=2,k=2}(r,\theta,\pi/2)
=a_{9,2,2}r^{\zeta_2^{(2)}}[-2\zeta_2^{(2)}(\zeta_2^{(2)}-2)
\cos^2\theta\sin^2\theta]
+a_{1,2,2}r^{\zeta_2^{(2)}}[-2\zeta_2^{(2)}(\zeta_2^{(2)}-2)\sin^2\theta]
\nonumber
\end{eqnarray}
We see that choosing a particular geometry eliminates certain
tensor contributions. In the case of set I we are left with 3
independent coefficients for $m\ne0$, the 2 coefficients from the
$m=0$ contribution (Eq. \ref{m0}), and the single coefficient
from the isotropic sector \ref{Siso}, giving a total of 6 fit
parameters. The general forms in \ref{ktens} can be used along
with the $k=0$ (axisymmetric) contribution \ref{Siso} to fit to
any second order tensor object. For convenience, the table shows
the number of independent coefficients that a few different
experimental geometries we have will allow in the $j=2$ sector.
It must be kept in mind that these forms are to be used {\em
only} when there is known to be homogeneity. If there is
inhomogeneity, then we cannot apply the incompressibility
condition to provide constraints in the various parity and
symmetry sectors and we must in general mix different parity
objects, using only the geometry of the experiment itself to
eliminate any terms.
\begin{table}
\begin{tabular}{|c|c|c|c|c|c|c|c|c|}
\hline
&\multicolumn{2}{c|}{$\phi=\pi/2,\alpha=\beta=3$} &
\multicolumn{2}{c|}{$\phi = 0,\alpha=\beta=3$} &
\multicolumn{2}{c|}{$\phi = 0,\alpha=\beta=1$}&
\multicolumn{2}{c|}{$\phi = 0,\alpha=3,\beta=1$}\\ \cline {2-9}
$k$&$\theta \ne 0$ &$\theta = 0$ & $\theta \ne 0$ &$ \theta = 0$
& $\theta \ne 0$ & $\theta = 0$ & $\theta \ne 0 $&$\theta = 0$ \\
\hline 0    & 2 & 2 & 2 & 2 & 2 & 2 & 2 & 0 \\ \hline -1 & 0 & 0
& 1 & 0 & 1 & 0 & 2 & 2 \\ \hline 1  & 1 & 0 & 0 & 0 & 0 & 0 & 0
& 0 \\ \hline -2) & 0 & 0 & 0 & 0 & 0 & 0 & 0 & 0 \\ \hline 2 & 2
& 0 & 2 & 0 & 2 & 2 & 2 & 0 \\ \hline \hline
Total   & 5 & 2 & 5 & 2 & 5 & 4 & 6 & 2 \\
\hline
\end{tabular}
\caption{The number of free coefficients in the $j=2$ sector for
homogeneous turbulence and for different geometries}
\end{table}

%
\section{The j=1 Component in the Inhomogeneous Case}
\label{app:fullj1}
\subsection{Antisymmetric Contribution}
    We consider the tensor
$$
T^{\alpha\beta}(\B{  r}) = \langle u^\alpha({\B  x} + {\B  r}) -
u^\alpha({\B  x})) (u^\beta({\B  x} + {\B  r}) + u^\beta({\B
x}))\rangle. $$ This object is trivially zero for
$\alpha=\beta$. In the  experimental setup, we measure at points
separated in the shear direction and therefore have inhomogeneity
which makes the object of mixed parity and symmetry.  We cannot
apply the incompressibility condition in same parity/symmetry
sectors as before to provide constraints. We must in general use
all 7 irreducible tensor forms. This would mean fitting for $7 \times
3 = 21$ independent coefficients plus 1 exponent $\zeta_1^{(2)}$
in the anisotropic sector, together with 2 coefficients in the
isotropic sector. In order to pare down the number of parameter
we are fitting for, we look at the antisymmetric part of
$T^{\alpha\beta}({\B  r})$
$$
{\widetilde T}^{\alpha\beta}({\B  r}) = {T^{\alpha\beta}({\B  r})
- T^{\beta\alpha}({\B  r}) \over 2} = \langle u^\alpha({\B
x})u^\beta({\B  x} + {\B  r})\rangle - \langle u^\beta({\B
x})u^\alpha({\B  x} + {\B  r})\rangle $$
which will only have contributions from the antisymmetric $j=1$ basis
tensors. These are \\
$\bullet$ Antisymmetric, odd parity
\begin{equation}
B_{3,1,m}^{\alpha\beta} = r^{-1}[r^\alpha\partial^\beta-
r^\beta\partial^\alpha] rY_{1,m}(\B  {\hat r})
\end{equation}
$\bullet$ Antisymmetric, even parity
\begin{eqnarray}
B_{4,1,m}^{\alpha\beta} &=& r^{-2}\epsilon^{\alpha\beta\mu}r_\mu
r Y_{1,m}(\B  {\hat r}) \nonumber\\ B_{2,1,m}^{\alpha\beta} &=&
r^{-2}\epsilon^{\alpha\beta\mu}\partial_\mu r Y_{1,m}(\B  {\hat
r}) \end{eqnarray}
As with the $j=2$ case we form a real basis $r {\tilde
Y}_{1,k}(\B  {\hat r})$ from the (in general) complex $r
Y_{1,m}(\B  {\hat r})$ in order to obtain real coefficients in
the  fits.
\begin{eqnarray}
r {\tilde Y}_{1,k=0}(\B  {\hat r}) &=& r Y_{1,0}(\B  {\hat r})= r\cos\theta
= r_3 \nonumber\\
r {\tilde Y}_{1,k=1}(\B  {\hat r}) &=& r {Y_{1,1}(\B  {\hat r})
+Y_{1,1} (\B  {\hat r}) \over 2i }
=r\sin\theta\sin\phi = r_2 \nonumber\\ r {\tilde
Y}_{1,k=-1}(\B  {\hat r}) &=& r
{Y_{1,-1}(\B  {\hat r}) - Y_{1,1}(\B  {\hat r}) \over 2 }
=r\sin\theta\cos\phi = r_1 \nonumber
\end{eqnarray}
And the final forms are
\begin{eqnarray}
B_{3,1,0}^{\alpha\beta}(\B  {\hat r}) &=& r^{-1}[r^\alpha
n^\beta- r^\beta n^\alpha] \nonumber\\
B_{4,1,0}^{\alpha\beta}(\B  {\hat r})
&=&r^{-2}\epsilon^{\alpha\beta\mu}r_\mu (\B{  r \cdot n}) \nonumber\\
B_{2,1,0}^{\alpha\beta}(\B  {\hat r}) &=&
r^{-2}\epsilon^{\alpha\beta\mu}n_\mu \nonumber\\
B_{3,1,1}^{\alpha\beta}(\B  {\hat r}) &=& r^{-1}[r^\alpha
p^\beta- r^\beta p^\alpha] \nonumber\\
B_{4,1,1}^{\alpha\beta}(\B  {\hat r})
&=&r^{-2}\epsilon^{\alpha\beta\mu}r_\mu (\B{  r \cdot p}) \nonumber\\
B_{2,1,1}^{\alpha\beta}(\B  {\hat r}) &=&
r^{-2}\epsilon^{\alpha\beta\mu}p_\mu \nonumber\\
B_{3,1,-1}^{\alpha\beta}(\B  {\hat r}) &=& r^{-1}[r^\alpha
m^\beta- r^\beta m^\alpha] \nonumber\\
B_{4,1,-1}^{\alpha\beta}(\B  {\hat r})
&=&r^{-2}\epsilon^{\alpha\beta\mu}r_\mu (\B{  r \cdot m}) \nonumber\\
B_{2,1,-1}^{\alpha\beta}(\B  {\hat r}) &=&
r^{-2}\epsilon^{\alpha\beta\mu}m_\mu \end{eqnarray}
Note: For a given $k$ the representations is symmetric about a
particular axis in  the  coordinate system chosen  (1=m (shear), 2=p
(horizontal), 3=n (mean-wind))
We now have 9 independent terms and we cannot apply
incompressibility in order to reduce the number of independent
coefficients in the  fitting procedure. We use the geometric
constraints of the  experiment to do this.\\
$\bullet$ $\phi = 0$ (vertical separation), $\alpha = 3, \beta =
3$ \begin{eqnarray} B_{3,1,0}^{31}(r,\theta,\phi=0) = -\sin\theta
\nonumber
\\ B_{2,1,1}^{31}(r,\theta,\phi=0) = 1 \nonumber\\
B_{3,1,-1}^{31}(r,\theta,\phi=0) = \cos\theta \end{eqnarray}
There are no contributions from the reflection-symmetric terms in
the $j=0$ isotropic sector since these are symmetric in the
indices. The helicity term in $j=0$ also doesn't contribute
because of the geometry. So, to lowest order
$$
{\widetilde T}^{\alpha\beta}({\B  r}) = {\widetilde
T}_{j=1}^{\alpha\beta}({\B  r})
 = a_{3,1,0}(r)(-\sin\theta) + a_{2,1,1}(r) + a_{3,1,-1}(r)\cos\theta
$$
We have 3 unknown independent coefficients and 1 unknown exponent
to fit for in the  data.
\subsection{Symmetric Contribution}
    We consider the structure function
$$
S^{\alpha\beta}({\B  r}) = \langle (u^\alpha({\B  x} + {\B  r}) -
u^\alpha({\B  x})) (u^\beta({\B  x} + {\B  r}) - u^\beta({\B
x}))\rangle $$ in the case where we have homogeneous flow.
This object is symmetric in the indices by construction, and it
is easily seen that homogeneity implies even parity in $r$:
$
    S^{\alpha\beta}({\B  r}) = S^{\beta\alpha}({\B  r})$ and
$S^{\alpha\beta}({\B  -r}) = S^{\alpha\beta}({\B  r}).$
 We reason that this object cannot exhibit a $j=1$
contribution from the $SO(3)$ representation in the following
manner. Homogeneity allows us to use the incompressibility
condition: $  \partial_\alpha S^{\alpha\beta} = 0$ and $ \partial_\beta
S^{\alpha\beta} = 0$,
separately on the basis tensors of a given parity and symmetry in
order to give relationships between their coefficients. For the
even parity, symmetric case we have for general $j \geq 2$ just
two basis tensors and they must occur in some linear combination
with incompressibility providing a constraint between the two
coefficients. However, for $j=1$ we only have one such tensor in
the even parity, symmetric group. Therefore, by
incompressibility, its coefficient must vanish. Consequently, we
cannot have a $j=1$ contribution for the even parity
(homogeneous), symmetric structure function.     Now, we consider
the case as available in experiment when ${\B  r}$ has some
component in the inhomogeneous direction. Now, it is no longer
true that $S^{\alpha\beta}({\B  r})$ is of even parity and
moreover it is also not possible to use incompressibility as
above to exclude the existence of a $j=1$ contribution. We must
look at all $j=1$ basis tensors that are symmetric, but not
confined to even
parity. These are \\
$\bullet$ Odd parity, symmetric
\begin{eqnarray}
B_{1,1,k}^{\alpha\beta}({\B  {\hat r}}) &\equiv&
r^{-1}\delta^{\alpha\beta}r {\tilde Y}_{1k}({\B  {\hat
r}})\nonumber\\ B_{7,1,k}^{\alpha\beta}({\B  {\hat r}}) &\equiv&
r^{-1}[r^\alpha \partial^\beta + r^\beta \partial^\alpha]
r{\tilde Y}_{1k}(\B  {\hat r}) \nonumber \\
B_{9,1,k}^{\alpha\beta}({\B  {\hat r}}) &\equiv& r^{-3}r^\alpha
r^\beta r {\tilde Y}_{1k}({\B  {\hat r}})\nonumber \\
B_{5,1,k}^{\alpha\beta}({\B  {\hat r}}) &\equiv& r
\partial^\alpha \partial^\beta r{\tilde Y}_{1k}({\B  {\hat r}})
\equiv 0
\end{eqnarray}
$\bullet$ Even parity, symmetric
\begin{eqnarray}
B_{8,1,k}^{\alpha\beta}({\B  {\hat r}}) &\equiv& r^-2[r^\alpha
\epsilon^{\beta\mu\nu} r_\mu \partial_\nu + r^\beta
\epsilon^{\alpha\mu\nu} r_\mu \partial_\nu] r{\tilde Y}_{1k}(\B
{\hat r}) \nonumber\\ B_{6,1,k}^{\alpha\beta}({\B  {\hat r}})
&\equiv& [\epsilon^{\beta\mu\nu} r_\mu \partial_\nu
\partial_\alpha + \epsilon^{\beta\mu\nu} r_\mu \partial_\nu
\partial_\beta] r{\tilde Y}_{1k}(\B  {\hat r}) \equiv 0
\end{eqnarray}
We use the real basis of $r^{-1}{\tilde Y}_{1k}({\B  {\hat r}})$
which are formed from the $r^{-1}Y_{1m}({\B  {\hat r}})$. Both
$B_{5,1,k}^{\alpha\beta}({\B  {\hat r}})$ and
$B_{6,1,k}^{\alpha\beta}({\B  {\hat r}})$ vanish because of the
taking of the double derivative of an object of single power in
$r$. We thus have 4 different contributions to symmetric $j=1$
and each of these is of 3 dimensions $(k= -1,0,1)$ giving in
general 12 terms in all. \begin{eqnarray}
B_{1,1,0}^{\alpha\beta}(\B{  {\hat r}}) &=&
r^{-1}\delta^{\alpha\beta}(\B{  r \cdot n})\nonumber\\
B_{7,1,0}^{\alpha\beta}(\B{  {\hat r}}) &=& r^{-1}[r^\alpha
n^\beta + r^\beta n^\alpha] \nonumber\\
B_{9,1,0}^{\alpha\beta}(\B{  {\hat r}}) &=&r^{-3}r^\alpha r^\beta
(\B{  r \cdot n})\nonumber\\ B_{8,1,0}^{\alpha\beta}(\B{  {\hat
r}}) &\equiv& r^{-2}[(r^\alpha m^\beta + r^\beta m^\alpha)(\B{  r
\cdot p}) -(r^\alpha p^\beta + r^\beta p^\alpha)(\B{  r \cdot
m})]\nonumber \\ B_{1,1,1}^{\alpha\beta}(\B{  {\hat r}}) &=&
r^{-1}\delta^{\alpha\beta}(\B{  r \cdot p})\nonumber\\
B_{7,1,1}^{\alpha\beta}(\B{  {\hat r}}) &=& r^{-1}[r^\alpha
p^\beta + r^\beta p^\alpha] \nonumber\\
B_{9,1,1}^{\alpha\beta}(\B{  {\hat r}}) &=&r^{-3}r^\alpha r^\beta
(\B{  r \cdot p}) \nonumber\\ B_{8,1,1}^{\alpha\beta}(\B{  {\hat
r}}) &\equiv& r^{-2}[(r^\alpha m^\beta + r^\beta m^\alpha)(\B{  r
\cdot n}) -(r^\alpha n^\beta + r^\beta n^\alpha)(\B{  r \cdot
m})] \nonumber \\ B_{1,1,-1}^{\alpha\beta}(\B{  {\hat r}}) &=&
r^{-1}\delta^{\alpha\beta}(\B{  r \cdot m})\nonumber\\
B_{7,1,-1}^{\alpha\beta}(\B{  {\hat r}}) &=& r^{-1}[r^\alpha
m^\beta + r^\beta m^\alpha] \nonumber\\
B_{9,1,-1}^{\alpha\beta}(\B{  {\hat r}}) &=&r^{-3}r^\alpha
r^\beta (\B{  r \cdot m}) \nonumber\\
B_{8,1,-1}^{\alpha\beta}(\B{  {\hat r}}) &\equiv&
r^{-2}[(r^\alpha p^\beta + r^\beta p^\alpha)(\B{  r \cdot n})
-(r^\alpha n^\beta + r^\beta n^\alpha)(\B{  r \cdot p})]
\end{eqnarray}

These are all the possible $j=1$ contributions to the symmetric,
mixed parity (inhomogeneous) structure function.

For the  experimental setup II, we want to analyze the
inhomogeneous structure function in the case $\alpha = \beta =
3$, and azimuthal angle $\phi=0$ (which corresponds to vertical
separation) and we obtain the basis tensors
\begin{eqnarray}
B_{1,1,0}^{33}(\theta) &=& \cos\theta \nonumber\\
B_{7,1,0}^{33}(\theta) &=& 2\cos\theta \nonumber\\
B_{9,1,0}^{33}(\theta) &=& \cos^3\theta \nonumber\\
B_{8,1,1}^{33}(\theta) &=& -2\cos\theta\sin\theta \nonumber\\
B_{1,1,-1}^{33}(\theta) &=& \sin\theta \nonumber\\
B_{9,1,-1}^{33}(\theta) &=& \cos^2\theta\sin\theta
\label{b15}
\end{eqnarray}
Table~D.1 gives the number of free coefficients in the symmetric
$j=1$ sector in the fit to the inhomogeneous structure function
for various geometric configurations.
\begin{table}
\begin{tabular}{|c|c|c|c|c|c|c|}
\hline
&\multicolumn{2}{c|}{$\phi= 0,\alpha=\beta=3$} &
\multicolumn{2}{c|}{$\phi = 0,\alpha=\beta=1$} &
\multicolumn{2}{c|}{$\phi = 0,\alpha=3,\beta=1$}\\ \cline {2-7}
$k$&$\theta \ne 0$ &$\theta = 0$ & $\theta \ne 0$ &$ \theta = 0$
& $\theta \ne 0$ & $\theta = 0$ \\ \hline 0   & 3 & 3 & 2 & 1 & 2
& 0 \\ \hline 1   & 1 & 0 & 1 & 0 & 0 & 0 \\ \hline -1  & 2 & 0 &
3 & 0 & 2 & 1 \\ \hline \hline Total & 6 & 3 & 6 & 1 & 4 & 1\\
\hline
\end{tabular}
\caption{The number of free coefficients in the symmetric $j=1$
sector for inhomogeneous turbulence and for different
geometries.} \end{table}
\section{The Matrix Form of the Operator of the Linear Pressure Model}
\label{app:lp1}
Using the basic identities of the $\Phi_{j m}(\B r)$ functions
(see \cite{ara99b}),
\begin{eqnarray*}
  \partial^2 \Phi_{j m}(\B r) &=& 0 \ , \\
  r^\mu \partial_\mu \Phi_{j m}(\B r) &=& j \Phi_{j m}(\B r) \ ,
\end{eqnarray*}
a short calculation yields:
\begin{eqnarray}
   && \KO C^\alpha(\B r) \equiv
    K^{\mu\nu}(\B r)\partial_\mu\partial_\nu C^\alpha(\B r) =
Dx^\epsilon\Big[
2c''_1 +
2(2+\epsilon)\frac{c'_1}{r} -
                (2+\epsilon)(j+1)(j+2)\frac{c_1}{r^2}
              \Big]B_{1jm}^\alpha(\B{\hat r}) \nonumber \\
     && \quad + Dx^\epsilon\Big[ 2c''_2 + 2(2+\epsilon)\frac{c'_2}{r}
        + 2(2+\epsilon)\frac{c_1}{r^2}  - 2(2+\epsilon)j(j-1)\frac{c_2}{r^2}
\Big]
     B_{2jm}^\alpha(\B{ \hat  r}) \nonumber \ .
\end{eqnarray}
Therefore, in matrix notation, the Kraichnan operator can be
written as:
\begin{eqnarray}
 && \KO \VecII{c_1}{c_2} = 2Dr^\epsilon
\MatII{1}{0}{0}{1}\VecII{c''_1}{c''_2}
  + 2D(2+\epsilon)r^{\epsilon-1} \MatII{1}{0}{0}{1}\VecII{c'_1}{c'_2}
     \nonumber \\
  && \quad - D(2+\epsilon)r^{\epsilon-2} \MatII{(j+1)(j+2)}{0}{-2}{j(j-1)}
      \VecII{c_1}{c_2} \nonumber \\
  && \equiv r^\epsilon \KM_2\VecII{c''_1}{c''_2} +
           r^{\epsilon-1}\KM_1\VecII{c'_1}{c'_2} +
           r^{\epsilon-2}\KM_0\VecII{c_1}{c_2} \ .
\label{def:K-mat}
\end{eqnarray}
Letting
\begin{equation}
  T^\alpha(\B r) = t_1(r)B_{1jm}^\alpha(\B{\hat  r}) +
    t_2(r)B_{2jm}^\alpha(\B{ \hat r}) \ ,
\end{equation}
and applying a Laplacian to $\PO T^\alpha$, we get,
\begin{eqnarray}
\label{eq:explicit-P}
  && \partial^2\PO T^\alpha =
    \Big[ -j t''_2 + j\frac{t'_1}{r} + j(2j-1)\frac{t'_2}{r}-
j(j+1)\frac{t_1}{r^2}
     - j(j-1)(j+1)\frac{t_2}{r^2}\Big]B_{1jm}^\alpha \nonumber\\&&+ \left[
t''_2
-
\frac{t'_1}{r} +
(2-j)\frac{t'_2}{r}\right]
     B_{2jm}^\alpha \ . \nonumber
\end{eqnarray}
Hence in matrix notation,
\begin{eqnarray}
\label{def:P-mat}
&& \partial^2\PO\VecII{t_1}{t_2} = \MatII{0}{-j}{0}{1}
        \VecII{t''_1}{t''_2}  +
\frac{1}{r}\MatII{j}{j(2j-1)}{-1}{2-j}
      \VecII{t'_1}{t'_2} \\&&
    \quad - \frac{1}{r^2}\MatII{j(j+1)}{j(j-1)(j+1)}{0}{0}
      \VecII{t_1}{t_2}  \equiv \PM_2\VecII{t^{''}_1}{t^{''}_2} +
\frac{1}{r}\PM_1
      \VecII{t'_1}{t'_2} + \frac{1}{r^2}\PM_0\VecII{t_1}{t_2} \ .\nonumber
\end{eqnarray}
Now that the matrix forms of the Kraichnan operator and of the
Laplacian of the projection operator have been found, we can
combine these two results to find the matrix form of the LHS of
\Eq{eq:diff-C}. To this aim let us define,
$$
  \VecII{t_1}{t_2} = \KO \VecII{c_1}{c_2} \ ,$$
and from Eq.~(\ref{def:K-mat},\ref{def:P-mat}) we get,
\begin{eqnarray}
  && \partial^2 \PO \KO \VecII{c_1}{c_2} =
    r^\epsilon \MM_4 \VecII{c^{(4)}_1}{c^{(4)}_2} +
    r^{\epsilon-1} \MM_3 \VecII{c^{(3)}_1}{c^{(3)}_2} +\ r^{\epsilon-2}
\MM_2
\VecII{c^{(2)}_1}{c^{(2)}_2}
   \nonumber\\&&+ r^{\epsilon-3} \MM_1 \VecII{c^{(1)}_1}{c^{(1)}_2}  +\
r^{\epsilon-4} \MM_0
\VecII{c_1}{c_2} \ ,
\nonumber
\end{eqnarray}
where the number in parenthesis denotes the order of the
derivative. The matrices $\MM_i$ are given by:
\begin{eqnarray}
\label{def:M-mat}
\MM_4 &\equiv& \PM_2\KM_2 \ , \\
\MM_3 &\equiv& 2\epsilon\PM_2\KM_2 + \PM_2\KM_1 + \PM_1\KM_2 \ , \nonumber
\\
\MM_2 &\equiv& \epsilon(\epsilon-1)\PM_2\KM_2 + 2(\epsilon-1)\PM_2\KM_1 +
\PM_2\KM_0
         \nonumber \\
      && \quad \ + \epsilon\PM_1\KM_2 + \PM_1\KM_1 + \PM_0\KM_2 \ ,
\nonumber \\
\MM_1 &\equiv& (\epsilon-1)(\epsilon-2)\PM_2\KM_1 + 2(\epsilon-2)\PM_2\KM_0
\nonumber \\
      && \quad \ + (\epsilon -1)\PM_1\KM_1 +\PM_1\KM_0 + \PM_0\KM_1 \
,\nonumber
\\
\MM_0 &\equiv& (\epsilon-2)(\epsilon-3)\PM_2\KM_0 + (\epsilon-2)\PM_1\KM_0
          + \PM_0\KM_0  \ . \nonumber
\end{eqnarray}
To find the RHS of \Eq{eq:diff-C} we expand the ``forcing''
$A^\alpha(\B r)$ in terms of the spherical vectors $\B B_{1jm}, \B
B_{2jm}$,
\begin{equation}
A^\alpha(\B r) = f_1(r) B^\alpha_{1jm}(\B{ \hat r}) +
                  f_2(r) B^\alpha_{2jm}(\B{ \hat r}) \ ,
\end{equation}
and applying a Laplacian we find the matrix form of $\partial^2
A^\alpha(\B r)$:
\begin{eqnarray}
&& \partial^2 \VecII{f_1}{f_2} =
   \left( \begin{array}{c}
    f''_1 + \frac{2}{r}f'_1 - (j+1)(j+2)\frac{1}{r^2}f_1 \\ \ \\
   f''_2 + \frac{2}{r}f'_2 + \frac{2}{r^2}f_1 - j(j-1)\frac{1}{r^2}f_2
   \end{array} \right)  \equiv \VecII{\rho_1}{\rho_2} \ .
\end{eqnarray}
At this point it is worthwhile to remember that the forcing term
$A^\alpha(\B r/L)$ is assumed to be analytic. As a result for
$r/L \ll 1$ its leading contribution in the $(j,m)$ sector is
proportional to $\partial^\alpha r^j Y_{j m}(\hat{\B r})
\sim r^{j-1}$. However $\partial^2 A^{\alpha}(\B r/L)$ is also
analytic, and must therefore also scale like $r^{j-1}$ for
small $r$, instead of like $r^{j-3}$ which could be the naive
dimensional guess.

To proceed we restrict ourselves to finding the solution in the
inertial range and beyond. In these ranges the dissipative term
$\kappa
\partial^2\partial^2 C^\alpha(\B r)$ is negligible and can be omitted,
thus reaching Eq.~(\ref{eq:C-mat}) for $c_1(r)$ and $c_2(r)$.



\end{document}